%% file: main.tex
\documentclass[preprint]{emulateapj}
\usepackage{amsmath,amssymb,amstext,comment}

\usepackage[breaklinks,colorlinks,citecolor=blue,linkcolor=magenta]{hyperref}

\usepackage{color} 
\usepackage{ulem} 

\newcommand{\mearth}{M_{\earth*}}

\usepackage[all]{hypcap} 
\usepackage{aas_macros}
\usepackage{natbib}
\usepackage{threeparttablex}
\usepackage{longtable}

\shorttitle{Numerical and Analytical TTV}
\shortauthors{Hadden \& Lithwick}

\begin{document}
\title{
{\it Kepler} Planet Masses and Eccentricities from TTV Analysis
} 
\author{Sam Hadden,Yoram Lithwick}
\affil{Department of Physics \& Astronomy, Northwestern University, Evanston, IL 60208, USA \& Center for Interdisciplinary Exploration and Research in Astrophysics (CIERA)}

\begin{abstract}
We conduct a uniform analysis of the transit timing variations (TTVs) of 145 planets from 55 {\it Kepler} multiplanet systems to infer planet masses and eccentricities.
Eighty of these planets do not have previously reported mass and eccentricity measurements.
We employ two complementary methods to fit TTVs: Markov chain Monte Carlo simulations  based on $N$-body integration and an analytic fitting approach.
Mass measurements of 49 planets,  including 12 without previously reported masses,  meet our criterion for classification as robust.
Using mass and radius measurements, we infer the masses of planets' gaseous envelopes for both our TTV sample as well as transiting planets with radial velocity observations.
Insight from analytic TTV formulae allows us to partially circumvent degeneracies inherent to inferring eccentricities from TTV observations. 
We find that planet eccentricities are generally small, typically a few percent, but in many instances are  non-zero.
\end{abstract}
\maketitle

\section{Introduction}
\label{sec:intro}	
	The {\it Kepler} mission's census of exoplanetary systems has provided us with a statistical picture of the properties of planetary systems
	\citep{2010Sci...327..977B}. 
	Small planets with radii in the range $R_p=1-4~R_\earth$ on short-period orbits, $P< 100$ days, are among the most abundant, occurring around roughly half 
	of Sun-like stars \citep[e.g.,][]{2013PNAS..11019273P,2013ApJ...766...81F}.
	The distributions of planet sizes and periods measured by {\it Kepler} provide important constraints for theories of planet formation and evolution.
	Mass measurements, which constrain planets' compositions, and eccentricity measurements, which reveal the current dynamical states of planetary systems,
	are also essential clues for understanding formation and evolution processes. 
	
The radial velocity (RV) method has provided the majority of  exoplanet mass and eccentricity measurements to date \citep[e.g., exoplanets.org;][]{2011PASP..123..412W}.
Mass measurements of transiting super-Earth and sub-Neptune planets are of particular interest given the ubiquity of such planets and their absence from our own solar system.
Many sub-Jovian transiting planets have had their masses measured through radial velocity follow-up \citep[e.g.,][]{
2010Sci...327..977B,2010AandA...520A..66B,2010ApJ...710.1724B,Hartman:2011ei,2012ApJ...749...15G,Gilliland:2013ep,Pepe:2013hs,2014AandA...569A..74B,2014ApJ...789..154D,2014MNRAS.444.2783M,2014AandA...567A.112A,2014ApJ...784...14K,Marcy:2014hr,2015ApJ...804..150E,Dressing:2015je,2016ApJ...816...95G,Sinukoff:2016cf}.
These efforts have yielded important constraints on the mass-radius relationship of super-Earth and sub-Neptune planets:
\citet{Weiss:2014ef}  infer a mass-radius relationship of planets smaller than $4~R_\earth$ using available RV mass determinations, supplemented with a handful of masses determined from transit timing.
 \citet{2015ApJ...801...41R} uses the sample of {\it Kepler} planets with Keck HIRES RV  follow-up to infer that planets transition from mainly rocky to volatile-rich compositions above a size of $1.6~R_\earth$.
However, the RV method is of limited applicability to {\it Kepler}'s many sub-Neptune planets since the radial velocities induced by such planets often require intense follow-up efforts to detect.
With a handful of exceptions, RV mass determinations of sub-Neptunes have been limited to planets with orbital periods shorter than 20 days.

	The limitations of RV are even more acute for measuring sub-Neptunes' eccentricities since high signal-to-noise ratios are required to 
	accurately infer eccentricities with RV \citep{2008ApJ...685..553S,2011MNRAS.410.1895Z}.
	Eccentricities of transiting planets can also be measured by modeling transit light curves.
	However, light curve modeling yields useful constraints only in special circumstances:
	giant planets with large eccentricities \citep{2012ApJ...756..122D},
	planets with occultation detections \citep{Shabram:2016gb},
	or host stars with strong density constraints from asteroseismology \citep{2014MNRAS.440.2164K}. 
	Statistical analyses of the transit durations of planets around well-characterized host stars have been used to
	constrain planet samples' overall eccentricity distributions \citep{VanEylen:2015je,Xie:2016dp},
	but inferring the eccentricities of individual sub-Neptune planets from light curve modeling remains difficult.

	Transit timing variations (TTVs) are a powerful tool for measuring masses and eccentricities in multi-transiting systems \citep{2005MNRAS.359..567A,Holman:2005jf}. 
	The large TTV amplitudes induced in planets near mean motion resonances (MMRs) can probe the masses and eccentricities of small planets at relatively long orbital periods which would otherwise be difficult or impossible to measure via the RV method.  
	However, inverting TTVs to infer planet properties poses a difficult parameter inference problem:
	it requires fitting a large number parameters, often with strong degeneracies, to noisy data.
 Statistical analyses of samples of TTV systems can overcome some of these difficulties \citep{Wu:2013cp,2014ApJ...787...80H}.
Alternatively, the parameter inference challenge can be addressed with Markov chain Monte Carlo (MCMC) simulations when fitting TTVs of individual systems.
	MCMC is well suited for high-dimensional parameter inference problems and has been used frequently in TTV studies
	\citep[e.g.,][]{2012Natur.487..449S,Huber:2013ha,2013ApJ...778..185M,2014ApJ...795..167S,2015Natur.522..321J,JontofHutter:2016ch,Hadden:2016ki,Mills:2016gi}.

	While MCMC can efficiently sample planet masses and orbits consistent with TTV observations,
	interpreting MCMC results is often complicated by strong parameter correlations and sensitivity to priors.
	Analytic TTV formulae identify degeneracies inherent to inverting TTVs and aid the interpretation of $N$-body MCMC results.
	Components of the TTV signal responsible for mass and eccentricity constraints can be identified with analytic formulae 
	and lend support to the robustness of $N$-body results to different prior assumptions.
     
	In this paper we compute MCMC fits to the TTVs of 55 {\it Kepler} multiplanet systems exhibiting significant TTVs,
	33 of which do not have $N$-body TTV fits reported previously in the literature. 
	In addition, our work provides a uniform treatment of TTV systems that have previously been analyzed elsewhere.
	We  {complement} our MCMC fits, which rely on $N$-body integrations, with an analytic approach to TTV modeling. 
	The paper is organized as follows: we review analytic TTV formulae in Section \ref{sec:analytic} and describe our 
	 fitting methods in Section \ref{sec:fit_methods}. 
	 Results of our fits are described in Section \ref{sec:results} and discussed in Section \ref{sec:discussion}: 
	 in Section \ref{sec:gas_fraction} we use planet radii and masses derived from our TTV fits to infer the masses of planets' gaseous envelopes.
	In Section \ref{sec:ecc_damping} we briefly discuss implications of the eccentricities inferred from TTVs.
	We conclude in Section \ref{sec:conclusion}.

\section{analytic TTV}
\label{sec:analytic}
\subsection{The Analytic TTV Formula}
\label{sec:formula}

A number of authors have derived analytic formulae to approximate TTVs  using perturbative methods \citep[e.g.,][]{2005MNRAS.359..567A,2008ApJ...688..636N,Nesvorny:2009bm,2012ApJ...761..122L,Deck:2015ed,Deck:2015tx,2016ApJ...818..177A,Hadden:2016ki,2016ApJ...823...72N}. 
Analytic formulae aid the interpretation of $N$-body fitting results by elucidating  degeneracies 
and identifying TTV features that constrain planet masses and eccentricities.
In this paper we apply the analytic formulae derived in \citet[][hereafter Paper I]{Hadden:2016ki}
as part of our analysis of each system's TTVs. 
The main features of these formulae are summarized below.

For clarity, we  focus our discussion on a planet near a $j$:$j$-1 first-order MMR with an exterior perturber.
{Other configurations are discussed in Section \ref{sec:other_config}} 
The analytic formulae express a planet's TTV as a sum of harmonic terms:
\begin{eqnarray}
\delta t(t) = \hat{\delta t_{\cal F}}{\text e}^{2\pi i t/P_{sup}}+
\hat{\delta t_{\cal S}}{\text e}^{4\pi i t/P_{sup}} + \nonumber\\
\hat{\delta t_{\cal C}}\sum_{k=1,k\ne j}^{\infty} C_k {\text e}^{k(2\pi i t/P_{syn})} + c.c.
\label{eq:the_ttv}
\end{eqnarray}
 where  {  the three successive terms are called `fundamental', `second-harmonic', and 
 `chopping' TTV's, and} `${c.c.}$' denotes complex conjugate. 
The { frequencies} of the harmonics 
are expressed in terms of the `super-period', $P_{sup}$, and synodic period, $P_{syn}$.  
These depend only on the period of the planet and its perturber,  and hence may be considered to be ``known''  for transiting planets.
 The dependence on masses and eccentricities is contained entirely in the amplitudes
 $\delta {\hat t}_{\cal F}$, $\delta {\hat t}_{\cal S}$, and $\delta {\hat t}_{\cal C}$.  
 Since the first two of these are complex numbers, and
 the third real, there are in total five observables that can be used to infer masses and eccentricities.
 (The coefficients $C_k$  depend only on the period ratio of the two planets).

The fundamental TTV is typically the dominant component.
Its complex amplitude depends on mass and eccentricity and is given by
\begin{equation}
\hat{\delta t_{\cal F}}=\mu' \left(A+B{\cal Z}^*\right) \label{eq:ttvF}
\end{equation}
where $A$ and $B$  are coefficients that depend only on the planets' periods, $\mu'$ is the perturber's planet-star mass ratio, and 
\begin{eqnarray}
{\cal Z}  &\equiv& \frac{f_{27} z+ f_{31} z'}{\sqrt{f_{27}^2+f_{31}^2}} \label{eq:zdef} \\
            &\approx& {1\over\sqrt{2}}(z - z') \label{eq:zapprox}
\end{eqnarray}
where $z$ and $z'$ are the {\it free} complex 
eccentricities\footnote{A planet's total complex eccentricity, $e{\text e}^{i\varpi}$, 
is the sum of its free eccentricity plus a forced component induced by the forcing of perturbing companions.
Forced components of the total eccentricity are typically much smaller than the free eccentricities we infer so that the difference between 
free  and total complex eccentricities is usually negligible.} of the inner and outer planet, and the $f_{i}$ are coefficients that depend only on the planets' period ratio (see Paper I). 

\subsection{Degeneracies}
In attempting to extract planet parameters from the dominant (fundamental) harmonic of the TTV,  there are at least two
kinds of degeneracies \citep{2012ApJ...761..122L}.
The first is between the two planets' individual complex eccentricities --- the {\it eccentricity-eccentricity degeneracy}.
Since the fundamental TTV depends  on the  planets' eccentricities only through the single linear combination ${\cal Z}$, 
the  $z$ and $z'$ of each planet is not separately measurable.
This degeneracy persists even with the inclusion of the chopping and second harmonic terms.

The  second degeneracy is between $\mu'$ and $|{\cal Z}|$,  which we call the {\it mass-eccentricity degeneracy}:
a smaller $\mu'$ can be compensated for by a larger $|{\cal Z}|$ (Eq. \eqref{eq:ttvF}).
Usually $B\gg A$ in Equation \eqref{eq:ttvF} so that the fundamental TTV amplitude is
sensitive to { $|{\cal Z}|$} and the degeneracy spans a large range in $\mu'$.

The mass-eccentricity degeneracy can be broken if the chopping component can be resolved in the TTV signal
\citep{2014ApJ...790...58N,Deck:2015ed}.
The chopping TTV is insensitive to eccentricities
and provides a measurement of the perturbing planet's mass.
In fact, the chopping TTV amplitude can simply be equated to the perturber's planet-star mass ratio,
\begin{equation}
\hat{\delta t_{\cal C}} =\mu' ,\label{eq:ttvC}
\end{equation}
 with appropriate choice of scaling for the $C_k$ coefficients in Equation \eqref{eq:the_ttv}.
If a perturbing planet's mass can be measured from the chopping TTV, 
then the fundamental TTV amplitude gives the planets' combined eccentricity (Eq.
\eqref{eq:ttvF}).
 
The mass-eccentricity degeneracy may alternatively be broken by detecting the second-harmonic component, the amplitude of which is given by 
\begin{equation}
\hat{\delta t_{\cal S}} = \mu'\left(D{\cal Z}^* + E{\cal Z}^{*2}\right) \label{eq:ttvS}
\end{equation}
with coefficients $D$ and $E$ depending only on planet periods.\footnote{\label{fnt:2to1MMR}
As shown in the Appendix of Paper I, the fact that Eq. \ref{eq:ttvS} depends on $z$ and $z'$ only through the combination
${\cal Z}$ is an approximation, albeit an excellent one. The approximation breaks down in the exceptional case of planets near the 2:1 MMR, allowing second-harmonic TTVs to constrain individual eccentricities.
}
Measuring both a planet's fundamental and second-harmonic TTV can break the degeneracy between mass and eccentricity since the functional dependence of each component on $\cal Z$ is different.
Most of the TTV systems analyzed in this paper have small eccentricities so that we typically infer upper limits on combined eccentricity from the lack of substantial second-harmonic TTVs.

\subsection{Other configurations}
\label{sec:other_config}
With minor modifications, Equation \eqref{eq:the_ttv} also describes the TTVs of planets subject to an interior perturber. Thus, breaking the mass-eccentricity degeneracy for one planet determines the pair's ${\cal Z}$ and thereby automatically breaks the degeneracy for both planets.

Equation \eqref{eq:the_ttv} without the fundamental TTV term also describes the TTVs of planets near a $j$:$j-2$ second-order MMR.
Proximity to a second-order resonance enhances the $\hat{\delta t_{\cal S}}$ term in Equation \eqref{eq:the_ttv}, which in this context we refer to as the `second-order resonance' component rather than second-harmonic component.
TTVs near second-order MMRs exhibit essentially the same mass-eccentricity and eccentricity-eccentricity degeneracies as planets near first-order MMRs and the mass-eccentricity degeneracy can be broken by measuring a chopping component in either planet's TTV.

We find that, in practice, measuring (only) fundamental TTVs of three or more planets generally does not resolve any of the degeneracies inherent to the two-planet case.

\section{TTV Fit Methods}
\label{sec:fit_methods}
\begin{figure*}[htbp]
\begin{center}
\includegraphics[width=0.95\textwidth]{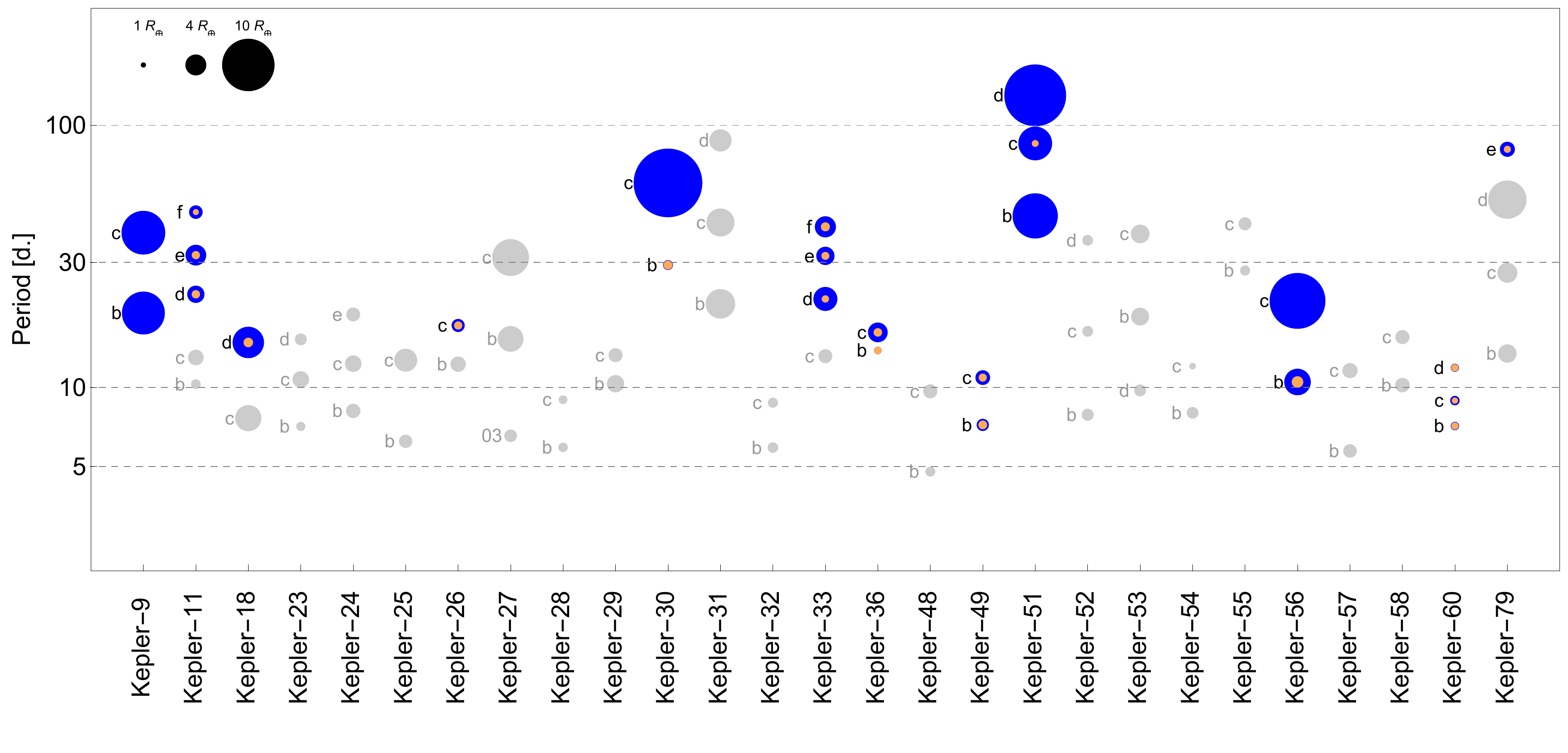}
\includegraphics[width=0.95\textwidth]{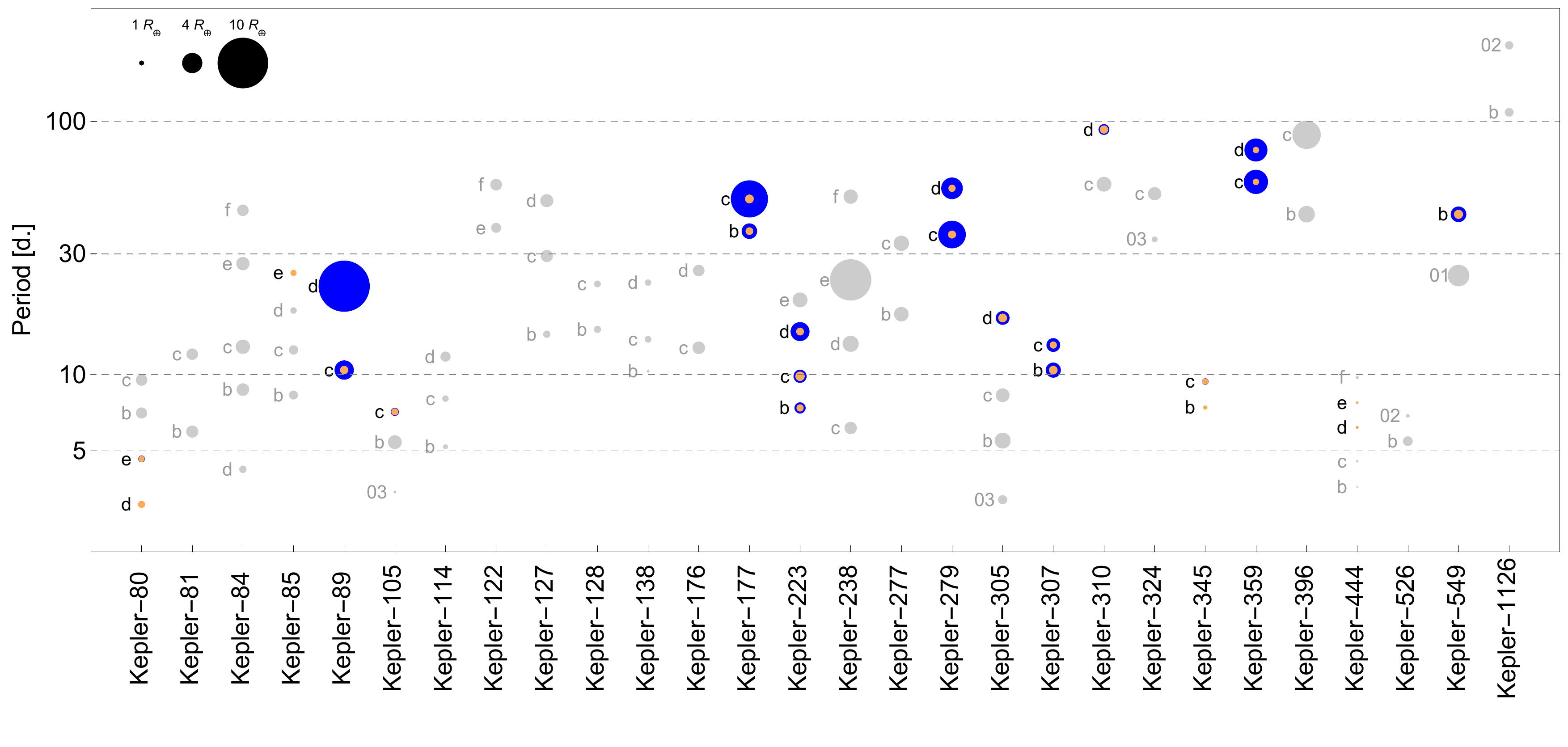}
\caption{
Multiplanet TTV systems fit in this paper.
Planets in each system are plotted along the vertical axis according to their periods.  
The (outer) radii of the plotted circles are proportional to observed radii.
The radii of planets with robust masses greater than $1~M_\earth$ and radii $R_p<8~R_\earth$ are shown decomposed into an Earth-composition core (orange) and H/He envelope (blue) using the results of Section \ref{sec:gas_fraction}.
Planets larger than $R_p>8~R_\earth$ with robust masses are shown in blue without a core and planets with robust masses smaller than $1~M_\earth$ are entirely orange.
Planets that do not have robustly constrained masses are shown in gray.
}
\label{fig:systems}
\end{center}
\end{figure*}

We analyze TTVs of 145  planets from  55 different {\it Kepler} systems.
Figure \ref{fig:systems} shows the TTV systems fit in this paper.  
Our analysis is based on transit times computed by \citet{2015ApJS..217...16R} from long cadence {\it Kepler} data spanning Quarters 1-17.
Each system is fit with both $N$-body MCMC simulations and the analytic TTV formulae.
Systems are discussed individually in Appendix \ref{sec:appendix}.
 Results for their masses and eccentricities are summarized in Tables \ref{tab:mass} and \ref{tab:ecc}.

Our MCMC simulations sample the most likely planet masses and orbital elements of a multi-transiting system given the planets' transit times by using $N$-body integrations.
Specifically, for a system of $N$ planets our MCMC samples the posterior distribution of each planet's planet-to-star mass ratio, $\mu_i$, 
eccentricity vector components $h_i \equiv e_i\cos(\varpi_i)$ and $k_i \equiv e_i\sin(\varpi_i)$, initial osculating period, $P_i$, and time of first transit, $T_i$ where $i=1,2,...,N$.
The complete details of our MCMC implementation are given in Paper I.

As in Paper I, in order to assess how robustly planet properties are constrained by the TTVs
we run two MCMC simulations for each system we fit with two different priors for masses and eccentricities.
Our first (default) prior is logarithmic in planet masses ($dP/dM\propto M^{-1}$) and uniform in eccentricities  ($dP/de\propto \text{const.}$).
Our second (high mass) prior is uniform in planet masses ($dP/dM\propto \text{const.}$) and logarithmic in eccentricity ($dP/de\propto e^{-1}$).
We will refer to the posterior distributions computed with the respective priors as the default and high mass posteriors.
Inferred planet masses are classified as `robust'  if the 1$\sigma$ credible region of the default posterior excludes $\mu=0$ and includes the peak of the high mass posterior.
Masses of 49 of the 145 planets are classified as robust. 
Twelve planets with robust masses do not have $N$-body fits previously reported in the literature.

We have chosen our two MCMC priors to weight toward opposite extremes of the mass-eccentricity degeneracy 
in order to assess the significance of this degeneracy in each system. 
Additionally, both priors are chosen to be ``uninformative" or broad and do not rely on any theoretical predictions for planet compositions or eccentricities. 
Aside from these considerations, our adopted priors are arbitrary and any number of other priors could be reasonably advocated.
It is possible that a different choice of priors could affect the inferred masses and eccentricities significantly.
Instead of exploring a wide variety of priors with $N$-body MCMC, which is not computationally feasible, we complement MCMC simulations with fits using the analytic formulae.
The analytic fits convert the amplitudes $\hat{\delta t_{\cal F}}$, $\hat{\delta t_{\cal S}}$, and $\hat{\delta t_{\cal C}}$ measured from TTVs to constraints in the $\mu'$-$|{\cal Z}|$ plane.
Constraints derived from our analytic fits are compared to the results of the $N$-body MCMC simulations for each system in Appendix \ref{sec:appendix}.
The analytic fits lend further confidence to the robustness of MCMC results by showing the degree to which various components of each planet's TTV constrain planet parameters.

TTVs do not probe planet masses directly but rather planet-star mass ratios.
Therefore we will often find it convenient to give planet masses in units of
\begin{equation}
\mearth\equiv M_\earth\times\left({M_*\over M_\sun}\right) \label{eq:mearth_def}
\end{equation}
 where $M_*/M_\odot$ is the ratio of host star mass to solar mass.  Employing $\mearth$ as the unit highlights mass uncertainties inherent to the TTVs separately from those arising from uncertain stellar properties.

\begin{figure}[htbp]
\begin{center}
\includegraphics[width=0.47\textwidth]{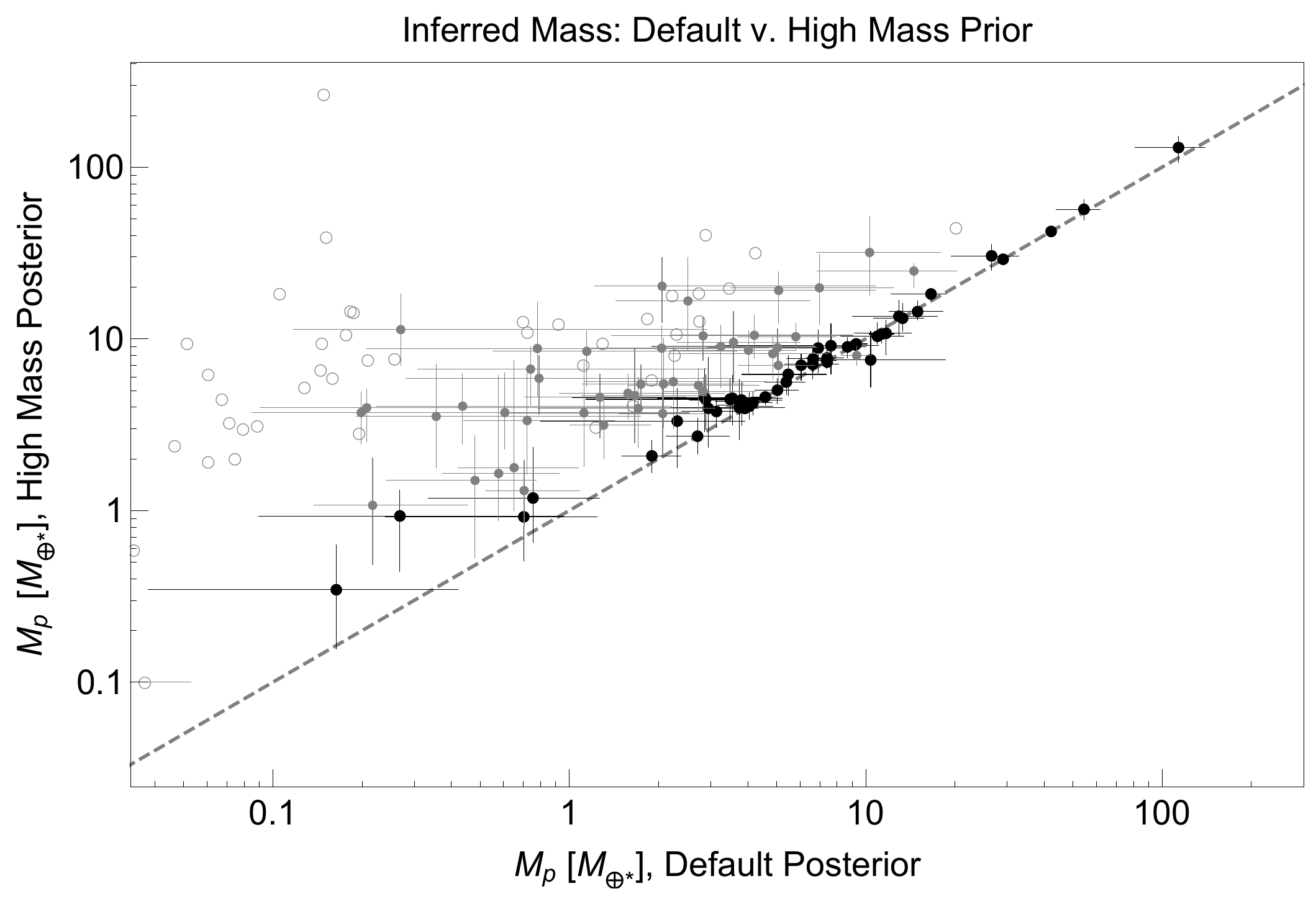}
\caption{
A comparison of planet masses inferred using the high mass versus default priors.
The dashed line indicates equal mass for both priors. 
Inferred masses that agree within 1$\sigma$ (2$\sigma$) are shown
as black (gray) points.
Error bars show 1$\sigma$ uncertainties.
Inferred masses which disagree at $>2\sigma$ are shown as empty circles with error bars omitted for clarity.
}
\label{fig:priors_compare}
\end{center}
\end{figure}

Figure \ref{fig:priors_compare} compares planet masses inferred using our default versus high mass priors.
Many mass inferences are seen to depend sensitively on the assumed prior, reflecting the degeneracies often inherent to inverting TTV observations.

\section{Results: Masses and Eccentricities}
\label{sec:results}
\begin{figure*}[htbp]
\begin{center}
\includegraphics[width=0.9\textwidth]{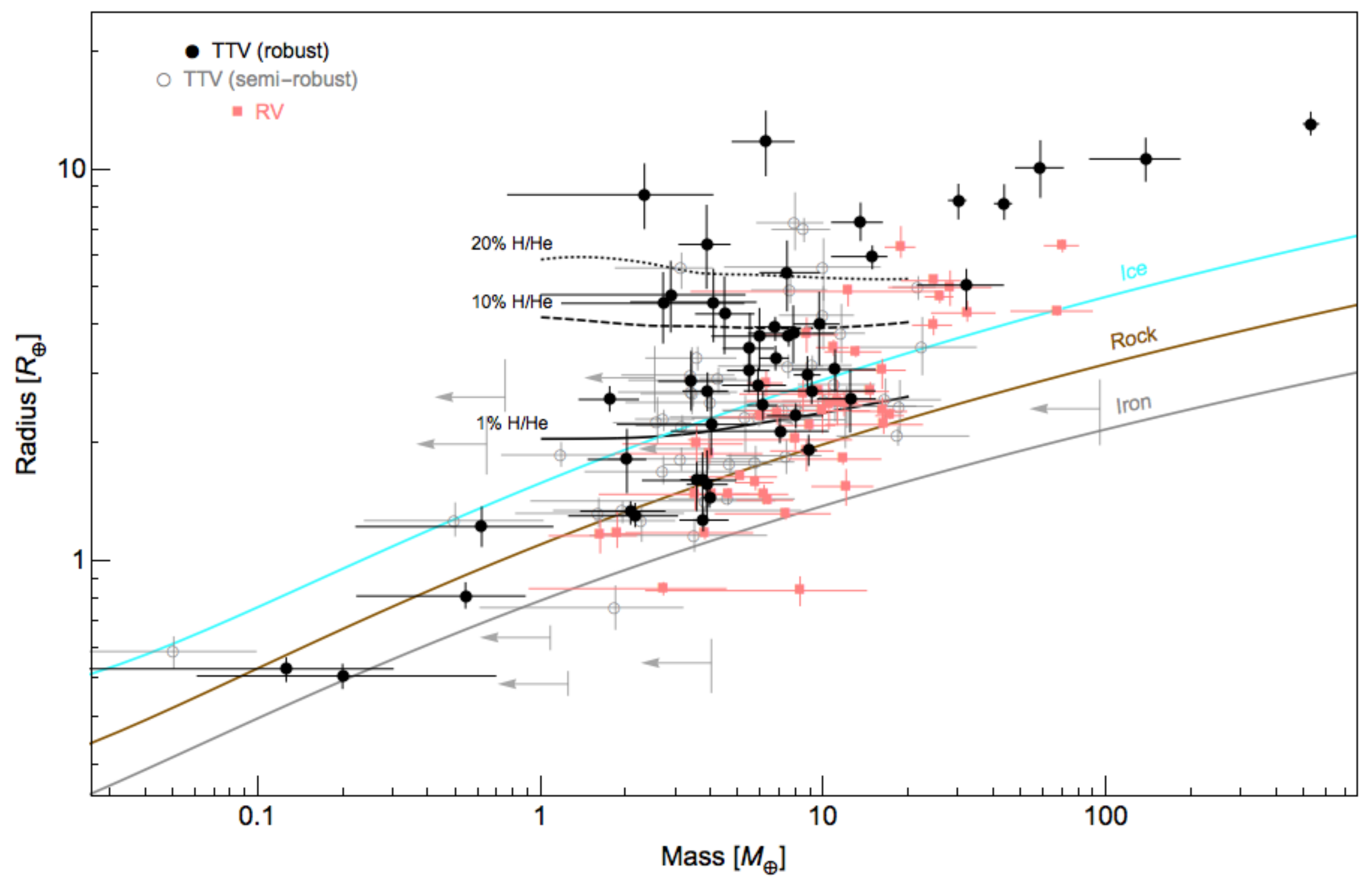}
\caption{
Planet radius versus mass. 
Masses from TTVs fit in Appendix \ref{sec:appendix} for planets classified as robust as defined in Section \ref{sec:fit_methods} are plotted as black circles.
Also shown as grey circles are planets whose masses under the two posteriors differ by up to  2$\sigma$ (``semi-robust'').
Error bars indicate the 1$\sigma$ credible regions from the default posteriors. 
Planets with RV mass measurements, listed in Table \ref{tab:rvdata}, are shown as pink squares.
For planets with masses that are consistent with zero, 1$\sigma$ upper bounds from the high mass posterior are indicated with an arrow.
Planet masses have been converted
to units of $M_\earth$ by multiplying stellar mass and planet-star mass ratios and accounting for stellar mass uncertainties (see Table \ref{tab:mass} caption for details).
Theoretical mass-radius relationships for pure ice, rock, and iron compositions from \citet{Fortney:2007hf} are plotted as colored curves.
Mass-radius relationships for planets with Earth-composition cores and H/He envelopes that make up 1,10, and 20\% of the total planet mass
are plotted as black curves which are interpolated from Tables 2 and 3 of \citet{2014ApJ...792....1L} assuming an age of 5 Gyr
and an incident stellar flux 100 times greater than that of Earth.
}
\label{fig:mass}
\end{center}
\end{figure*}

Figure \ref{fig:mass} shows inferred TTV planet masses on a mass-radius plot.
A sample of transiting planets with RV-measured masses are shown as well for comparison.
Most of the planets are less massive than $\lesssim10~M_\earth$ and exhibit a wide diversity of radii.
Many of the planets are less dense than a hypothetical pure water-composition planet, necessitating the presence of substantial gaseous envelopes to explain their low densities.

\begin{figure}[htbp]
\begin{center}
\includegraphics[width=0.45\textwidth]{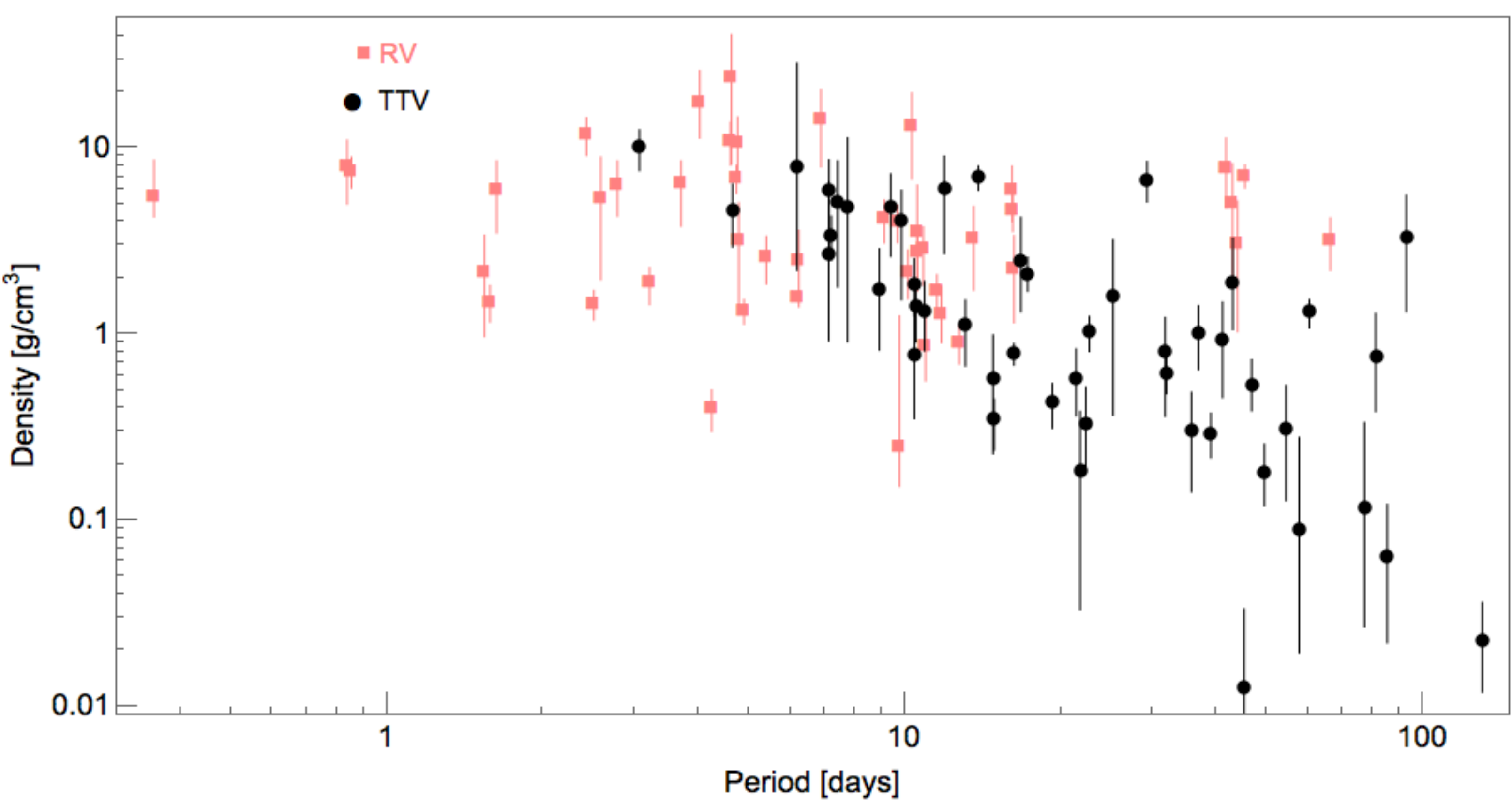}
\caption{
Planet densities versus orbital period for planets with masses measured via TTV (black circles) and RV (pink squares). 
TTV planet densities are  from the ``robust'' sample, with default posteriors.
}
\label{fig:period_density}
\end{center}
\end{figure}

Compared to small transiting planets with RV mass measurements, our TTV sample typically finds lower density planets as illustrated by Figure \ref{fig:period_density}. 
 Although this trend has been noted before, \citep[e.g.,][]{Weiss:2014ef,2015ApJ...813L...9D}, it is sometimes suggested that this discrepancy
is due to mass errors in either the TTV or RV measurements. 
But Figure  \ref{fig:period_density} demonstrates that much of the discrepancy 
can be explained if planets farther from the star tend to be less dense---perhaps because
they are less affected by photoevaporation. 
One should note, however, that the distribution of planets in Figure \ref{fig:period_density} is also affected by a variety of selection effects.  For example, 
 the lack of dense planets at long orbital periods could reflect the decreased transit detectability of small planets with long periods \citep[e.g.,][]{Gaidos:2012df}. 
 Also,  differences between RV and TTV populations could reflect the different dependence of the two techniques on planet mass and radius \citep[e.g.,][]{2016MNRAS.457.4384S}. 
Nonetheless, both TTVs and RVs find planets of similar density between periods of $3~\text{days}<P<20~\text{days}$ where there is significant overlap between the two samples.
A simple two-sample Kolmogorov-Smirnov test comparing all densities measured with the two methods gives a probability
of $p=2\times10^{-4}$ that they are drawn from the same underlying distribution, while the probability increases to $p=0.26$ if both samples are restricted to the period range 
$3~\text{days}<P<20~\text{days}$.

\begin{figure}[htbp]
\begin{center}
\includegraphics[width=0.5\textwidth]{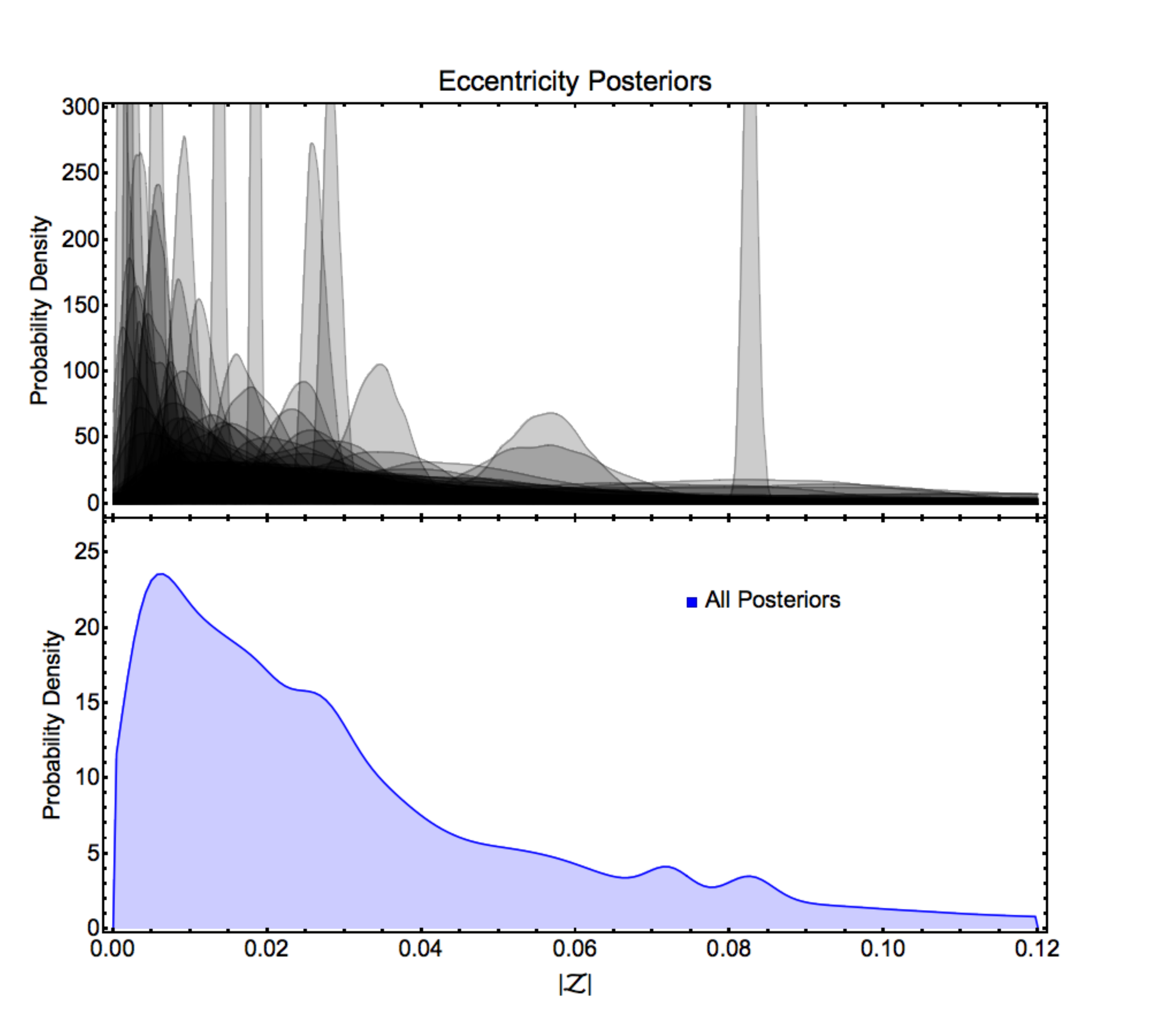}
\caption{
{\bf Top panel}: 
The posterior probability distributions of combined eccentricities, $|{\cal Z}|$, for all adjacent planet pairs (both robust and otherwise) from $N$-body MCMC fits.
The probability distributions are computed by applying a Gaussian kernel density estimate to the $N$-body MCMC default posterior samples.
Kernel bandwidths, $h$, are chosen using the `Silverman rule', $h = 1.06 \sigma N^{-1/5}$, where 
$\sigma$ is the sample variance and $N$ is the number of samples \citep{Silverman:1986uh}.
Combined eccentricities are shown only for adjacent planet pairs; combined eccentricities of non-adjacent planets are typically poorly constrained.
{\bf Bottom panel}: A smoothed histogram computed by combining all posterior samples shown in the top panel, {using a bandwidth 
$h=0.003$}.
The resulting `distribution' illustrates the typical magnitudes of combined eccentricities shown in the top panel
though it does not represent a true probability distribution.
}
\label{fig:ecc}
\end{center}
\end{figure}

We turn  now to the planets' eccentricities. 
As discussed in Section \ref{sec:analytic}, although TTV observations do not strongly constrain individual planet eccentricities, they can constrain the combination ${\cal Z}\approx (z-z')/\sqrt{2}$.
This combination may be considered as a surrogate for the individual planets' complex eccentricities unless planets have 
 $z\approx z'$, that is, comparable eccentricities and aligned orbits.
We expect planets' complex eccentricities have random relative orientations (see discussion in Paper I) so that, overall, $|{\cal Z}|$ values are a reliable surrogate for eccentricities.
The inferred values of $|{\cal Z}|$ are summarized in Figure \ref{fig:ecc}.
The majority of eccentricities are inferred to be small: the median of all the posterior samples shown in Figure \ref{fig:ecc} is $|{\cal Z}|=0.025$.
While many planet pairs' combined eccentricities are small, they are frequently inconsistent with zero.

Table  \ref{tab:ecc} lists combined eccentricities for individual planet pairs.  In addition to $|{\cal Z}|$, 
it is of interest to know which planets are consistent with ${\cal Z}=0$; such planets might have experienced
significant damping by tides or other effects.  Credible regions in $|{\cal Z}|$ cannot be used to address this question because $|{\cal Z}|$ must be non-negative.
Therefore, following \citet{2011MNRAS.410.1895Z}, we define the signed quantity, ${\cal Z}_\text{proj}$, which is the projection of the ${\cal Z}$s from the MCMC posterior onto the median of their distribution. 
More precisely, we define the median  ${\cal Z}_\text{med}$, by computing the median real and imaginary components of ${\cal Z}$.
Then, given ${\cal Z}$, the value of ${\cal Z}_\text{proj}$ is defined as 
\begin{equation}
{\cal Z}_\text{proj} = \frac{{\cal Z}{\cal Z}_\text{med}^*}{|{\cal Z}_\text{med}|}
\label{eq:zhat}
\end{equation}
where the `$^*$' indicates complex conjugate.
\citet{2011MNRAS.410.1895Z} show that an analogous quantity is useful for recovering $e=0$ solutions in the analysis of radial velocity data generated from circular orbits.
Of the 90 adjacent planet pairs in our TTV sample, more than 60\% have ${\cal Z}_\text{proj}$s inconsistent with 0 at 2$\sigma$ confidence for both the default and high mass posteriors.

\section{Discussion}
\label{sec:discussion}
\subsection{Gaseous Envelopes}
\label{sec:gas_fraction}	
\begin{figure*}[htbp]
\begin{center}
\includegraphics[width=0.95\textwidth]{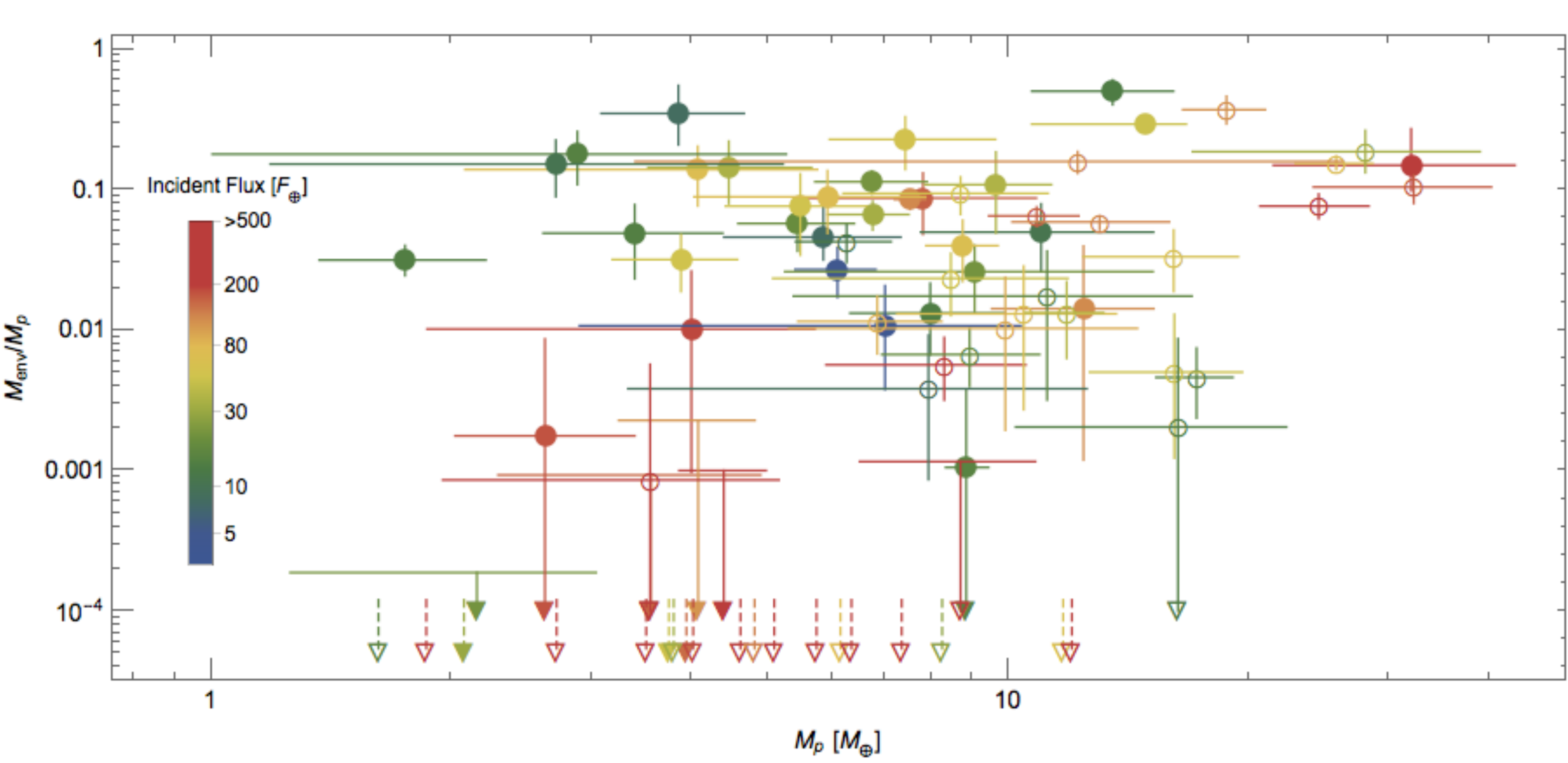}
\caption{
Envelope mass fraction versus planet mass.
TTV planets and RV planets are shown with shaded and open symbols, respectively.
Planets' incident fluxes are indicated by color scale.
Envelope mass fractions are computed assuming an age of 5 Gyr  from \citet{2014ApJ...792....1L}'s Tables 2 and 3 via interpolation.
Error bars show 1$\sigma$ uncertainties.
Error bars extending below $M_{env}/M_p<10^{-4}$ are indicated with arrows.   
Points with error bars entirely below $M_{env}/M_p<10^{-4}$ are indicated with dashed arrows and mass error bars are omitted for clarity. 
}
\label{fig:gas_frac}
\end{center}
\end{figure*}
Many of the planets shown in Figure \ref{fig:mass} 
have such low densities that they must have massive gaseous envelopes. 
 Such massive envelopes were likely accreted when the protoplanetary disk was still present; outgassing can only produce envelopes less massive than $\sim$2\% of a planet's total mass \citep{Rogers:2011gz}.
Here we convert observed masses and radii to envelope masses by interpolating the tables of
\cite{2014ApJ...792....1L}.   
Those authors simulate envelopes on top of Earth-composition cores; as time proceeds, their envelopes cool, while being irradiated by the star.
  Thus the masses and radii of 
 of their envelopes are essentially related by the fact that the cooling time should 
 be comparable to the age. Although there are multiple unknowns that will affect the predicted envelope 
 mass
 (especially atmospheric opacity and composition and, to a lesser extent, the core composition and heating/cooling)
 \citep[see, e.g.][]{Rogers:2011gz,2014ApJ...787..173H,2014ApJ...792....1L},
 the results of their simulations are adequate for our purposes.

Inferred envelope mass fractions, $M_{env}/M_p$, are plotted in Figure \ref{fig:gas_frac}
for  planets more massive than $1~M_\earth$ and smaller than $8~R_\earth$ with either RV or robust TTV mass measurements. 
 \citet{2014ApJ...792....1L} find that the radii of planets with envelope mass fractions $\gtrsim$1\% are quite insensitive to core mass.
Therefore, the diversity of gas fractions  in Figure \ref{fig:gas_frac} reflects the diversity of radii seen in Figure \ref{fig:mass}.
On the other hand, inferred envelope mass fractions  $\lesssim$1\% are likely sensitive to our assumption of Earth-composition cores and some of the envelope mass fractions in Figure \ref{fig:gas_frac} may actually reflect a diversity of core compositions. 

How do the inferred envelopes shown in Figure \ref{fig:gas_frac} compare to theoretical predictions of gas envelope accretion?
A number of studies have examined envelope accretion by planetary cores embedded in a gaseous protoplanetary disk  \citep[e.g.][]{2012ApJ...753...66I,2014ApJ...791..103B,2016ApJ...825...29G,2016ApJ...817...90L}.
These studies find that super-Earth cores readily accrete envelopes between a few to tens of percent of their total mass over the lifetime of a protoplanetary disk, similar to those 
shown in Figure \ref{fig:gas_frac}.
Planets in Figure \ref{fig:gas_frac} that are consistent with having no envelope generally receive higher fluxes than similar-mass planets covered by significant envelopes.
This trend could result from photoevaporation removing the envelopes initially accreted by planets close to their stars \citep[e.g.][]{Lopez:2012hi,2013ApJ...775..105O}.
Among planets that do have significant envelopes, there is no obvious trend in envelope fraction with planet mass.
This suggests a wide diversity of factors that influence envelope accretion and retention. 
We plan to compare observed envelope masses with theoretical predictions of accretion and mass-loss in a future work.

\subsection{Eccentricities}
\label{sec:ecc_damping}
Overall, our results show that most planets have relatively small eccentricities of a few percent.
This is consistent with the results of past TTV studies \citep{Wu:2013cp,2014ApJ...787...80H} and transit duration analyses \citep{VanEylen:2015je,Xie:2016dp} of multiplanet systems.
These studies infer the overall eccentricity distribution of their samples by fitting Rayleigh distributions:
\citet{Wu:2013cp} find a mean eccentricity $\bar{e}\sim 0.01$ for a sample of 44 planets,
\citet{2014ApJ...787...80H} find  $0.02<\bar{e}<0.03$ for a sample of 139 planets,
\citet{VanEylen:2015je} find  $0.05<\bar{e}<0.08$ for a sample of 74 planets, and
\citet{Xie:2016dp} find $\bar{e}<0.07$ for a sample of 330 planets.
In contrast to these population-level studies, we are able to measure the combined eccentricities of individual planet pairs, often with uncertainties on the order of a percent or better.
Both \citet{VanEylen:2015je} and \citet{Xie:2016dp} suggest that TTV systems' proximity to resonance may imply eccentricities  that are distinct from the larger population of multiplanet systems.
While it is true that TTV systems are preferentially closer to resonance (see Figure \ref{fig:sample_compare} in Appendix \ref{sec:appendix}),
they are not as radically distinct from the larger multiplanet population as is often suggested.

As noted above, a majority of our TTV sample (roughly $60\%$)  have eccentricities that are non-zero at 2$\sigma$ confidence.
This seemingly contradicts the inference that the planets were born in a disk of gas, which would presumably damp eccentricities.
The non-zero eccentricities suggest that planets may have accreted their envelopes from a depleted disk in which gas damping was ineffective \citep{2016ApJ...817...90L}.

\begin{figure}[htbp]
\begin{center}
\includegraphics[width=0.45\textwidth]{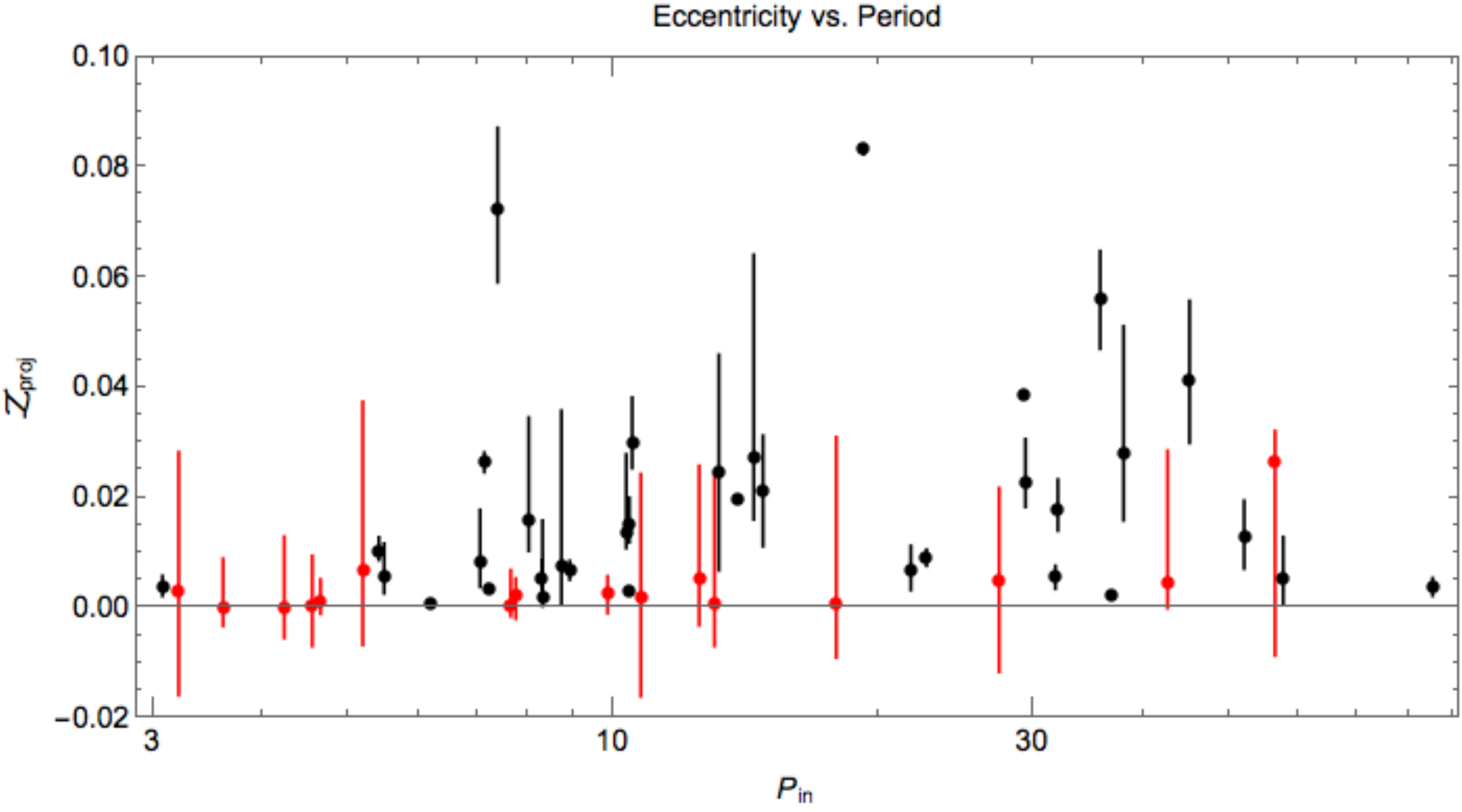}
\caption{
Signed combined eccentricities, ${\cal Z}_\text{proj}$ (Eq. \eqref{eq:zhat}), versus period for adjacent planet pairs.
Error bars indicate 1$\sigma$ credible regions computed with the default prior.
${\cal Z}_\text{proj}$  values that are consistent with 0 at the 1$\sigma$ level are emphasized as red points.
Only combined eccentricities that are robustly measured, meaning the default posterior 1$\sigma$ credible region contains the high mass posterior peak likelihood value, are plotted.
We have also excluded points with 1$\sigma$ credible regions larger than 0.05 for clarity.
}
\label{fig:ecc_period_damping}
\end{center}
\end{figure}

Some of the planets in Table \ref{tab:mass} orbit their host star at close enough separations that their eccentricities may be affected by tidal dissipation. 
Tidal dissipation depends steeply on planet period and the shortest period planets are expected to have experienced significant tidal damping.
Figure \ref{fig:ecc_period_damping} plots ${\cal Z}_\text{proj}$ (Equation \eqref{eq:zhat}) versus the periods of planet pairs' inner planets.
All but one of the seven shortest period planets within $P\lesssim 5$ days have combined eccentricities consistent with 0, suggesting circularization by tides.
The shortest period planet, Kepler-80 d, has a small but non-zero ${\cal Z}_\text{proj}$.
Beyond periods of 5 days there is a mixture of combined eccentricities that are consistent with 0 or very nearly so, as well as combined eccentricities that are definitively non-zero.

\section{Summary and Conclusion}
\label{sec:conclusion}
We have presented a uniform analysis of the TTVs of 55 multiplanet systems in order to infer planet masses and eccentricities.
We employ both $N$-body and analytic fitting in a complementary approach in order identify the degeneracies inherent to TTV inversion
and understand when and how they are broken by components of the TTV signal. 
We use two sets of MCMC simulations for each systems, each of which uses a prior weighted to opposite extremes of the mass-eccentricity
degeneracy predicted by the analytic formulae, in order to classify inferred planet masses as robust or not.

 Low mass ($\lesssim10 M_\earth$) planets exhibit a wide diversity of sizes,
with many of these planets less dense than a hypothetical pure ice composition planet, indicating the presence of significant gaseous envelopes. 
The wide diversity of planet sizes can be attributed to a diversity of planet envelope mass fractions.
We use our TTV fits together with the evolutionary models of \citet{2014ApJ...792....1L}, to convert planet masses and radii to envelope mass fractions.
We plan to use our sample of TTV-characterized planets in a future work examining planets' accretion and retention of gas envelopes during and after dispersal of the protoplanetary disk.

With guidance from the analytic TTV model, we have focused on planets' combined eccentricities, ${\cal Z}$, rather than individual eccentricities, which
are largely unconstrained by TTVs. 
We find that planets typically have eccentricities of a few percent or less, in agreement with past statistical studies of multiplanet systems \citep{Wu:2013cp,2014ApJ...787...80H,VanEylen:2015je,Xie:2016dp}.
The shortest period planets have eccentricities consistent with 0 and thus may have experienced significant tidal eccentricity damping. 
Detailed modeling of individual TTV systems can potentially shed light on how dynamical processes such as migration and eccentricity damping may have shaped the systems we observe today. 

We measure a number of planet masses for planets in the super-Earth/mini-Neptune size range where there is great theoretical interest in understanding the mass-radius relationship.
The observational biases of our TTV sample are difficult to assess given the TTV signal's complicated dependence on periods and eccentricities. 
We see incorporating TTV non-detections and quantifying selection effects as important directions for future work in understanding the mass-radius relationship.

\acknowledgments{
  Acknowledgments.
  We thank Lauren Weiss for helpful discussion.
  We are grateful to the Kepler team for acquiring and publicly releasing such spectacular results.
  This research has made use of the NASA Exoplanet Archive, which is operated by the California Institute of Technology,
   under contract with the National Aeronautics and Space Administration under the Exoplanet Exploration Program.
 SH acknowledges support from the NASA Earth and Space Science Fellowship program, grant number NNX15AT51H. 
 YL acknowledges grants AST-1109776 and AST-1352369 from NSF, and NNX14AD21G from NASA.
}

\bibliographystyle{apj}
\bibliography{Untitled}
\clearpage

\defcitealias{Holman:2010db}{Ho10} 
\defcitealias{2014arXiv1403.1372D}{Dr14} 
\defcitealias{Lissauer:2011el}{Li11} 
\defcitealias{2013ApJ...770..131L}{Li13} 
\defcitealias{Borsato:2014it}{Bo14} 
\defcitealias{Cochran:2011kg}{Co11} 
\defcitealias{Marcy:2014hr}{Ma14} 
\defcitealias{Hadden:2016ki}{Ha16} 
\defcitealias{JontofHutter:2016ch}{JH16} 
\defcitealias{2016arXiv160909135M}{Mig16} 
\defcitealias{2012Natur.487..449S}{SO12} 
\defcitealias{Carter:2012gq}{Ca12} 
\defcitealias{Masuda:2014dm}{Ma14} 
\defcitealias{Huber:2013ha}{Hu13} 
\defcitealias{Gozdziewski:2015ws}{Go15} 
\defcitealias{2014ApJ...785...15J}{JH14} 
\defcitealias{MacDonald:2016vo}{Mac16} 
\defcitealias{2013ApJ...778..185M}{Ma13} 
\defcitealias{2012ApJ...759L..36H}{Hi12} 
\defcitealias{2013ApJ...768...14W}{We13} 
\defcitealias{2015Natur.522..321J}{JH15a} 
\defcitealias{Mills:2016gi}{Mi16} 

\begin{ThreePartTable}
\begin{TableNotes}
\item [a] \label{tn:24c} Planet-star radius ratio from \citet{2011ApJ...736...19B}.
\item [b] \label{tn:29} Planet-star radius ratio from \citet{Fabrycky:2012ed}.
\item [c] \label{tn:51c} Planet-star radius ratio from \citet{Masuda:2014dm}.
\item [d] \label{tn:55c} Planet-star radius ratio from \citet{2013MNRAS.428.1077S}.
\item [e] \label{tn:223e} Planet-star radius ratio from \citet{Mills:2016gi}.
\item [f] \label{tn:359d} Planet-star radius ratio from \citet{Rowe:2014jq}.
\item [g] \label{tn:1126} Planet-star radius ratio from {\it Kepler} KOI Q1-17 data release DR24, hosted on the Exoplanet Archive.
\end{TableNotes}
\begin{longtable}{|c | c  c | c | c c  | c  c| c |}
\caption*{ {\bf Masses from TTVs.}
	 Values and uncertainties reflect the peak posterior probabilities and 68.3\% credible regions. 
	The peak posterior probabilities are computed by finding the maximum likelihood of a
	kernel density estimate computed from the posterior sample.
	Credible regions are so-called ``highest posterior density intervals": 
	the smallest parameter range containing 68.3\% of the posterior sample.
	In columns 5--8, 68.3\% upper limits are listed for planets with masses that are consistent with 0.
	Planets with robustly inferred masses are indicated with a `*' (see Section \ref{sec:fit_methods}).
	References are listed in column 9 for planets with masses previously inferred from $N$-body TTV fits or RV observations.
	Planet radii, masses, and densities in columns 3, 5--7 incorporate the following: planet-star mass ratios sampled from our MCMC posteriors;
	planet-star radius ratios from the light curve fit posteriors of \citet{2015ApJS..217...16R};
	and randomly generated samples of host star  properties.
	For the latter, samples of host star radii, masses, and densities are generated based on values reported in the {\it Kepler} Stellar Q1-17 data release DR25, hosted on the Exoplanet Archive.
	For each {\it Kepler} system, random samples of stellar mass, radius, and density are drawn from skew-normal distributions \citep{Azzalini:1985fe}
	with scale and shape parameters chosen to match the reported $\pm1\sigma$ error bars.
	For some planets, \citet{2015ApJS..217...16R}'s light curve fit posteriors are missing or contain a large number of points with impact parameters $b > 1$ and are clearly pathological.
	The radii of these planets are computed using planet-star radius ratios from other sources as indicated.
 } 
\label{tab:mass} \\
\hline 
\multicolumn{1}{| c |} {\bf  Planet} &
\multicolumn{1}{| c } {\bf  Period} &
\multicolumn{1}{c |} {\bf  Radius}        &
\multicolumn{1}{c |} {\bf Star Mass}        &
\multicolumn{1}{c }  {\bf  Mass }          & 
\multicolumn{1}{c |}  {\bf Density} &
\multicolumn{1}{c }  {\bf  Mass}          & 
\multicolumn{1}{c |}  {\bf Density} 	&
\multicolumn{1}{c |}   {\bf Ref.}	\\  
{}& 
{}& 
{ }   &
{ }   &
{(default prior)} &
{(default prior)} &
{(high mass prior)} &
{(high mass prior)} &
 {} \\ 
{}& 
{[days]}& 
{ [$R_\oplus$]}   &
{ [$M_\odot$]} &
{ [$M_\oplus$]} &
{ [g/$\text{cm}^3$]} &
{ [$M_\oplus$]} &
{ [g/$\text{cm}^3$]} &
 {} \\ 
 \hline 
\endfirsthead
\multicolumn{9}{c}%
{{\bfseries \tablename\ \thetable{} -- continued from previous page}}\\ \hline
\multicolumn{1}{| c |} {\bf  Planet} &
\multicolumn{1}{| c } {\bf  Period} &
\multicolumn{1}{c |} {\bf  Radius}        &
\multicolumn{1}{c |}  {\bf  Star Mass }          & 
\multicolumn{1}{c }  {\bf  Mass }          & 
\multicolumn{1}{c |}  {\bf Density} &
\multicolumn{1}{c}  {\bf  Mass}          & 
\multicolumn{1}{c |}  {\bf Density} &
\multicolumn{1}{c |}   {\bf Ref.}	\\  
{}& 
{}& 
{ }   &
{ }   &
{(default prior)} &
{(default prior)} &
{(high mass prior)} &
{(high mass prior)} &
 {} \\ 
{}& 
{[days]}& 
{ [$R_\oplus$]}   &
{ [$M_\odot$]} &
{ [$M_\oplus$]} &
{ [g/$\text{cm}^3$]} &
{ [$M_\oplus$]} &
{ [g/$\text{cm}^3$]} &
 {} \\ 
 \hline 
\endhead
\hline \multicolumn{9}{|r|}{{Continued on next page}} \\ \hline
\endfoot
\hline
 \hline
\insertTableNotes
\endlastfoot
 \input{masstable.tex}
\end{longtable}
\end{ThreePartTable}

\clearpage

\begin{longtable}{| c c c| c c | c c |}
\caption*{
{\bf Combined Eccentricities of Adjacent TTV Planet Pairs.}
\footnotesize
Column two lists each planet pair's nearest first- or second-order MMR. Column three lists planet pairs' normalized distance to resonance, $\Delta = (j-k)P'/(jP)-1$, where $P$ and $P'$ are the periods of the inner and outer planet, respectively and $k=1$ or $2$ as appropriate for a first- or second-order resonance. Values and uncertainties reflect the peak posterior probabilities and 68.3\% credible regions, computed as described in the caption of Table \ref{tab:mass}. Our ${\cal Z}$ is defined in terms of {\it free} eccentricities  (Eq. \ref{eq:zdef}), whereas the MCMC  outputs total (free+forced) eccentricity; we convert to free eccentricity for this table by subtracting off the analytically calculated forced components.
} \\
\hline 
\multicolumn{1}{| c } {\bf  Planet Pair} &
\multicolumn{1}{ c } {\bf  Resonance} &
\multicolumn{1}{ c |} {  $\Delta$}        &
\multicolumn{1}{ c }  { $|{\cal Z}|$}          & 
\multicolumn{1}{ c |}  {${\cal Z}_\text{proj}$} &
\multicolumn{1}{ c }  { $|{\cal Z}|$}          & 
\multicolumn{1}{ c |}  {${\cal Z}_\text{proj}$}  \\
{}& 
{}& 
{ }   &
{(default prior)} &
{(default prior)} &
{(high mass prior)}  &
{(high mass prior)}  
\\ 
 \hline 
\endfirsthead
\multicolumn{7}{c}%
{{\bfseries \tablename\ \thetable{} -- continued from previous page}}\\ \hline
\multicolumn{1}{| c } {\bf  Planet Pair} &
\multicolumn{1}{ c } {\bf  Resonance} &
\multicolumn{1}{ c } {  $\Delta$}        &
\multicolumn{1}{ c }  { $|{\cal Z}|$}          & 
\multicolumn{1}{ c |}  {${\cal Z}_\text{proj}$} &
\multicolumn{1}{ c }  { $|{\cal Z}|$}          & 
\multicolumn{1}{ c |}  {${\cal Z}_\text{proj}$}  \\
{}& 
{}& 
{ }   &
{(default prior)} &
{(default prior)} &
{(high mass prior)}  &
{(high mass prior)}  \\
\hline 
\endhead
\hline \multicolumn{7}{|r|}{{Continued on next page}} \\ \hline
\endfoot
\hline
\hline
\endlastfoot
\input{ecctable.tex}
\label{tab:ecc}
\end{longtable}

\clearpage

\begin{table}[htdp]
\caption{Transiting RV Planets}
{\footnotesize
Periods, masses, and radii of transiting RV planets smaller that $8~R_\earth$.
All planets selected from the Exoplanet Archive.  Only planets with $1\sigma$ mass uncertainties inconsistent with $0$ are included.
We exclude RV measurements of planets from the Kepler-18, 25, 48, and 89 systems since they are also in our TTV sample.
}
\begin{center}
\begin{tabular}{|c|c|c|c|c|}
\hline
{\bf Name }& {\bf Period} & {\bf Mass }& {\bf Radius }& { \bf Reference }\\
		&[days]		&	$[M_\earth]$& $[R_\earth]$& \\ \hline
\input{rvtable.tex}\\ \hline
\end{tabular}
\end{center}
\label{tab:rvdata}
\end{table}%

\appendix

\section{A: Description of Individual Systems}
\label{sec:appendix}
\begin{figure}[htbp]
\begin{center}
\includegraphics[width=0.45\textwidth]{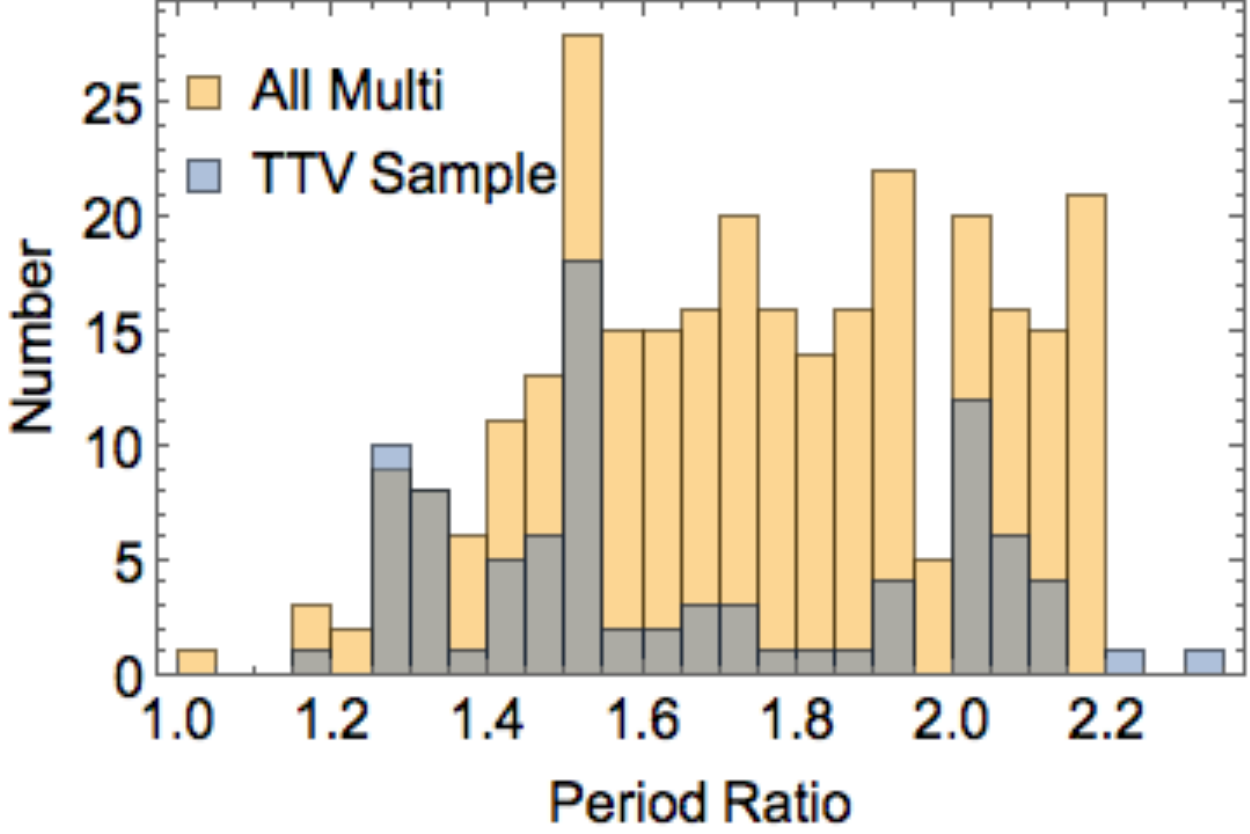}
\includegraphics[width=0.45\textwidth]{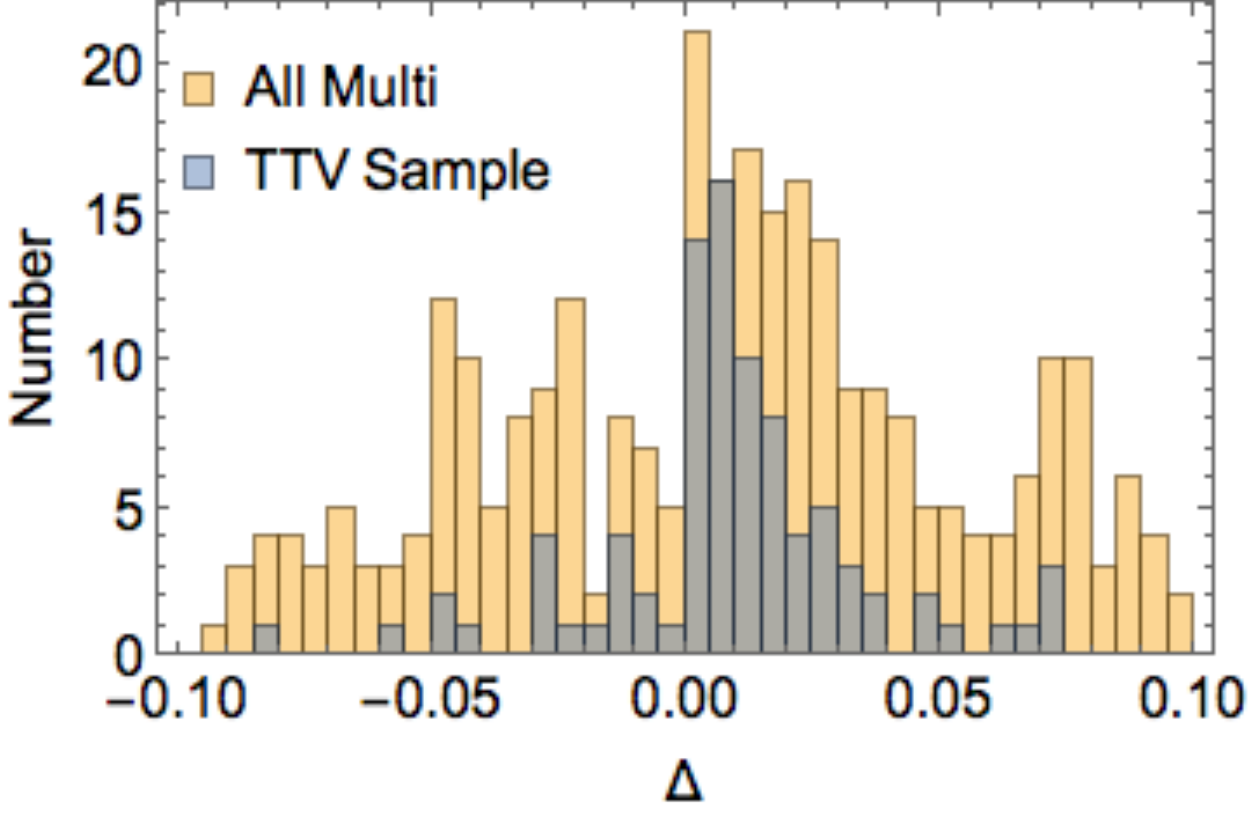}
\caption{
Comparison of our selected TTV sample to all {\it Kepler} multi-transiting planets. 
The ``All Multi" sample contains all confirmed pairs of planets with period ratios less than $P'/P<2.2$ taken from the Exoplanet Archive on November 4th, 2016. 
{\bf Left panel:} Histogram of the period ratios of adjacent planet pairs.
{\bf Right panel:} Histogram of distances to the nearest first- or second-order resonance, $\Delta$, defined in Equation \eqref{eq:delta_def}.
}
\label{fig:sample_compare}
\end{center}
\end{figure}

Our initial selection of systems is made from TTVs identified in the Q1-17 catalogue of \citet{2016ApJS..225....9H}.
We limit our selection to multiplanet systems hosting at least one planet pair with a period ratio smaller than $P'/P<2.2$.
After selecting systems on the basis of period ratios, we identify planets with significant TTVs as follows: 
\begin{itemize}
\item Each planet's transit times are fit with two linear models:
the first is a simple linear trend corresponding to a constant orbital period and
the second fits each planet's transit times as a linear trend plus a TTV induced by the other planets in the system.
The TTV is assumed to  be described by Equation \eqref{eq:the_ttv} and parameterized by the amplitudes $\hat{\delta t_{\cal F}}$, $\hat{\delta t_{\cal S}}$, and $\hat{\delta t_{\cal C}}$.
Since $P_{sup}$ and $P_{syn}$ in Equation \eqref{eq:the_ttv} are defined in terms of ``average" orbital periods that differ from the orbital period fit without accounting for TTVs, 
fits are iterated to achieve convergence in average orbital periods.
\item  we compute the Bayesian Information Criterion \citep[BIC;][]{Schwarz:1978kf}, defined as 
\begin{equation}
\text{BIC}=\chi^2+k\ln(N_\text{trans.})
\end{equation}
for both linear fits of a planet's transit times where $\chi^2$ has the standard definition, $k$ is the number of fit parameters,
and $N_\text{trans.}$ is the number of transit times that are fit.
We select TTVs for which the sinusoidal fit improves the BIC by $>$10 relative to the simple linear trend model, indicating very 
strong evidence for the analytic TTV model over the simple linear trend \citep{Kass:2012eh}.
\item Finally, we remove a handful of systems after visual inspection of the analytic fit
(KOI-0295, KOI-0571, KOI-1873, KOI-2029) 
or because we are unable to find a satisfactory initial $N$-body fit (KOI-262, KOI-880, KOI-1426, KOI-2693), potentially because of  
non-transiting perturbers.
\end{itemize}
In systems with three or more planets, any planets that are separated from the next adjacent planet by a period ratio greater than $P'/P>2.2$ are ignored by our analysis.\footnote{
Two exceptions to this cut, Kepler-52 d and Kepler-27.03, were included in our fits before we arrived at our final selection criteria.  
These planets are included in our posterior data though neither contributes useful constraints.}
Figure \ref{fig:sample_compare} compares our selected sample to all confirmed {\it Kepler} planet pairs with $P'/P<2.2$, showing both period ratios and normalized distance
to the nearest first- or second-order MMR, 
\begin{equation}
\Delta = \frac{j-k}{j}\frac{P'}{P}-1\label{eq:delta_def}
\end{equation}
with $k=1, 2$ or as appropriate for first- or second-order MMRs.
The TTV sample is biased towards closer period ratios compared to the complete multiplanet sample, largely because it lacks planets between the 3:2 and 2:1 MMRs.
It is often claimed that because TTV systems contain planets close to mean-motion resonances they may have unique formation channels not shared with the broader population of multiplanet systems.
The proximity to resonance of our sample, as measured by $\Delta$, is compared to the full sample of multiplanet systems in Figure \ref{fig:sample_compare}. 
Pairs with $|\Delta|>0.05$, which constitute $\sim$30\% of the total sample, are essentially absent from the TTV sample.
Otherwise, the TTV sample is not radically distinct in its proximity to resonance.

The posterior samples from our MCMC fits along with plots showing the TTVs of each system are available online at \href{https://doi.org/10.5281/zenodo.162965}{doi.org/10.5281/zenodo.162965}.
Each system is discussed below (see Paper I for discussion of the Kepler-26, Kepler-33, Kepler-128, and Kepler-307 systems).
Systems have been categorized into four groups based on the analytic fits:
 \begin{itemize}
 \item	{\bf Chopping Systems} are presented in Section \ref{sec:Chopping}. 
 		These systems host at least one planet with a chopping signal that allows the mass-eccentricity degeneracy to be broken, 
 		resulting in constraints on planet masses and combined eccentricities.
		The chopping signal constrains the mass of a perturbing companion.
		With the perturber's mass constrained, the TTV signal component caused by a first- or second-order 
		MMR uniquely determines the planets' combined eccentricity. 
		Some systems where a lack of chopping places a strong upper bound on planet masses
		are also included in this category.
 \item	{\bf Second Harmonic Systems} are presented in Section \ref{sec:SecondHarmonic}. 
 		This group is comprised of systems with a pair of planets near a first-order MMR.
		The second harmonic TTV of at least one of the planets  helps
		break the  mass-eccentricity degeneracy. 
		The second harmonic TTV often does not provide as strong a constraint on planet mass as the chopping TTV.
		In most cases, the second-harmonic TTV sets an upper limit to the 
		combined eccentricity, or equivalently, a lower limit on planet masses.
\item	{\bf Resonant \& Massive Systems} include systems that have planets either extremely close to or librating in resonance
		or systems with especially massive planets.
		The analytic formulae are not expected to provide accurate approximations of the TTVs of such planets. 
\item		{\bf Degenerate Systems} do not have significant chopping or second-harmonic signals
 		and show a strong degeneracy between masses or eccentricities. 
		These systems are summarized in \ref{sec:Degenerate}.
 \end{itemize}

	Below, analytic fits are compared to $N$-body results in a series of `analytic constraint plots'.
	To generate these plots, we convert the 1$\sigma$ uncertainties in $\hat{\delta t_{\cal F}}$, $\hat{\delta t_{\cal S}}$, and $\hat{\delta t_{\cal C}}$ 
	obtained from least-squares fitting to 1$\sigma$ uncertainty bands in the $\mu'$-$|\cal Z|$ 
	plane.\footnote{Chopping amplitude constraints are computed by incorporating the prior assumption that masses are non-negative.
	This constraint affects the 1$\sigma$ upper limits computed for $\hat{\delta t_{\cal C}}$'s that are consistent with negative masses.}
	The details of this procedure are described in Paper I.
	Planets with robustly inferred masses, as defined in Section \ref{sec:fit_methods}, are indicated with a `*'.
	
\subsection{\ref{sec:Chopping}: Chopping Systems}
\label{sec:Chopping}

\begin{figure}[htbp]
 	\begin{center}
 	\includegraphics[width=0.45\textwidth]{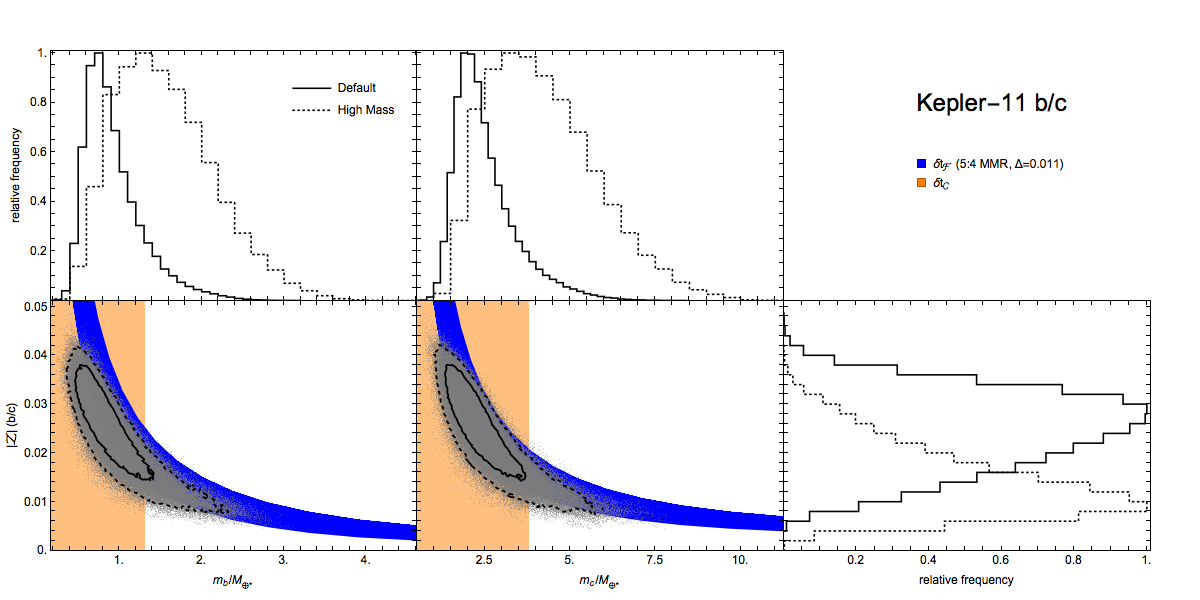}
	 \includegraphics[width=0.45\textwidth]{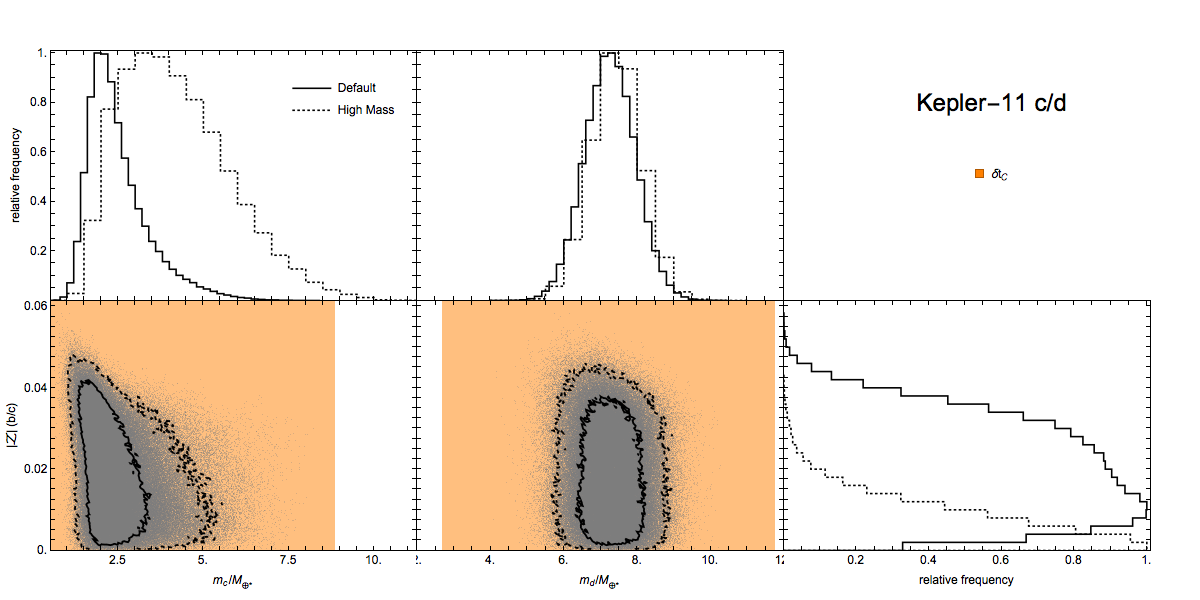}
	\includegraphics[width=0.45\textwidth]{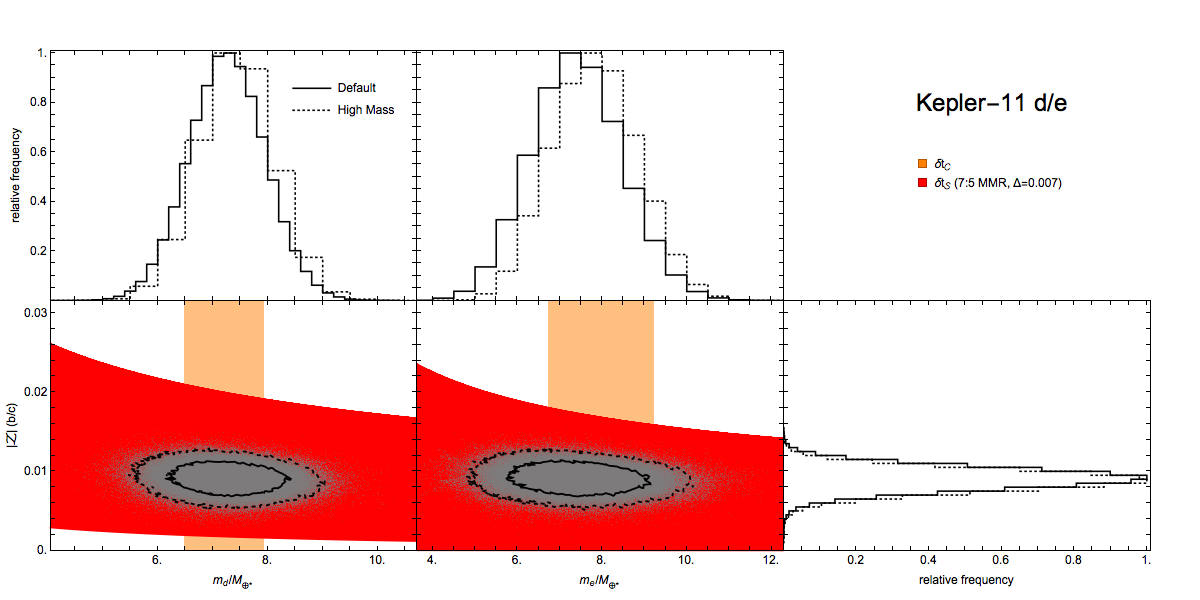}
	\includegraphics[width=0.45\textwidth]{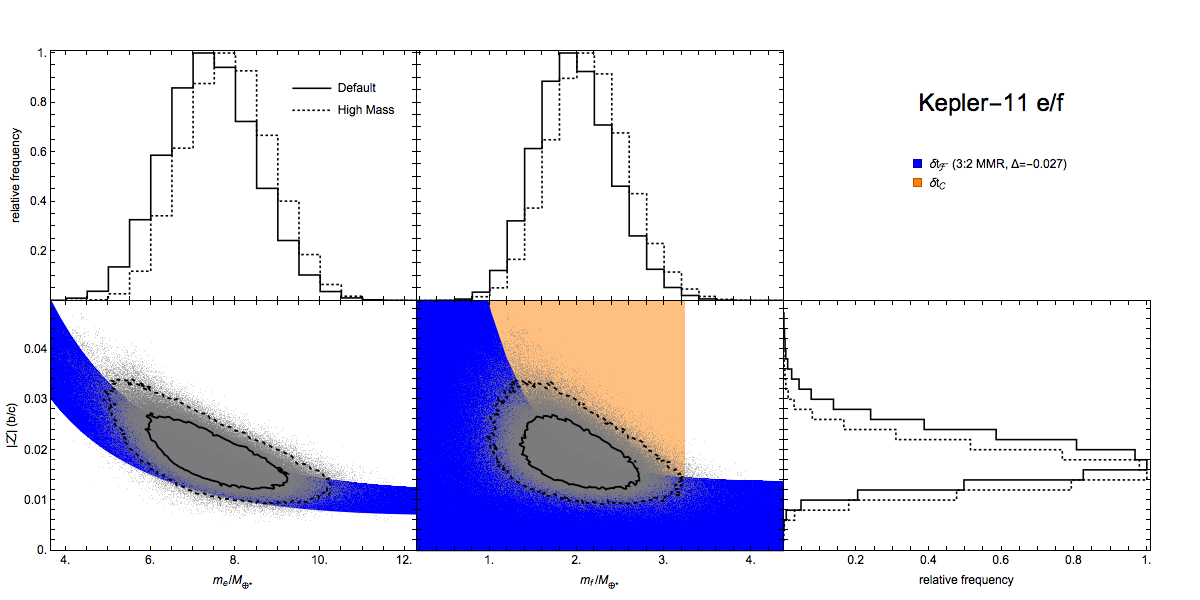}
 \caption{
Constraint plots for the Kepler-11 system.
Each panel summarizes the constraints derived from the interactions of a particular planet pair. 
Specifically, the constraints placed {\it by} the TTV of the outer planet {\it on} the mass of the inner planet 
plus the combined eccentricity are shown in the bottom left frame and the constraints of the inner planet's
TTV on the outer planet's properties are shown in the bottom middle frame.
Posterior samples from the default prior MCMC are plotted as gray points with black lines indicating
the 68\% (solid) and 95\% (dashed) credible regions.
Histograms show the marginalized mass and combined eccentricity posterior distributions from the 
default prior (solid) and high mass prior (dotted) $N$-body MCMCs.
The 1$\sigma$ constraints from fundamental TTVs are shown in blue, chopping TTVs in orange,
and second-harmonic/second-order resonance TTVs in red.
Note that planet masses are plotted relative to the host star mass in units of $\mearth$
defined in Equation \eqref{eq:mearth_def}.
 }
 	\label{fig:kep11cons}
 	\end{center}
 \end{figure}

{\it Kepler-11 b, c, d*, e*, f*} (Fig. \ref{fig:kep11cons}): 
The inner five  planets of Kepler-11 are near a series of first and second-order MMRs.
The TTVs robustly constrain the masses of planets d, e, and f. 
The bottom left panel of Figure \ref{fig:kep11cons} shows that planet d and e's masses are strongly constrained by their mutual chopping TTVs.
The bottom right panel of Figure \ref{fig:kep11cons} summarizes the constraints from planet e and f's interaction. 
Since the mass of planet e is already constrained by its interaction with planet d, planet f's fundamental TTV constrains the e/f pair's combined eccentricity. 
The combination of planet e's fundamental and chopping TTV constrains the mass of planet f.
	
The masses of the innermost pair, b and c, are less robust to the choice of priors.  
The top left panel of Figure \ref{fig:kep11cons} shows their inferred properties are consistent with their fundamental TTVs
and that their lack of chopping TTVs imposes upper bounds on their masses.

The upper right panel of Figure \ref{fig:kep11cons} shows that planet c's chopping TTV gives a weak constraint on planet d's mass and the lack of chopping in planet d's TTV gives an upper limit to the mass of planet c.
Planet c and d are not close to any first or second-order MMRs so there is little correlation between their masses and combined eccentricity.

TTV analyses of the Kepler-11 system have been conducted by number of authors \citep{Lissauer:2011el,2012MNRAS.427..770M,2013ApJ...770..131L,Borsato:2014it} and
our results agree well with previous work.
We omit the outermost planet of the system, Kepler-11 g, from our analysis since it does significantly influence the TTVs of the other planets  \citep{Lissauer:2011el}.

\begin{figure}[htbp]
	\begin{center}
	\includegraphics[width=0.9\textwidth]{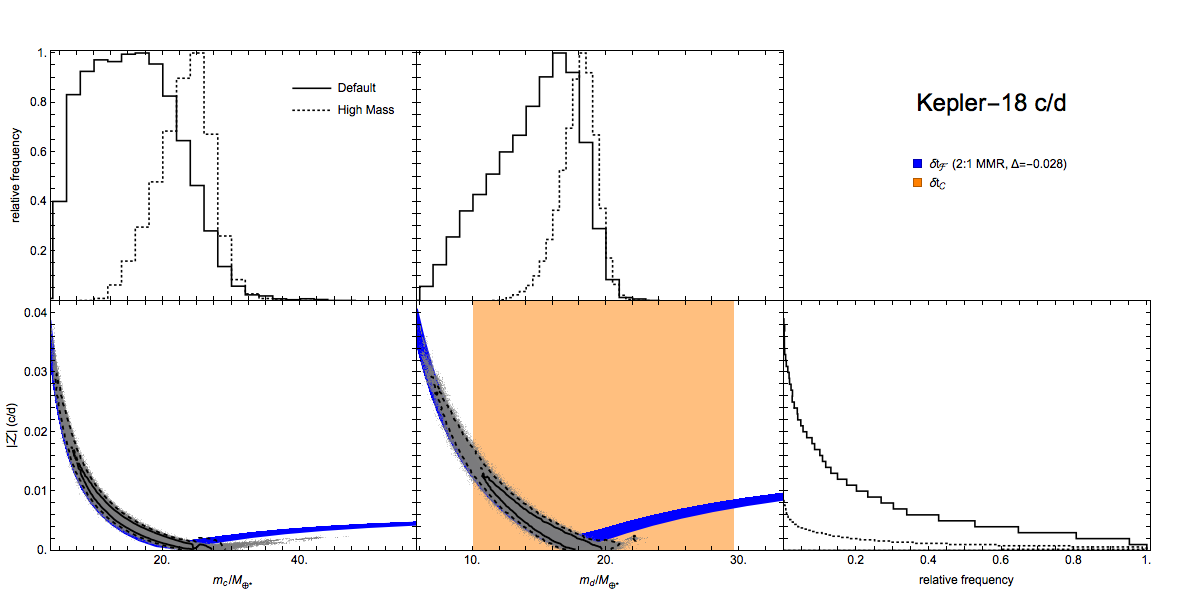}
\caption{
Constraint plots for Kepler 18 (see Figure \ref{fig:kep11cons} for description).
}
	\label{fig:kep18cons}
	\end{center}
\end{figure}

{\it Kepler-18 c, d*} (Fig. \ref{fig:kep18cons}):
 Planet d's inferred mass agrees well with the analytic chopping constraint.
The chopping amplitude measurement has a fractional uncertainty of $\sim$50\%, 
so there is still a large degree of mass-eccentricity degeneracy.
Consequently, planet c's mass does not meet our standard for classification as robust.

Kepler-18 was previously analyzed by \citet{Cochran:2011kg}, who fit both RV and TTV observations of Kepler-18 spanning Quarters 0--6. 
They find planet masses of $[m_c,m_d]=[17.3 \pm 1.9,16.4 \pm 1.4]M_\earth$ using a combination of RV and TTV data. 
Their results are consistent with our $N$-body MCMC fits.

Kepler-18 hosts an additional planet, b, interior to planet c. The TTVs of planet b appear flat and we find that including it in our $N$-body MCMC fitting does not yield a strong constraint on its properties or affect the inferred properties of planets c and d.

\begin{figure}[htbp]
\begin{center}
	\includegraphics[width=0.9\textwidth]{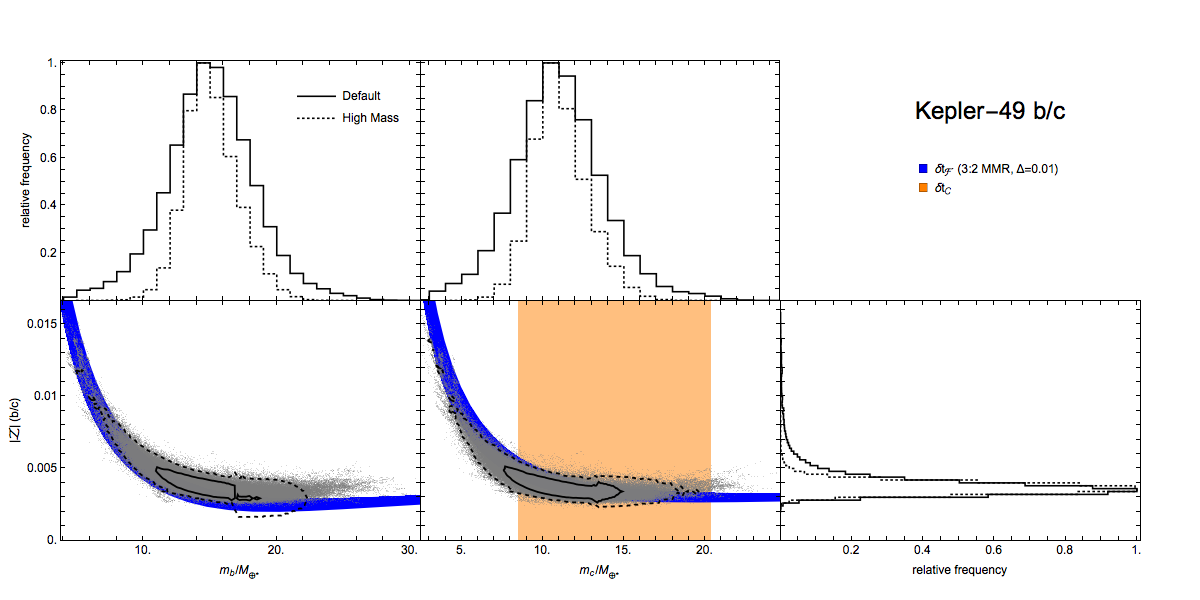}
\caption{
Constraint plots for Kepler-49 (see Figure \ref{fig:kep11cons} for description).
The combined eccentricity has been corrected by subtracting the forced component.
}
	\label{fig:kep49cons}
	\end{center}
\end{figure}

{\it  Kepler-49 b*,c*} (Fig. \ref{fig:kep49cons}): 
Planet b's chopping TTV breaks the mass-eccentricity degeneracy for this planet pair, resulting in two robustly inferred masses.
\citet{JontofHutter:2016ch} fit masses $[m_b,m_c]=[9.2^{+3.7}_{-3.5},5.9^{+2.7}_{-2.3}]M_\earth$ analyzing Q1-17 transit times of Kepler-49 b and c, in good agreement with the default prior MCMC results,  $[m_b,m_c]=[8.1^{+1.8}_{-1.8},5.8^{+1.5}_{-1.4}]M_\earth$.

\begin{figure}[htbp]
\begin{center}
	\includegraphics[width=0.9\textwidth]{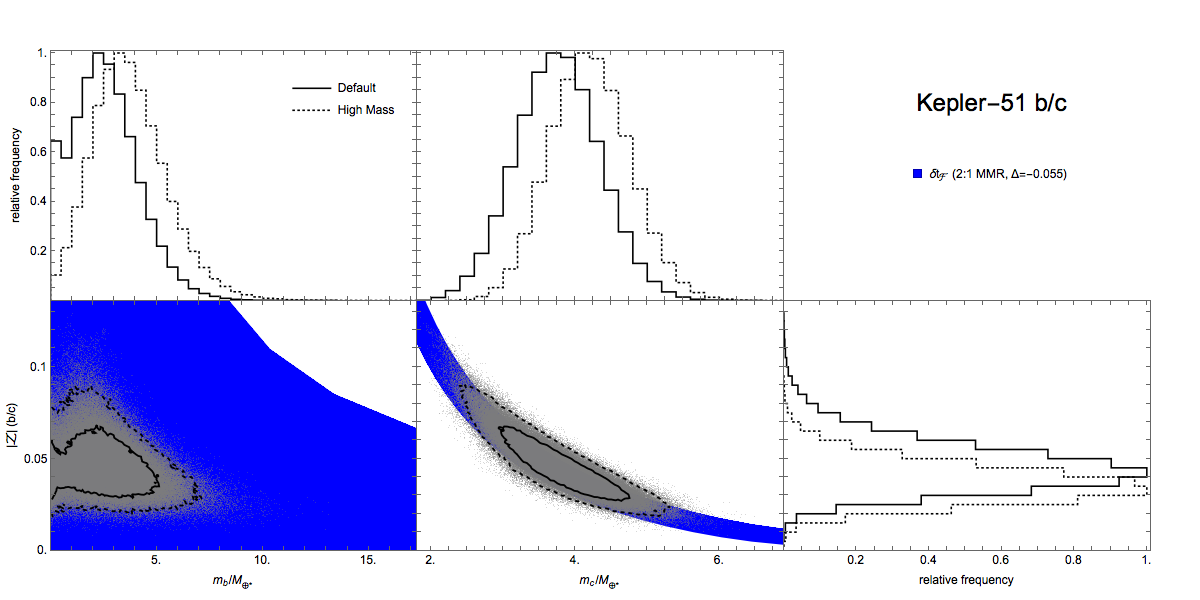}
	\includegraphics[width=0.9\textwidth]{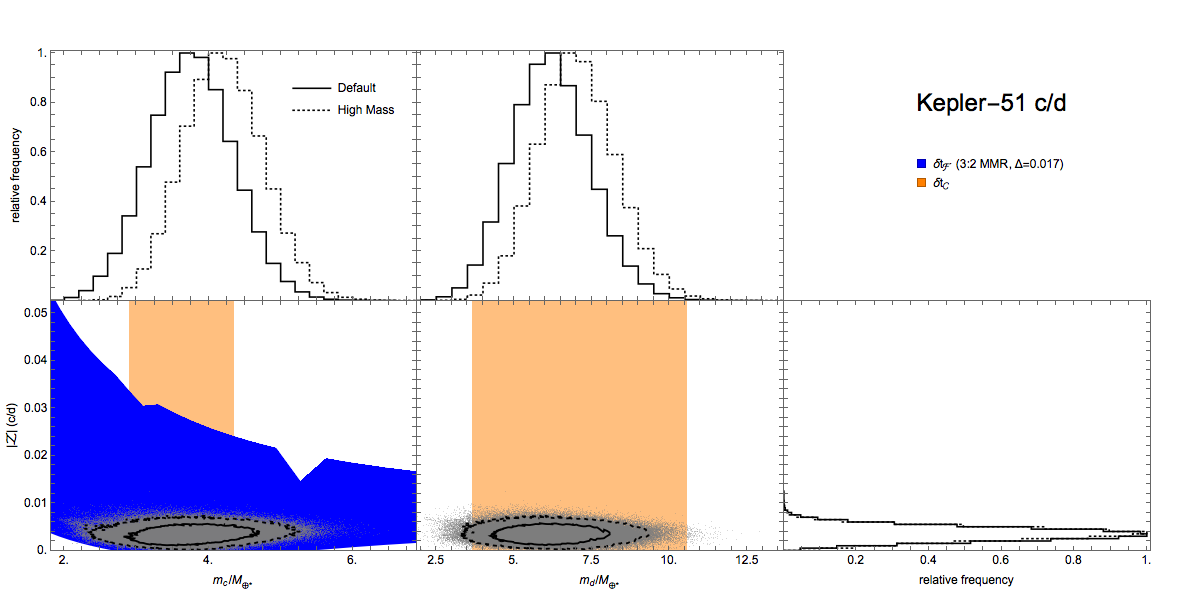}
\caption{Constraint plots for Kepler-51 (see Figure \ref{fig:kep11cons} for description).}
	\label{fig:kep51cons}
	\end{center}
\end{figure}

{\it  Kepler-51 b*,c*,d*} (Fig. \ref{fig:kep51cons}): 
Planet c and d's masses are well constrained by their mutual chopping TTVs.
Planet b and c's  combined eccentricity is constrained by the fundamental TTV of planet b since the mass of planet c is already constrained by its effect on planet d.
\citet{Masuda:2014dm} previously conducted an MCMC analysis of the Kepler-51 system's TTVs using {\it Kepler} data spanning Quarters 1-16
and fits masses $[m_b,m_c,m_d]=[2.1^{+1.5}_{-0.8},4.0\pm0.4,7.6\pm1.1]M_\earth$.
The  masses found by \citet{Masuda:2014dm}  agree well with our default prior results: $[m_b,m_c,m_d]=[2.4^{+1.7}_{-1.6},3.8^{+0.9}_{-0.7},6.2^{+1.6}_{-1.5}]M_\earth$.

\begin{figure}[htbp]
 	\begin{center}
 	\includegraphics[width=0.9\textwidth]{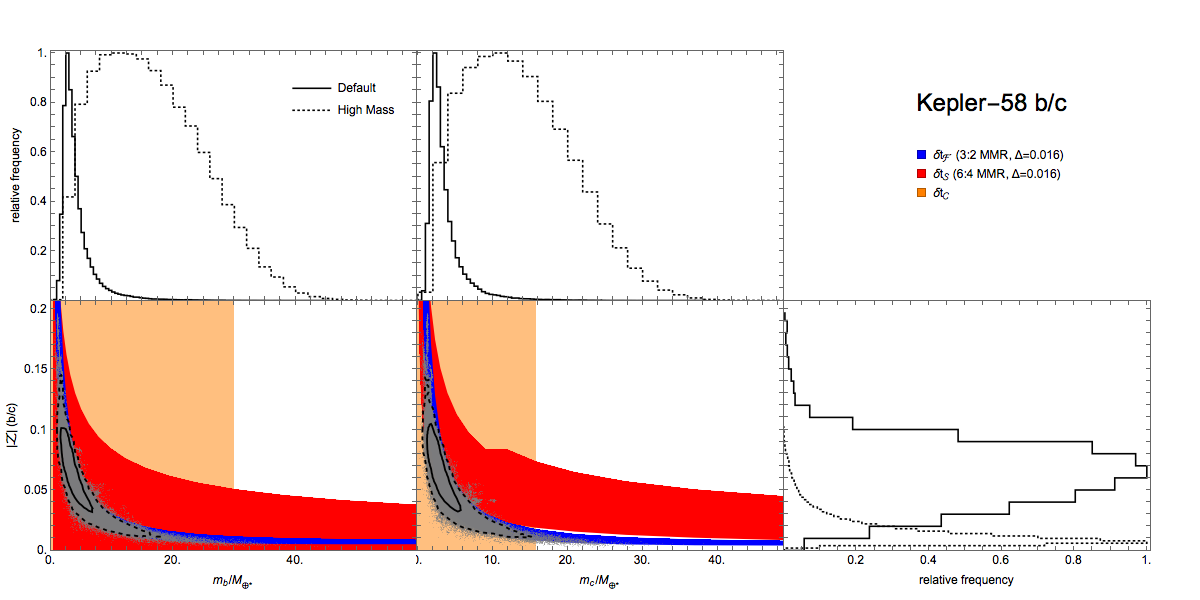}
 \caption{
 Constraint plots for Kepler-58  (see Figure \ref{fig:kep11cons} for description). 
 }
 	\label{fig:kep58cons}
 	\end{center}
 \end{figure}

{\it Kepler-58 b,c} (Fig. \ref{fig:kep58cons}):
Kepler-58 b and c's lack of chopping place upper bounds on the planets' masses that agree well with 
the upper limits inferred from the high mass prior $N$-body MCMC results.

\begin{figure}[htbp]
	\begin{center}
	\includegraphics[width=0.45\textwidth]{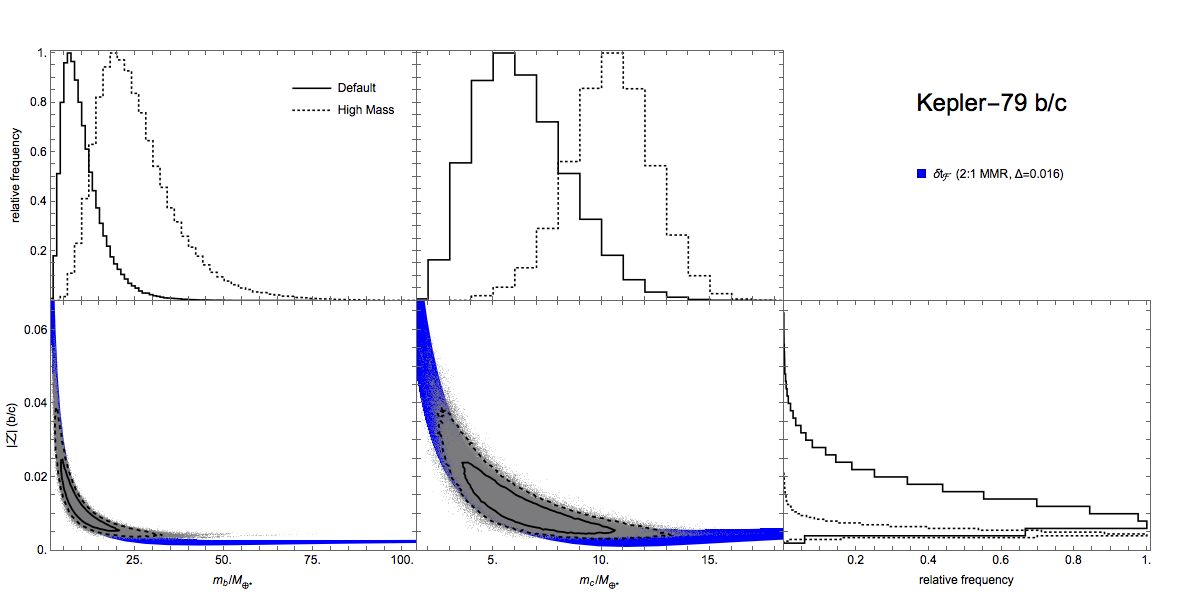}
	\includegraphics[width=0.45\textwidth]{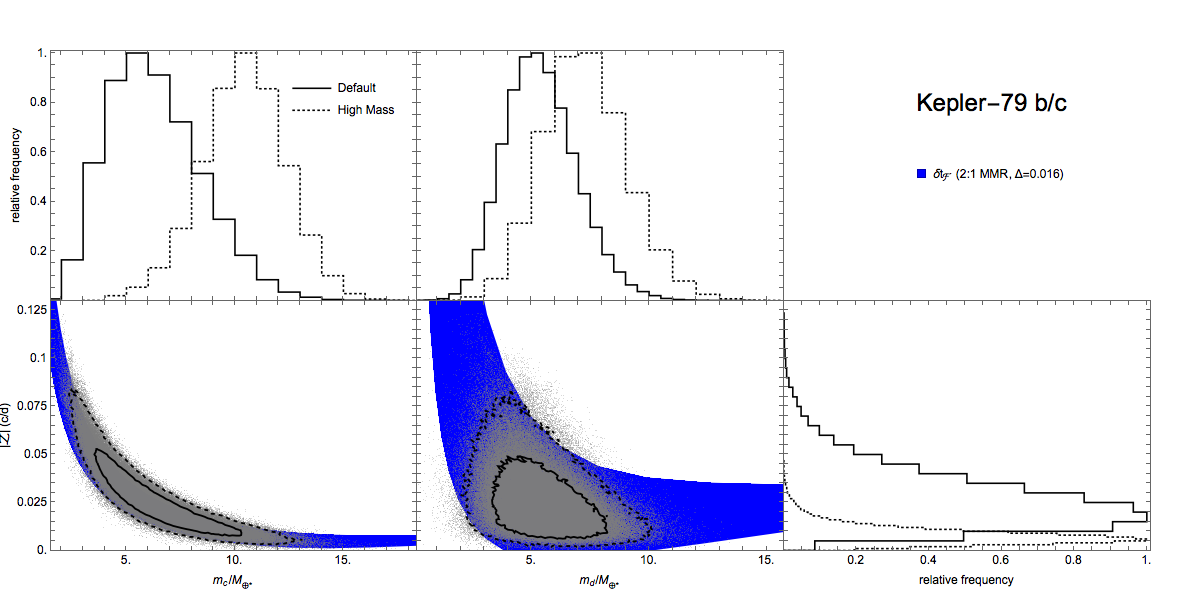}
	\includegraphics[width=0.45\textwidth]{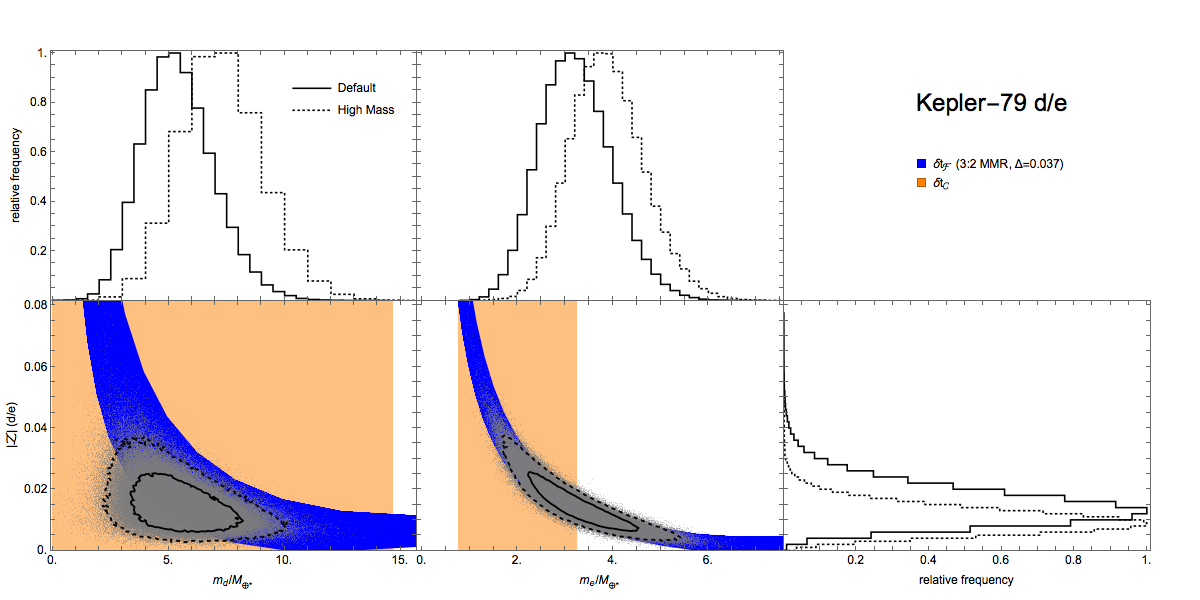}
\caption{
Constraint plots for Kepler-79 (see Figure \ref{fig:kep11cons} for description).}
\label{fig:kep79cons}
\end{center}
\end{figure}

{\it Kepler-79 b,c,d,e*} (Fig. \ref{fig:kep79cons}):
The planets of the Kepler-79 system are near a succession of first order MMRs: 2:1 (b/c), 2:1 (c/d), and 3:2 (d/e).
 Planet e's mass is robustly constrained and shows fair agreement with the analytic chopping constraint.
\citet{2014ApJ...785...15J} also analyze the TTVs of the Kepler-79 system using Quarters 1-14 short-cadence data.
Their results agree with ours for the robustly measured mass of planet e.

\begin{figure}[htbp]
	\begin{center}
	\includegraphics[width=0.45\textwidth]{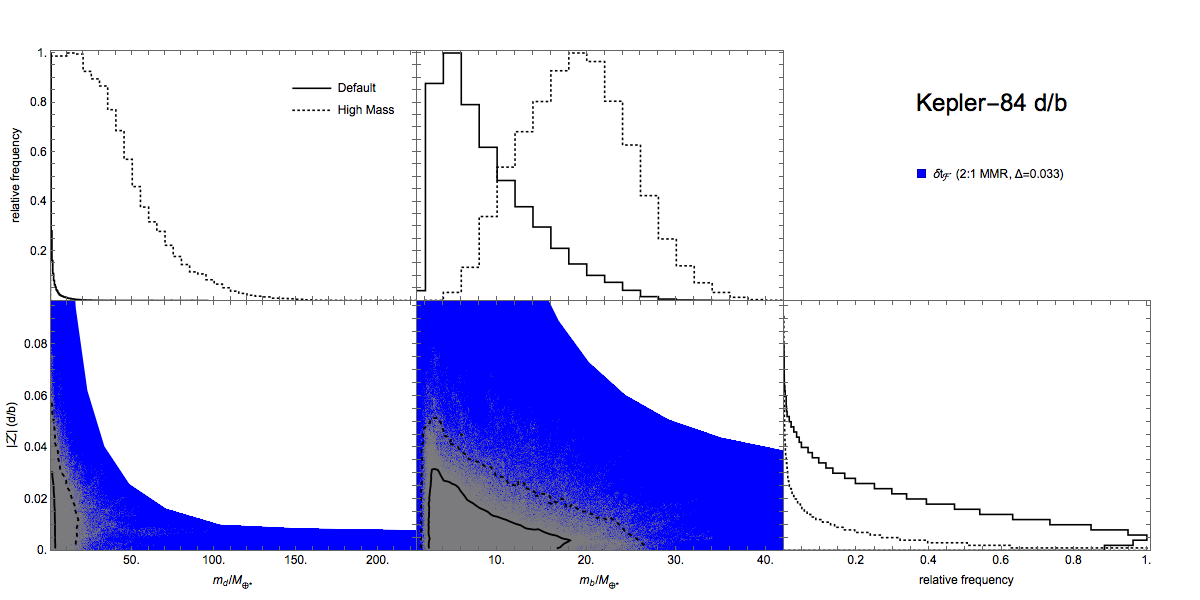}
	\includegraphics[width=0.45\textwidth]{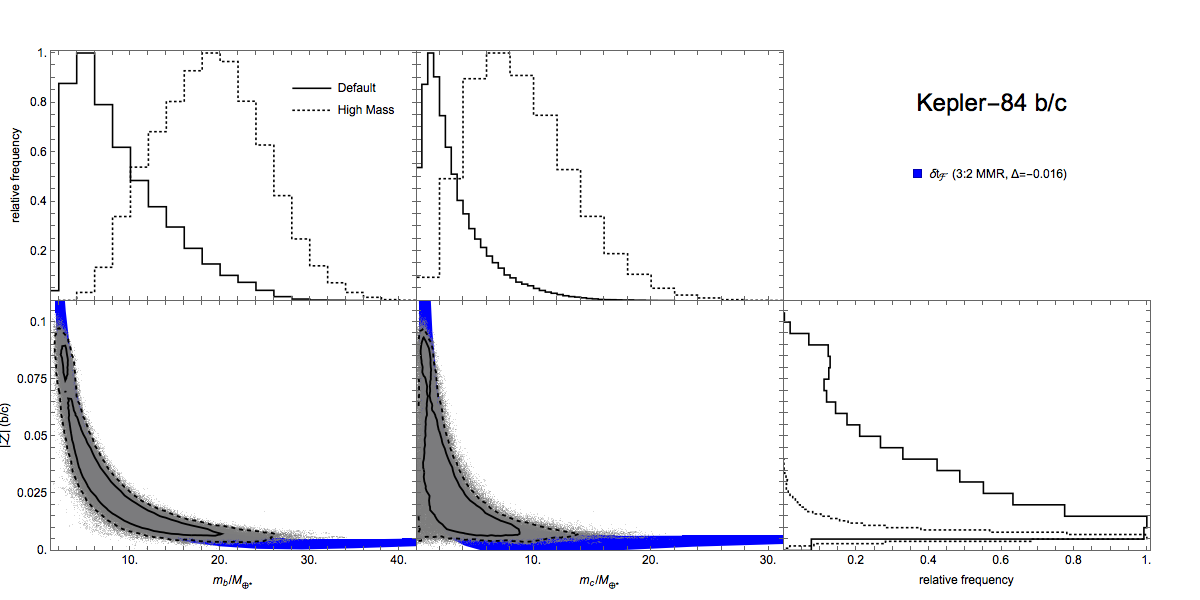}
	\includegraphics[width=0.45\textwidth]{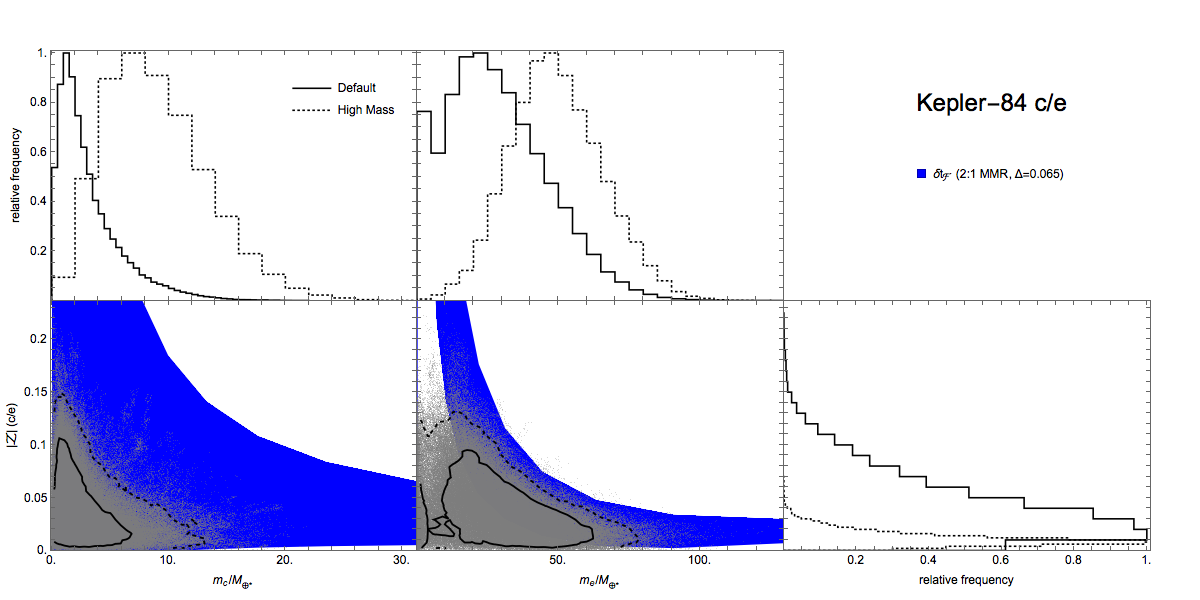}
	\includegraphics[width=0.45\textwidth]{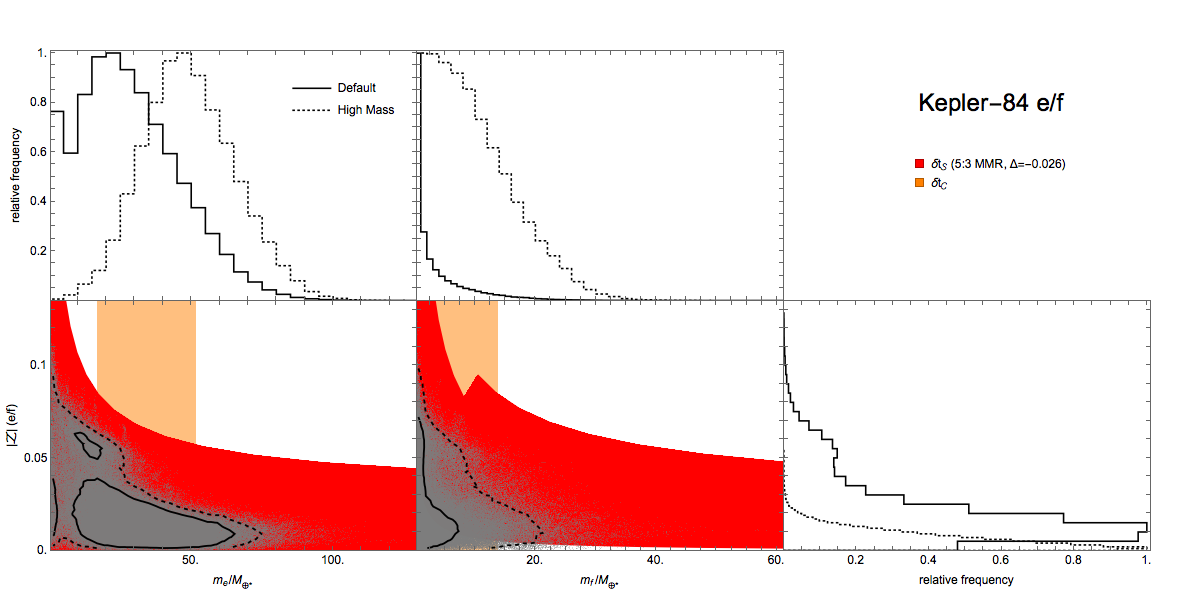}
\caption{
	Constraint plots for Kepler-84  (see Figure \ref{fig:kep11cons} for description).
}
	\label{fig:kep84cons}
	\end{center}
\end{figure}

{\it Kepler-84 d, b, c, e, f} (Fig. \ref{fig:kep84cons}):
Kepler-84 hosts 5 planets near a series of first and second-order MMRs.
Planet f's chopping TTV favors a planet e mass of $m_e\sim 40 \mearth$, though the default prior's weighting towards low masses results in a posterior that is consistent with $m_e=0$.
The other planets suffer strong mass-eccentricity degeneracies or have weak mass upper bounds from the non-detection of TTVs.

\begin{figure}[htbp]
	\begin{center}
	\includegraphics[width=0.45\textwidth]{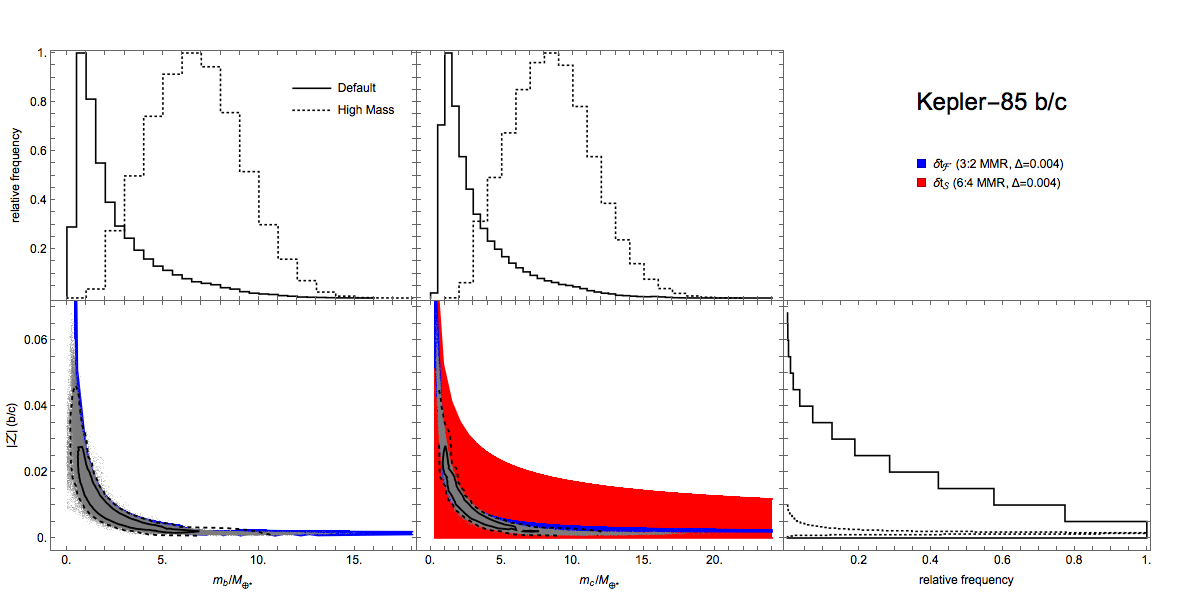}	
	\includegraphics[width=0.45\textwidth]{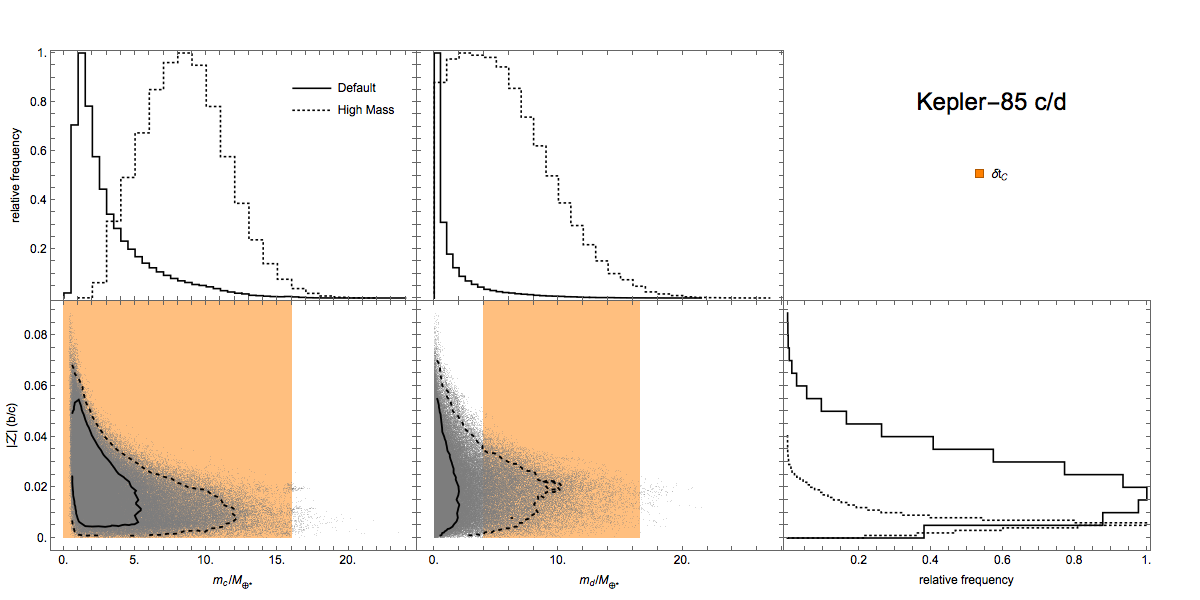}
	\includegraphics[width=0.45\textwidth]{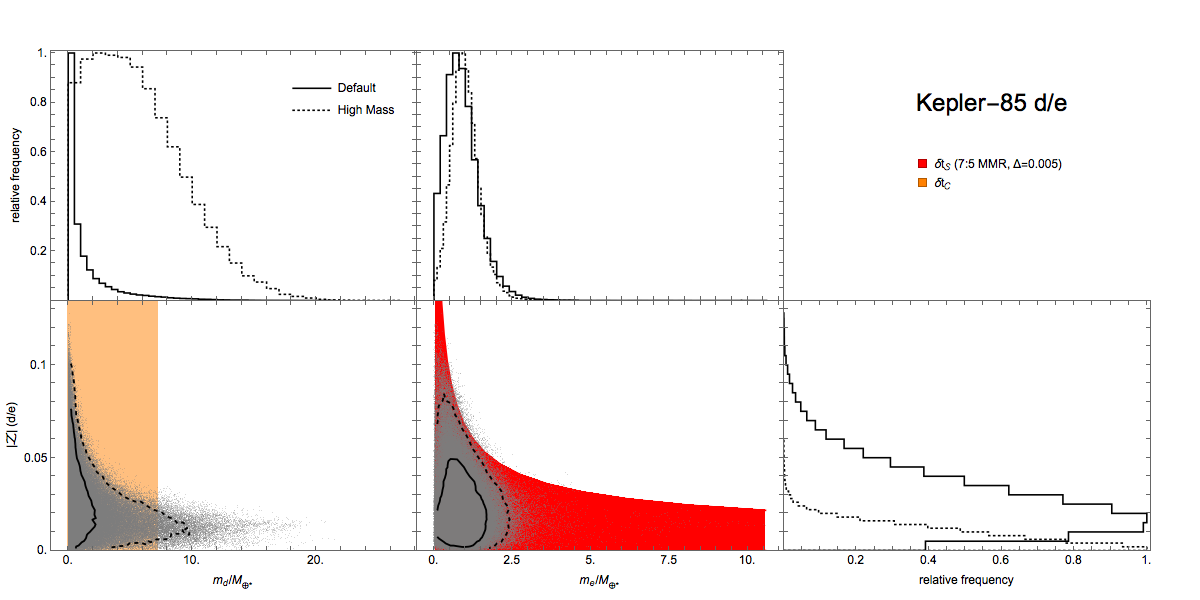}
	\includegraphics[width=0.45\textwidth]{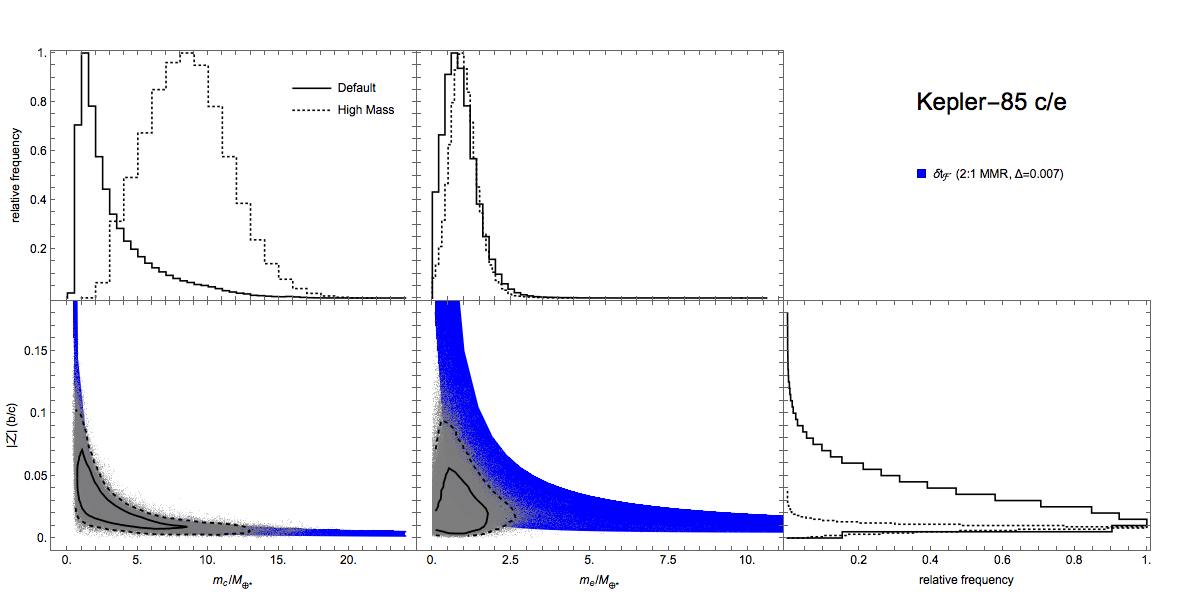}
\caption{
	Constraint plots for Kepler-85 (see Figure \ref{fig:kep11cons} for description).
	The combined eccentricity of the b/c pair has been corrected by subtracting the forced component.
}
	\label{fig:kep85cons}
	\end{center}
\end{figure}

{\it Kepler-85 b, c, d, e*} (Fig. \ref{fig:kep85cons}):
Kepler-85 c, d, and e's lack of strong chopping TTVs place upper bounds on planet c and d's masses. 
Planet b and c have significant fundamental TTVs from their near-3:2 commensurability.
Planet b and c's TTVs are subject to the mass-eccentricity degeneracy.
Planet e's mass is robust based on the MCMC results, though it is not clear from the analytic constraints 
shown in Figure \ref{fig:kep85cons} why this is the case. 
An analytic MCMC fit to Kepler-85's TTVs yields $m_e=2.8^{+1.6}_{-2.0}\mearth$,
consistent with, though more uncertain than, the $N$-body MCMC results.
We are unable to fully account for Kepler-85 e's well-constrained mass with the analytic model.

\begin{figure}[htbp]
	\begin{center}	
	\includegraphics[width=0.9\textwidth]{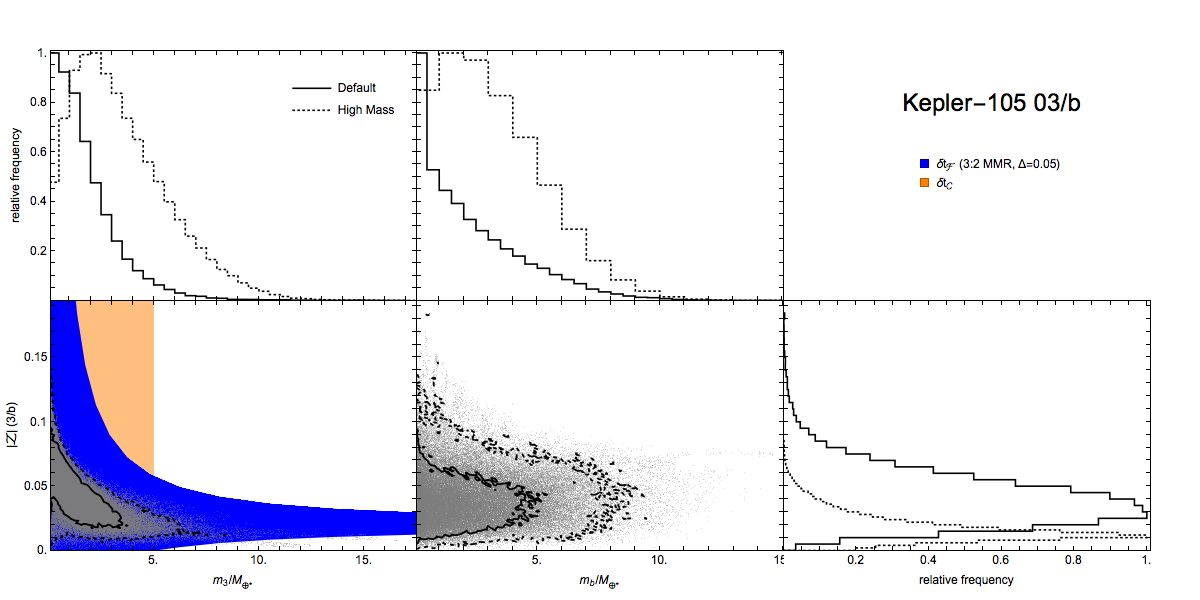}
	\includegraphics[width=0.9\textwidth]{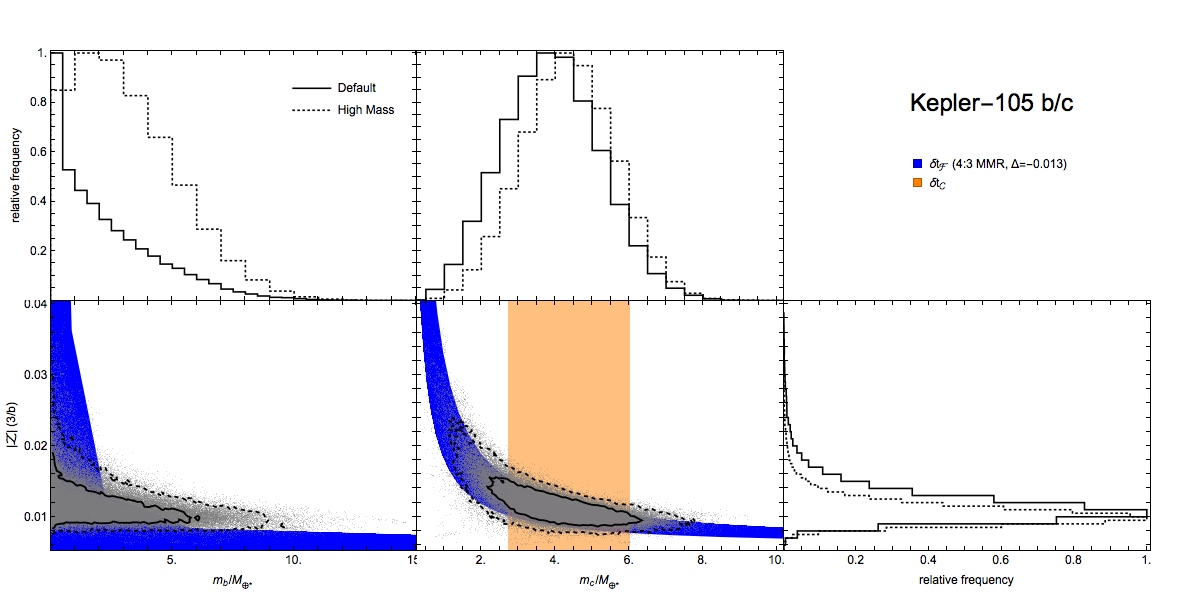}
\caption{
Constraint plot for Kepler-105 (see Figure \ref{fig:kep11cons} for description).
}
	\label{fig:kep105cons}
	\end{center}
\end{figure}

{\it Kepler-105 03,b,c*} (Fig. \ref{fig:kep105cons}):  
Planet c's mass is robustly constrained by chopping.
KOI-115.03 and Kepler-105 b do not induce a significant variations in each other's transit times, 
imposing an upper limit that agrees well with the high mass prior $N$-body results.
\citet{JontofHutter:2016ch} measure  planet c's mass to be $m_c=4.6\pm0.9\mearth$ from an analysis of  Q1-17 transit times,
consistent with our inferred value, $m_c=2.9\pm1.5\mearth$ at roughly 1$\sigma$ confidence.

\begin{figure}[htbp]
	\begin{center}
	\includegraphics[width=0.9\textwidth]{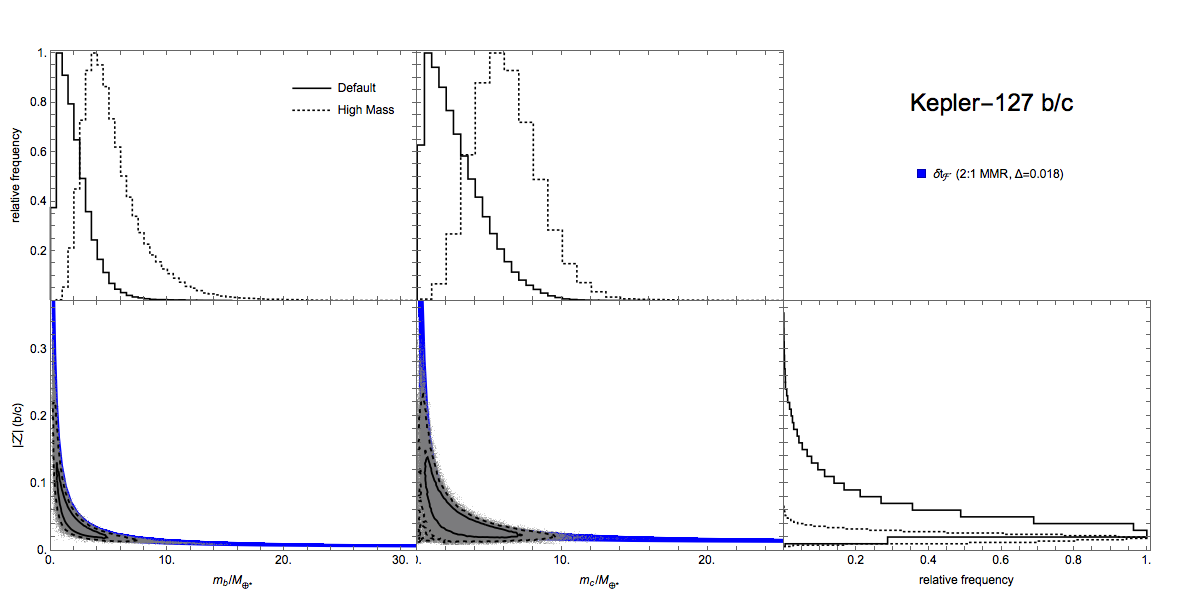}
	\includegraphics[width=0.9\textwidth]{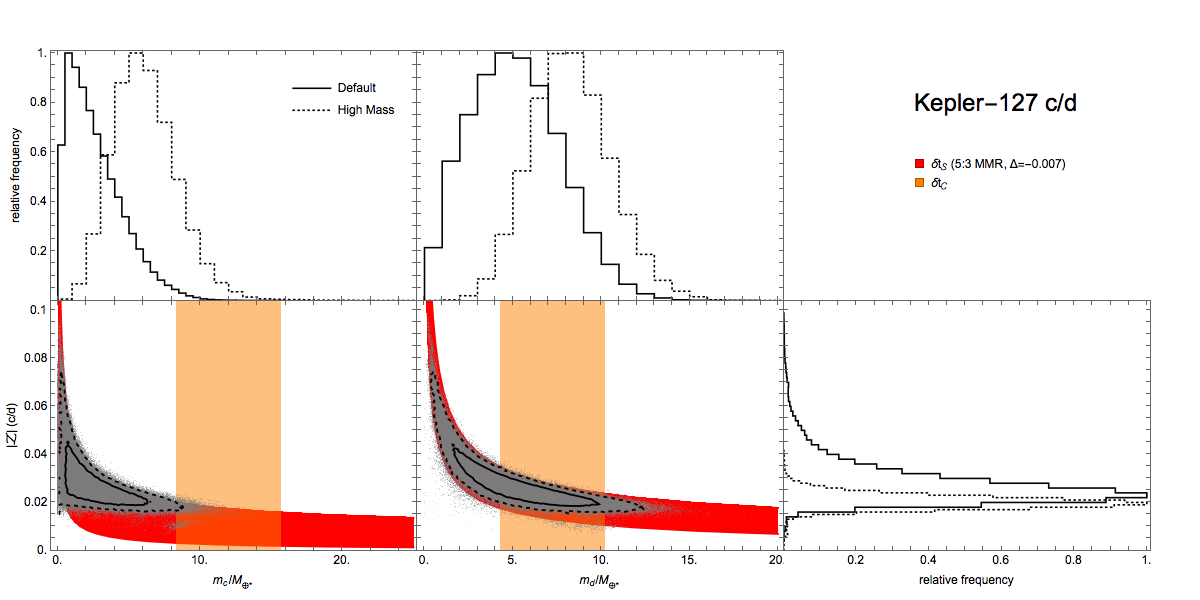}
\caption{
Constraint plots for Kepler-127 (see Figure \ref{fig:kep11cons} for description).
}
	\label{fig:kep127cons}
	\end{center}
\end{figure}

{\it Kepler-127 b, c, d }(Fig. \ref{fig:kep127cons}):
The bottom panel shows that planet d's mass is constrained by chopping.
Consequently, the combined eccentricity of c and d is well constrained by their 5:3 MMR TTVs.
Planet c's mass is inferred to be slightly lower than the chopping constraint, though
its inferred mass is not robust to the choice of priors.
The 2:1 fundamental TTVs of planets b and c suffer a strong mass-eccentricity degeneracy.

\begin{figure}[htbp]
	\begin{center}
	\includegraphics[width=0.9\textwidth]{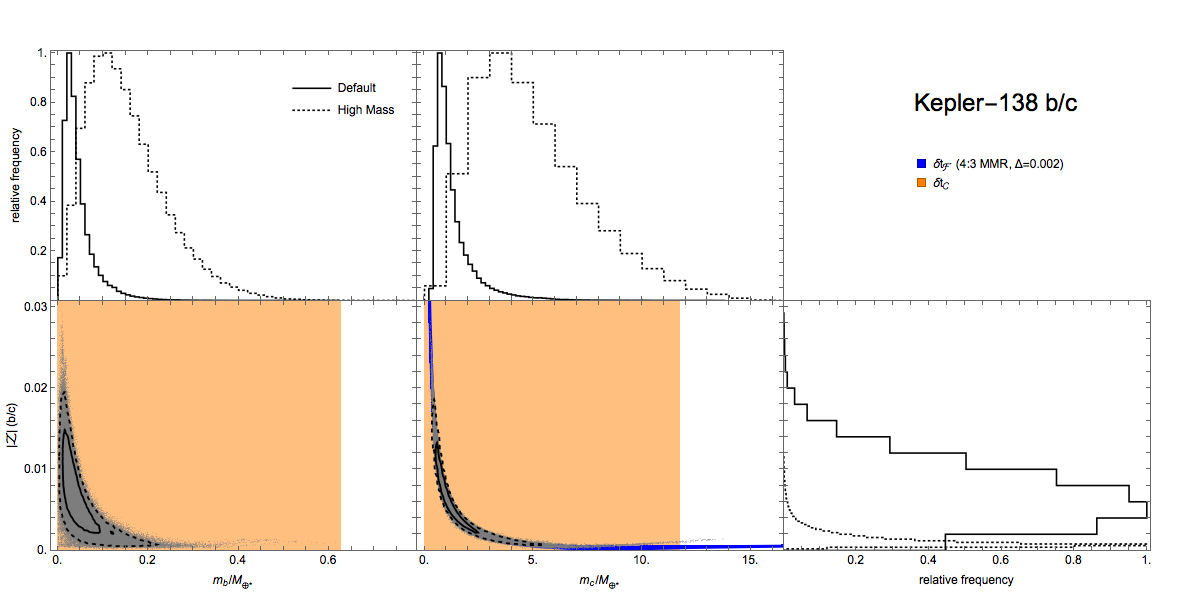}
	\includegraphics[width=0.9\textwidth]{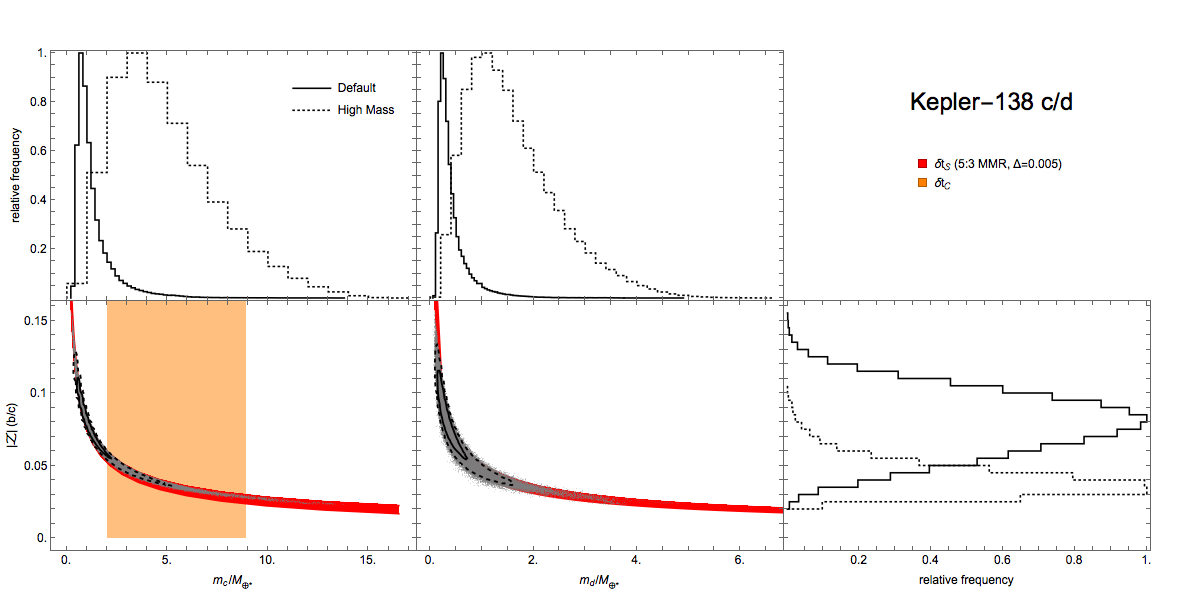}
\caption{
Constraint plots for  Kepler-138  (see Figure \ref{fig:kep11cons} for description).
The combined eccentricity of the b/c pair has been corrected by subtracting the forced component.
}
	\label{fig:kep138cons}
	\end{center}
\end{figure}

{\it Kepler-138 b, c, d} (Fig. \ref{fig:kep138cons}):
Planet d's chopping and second-order resonance TTV together constrain the mass of planet c and the pair's combined eccentricity, though
the chopping TTV amplitude has a large uncertainty, with $1 \mearth \lesssim m_c\lesssim 8 \mearth$ at 1$\sigma$ confidence. 
Consequently, the inferred masses are quite sensitive to the assumed priors. 

\citet{2015Natur.522..321J} measure masses $[m_b,m_c,m_d]=[0.13^{+0.12}_{-0.08},3.85^{+3.77}_{-2.30},1.28^{+1.36}_{-0.78}]\mearth$ from MCMC analysis of transit data up to Quarter 14. 
Our default priors result in lower planet masses (Table \ref{tab:mass}) than inferred by \citet{2015Natur.522..321J}, who adopt priors that are uniform in planet masses and eccentricities.
Our high mass priors give close agreement with the masses found by \citet{2015Natur.522..321J}.

\begin{figure}[htbp]
	\begin{center}
	\includegraphics[width=0.9\textwidth]{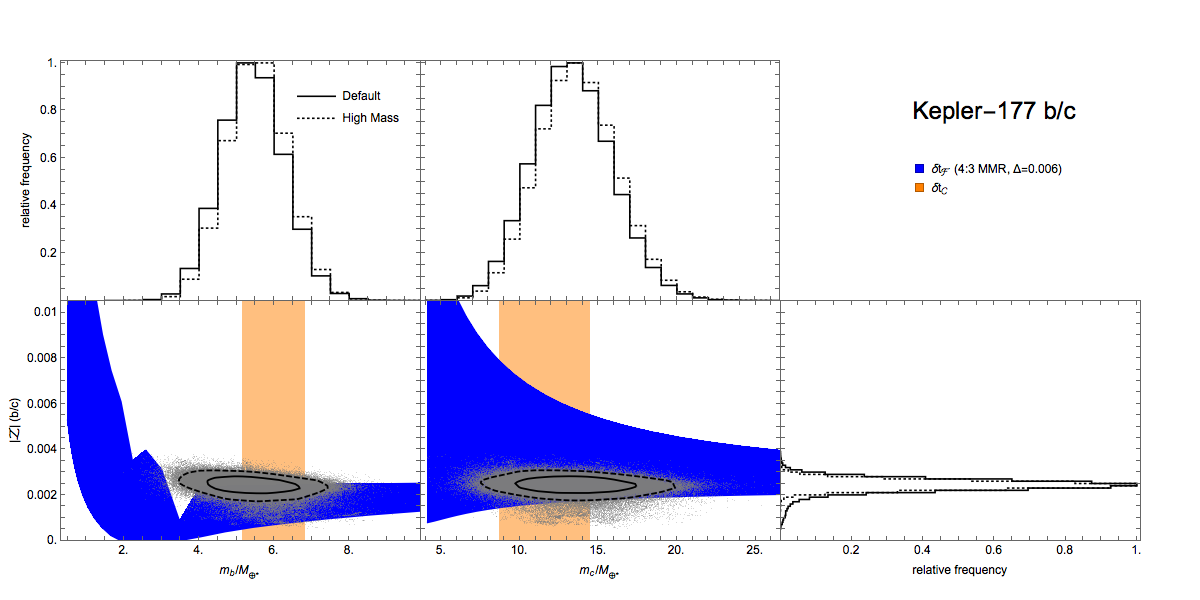}
\caption{
Constraint plots for Kepler-177 (see Figure \ref{fig:kep11cons} for description).
The combined eccentricity has been corrected by subtracting the forced component.
}
	\label{fig:kep177cons}
	\end{center}
\end{figure}

{\it Kepler-177 b*,c*} (Fig. \ref{fig:kep177cons}):
Kepler-177 b and c's mutual fundamental and chopping TTVs constrain the masses and combined eccentricity of the planet pair.
\citet{JontofHutter:2016ch} analyze the TTVs of Kepler-177 b and c and find planet masses $[m_b,m_c]=[5.7\pm0.8,14.6^{+2.7}_{-2.5}]\mearth$,
which agree well with our results: $[m_b,m_c]=[5.4\pm0.8,13.3^{+2.4}_{-2.7}]\mearth$ .
\citet{JontofHutter:2016ch} preform a second fit using an alternate set of transit times 
and find masses that are larger by more than 1$\sigma$ for both planets.

\begin{figure}[htbp]
	\begin{center}
	\includegraphics[width=0.9\textwidth]{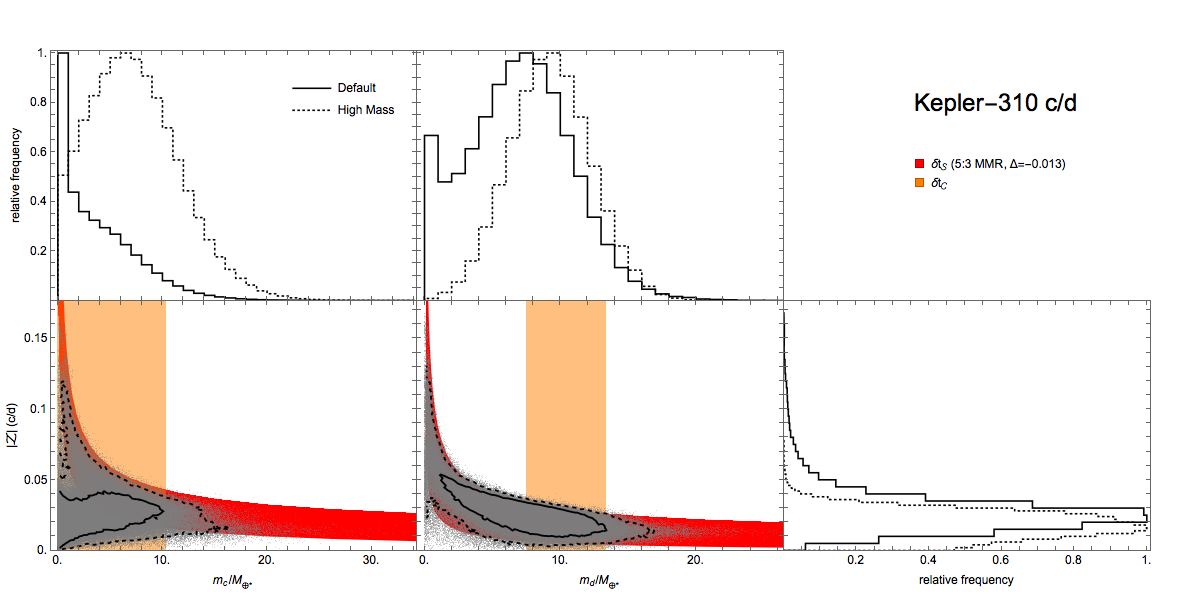}
\caption{
Constraint plot for Kepler-310  (see Figure \ref{fig:kep11cons} for description).
}
\label{fig:kep310cons}
	\end{center}
\end{figure}

{\it Kepler-310 c,d*} (Figure \ref{fig:kep310cons}):
Planet c's chopping and second-order resonance TTVs combine to constrain the mass of planet d and the pair's combined eccentricity. 
The non-detection of chopping in planet d's TTV gives an upper limit on planet c's mass that agrees well with the $N$-body results.

\begin{figure}[htbp]
	\begin{center}
	\includegraphics[width=0.9\textwidth]{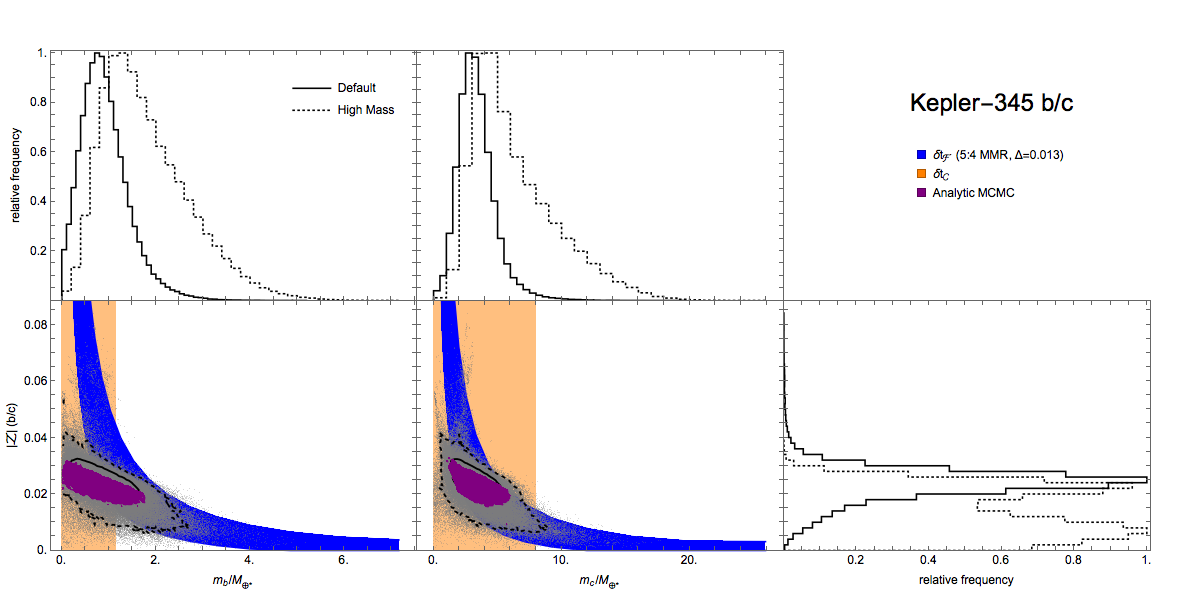}
\caption{
Constraint plots for Kepler-345 (see Figure \ref{fig:kep11cons} for description).
The analytic MCMC includes the effects of the fourth-order 19:15 MMR
in addition to the fundamental, second harmonic and chopping components.
}
	\label{fig:kep345cons}
	\end{center}
\end{figure}

{\it  Kepler-345 b*,c*} (Fig. \ref{fig:kep345cons}):
The absence of chopping in Kepler-345 b and c's TTVs places upper limits on their masses. 
Kepler-345 b and c are very near the fourth-order 19:15 MMR, with $|15P_c/19P_b - 1| = 6\times10^{-4}$.
We fit Kepler-345 b and c's TTVs with an analytic MCMC fit that, in addition to the fundamental, second-harmonic,
and chopping TTVs, includes the effects of the 19:15 MMR on each planet's TTV.\footnote{Terms in the analytic TTV formulae accounting for the fourth-order 19:15 MMR are derived by a straightforward extension of the derivation in Paper 1 \citep[see also][]{Deck:2015tx}.
}
By comparing the results of analytic MCMC fits with and without the effects of 19:15 MMR terms included, we find that these terms are 
crucial for constraining the maximum value of the planet pairs' combined eccentricity.

\begin{figure}[htbp]
	\begin{center}
	\includegraphics[width=0.9\textwidth]{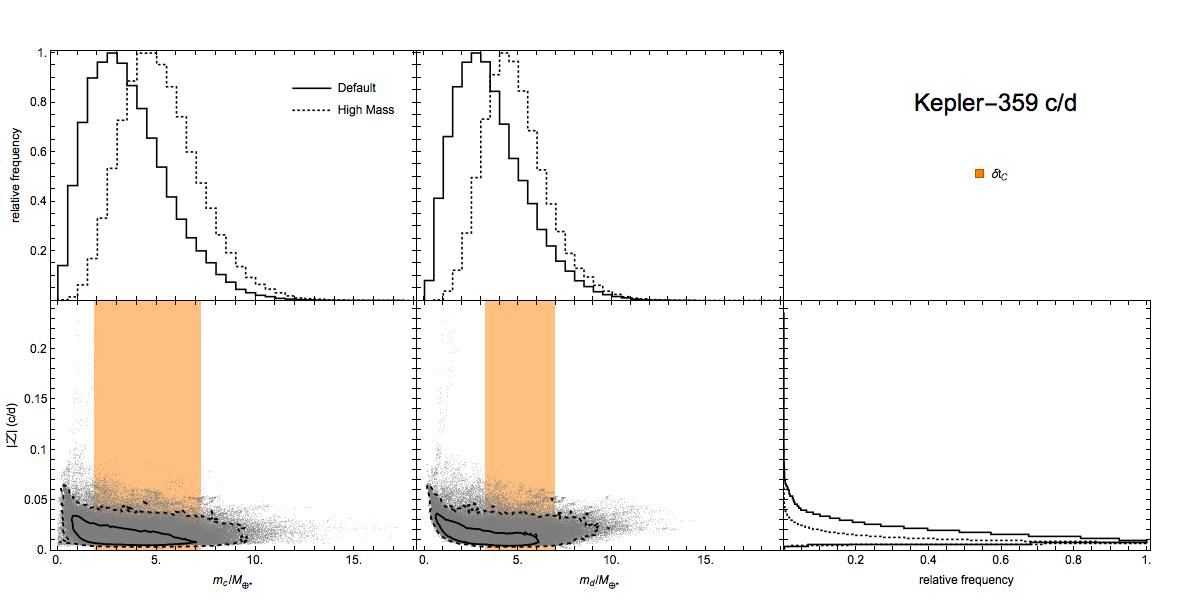}
\caption{
Constraint plots for Kepler-359 (see Figure \ref{fig:kep11cons} for description).
}
	\label{fig:kep359cons}
	\end{center}
\end{figure}

{\it Kepler-359 c*,d*} (Fig. \ref{fig:kep359cons}):
Kepler-359 c and d's mutual chopping TTVs constrain the both planets' masses.
The best-fit periods place the planet pair quite near the 4:3 MMR ($\Delta=-0.002$).
Because of Kepler-359 c and d's proximity to resonance, 
their transit time observations cover a relatively short portion of the super-period of their fundamental TTVs.
As a result, there is a strong covariance between the planets' precise periods and their fundamental TTV amplitudes which
complicates the application of our analytic formulae.
Constraints from the fundamental TTV have been omitted in Figure \ref{fig:kep359cons}.

\begin{figure}[htbp]
	\begin{center}
	\includegraphics[width=0.45\textwidth]{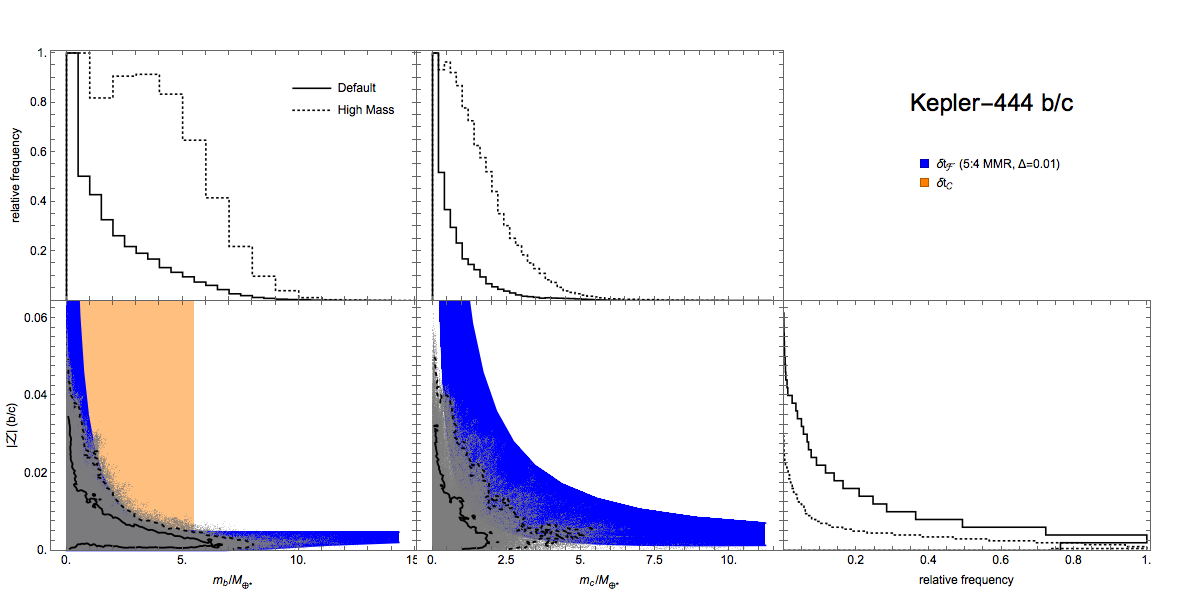}
	\includegraphics[width=0.45\textwidth]{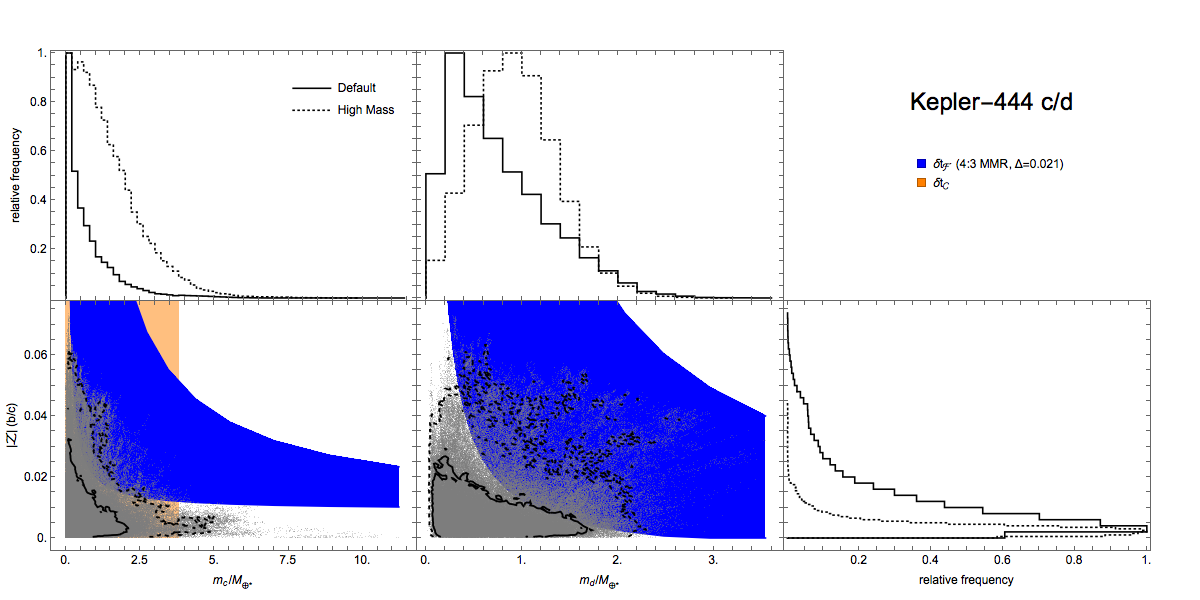}
	\includegraphics[width=0.45\textwidth]{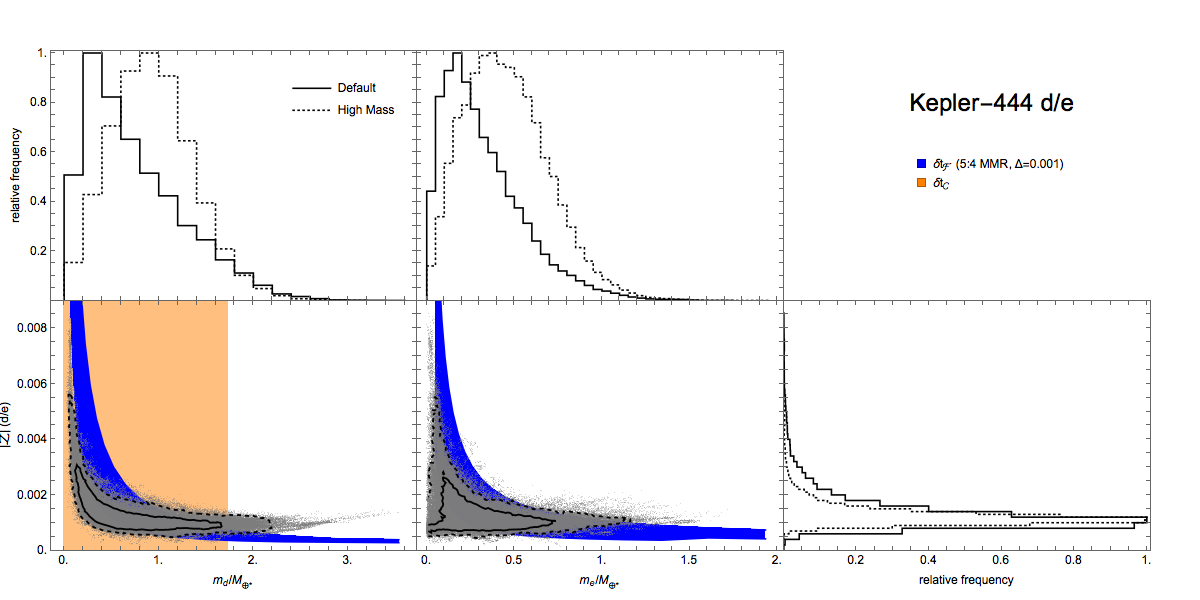}
	\includegraphics[width=0.45\textwidth]{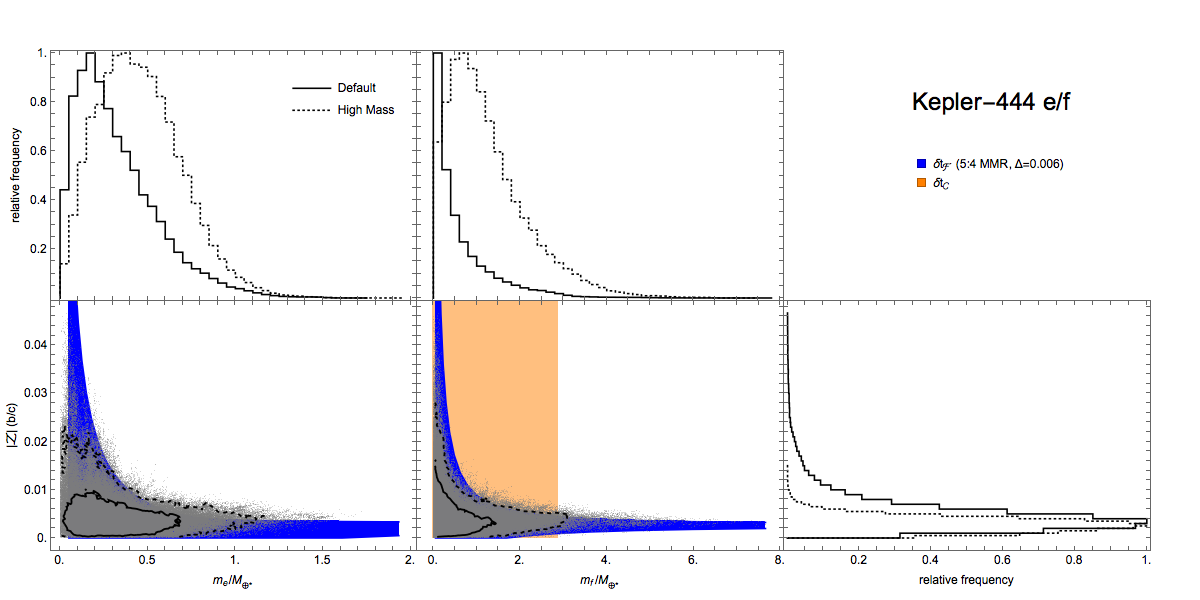}
\caption{
Constraint plots for Kepler-444 (see Figure \ref{fig:kep11cons} for description).
The combined eccentricities of the d/e and e/f pairs habe been corrected by subtracting the forced component.
}
	\label{fig:kep444cons}
	\end{center}
\end{figure}

{\it  Kepler-444 b,c,d*,e*,f} (Fig. \ref{fig:kep444cons}): 
Kepler-444 is a compact five-planet system with each adjacent pair of planets near a first order mean motion resonance.	 
\citet{Campante:2015ei} validate the planetary nature of the system and measure a precise stellar age of $11.2\pm1.0$ Gyr via astroseismology.
Planets b/c are near a 5:4 MMR, c/d are near a 4:3 MMR, d/e are near or in a 5:4 MMR, and e/f are near a 5:4 MMR.
None of planets' TTVs show large variations.
The planets' lack of chopping TTVs place stringent upper limits on their masses.
The masses of planets d and e are robust to the choice of priors. 
Their masses and combined eccentricity are consistent with the fundamental and chopping constraints,
though no upper bound on their eccentricity is apparent from the analytic constraints.
The planet pair is very near resonance ($\Delta=0.001$), where the assumptions of analytic model start to break down.
Their extreme proximity to resonance likely plays a role in limiting the maximum combined eccentricity consistent with their TTVs.

\begin{figure}[htbp]
	\begin{center}
	\includegraphics[width=0.9\textwidth]{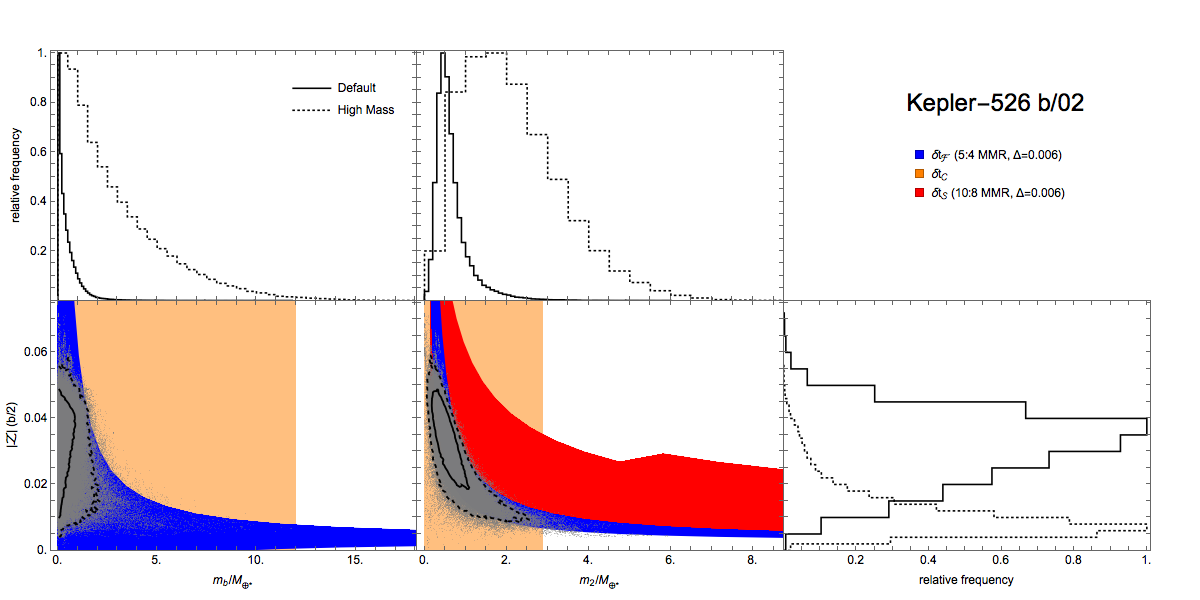}
\caption{
Constraint plots for Kepler-526 (see Figure \ref{fig:kep11cons} for description).
}
\label{fig:kep526cons}
	\end{center}
\end{figure}

 {\it Kepler-526 b, 02} (Fig. \ref{fig:kep526cons}):
Kepler-526 b and 02 are near a 5:4 MMR.
Planet b does not induce any detectable TTV in Kepler-526.02.
Lack of chopping TTVs provides upper limits on the masses of Kepler-526 b and 02.

\begin{figure}[htbp]
	\begin{center}
	\includegraphics[width=0.9\textwidth]{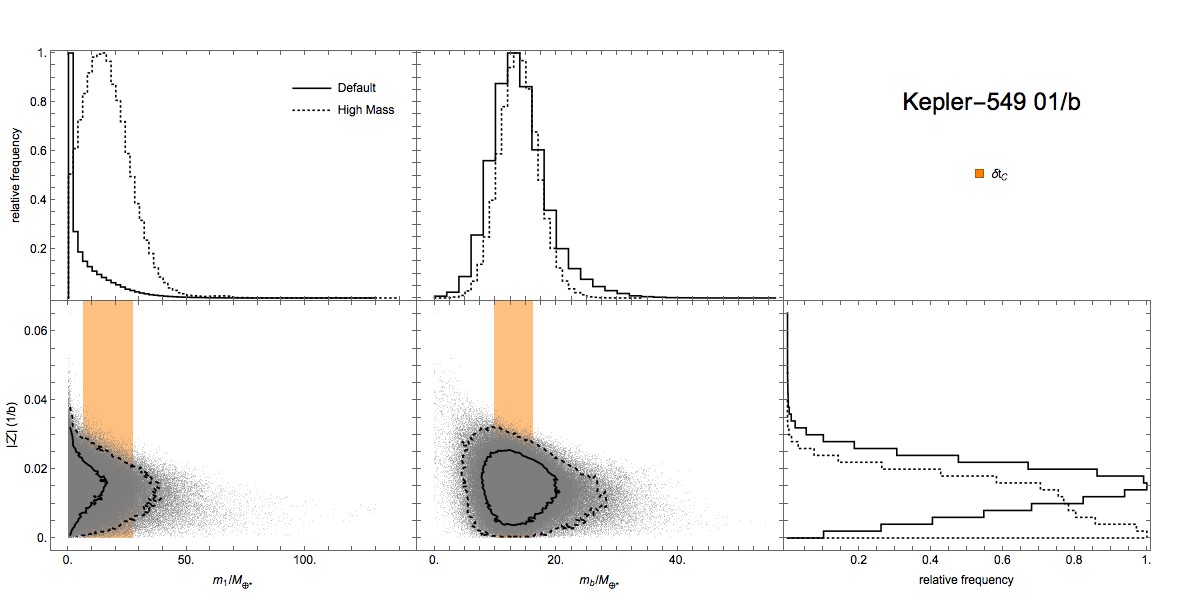}
\caption{
Constraint plots for Kepler-549 01/b (see Figure \ref{fig:kep11cons} for description).
}
	\label{fig:koi427cons}
	\end{center}
\end{figure}

{\it Kepler-549 01,b*} (Fig. \ref{fig:koi427cons}):
The masses of Kepler-549.01 and b are constrained by their mutual chopping components, especially planet b's mass.
The planets lack of TTV from the nearby 5:3 MMR ($\Delta = 0.047$) places a loose upper bound on their combined eccentricity (not shown in Figure \ref{fig:koi427cons}).
The $N$-body posterior shown in Figure \ref{fig:koi427cons} is confined to significantly smaller combined eccentricities
 than the upper bound derived from the lack of a 5:3 MMR signal.
 Kepler-549 01/b are very near the third-order 7:4 MMR ($\Delta = -0.003$) and this likely plays a role in further constraining the eccentricities of the planet pair.

\subsection{\ref{sec:SecondHarmonic}: Second Harmomic Systems}
\label{sec:SecondHarmonic}

\begin{figure}[htbp]
 	\begin{center}
 	\includegraphics[width=0.9\textwidth]{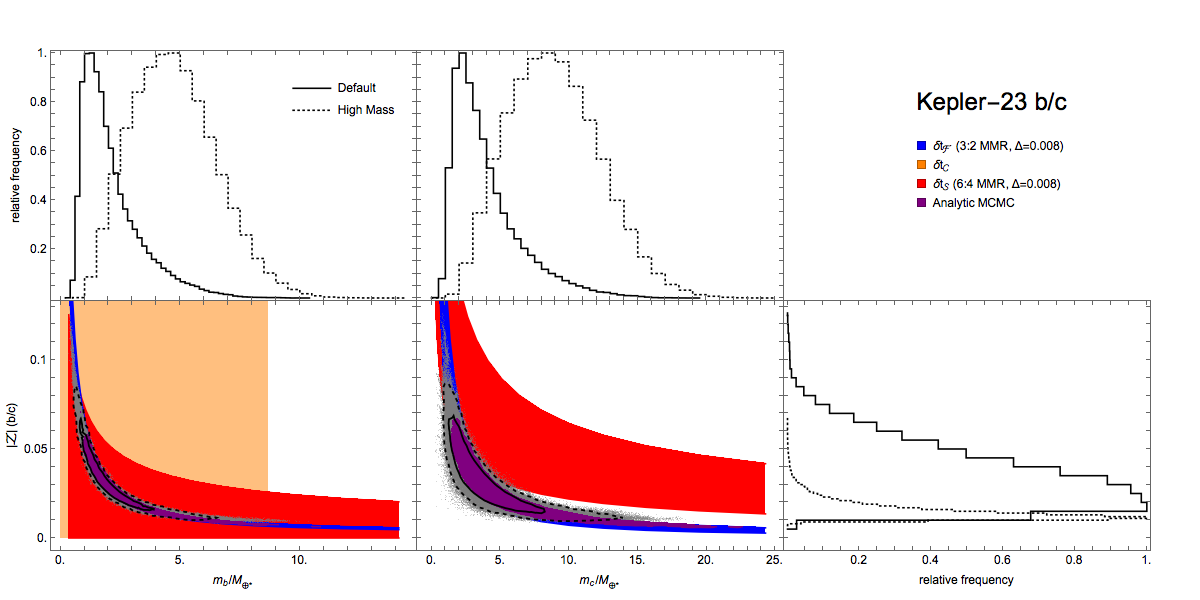}
     \includegraphics[width=0.9\textwidth]{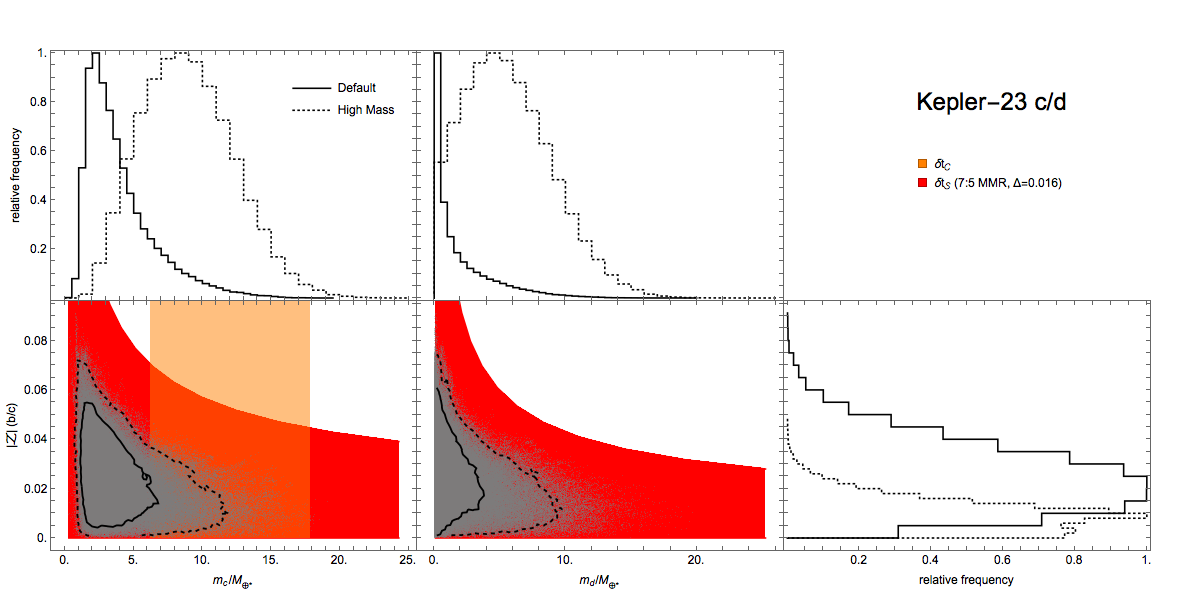}
 \caption{
 Constraint plots for Kepler-23 b, c and d (see Figure \ref{fig:kep11cons} for description).
 The analytic MCMC results shown for Kepler-23 b and c are fit to the fundamental and second-harmonic TTVs of 
 planet b and c, ignoring planet d and any chopping.
 }
 	\label{fig:kep23cons}
 	\end{center}
 \end{figure}

{\it Kepler-23 b,c,d} (Fig. \ref{fig:kep23cons}):  
Kepler-23 b and c are near the 3:2 MMR and have strong fundamental TTVs.
The constraints from their mutual interactions are summarized in the top panels of Figure \ref{fig:kep23cons}.
The planet's second-harmonic TTV constraints appear to be in tension with each other: the constraint in the right panel (planet b's TTV) favors large $|{\cal Z}|$ values that are ruled out by the left panel constraint (planet c's TTV).
The right-hand panel second-harmonic TTV constraint is consistent with 0 at the 1.5$\sigma$ level.
We use an analytic MCMC fit to further investigate the discrepancy between the two constraints. 
The analytic MCMC combines the constraints of both planets TTVs simultaneously so that  
the stronger of the two conflicting second-harmonic TTV constraints will dictate the inferred solution.
The analytic MCMC results in Figure \ref{fig:kep23cons} demonstrate the combination of
both planets' TTV constraints gives good agreement with the $N$-body results.

The constraints from planet c and d's mutual TTVs are summarized in the bottom panels of Figure \ref{fig:kep23cons}.
Planet c and d are near the 7:5 MMR  but do not induce significant second-order resonance TTVs in each other.
Planet c's mass inferred with the default prior MCMC is somewhat smaller than the mass inferred from the chopping constraint, indicating that the 
higher prior probability assigned to lower masses by the default prior outweighs the likelihood contribution of the chopping TTV.

\begin{figure}[htbp]
	\begin{center}
	\includegraphics[width=0.9\textwidth]{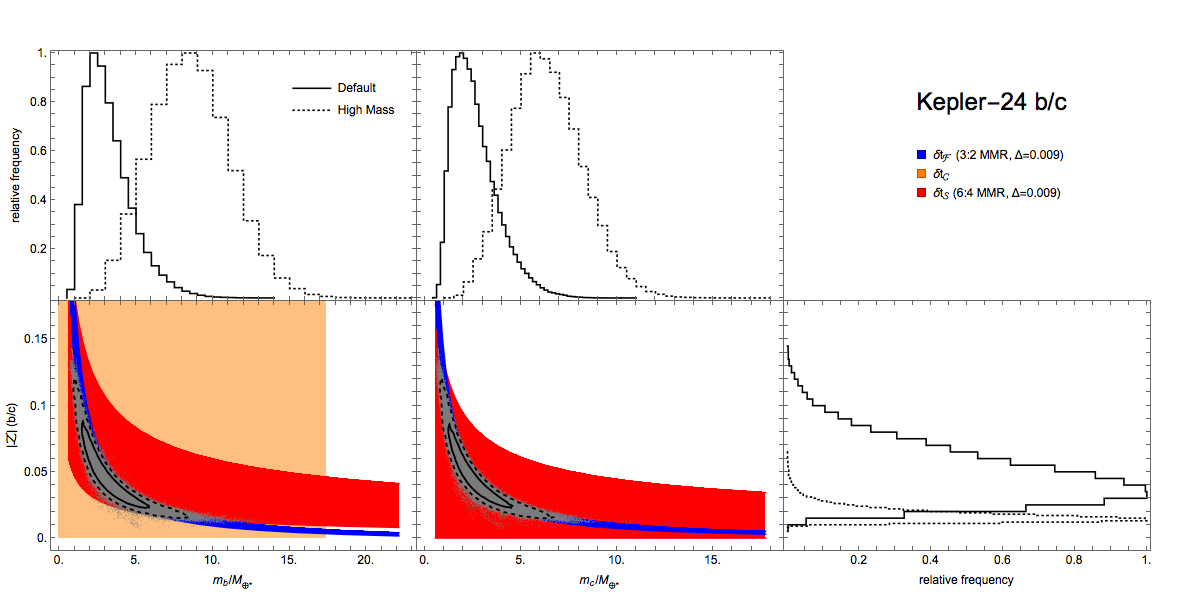}
	\includegraphics[width=0.9\textwidth]{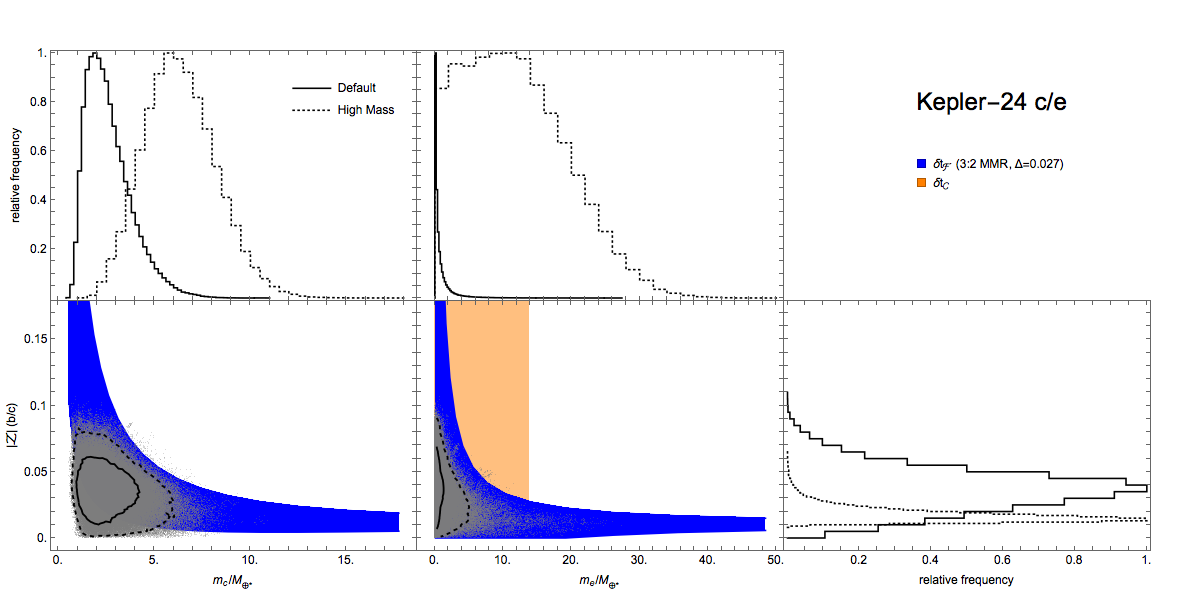}
\caption{
Constraint plots for the Kepler-24 system (see Figure \ref{fig:kep11cons} for description).
}
	\label{fig:kep24cons}
	\end{center}
\end{figure}

{\it Kepler-24 b,c,e} (Fig. \ref{fig:kep24cons}):
 Constraints from planet b and c's interactions are summarized in the top panels of Figure \ref{fig:kep24cons}.
The second-harmonic TTV constraint in the bottom left panel (from planet c's TTV) sets a lower limit on $|{\cal Z}|$.
The planet masses are not strongly constrained and their inferred masses are sensitive to the assumed priors. 
Planet e does not induce any detectable TTV in planet c and the interactions of planet c and e are not very constraining. 

Kepler-24 hosts an additional planet, d, with a radius of 1.7$R_\oplus$ on a 4.2 day orbit somewhat near the 2:1 MMR with planet c.
Planet d shows no significant TTV.  
We conducted $N$-body MCMC fits including planet d and find no significant difference in the posterior distributions of planet b, c and e's parameters.

\begin{figure}[htbp]
 	\begin{center}
     \includegraphics[width=0.9\textwidth]{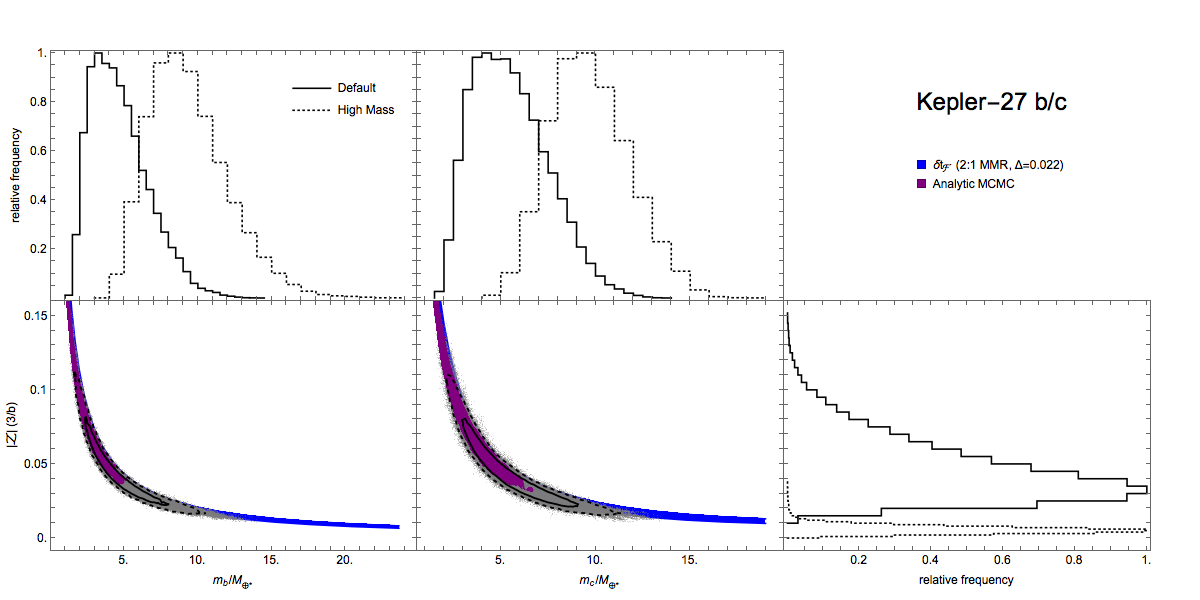}
 \caption{
 Constraint plots for Kepler-27 b and c (see Figure \ref{fig:kep11cons} for description). 
 The purple region shows the 68\% confidence regions derived from an MCMC simulation using the analytic model.
 }
 	\label{fig:kep27cons}
 	\end{center}
 \end{figure}

 {\it Kepler-27 03,b,c} (Fig. \ref{fig:kep27cons}):
In addition to large fundamental TTVs, planets b and c have non-zero second harmonic TTVs at  1$\sigma$ significance. 
The constraints from the second-harmonic TTV cannot easily be plotted in the $\mu-|{\cal Z}|$ plane because of the indirect terms in the analytic formulae for planets near the 2:1 MMR (see Paper I). Instead, we illustrate the contribution of the second-harmonic TTV to constraining planet parameters
by including in Figure \ref{fig:kep27cons} the results of an MCMC fit using the analytic formulae.
The analytic MCMC shows fair agreement, though it favors slightly larger eccentricities and smaller masses than the full $N$-body fit. 
The Kepler-27 system hosts an additional (candidate) planet, KOI-0841.03 interior to planets b and c that does not lie near any low order resonances or exhibit any significant TTVs. We include KOI-0841.03 in our $N$-body MCMC fits but find that its properties are largely unconstrained.
 
\begin{figure}[htbp]
\begin{center}
	\includegraphics[width=0.9\textwidth]{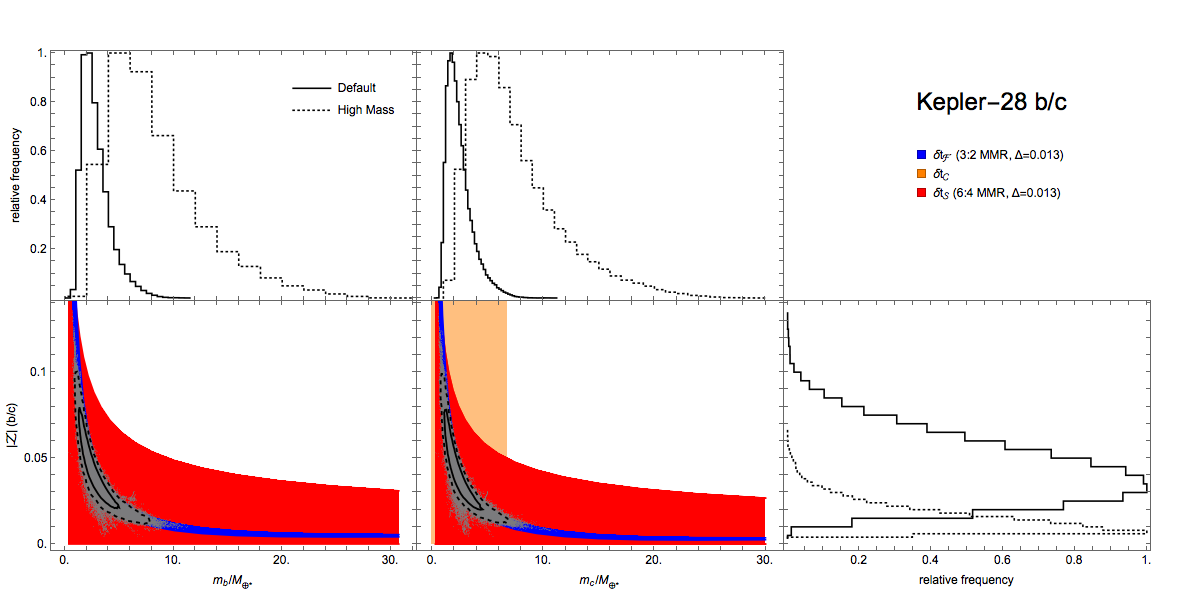}
\caption{Constraint plots for the Kepler-28 system (see Figure \ref{fig:kep11cons} for description).}
	\label{fig:kep28cons}
	\end{center}
\end{figure}

{\it Kepler-28 b, c} (Fig. \ref{fig:kep28cons}):
Neither Kepler-28 b nor c has a significant second-harmonic TTV. 
The absence of second-harmonic TTVs imposes an upper bound of $|{\cal Z}|\lesssim 0.1$ on their combined eccentricity.
A lack of chopping places a 1$\sigma$ upper bound $m_c\lesssim7\mearth$ on the mass of planet c.

\begin{figure}[htbp]
	\begin{center}
	\includegraphics[width=0.9\textwidth]{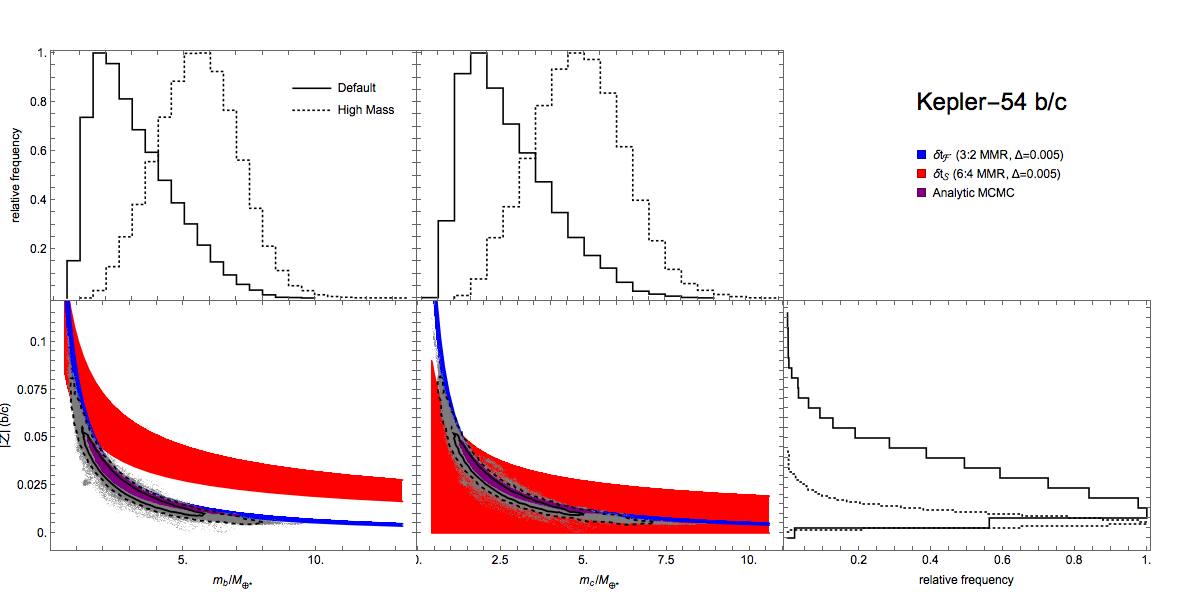}
\caption{
Comparison of MCMC priors for the Kepler-54 system (see Figure \ref{fig:kep11cons} for description). 
The purple regions indicate the 68\% credible region derived from an MCMC simulation with the analytic model.
}
\label{fig:kep54cons}
\end{center}
\end{figure}

 {\it Kepler-54 b, c} (Fig. \ref{fig:kep54cons}):
	Kepler-54 b and c show strong fundamental TTVs caused by their proximity to the 3:2 MMR.
	Figure \ref{fig:kep54cons} indicates that the two planet's second harmonic TTVs  give conflicting constraints.
	To determine which of the conflicting constraints
	more strongly influences the likelihood of inferred planet parameters we fit the TTVs of both planets simultaneously with an analytic MCMC.
	Analytic MCMC results, shown in purple in Figure \ref{fig:kep54cons}, 
	demonstrate that simultaneous fitting of both TTVs gives good agreement with the $N$-body results.	
		
\begin{figure}[htbp]
	\begin{center}
    \includegraphics[width=0.9\textwidth]{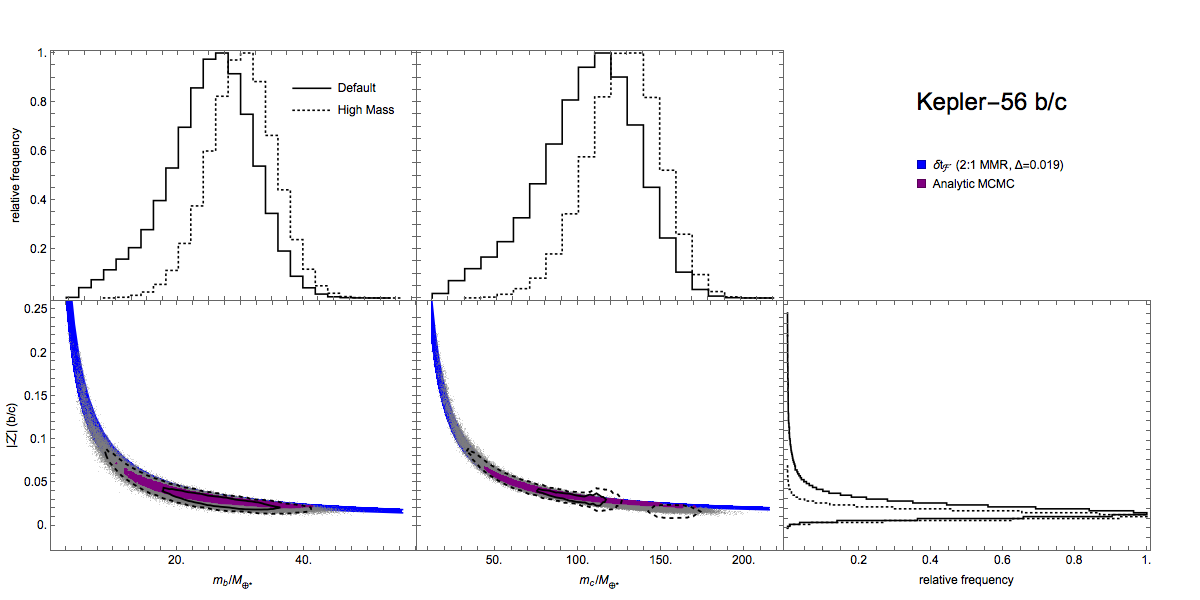}
\caption{
Constraint plots for Kepler-56 (see Figure \ref{fig:kep11cons} for description). 
The purple regions indicate 68\% credible regions derived from an MCMC simulation with the analytic model.
}
\label{fig:kep56cons}
\end{center}
\end{figure}

{\it Kepler-56 b*,c*} (Figure \ref{fig:kep56cons}):
Kepler 56 b and c are near a 2:1 MMR.
Planet b's second-harmonic TTV amplitude is non-zero at $>2\sigma$ significance.
As with Kepler-27 above, we include results from an analytic MCMC fit to illustrate the constraints contributed by the second-harmonic TTV.
The analytic MCMC shows good agreement with the $N$-body results.

\citet{Huber:2013ha} previously analyzed the Kepler-56 system, fitting 10 RV measurements plus a full ``photodynamical" model 
fit directly to the {\it Kepler} light curve.
\citet{Huber:2013ha} measure masses $[m_b,m_c]=[22.1^{+3.9}_{-3.6},181^{+21}_{-19}]M_\earth$, consistent
with our MCMC results in Table \ref{tab:mass}.

\begin{figure}[htbp]
	\begin{center}
	\includegraphics[width=0.9\textwidth]{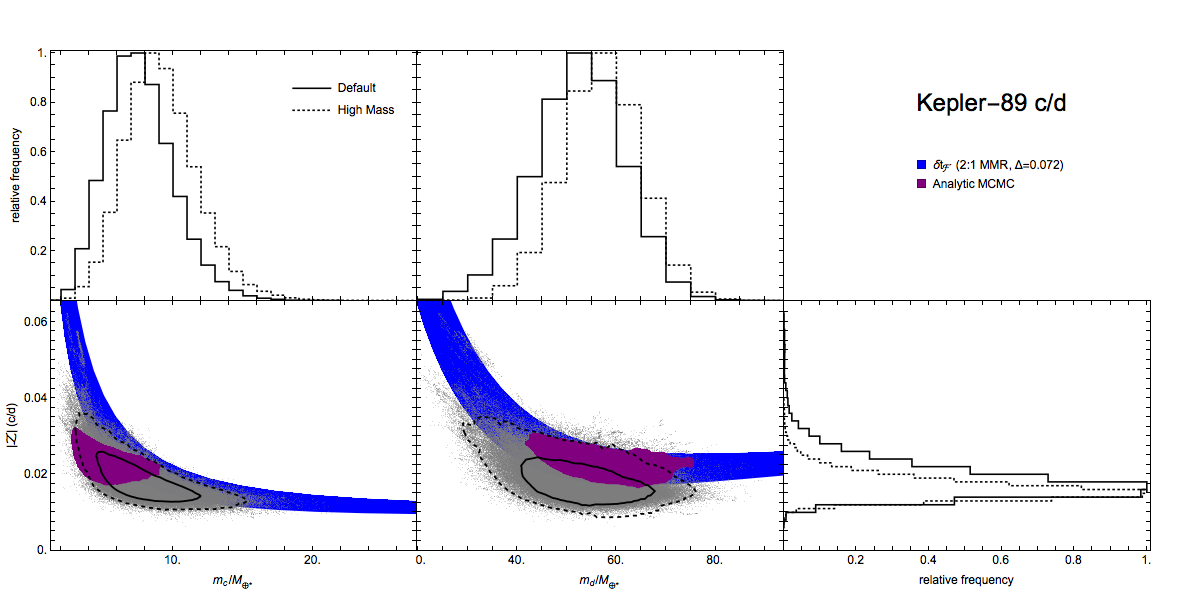}
\caption{
Constraint plot for Kepler-89 c and d (see Figure \ref{fig:kep11cons} for description).
}
	\label{fig:kep89cons}
	\end{center}
\end{figure}

{\it Kepler-89 c*,d*} (Fig. \ref{fig:kep89cons}):
Both Kepler-89 c and d have non-zero second-harmonic TTVs at $>$1$\sigma$ significance.
Figure \ref{fig:kep89cons} shows the constraints from the planets' fundamental TTVs along with the results of an analytic MCMC that includes second-harmonic TTV terms.
The analytic MCMC shows good agreement with the $N$-body posteriors, demonstrating that the second harmonic signals break the mass/eccentricity degeneracy for this system.
Kepler-89 hosts two additional planets, b and e, excluded from our analysis because their periods place 
them well away form any low-order MMRs with planet c and d.

Kepler-89 c and d are among the  few planets for which both RV and TTV mass measurements have been reported, with RV mass measurements reported by both \citet{2012ApJ...759L..36H}  and \citet{2013ApJ...768...14W} and a previous TTV measurement reported by \citet{2013ApJ...778..185M}.
The masses fit by both \citet{2013ApJ...778..185M} and \citet{2012ApJ...759L..36H} (who measure the mass of planet d only) agree well with the results of our TTV analysis.
\citet{2013ApJ...768...14W} find $m_d=106\pm11~M_\earth$ , in tension with our determination of planet d's mass even if we adopt their best-fit stellar mass, $M_*=1.3~M_\sun$.

\begin{figure}[htbp]
\begin{center}
\includegraphics[width=0.9\textwidth]{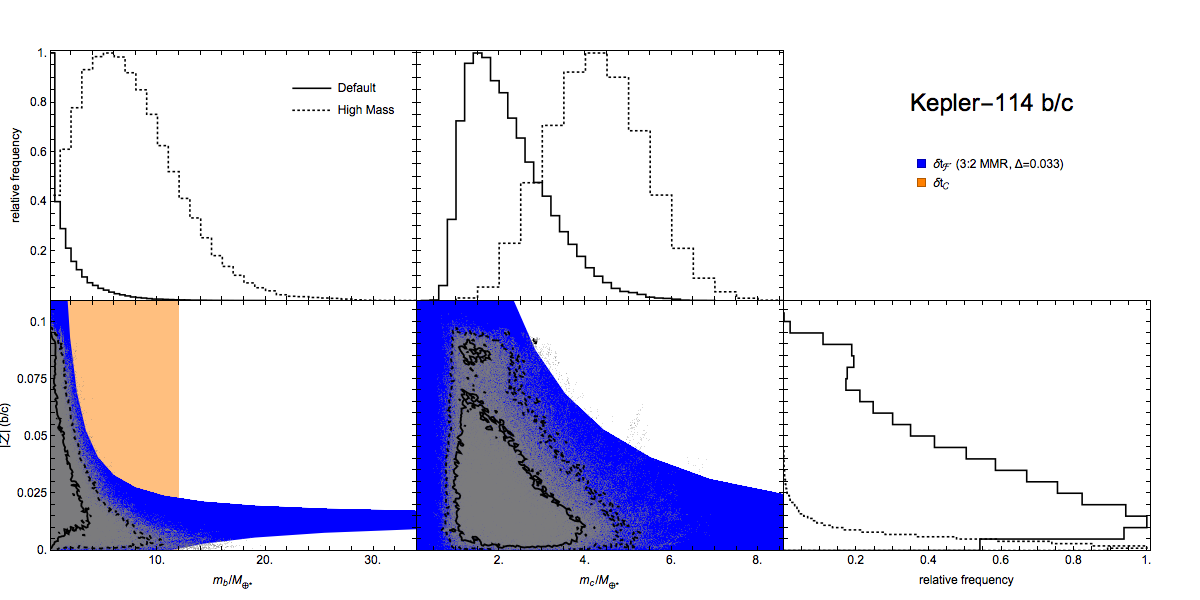}
\includegraphics[width=0.9\textwidth]{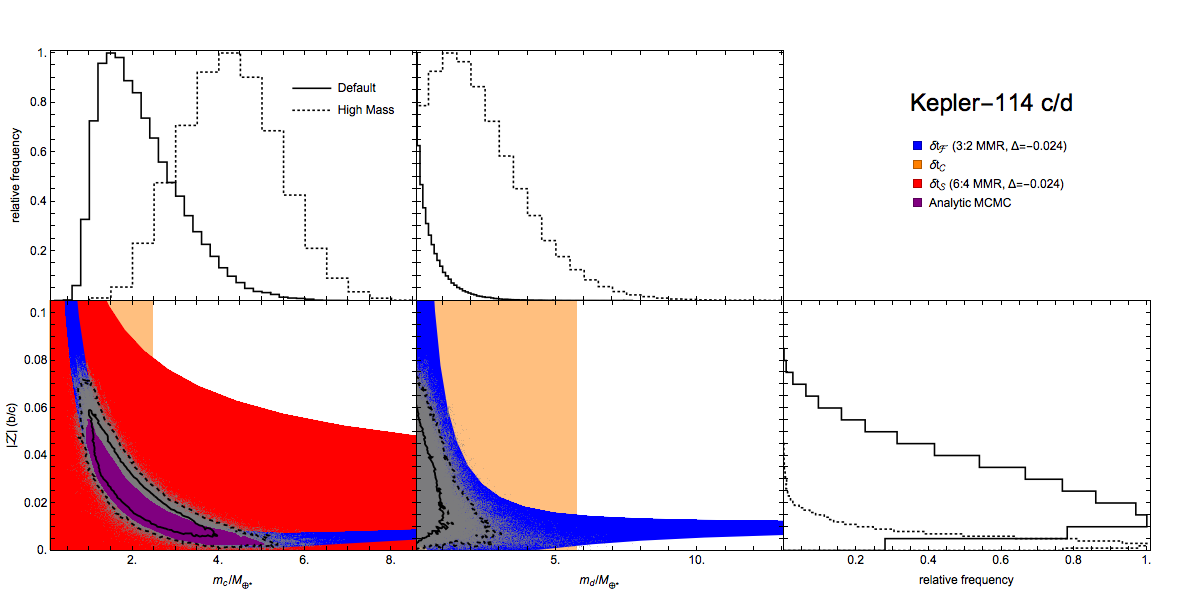}
\caption{
Constraint plots for Kepler-114 (see Figure \ref{fig:kep11cons} for description).
The purple region shows the 68\% credible region derived from an analytic MCMC fit of planet d's TTV.
The analytic MCMC fit only the TTV of planet d as a function of $m_c$  and the combined eccentricity, ${\cal Z}$,
of the c/d pair.
}
	\label{fig:kep114cons}
	\end{center}
\end{figure}

{\it Kepler-114 b,c,d} (Fig. \ref{fig:kep114cons}):
Kepler-114 is a three planet system with both the inner pair (b/c) and outer pair (c/d) of planets near 3:2 MMRs.
Interactions between b and c, summarized in the top panels of Figure \ref{fig:kep114cons}, do not provide strong constraints. 

The bottom panels of Figure \ref{fig:kep114cons} summarize the interactions of the c/d pair.
Planet c does not induce any significant TTV in planet d.
The bottom left panel shows a strong fundamental TTV constraint (from planet d's TTV) but the chopping and second-harmonic TTV amplitudes are consistent with zero at the 1$\sigma$ level.
Results of an analytic MCMC are included in Figure \ref{fig:kep114cons}.
We find, by turning on and off the contribution of the chopping and second-harmonic TTVs separately in the analytic MCMC, that both components help constrain planet c's mass. 

\begin{figure}[htbp]
	\begin{center}
	\includegraphics[width=0.9\textwidth]{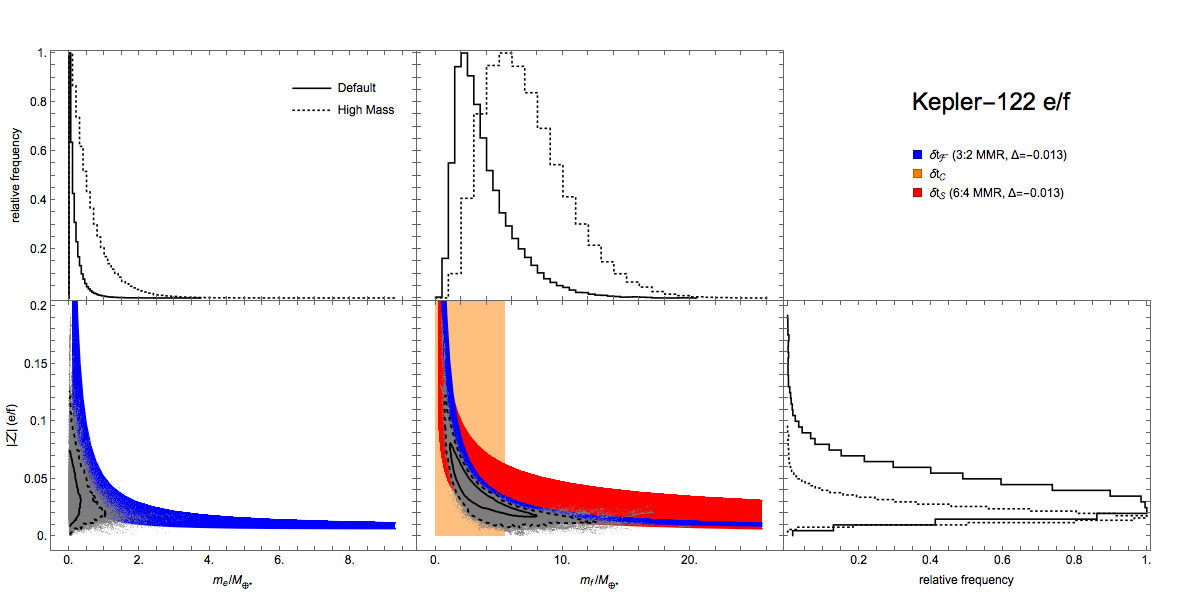}
\caption{
Constraint plots for Kepler-122 (see Figure \ref{fig:kep11cons} for description).
}
	\label{fig:kep122cons}
	\end{center}
\end{figure}

{\it Kepler-122 e,f} (Fig. \ref{fig:kep122cons}): 
Planet e's second-harmonic TTV and lack of chopping help break some of mass-eccentricity degeneracy, constraining $|{\cal Z}|\lesssim 0.1$ and $m_f\lesssim6\mearth$.

\begin{figure}[htbp]
 	\begin{center}
 	\includegraphics[width=0.9\textwidth]{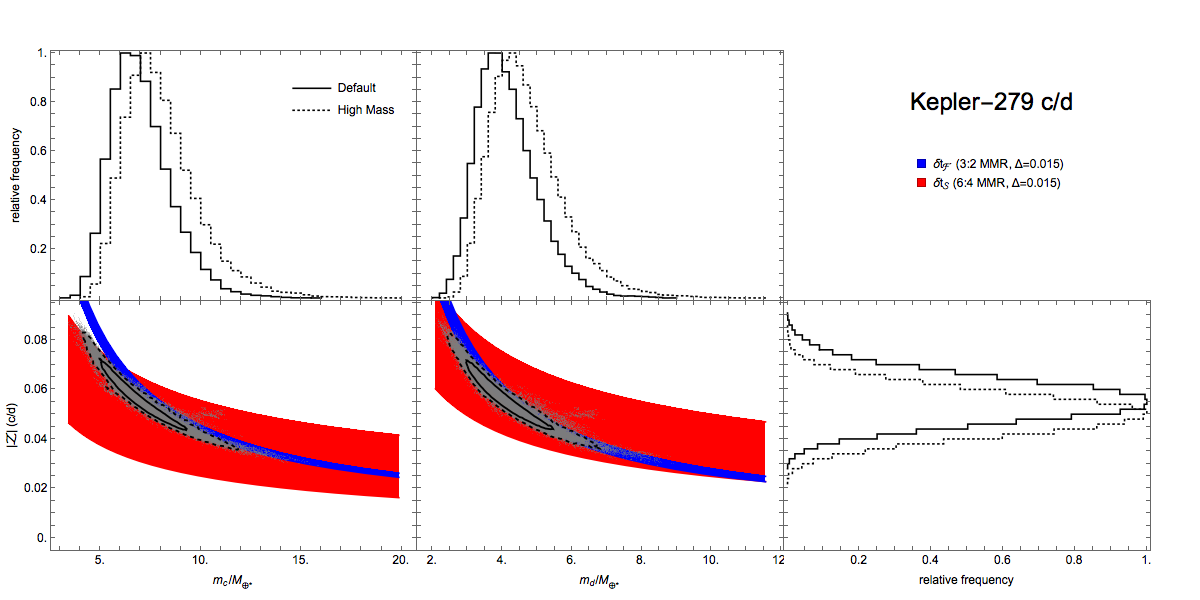}
 \caption{
 Constraint plots for Kepler-279 (see Figure \ref{fig:kep11cons} for description). 
 }
 	\label{fig:kep279cons}
 	\end{center}
 \end{figure}

{\it Kepler-279 c*,d*} (Fig. \ref{fig:kep279cons}): 
Kepler-279 c and d both have significant ($>3\sigma$) non-zero second-harmonic TTVs and
the combined constraints of the planets fundamental and second harmonic TTVs break the mass eccentricity degeneracy.
The inferred masses and combined eccentricity are not strongly affected by the choice of priors.
Kepler-279 hosts a third planet, b, interior to planets c and d with a  period of 12.3. Planet b's period places it far from planets c and d
and it is not expected to have an appreciable influence on either c or d's TTV.

\begin{figure}[htbp]
	\begin{center}
	\includegraphics[width=0.9\textwidth]{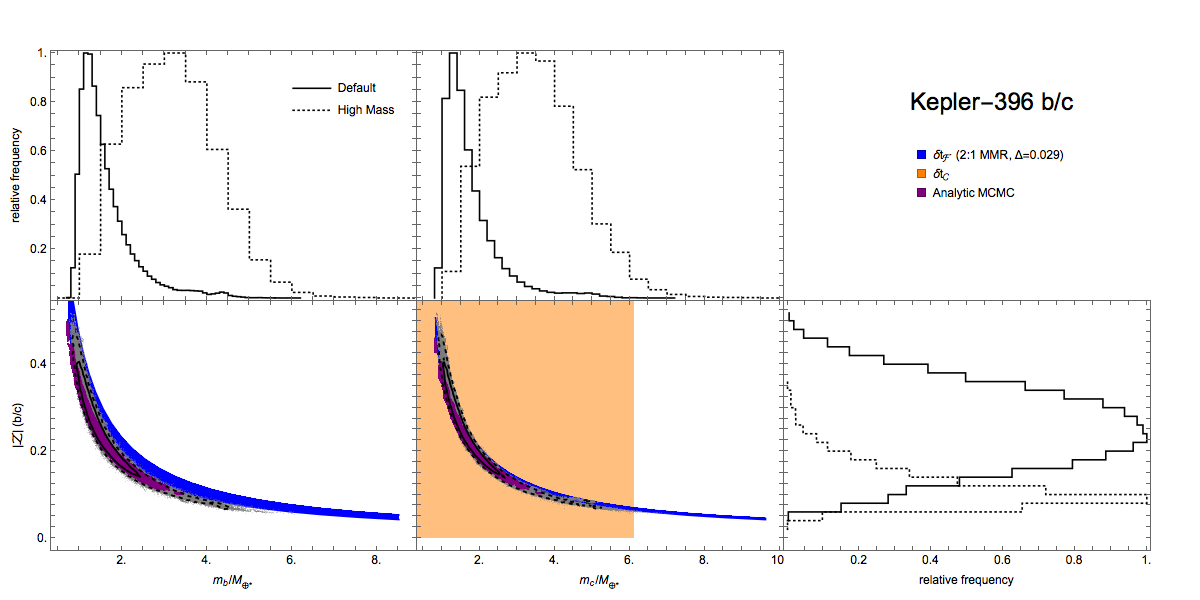}
\caption{
Constraint plots for Kepler-396 (see Figure \ref{fig:kep11cons} for description).
The purple regions indicate 68\% credible regions from an analytic MCMC fit.
}
	\label{fig:kep396cons}
	\end{center}
\end{figure}

{\it Kepler-396 b,c} (Fig. \ref{fig:kep396cons}):
Kepler-396 b's second-harmonic TTV amplitude is non-zero with $>2\sigma$ confidence.
We include the results of an analytic MCMC in 
Figure \ref{fig:kep396cons} since second-harmonic TTV constraints cannot be 
visualized in the $\mu-|{\cal Z}|$ plane for planets near the 2:1 MMR.
The analytic MCMC fit agrees well with the $N$-body results.
Lack of chopping  gives a 1$\sigma$ upper bound of $m_c\lesssim6\mearth$ on planet c's mass.

\subsection{\ref{sec:Other}: Resonant \& Massive Systems}
\label{sec:Other}

\begin{figure}[htbp]
	\begin{center}
	\includegraphics[width=0.9\textwidth]{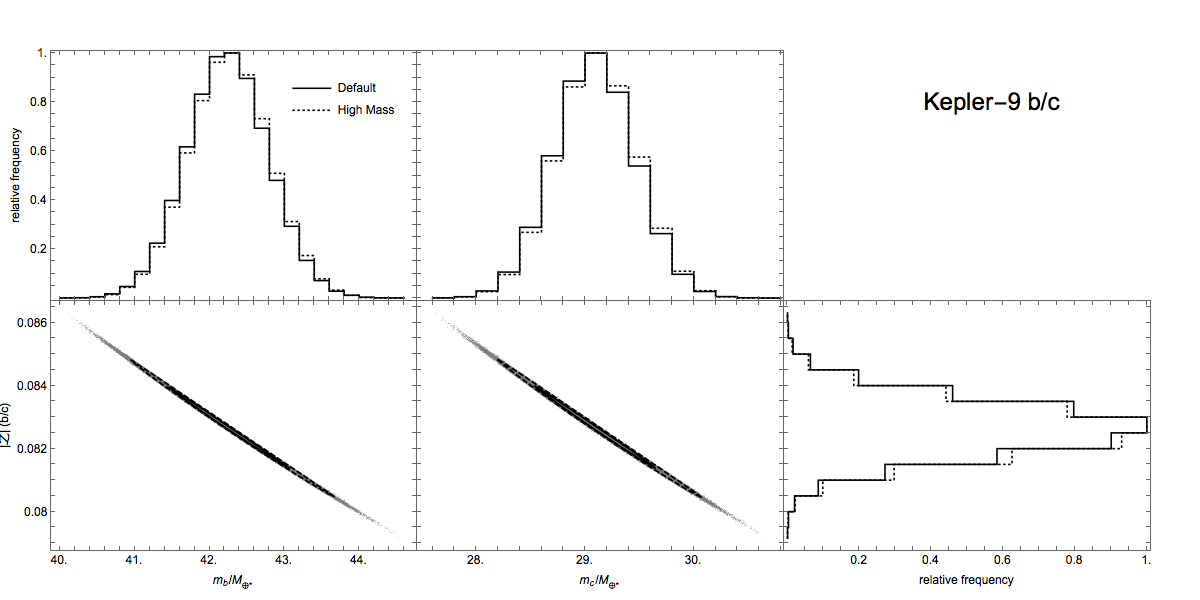} 
\caption{
Constraint plots for Kepler-9 (see Figure \ref{fig:kep11cons} for description).
Analytic fits omitted.
}
	\label{fig:kep9cons}
	\end{center}
\end{figure}

{\it Kepler-9 b*,c*} (Fig. \ref{fig:kep9cons}):
Kepler-9  b and c are a pair of Saturn-sized planets near a 2:1 MMR. 
Both planets' masses and their combined eccentricity are measured quite precisely.
Fits of Kepler-9 b and c's TTV with the analytic formulae under-predict the masses found by
$N$-body fitting by $\sim10\mearth$ for both planets. 
The analytic formulae fail to accurately recover the mass of Kepler-9 b and c because the formulae break down for planet pairs in or too near a mean motion resonance.
The width of a first-order MMR, as measured in terms of $\Delta$, scales as $\mu^{2/3}$  \citep[e.g.,][]{Henrard:1983cz} and Kepler-9 b and c's large masses place them quite close to being in the 2:1 resonance.
We find via $N$-body integrations that while the resonant angles $2\lambda_c-\lambda_b-\varpi_b$ and $2\lambda_c-\lambda_b-\varpi_c$
circulate for the best fit masses of Kepler-9 b and c, increasing the masses by only $\sim$30\% causes the angles to transition from circulation to libration. 

Kepler-9 has been studied previously by multiple authors: \citet{Holman:2010db,Borsato:2014it,2014arXiv1403.1372D}. 
Both \citet{Borsato:2014it} and \citet{2014arXiv1403.1372D} fit masses from TTV analysis that are $\sim$60\% smaller than those found by \citet{Holman:2010db}, who base their planet masses primarily on six RV observations.\footnote{\citet{Holman:2010db} also include limited number of TTV observations spanning only $\sim250$d in their fitting. They find that their TTV observations alone do not strongly constrain the planet masses.}
The masses and eccentricities that we fit agree well with those determined by \citet{Borsato:2014it} and \citet{2014arXiv1403.1372D}. 
The source of disagreement between the TTV- and RV-derived masses is unclear 
though, as the authors of both previous TTV studies note, the RV observation span a short time baseline, less than an entire orbit of planet c, and follow-up RV observations could shed light on the discrepancy.

\begin{figure}[htbp]
	\begin{center}
    \includegraphics[width=0.9\textwidth]{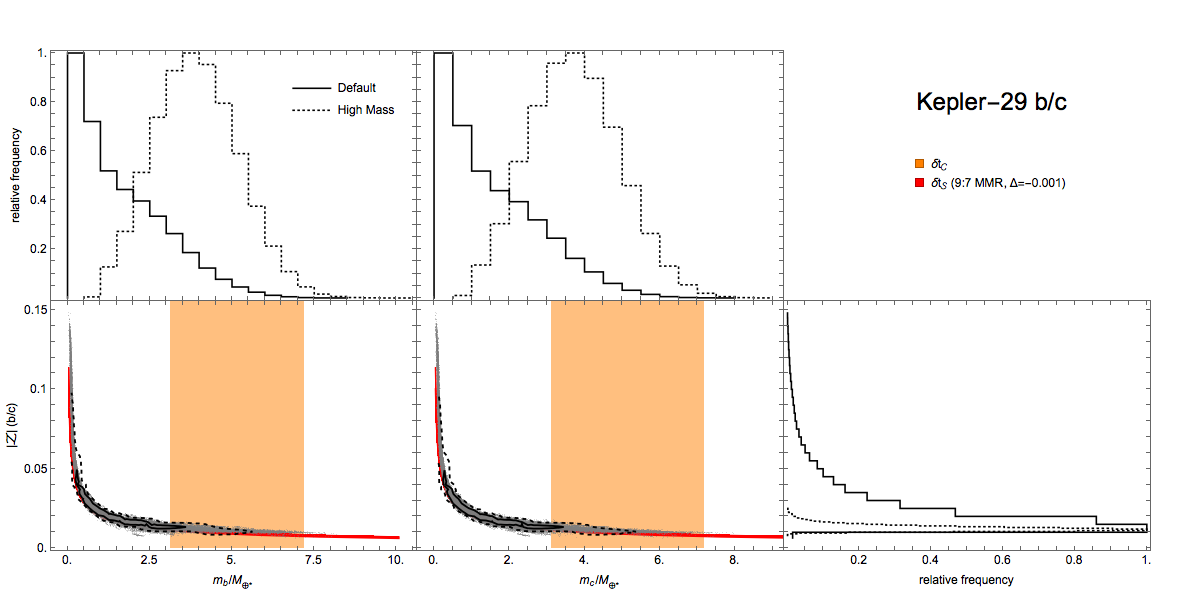}
\caption{
Constraint plots for Kepler-29 (see Figure \ref{fig:kep11cons} for description).
}
\label{fig:kep29cons}
\end{center}
\end{figure}

\begin{figure}[htbp]
	\begin{center}
	\includegraphics[width=0.65\textwidth]{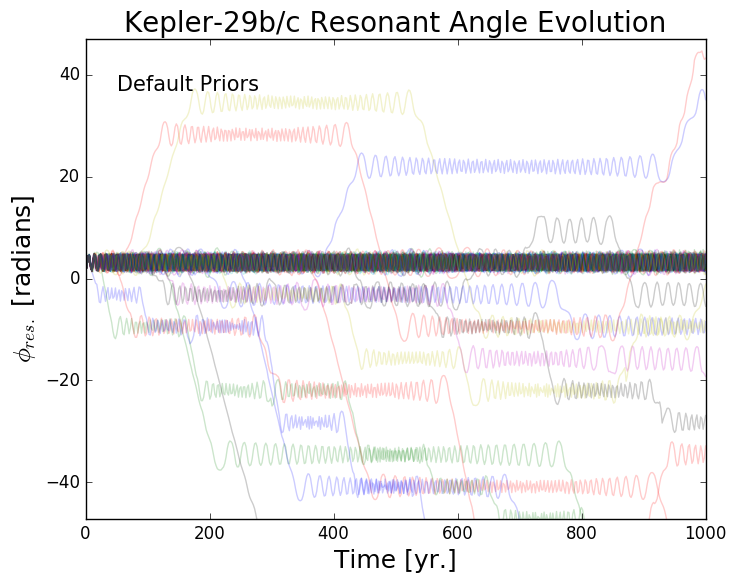}
	\includegraphics[width=0.65\textwidth]{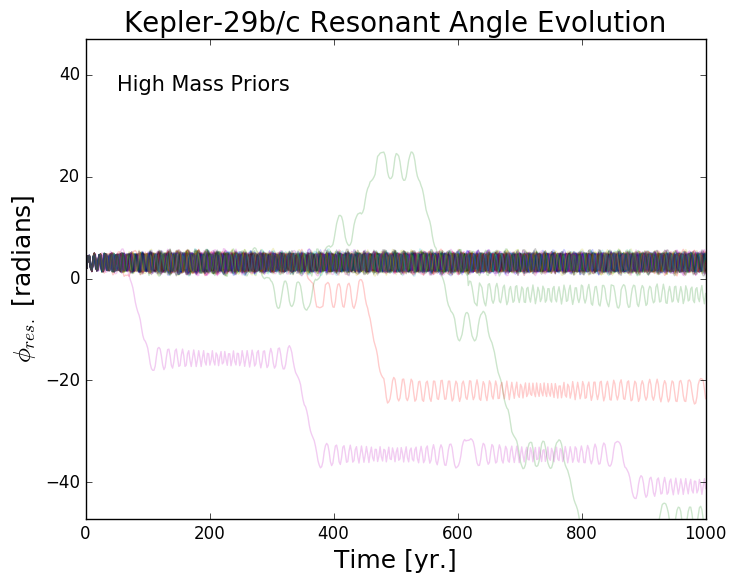}
\caption{
	{\bf Top panel:} Time-evolution of the Kepler-29b/c 9:7 resonant angle for 100 random 
	initial conditions drawn from the default  posteriors. 
	The resonant angle is defined as the complex phase of Equation \eqref{eq:res_ang2}.
	The resonant angles librate or alternates intermittently between libration and circulation.
	Alternating behavior is indicative of chaotic orbits.
	Note the vertical axis extends beyond $\pm \pi$ to account for windings of resonant angles that circulate.
	{\bf Bottom panel:} Same as top panel but with initial conditions drawn from the high mass posteriors.
}
\label{fig:kep29res}
\end{center}
\end{figure}

{\it Kepler-29 b,c} (Fig. \ref{fig:kep29cons}):
Our $N$-body fits indicate that Kepler 29 b and c's are librating in the 9:7 MMR.
The $N$-body MCMC results agree well with analytic constraints derived form the second-order resonance TTV despite the fact that the pair's  libration in resonance violates the assumptions of the formulae's derivation (see Paper I). 
We confirm that  Kepler 29 b and c are in the 9:7 MMR 
with a set of 100 $N$-body integrations using initial conditions drawn randomly from the MCMC posteriors. 
The $N$-body integrations are done with the REBOUND code's  WHFast integrator \citep{2015MNRAS.452..376R}.
Confirming that the planet pair is in resonance requires testing resonant angle(s) for libration.
There are three resonant angles associated with a 9:7 MMR: 
$9\lambda_c-7\lambda_b-2\varpi_b$,  $9\lambda_c-7\lambda_b-\varpi_b-\varpi_c$, and $9\lambda_c-7\lambda_b-2\varpi_c$,
each of which appears as a cosine term in the Laplacian expansion of the planets' disturbing function \citep{2000ssd..book.....M}.
The sum of the cosine terms can be written, to lowest order in eccentricities, in terms of the planets' complex eccentricities as
\begin{eqnarray}
{1\over 2}(f_{45} z_b^{*2} + f_{53} z_c^{*2} +f_{49} z^*_b z^*_c){\text e}^{i(9\lambda_c-7\lambda_b)} + c.c. \label{eq:res_ang2}
\end{eqnarray}
where the $f_{i}$ combinations of Laplace coefficients as defined in Appendix B of \citet{2000ssd..book.....M}, `*' indicates complex conjugation, and `$c.c.$' denotes the complex conjugate of the preceding term.
Instead of testing each of the three possibly resonant angles separately, 
we plot the complex phase of the term preceding `$c.c.$' in Equation \eqref{eq:res_ang2} in Figure \ref{fig:kep29res}.
Libration of this complex phase implies libration of at least one of the 9:7 MMR angles enumerated above.
Figure \ref{fig:kep29res} shows the time evolution of the resonant angle from $N$-body integrations drawn from both the default and high mass posteriors.
For all initial conditions, the resonant angle either librates  or chaotically alternates between libration and circulation.\footnote{We confirm the chaotic nature of the initial conditions that produce alternating evolution with the MEGNO chaos indicator \citep{2003PhyD..182..151C} implemented in the REBOUND code.}

\citet{JontofHutter:2016ch} and \citet{2016arXiv160909135M} also conduct an MCMC analysis of Kepler-29 b and c's Q1-17 transit times.
Our high-mass priors yield  masses and error bars similar to theirs.
Yet despite that, we do not consider this system as robust because the default and high mass priors disagree.

\begin{figure}[htbp]
 \begin{center}
 \includegraphics[width=0.9\textwidth]{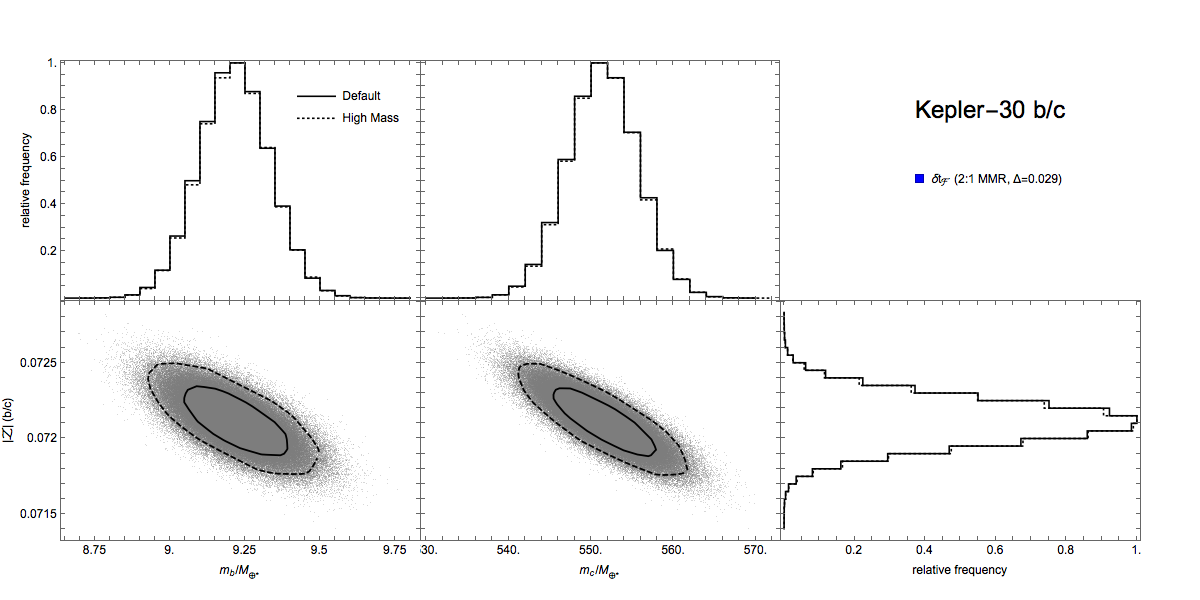}
 \caption{Constraint plot for Kepler-30 (see Figure \ref{fig:kep11cons} for description). 
 Analytic fits omitted.}
 	\label{fig:kep30cons}
 	\end{center}
 \end{figure}

{\it  Kepler-30 b*,c*} (Fig. \ref{fig:kep30cons}): 
Kepler-30 b and c  are near a 2:1 MMR. 
The analytic TTV formulae are not expected to be accurate given Kepler-30 c's large mass ($m_c\approx550\mearth$).
\citet{2012Natur.487..449S} fit masses $[m_b,m_c]=[11.3\pm1.4,640\pm50]\mearth$ 
based on a TTV analysis of 2.5 years of transit data. We find somewhat smaller masses than \citet{2012Natur.487..449S},
though our results are consistent within $\lesssim2\sigma$.
Our TTV analysis of Kepler-30 is based on the transit times computed by \citet{2016ApJS..225....9H} who note that the
transit times computed by \citet{2015ApJS..217...16R} for Kepler-30 b are erroneous.

\begin{figure}[htbp]
\begin{center}
	\includegraphics[width=0.9\textwidth]{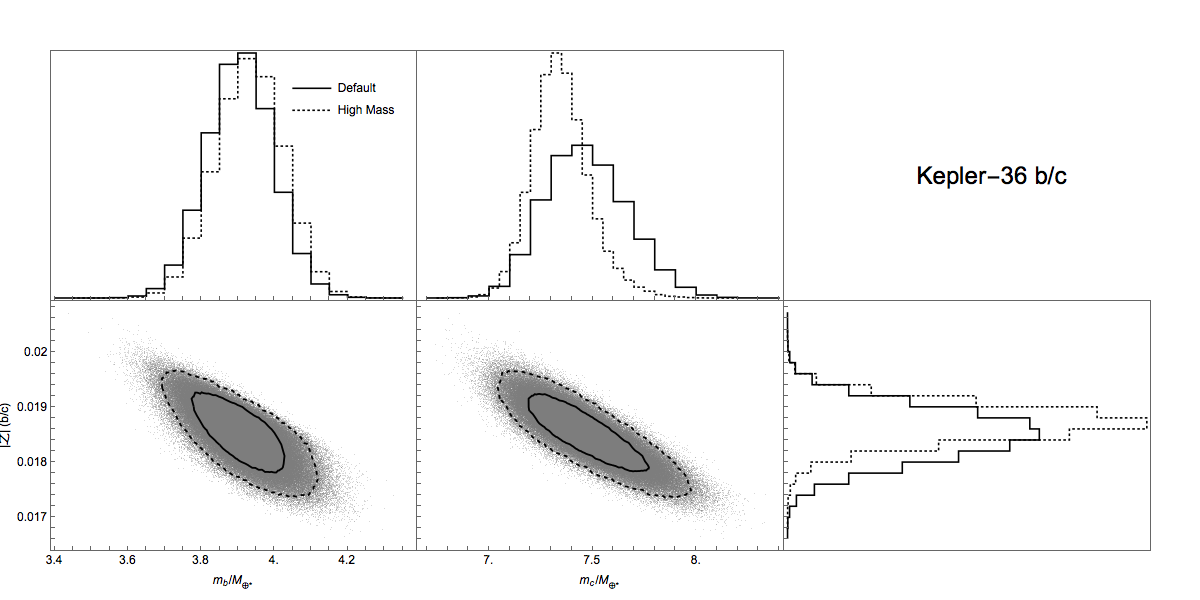}
\caption{Constraint plot for Kepler-36 (see Figure \ref{fig:kep11cons} for description).
Analytic fits omitted.}
	\label{fig:kep36cons}
	\end{center}
\end{figure}

\begin{figure}[htbp]
\begin{center}
	\includegraphics[width=0.6\textwidth]{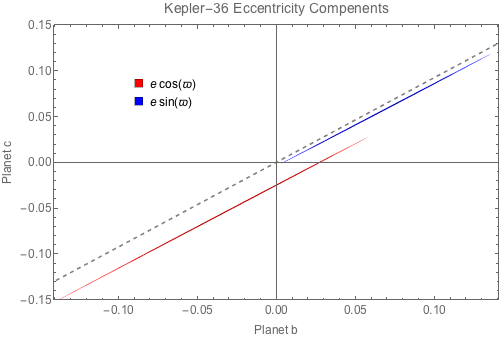}
\caption{
$N$-body MCMC posteriors for Kepler-36 b and c eccentricity components. 
The posterior distributions of $e_i\cos(\varpi_i)$ (red)  and $e_i\sin(\varpi_i)$ (blue) for $i=b,c$ are plotted with
68\% and 95\% credible regions indicated by dark and light shading, respectively.
The dashed line shows the expected correlation slope, $-f_{27}/f_{31}\approx0.9$, for a constant value of ${\cal Z}$ (Eq. \ref{eq:zdef}).
The posterior shows a strong correlation in the direction expected from the analytic TTV formulae. 
}
	\label{fig:kep36ecc}
	\end{center}
\end{figure}

{\it Kepler-36 b*,c*} (Fig. \ref{fig:kep36cons}):
Kepler-36 b and c are near a 7:6 MMR ($\Delta=0.005$).
We omit any analysis of this system with the analytic formulae because they provide poor approximations to the planets' TTVs due to both their extreme proximity and their small $\Delta$.
Despite the analytic model's poor preformance,
the planets' eccentricity vector components are still strongly correlated as predicted by the analytic model.  
Figure \ref{fig:kep36ecc} compares the eccentricity vector posteriors to the correlation predicted by the analytic model.
Kepler-36 b and c's eccentricities are poorly constrained by the TTVs and only their combination, ${\cal Z}$, is measured accurately.
Our analysis agrees with the previous TTV mass measurements of Kepler-36 b and c by \citet{Carter:2012gq}.

\begin{figure}[htbp]
\begin{center}
	\includegraphics[width=0.9\textwidth]{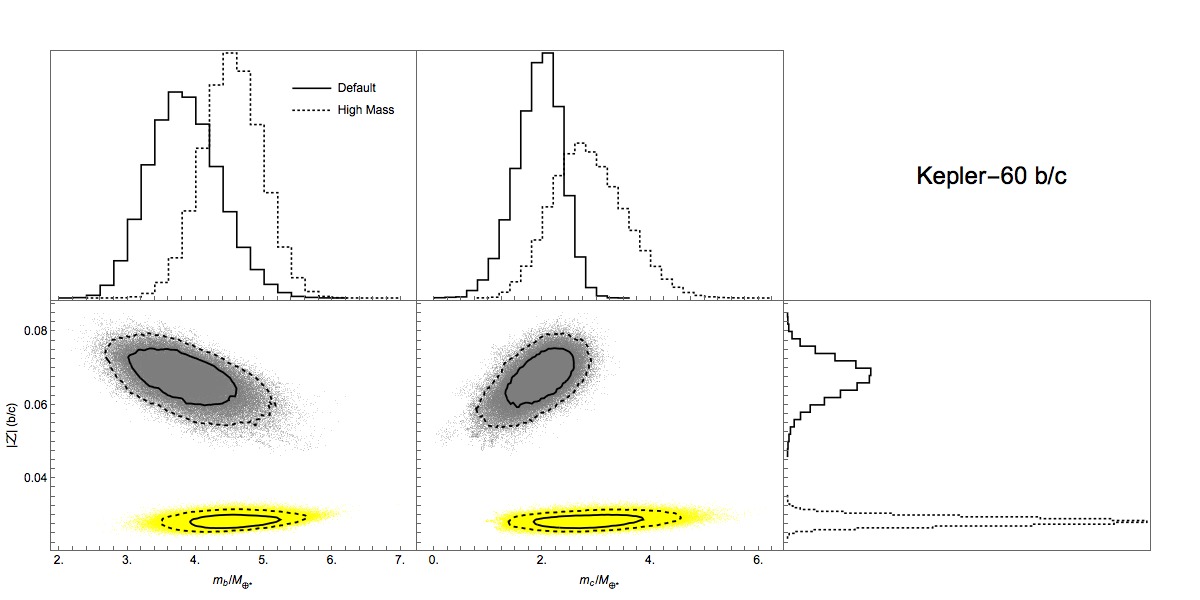}
	\includegraphics[width=0.9\textwidth]{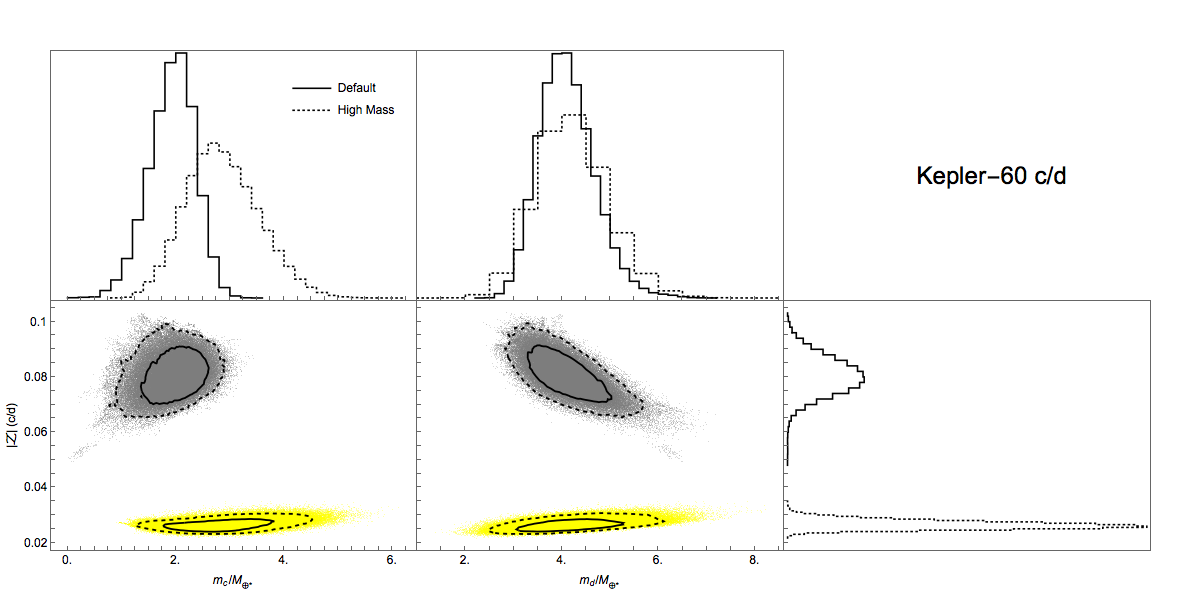}
\caption{Constraint plot for Kepler-60 (see Figure \ref{fig:kep11cons} for description).  Analytic fits are omitted.
The MCMC posterior samples computed with the high mass priors are shown in yellow.}
\label{fig:kep60cons}
	\end{center}
\end{figure}

{\it Kepler-60 b*,c*,d*} (Fig. \ref{fig:kep60cons}):
The Kepler-60 planets are in a chain of MMRs.
The inner b/c are in a 5:4 MMR and the outer  c/d pair is in a 4:3 MMR.\footnote{
The resonant angles of Kepler-60 b/c and c/d undergo libration.
Strictly speaking, `in resonance' also implies the existence of a separatrix in phase space \citep[e.g.,][]{Henrard:1983cz}.
Deducing the existence of a separatrix requires a more  detailed dynamical analysis beyond the scope of this work.
}
This configuration placed the b/d pair near a 5:3 MMR. 
Furthermore, the planets are in or near a three-body resonance satisfying $|n_b-2n_c+n_d|\approx0$.
Figure \ref{fig:kep60cons} shows that the default and high-mass posterior samples occupy disjoint regions of 
parameter spaces. 
This indicates that the likelihood function defined by the TTV observations has multiple local maxima of similar likelihood, and 
each prior weights more towards a different maximum.

\begin{figure}[htbp]
\begin{center}
	\includegraphics[width=0.45\textwidth]{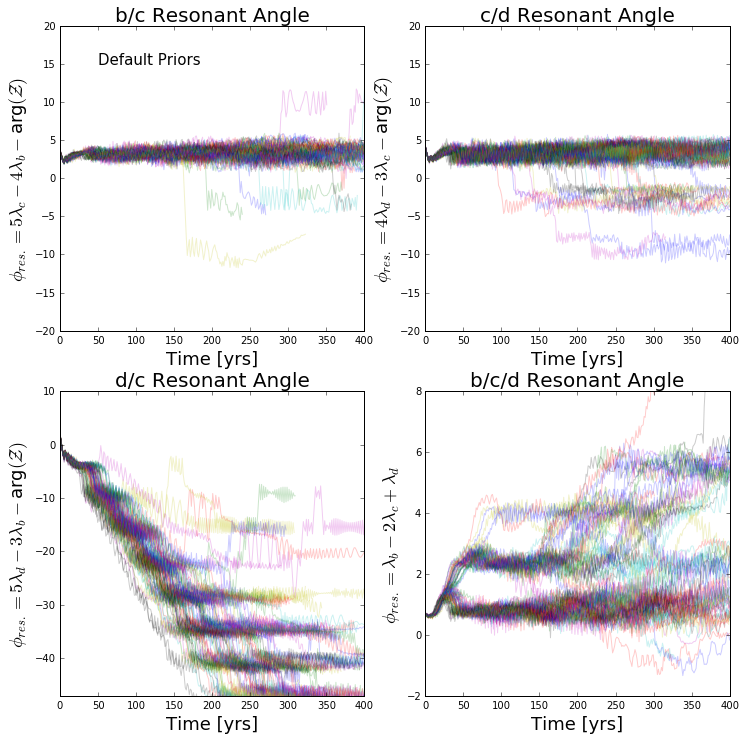}
	\includegraphics[width=0.45\textwidth]{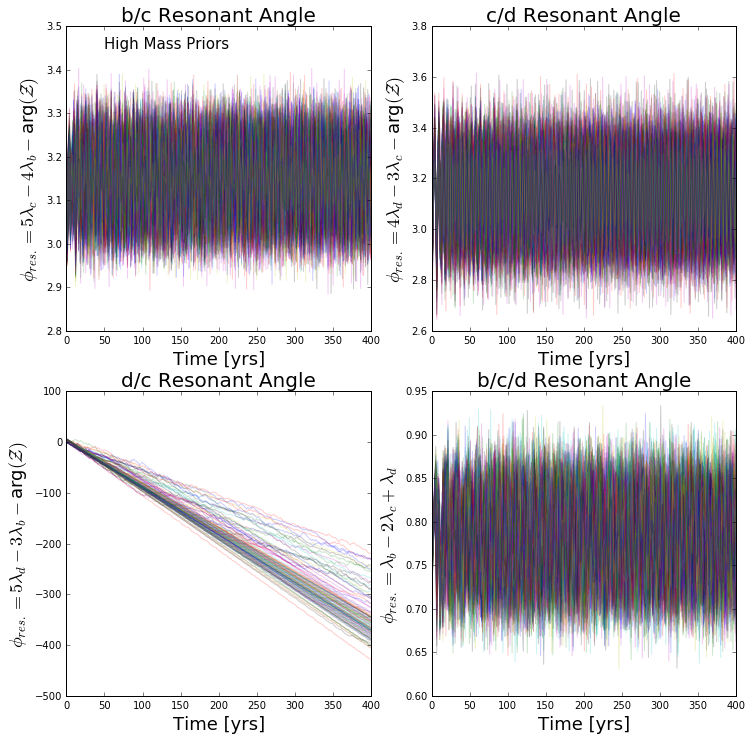}
\caption{
Resonant angle time evolution for the Kepler-60 system. 
Left panels show 100 random simulations with initial conditions drawn from the MCMC posterior computed using 
the default priors.
Right panels show the evolution of the same angles for initial conditions drawn from the high mass priors MCMC.
Many of the initial conditions taken from the default MCMC posteriors are unstable on the timescale of a few 100 years.
Vertical axes are extended beyond $\pm \pi$ to account for windings of resonant angles that circulate.
}
\label{fig:kep60res_ang}
\end{center}
\end{figure}

Figure \ref{fig:kep60res_ang} examines the time evolution of the resonant angles of 
each first-order MMR and the three-body resonance.
For each planet pair near a $j$:$j$-1 MMRs, Figure \ref{fig:kep60res_ang} plots the resonant angles
\begin{equation}
\phi_{res.}=j\lambda' -(j-1)\lambda - \arg{(\cal Z)}
\end{equation}
, where $\lambda$ and $\lambda'$ are the mean anomalies of the inner and outer planet, respectively.
The dynamics of a first-order resonance can be shown to depend only on this single resonant angle \citep[e.g.,][]{Batygin:2013en}.
We investigate resonant angle evolution for initial conditions drawn from both the high mass and default posteriors.
We find that most (95/100) of the initial conditions taken from the default posterior are unstable in less than  $10^3~\mathrm{ yr.}\approx 5\times10^4 P_b$, while all high-mass posterior initial conditions were stable for at least $10^3~\mathrm{ yr}$.
Given that most of the samples computed with the default prior were found to be unstable, we
conclude that the high mass posteriors more accurately reflect the true planet properties.
For Kepler-60 only, we use the high mass posteriors instead of the default priors to compute the masses, densities, and envelope fractions plotted of the figures in the body of the paper.

The high-mass prior MCMC results (Table \ref{tab:mass}) agree well with the masses measured by the two previous studies of \citet{Gozdziewski:2015ws} and \citet{JontofHutter:2016ch}.
Both \citet{Gozdziewski:2015ws} and \citet{JontofHutter:2016ch}  find similar values for the masses of the Kepler-60 system.
We find that the three-body resonance angle, $\lambda_b-2\lambda_c+\lambda_d$, librates for all initial conditions drawn from the MCMC posterior computed with the high mass prior.
\citet{JontofHutter:2016ch} find 20\% of the posterior samples they test have a three-body resonance angle that circulates while the system remains stable for at least 1 Myr.

\begin{figure}[htbp]
\begin{center}
\includegraphics[width=0.45\textwidth]{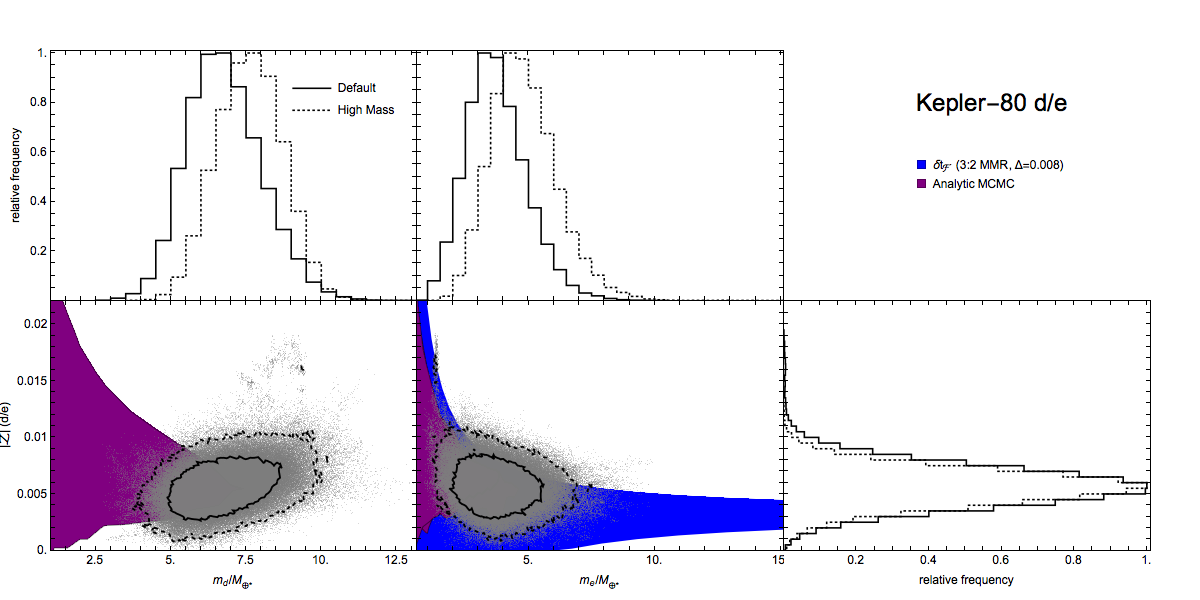}
\includegraphics[width=0.45\textwidth]{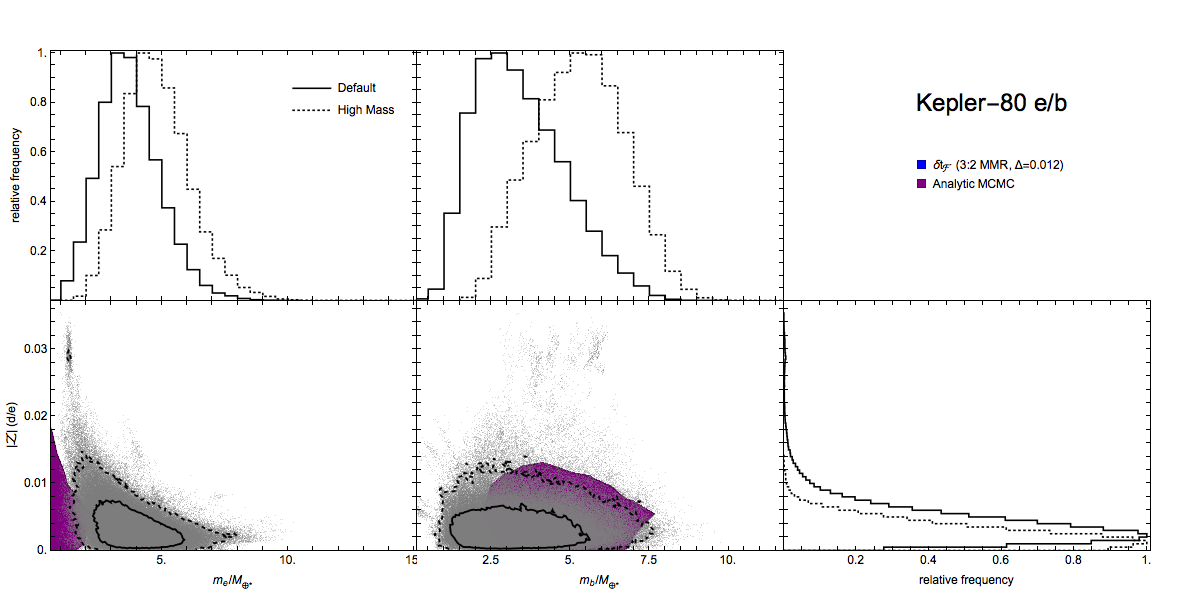}
\includegraphics[width=0.45\textwidth]{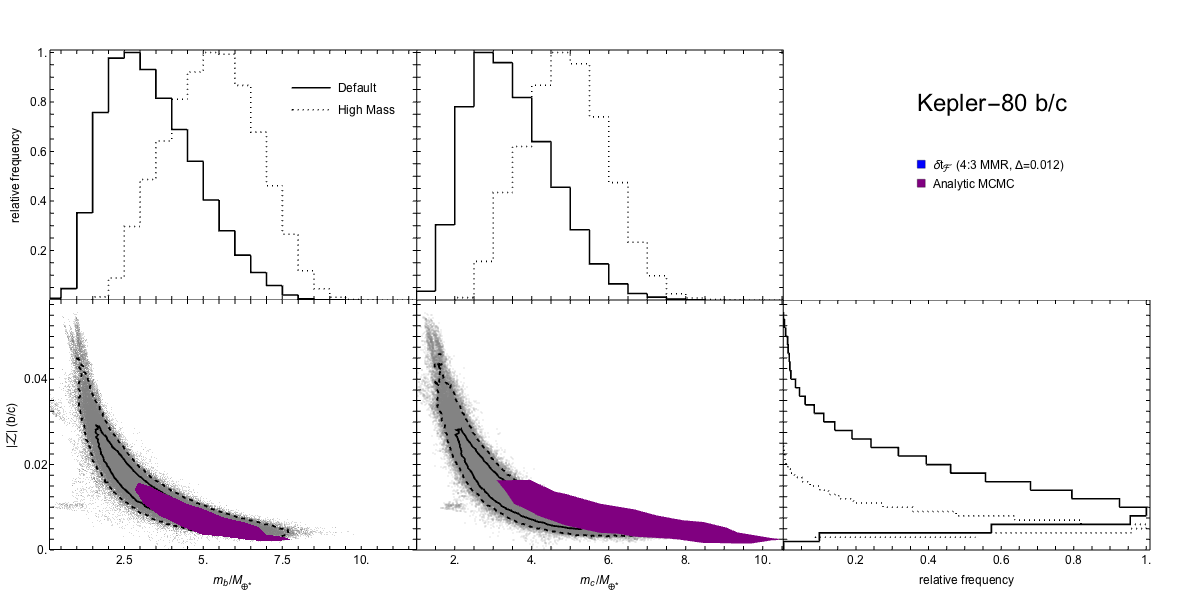}
\includegraphics[width=0.45\textwidth]{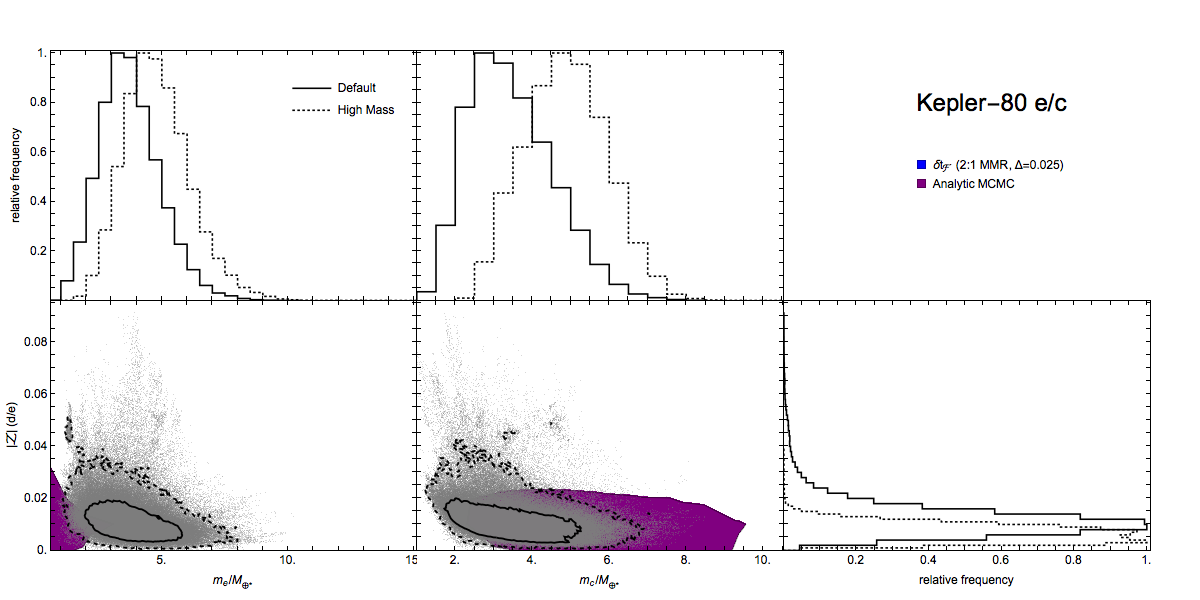}
\caption{
Constraint plots for Kepler-80 (see Figure \ref{fig:kep11cons} for description).
The purple regions indicate 68\% credible regions from an analytic MCMC fit.
}
\label{fig:kep80cons}
	\end{center}
\end{figure}

\begin{figure}[htbp]
\begin{center}
	\includegraphics[width=0.75\textwidth]{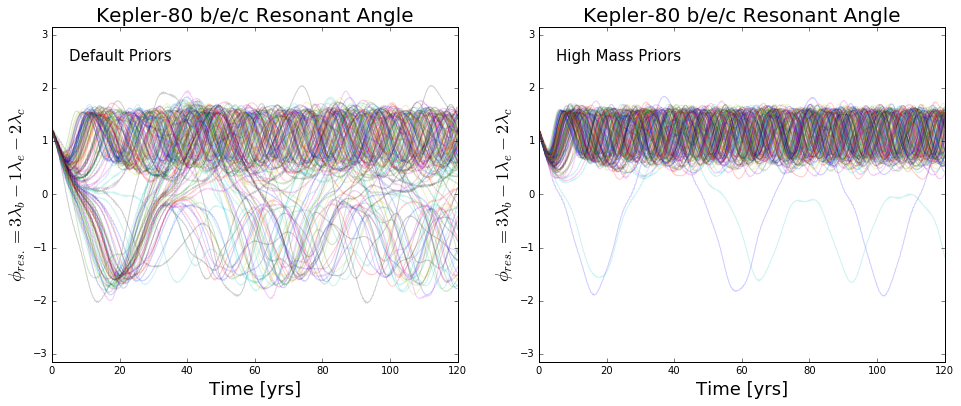}
	\includegraphics[width=0.75\textwidth]{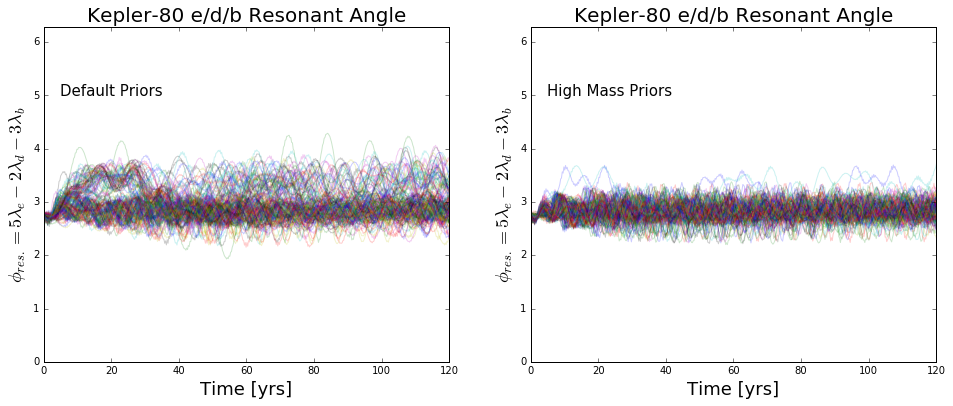}
\caption{
Resonant angle time evolution for the Kepler-80 system. 
The left panels shows 100 random simulations with initial conditions drawn from the default posterior.
The right panels shows the same angles with initial conditions drawn from the high mass posteriors.}
	\label{fig:kep80res_ang}
	\end{center}
\end{figure}

{\it Kepler-80 d*,e*,b,c} (Figure \ref{fig:kep80cons}):
The planets of Kepler-80 are arranged in a complex chain of resonances.
Each adjacent pair is near a first-order MMR, starting with the d/e pair near  a 3:2 MMR, 
followed by the e/b pair near a 3:2 MMR, and finally the b/c pair near a  4:3 MMR.
Additionally, the second and fourth planets, e  and c, are near a 2:1 MMR.
Based on $N$-body integrations initialized from our MCMC posteriors, all the resonant angles associated with these first-order MMRs circulate.
Three-body resonances occur when the frequencies of two two-body resonance angles are equal and
the circulation rates of each two-body resonant angle in the Kepler-80 system are nearly equal, i.e.
$3n_e -2n_d\approx3n_b-2n_e\approx 4n_c-3n_b$.
The two three-body resonance angles,
\begin{eqnarray}
\phi_{e,d,b} &= 5\lambda_e - 2\lambda_e - 3 \lambda_b \label{eq:ang_edb}\\
\phi_{b,e,c} &= 3\lambda_b - \lambda_e -2 \lambda_c \label{eq:ang_bec}
\end{eqnarray}
, librate in the Kepler-80 system.
Figure \ref{fig:kep80res_ang} plots the time evolution of the resonant angles,
Eqns. \eqref{eq:ang_edb} and \eqref{eq:ang_bec}, computed by $N$-body integrations initialized from the
MCMC posterior distributions.
Each angle librates for every initial condition that we integrate.

Because of the three-body resonances, planets e, b, and c  have multiple fundamental TTVs with the same super-period.
Planet e, for example, has fundamental TTVs induced by its proximity to the 3:2 MMRs with both  planet d and b,
both with super-periods of $\sim$192 days.
Multiple fundamental TTVs with the same frequency inhibits our linear fitting procedure for generating constraint plots 
since the sinusoidal basis functions used to fit fundamental TTVs are not independent.
To illustrate the constraints of the analytic model in Figure \ref{fig:kep80cons} we preform an MCMC that uses the analytic formulae to fit the planets' TTVs directly,
as opposed to fitting amplitudes measured by a least-squares fit.
The results of the analytic MCMC in Figure \ref{fig:kep80cons} show fair agreement for planed b and c but suggest that planet d and e's masses are consistent with 0, at odds with the $N$-body results.

Least-square fits to the transit times of planets b and c (and to a lesser extent  planet e) are significantly improved by allowing for a quadratic trend in the TTVs.
We suspect that this long-term quadratic trend arises from the dynamical influence of the three-body resonances on the TTVs  
and contributes to the robust constraints for planet d and e's masses found via $N$-body MCMC.

\citet{MacDonald:2016vo} recently conducted a TTV analysis of the Kepler-80 system.
Their mass determinations for the robustly constrained planets d and e are consistent with ours within 1$\sigma$ uncertainties after accounting for their use of a somewhat larger stellar mass of $M_*=0.73\pm0.02~M_\sun$ compared to our $M_*=0.6\pm0.03~M_\sun$.

\begin{figure}[htbp]
\begin{center}
\includegraphics[width=0.45\textwidth]{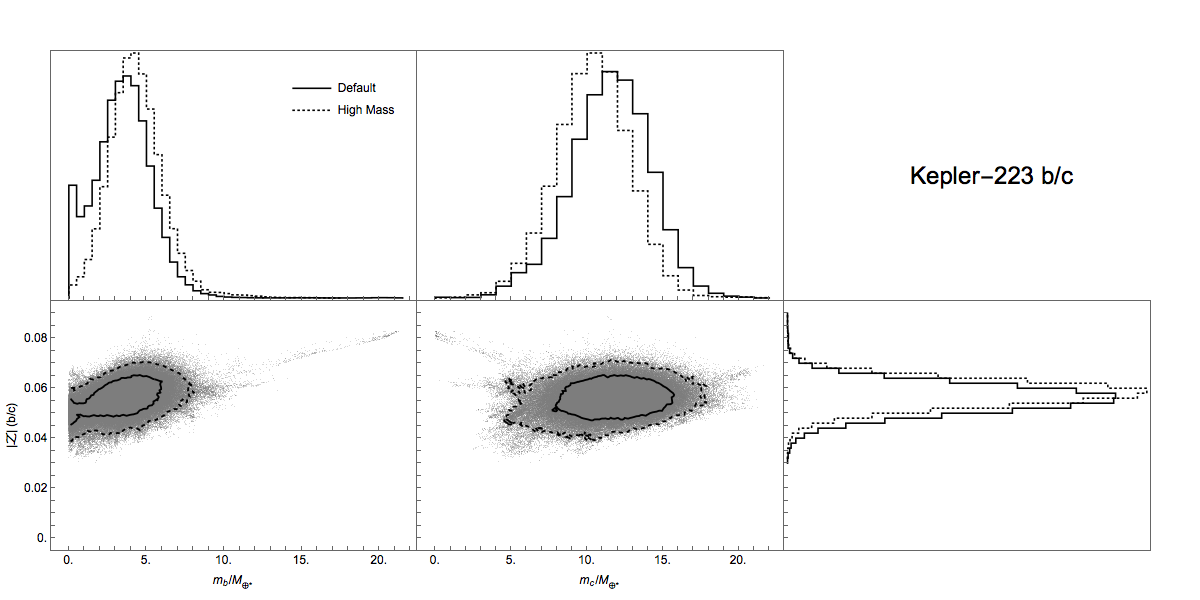}
\includegraphics[width=0.45\textwidth]{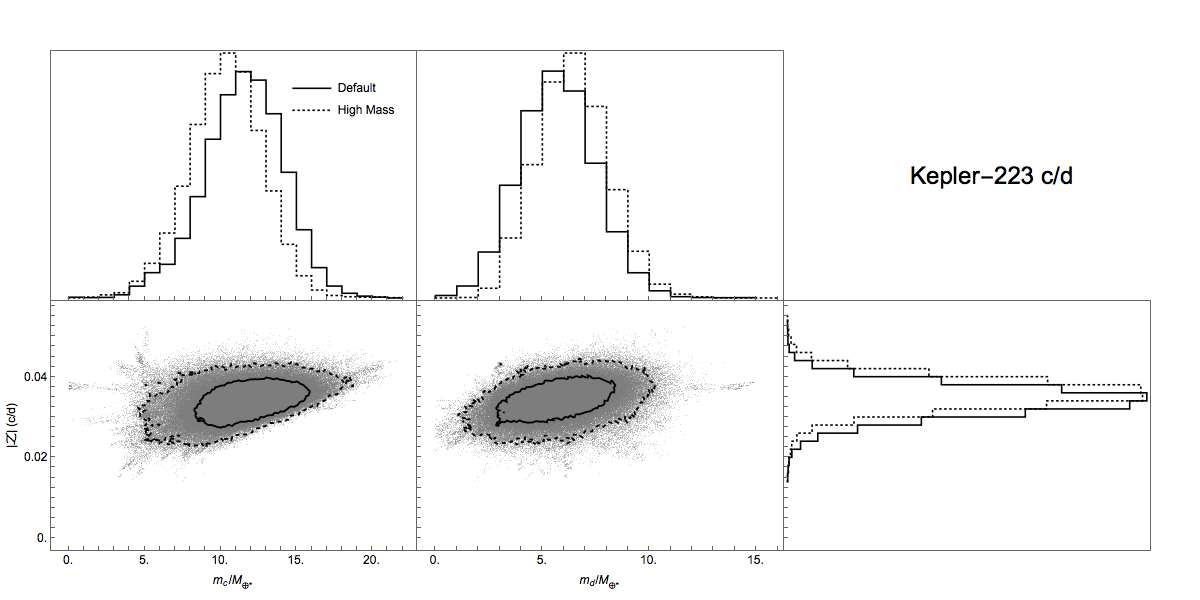}
\includegraphics[width=0.45\textwidth]{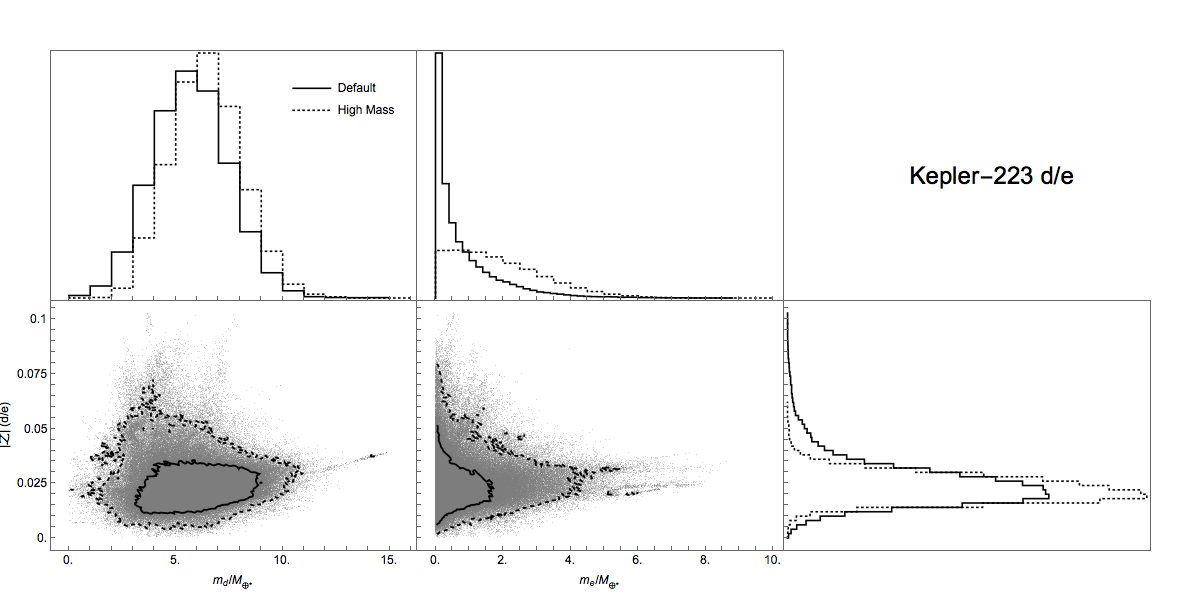}
\caption{Constraint plots for Kepler-223 (see Figure \ref{fig:kep11cons} for description). Analytic fits omitted.}
\label{fig:kep223cons}
\end{center}
\end{figure}

\begin{figure}[htbp]
\begin{center}
	\includegraphics[width=0.9\textwidth]{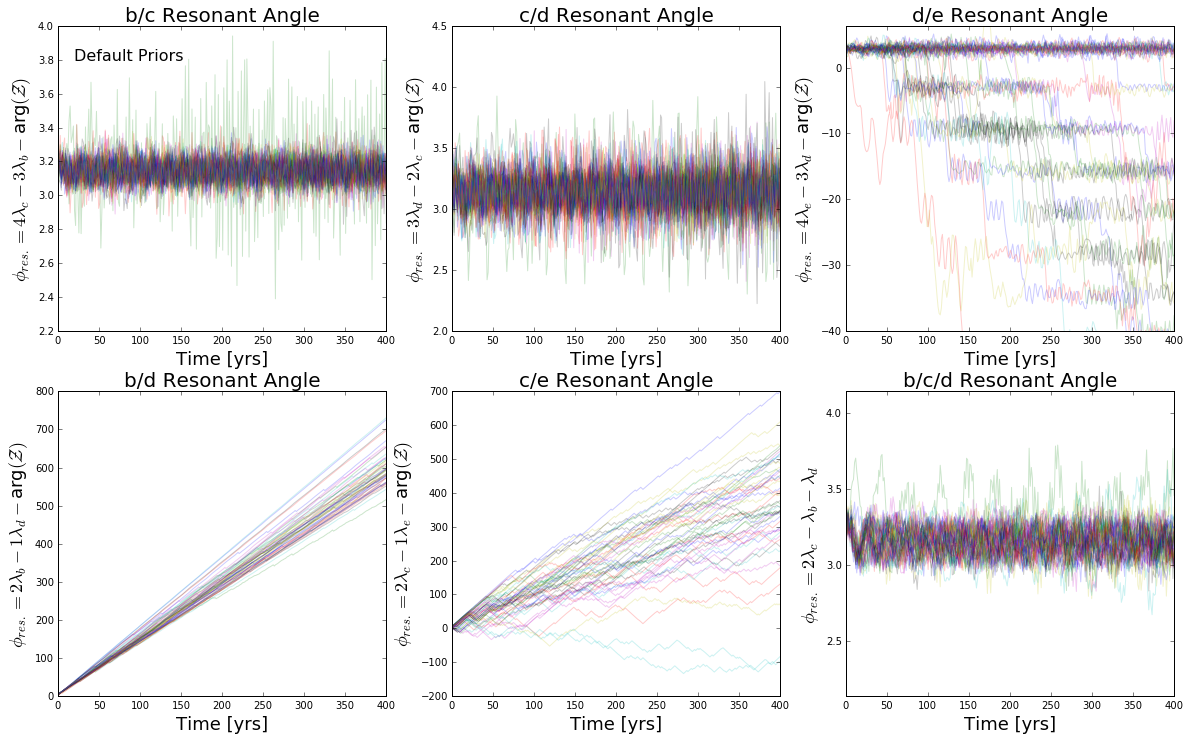}
	\includegraphics[width=0.9\textwidth]{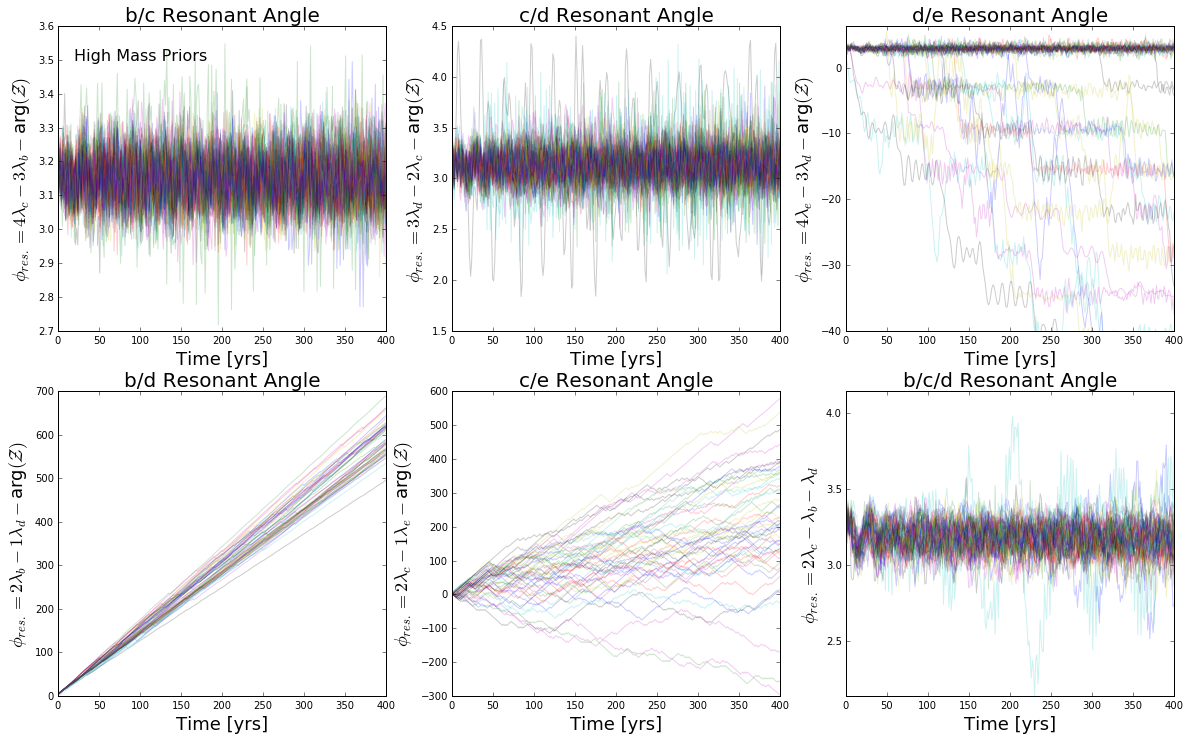}
\caption{
Resonant angle time evolution for  Kepler-223 system.
The top two rows show the time evolution for initial conditions drawn from the default posterior.
The bottom two rows show the time evolution of the same angles for initial conditions drawn form the high mass posterior.
Vertical axes are extended beyond $\pm \pi$ to account for windings of resonant angles that circulate.
}
	\label{fig:kep223res_ang}
	\end{center}
\end{figure}

{\it Kepler-223 b*,c*,d*,e} (Fig. \ref{fig:kep223cons}):
	The periods of the Kepler-223 planets place each successive adjacent pair of planets in or near a first-order MMR 
	with a 4:3 MMR between b and c, a 3:2 MMR between c ad d and a 4:3 MMR between d and e.
	We test for resonant angle librations in the Kepler-223 system with $N$-body integrations.
	The top row of Figure \ref{fig:kep223res_ang} plots the resonant angles of each adjacent planet pair. 
 	The resonant angles of the pairs b/c and c/d librate over the entire 400 year integration.
	The d/e resonant angle librates for 23 of the 50 simulations, while in the other 27 simulations it alternates 
	chaotically between libration and circulation.
	The left two panels of the bottom row show the 2:1 MMR resonant for planets b/d and c/e respectively, 
	both of which circulate.
	Finally the bottom right panel of Figure \ref{fig:kep223res_ang} shows that the 
	three-body resonant angle $2\lambda_c-\lambda_b-\lambda_d$ librates about $\pi$ all initial conditions.
	We do not apply the analytic TTV formulae to the Kepler-223 system since they are not expected to accurately represent
	the TTVs of planets librating in resonance.	
	
	\citet{Mills:2016gi} 
	infer masses $[m_b,m_c,m_d,m_e]=[7.4^{+1.3}_{-1.1},5.1^{+1.7}_{-1.1},8.0^{+1.5}_{-1.3},4.8^{+1.4}_{-1.2}]M_\earth$
	from a TTV analysis of the Kepler-223 system's Q1-17 transit times.
	Their results roughly agree with ours; the measurement of $m_c$ is the most in tension ($2.4\sigma$).
	Furthermore, our TTV fit gives only an upper bound for $m_e$, but the upper limit from the high-mass priors
	fit is consistent with the $m_e$ fit by \citet{Mills:2016gi}.

\begin{figure}[htbp]
\begin{center}
	\includegraphics[width=0.45\textwidth]{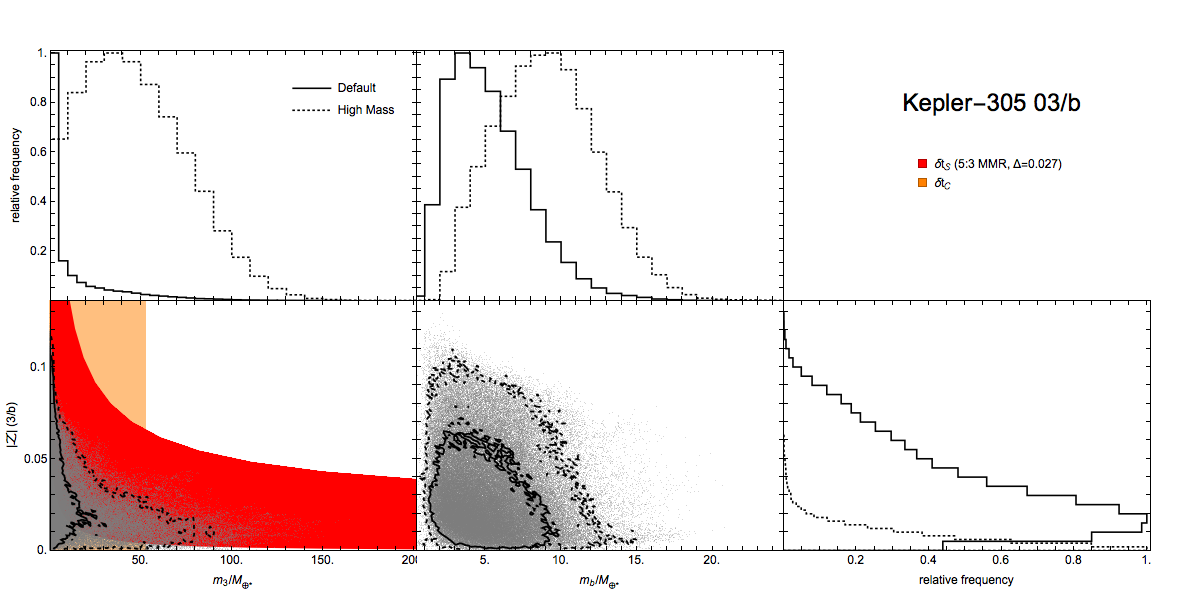}
	\includegraphics[width=0.45\textwidth]{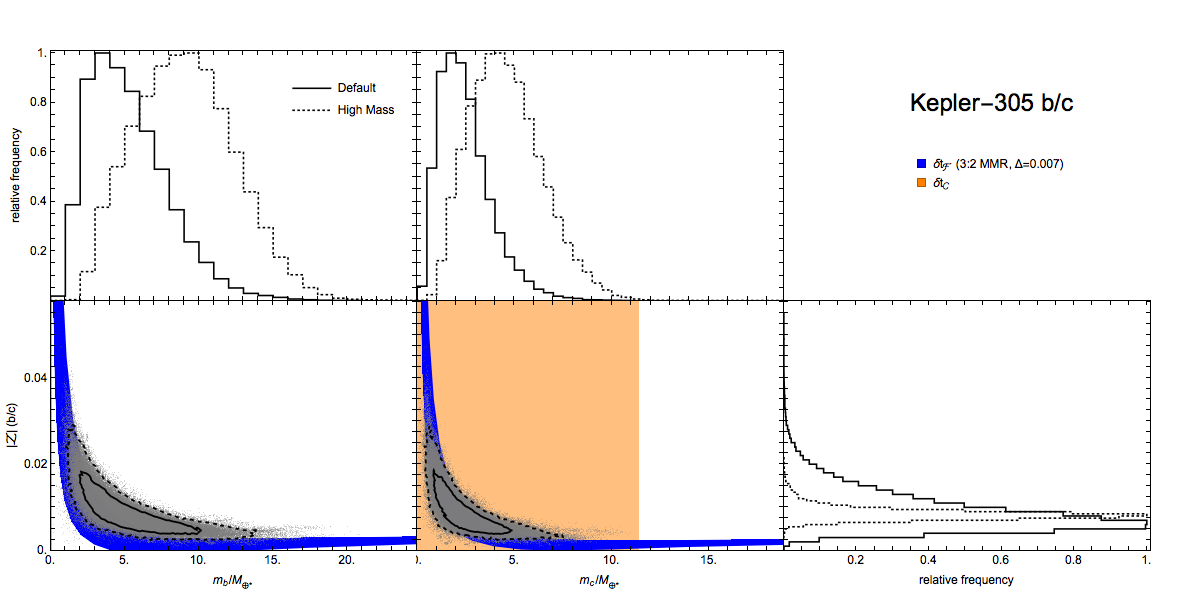}
	\includegraphics[width=0.45\textwidth]{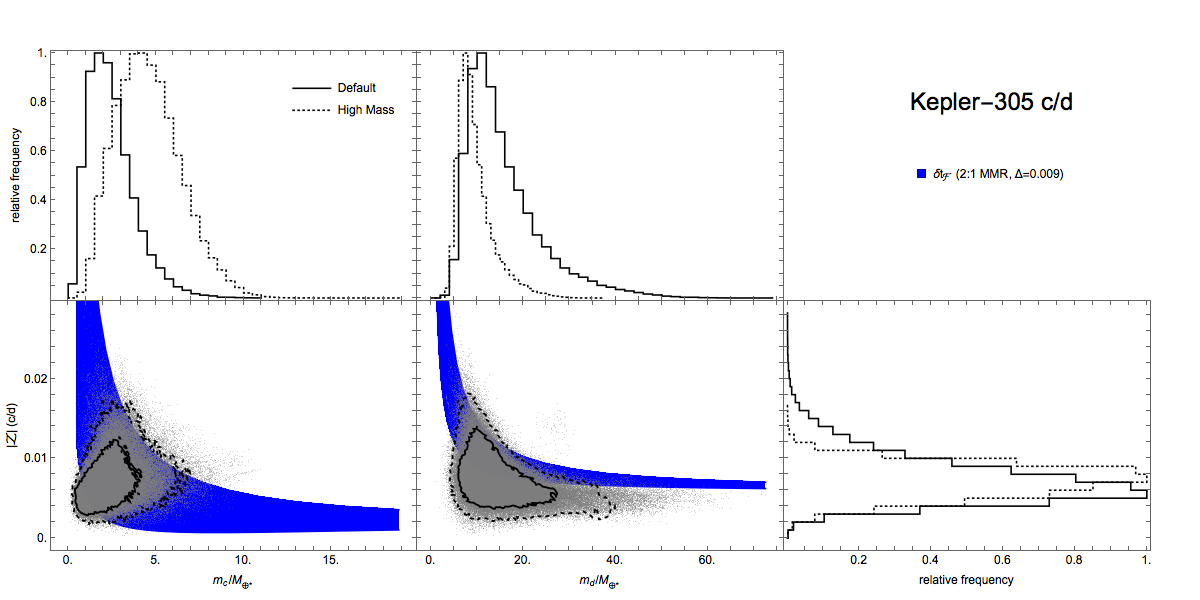}
\caption{
Constraint plots for Kepler-305 (see Figure \ref{fig:kep11cons} for description). 
The combined eccentricity of the b/c pair has been corrected by subtracting the forced component.
}
	\label{fig:kep305cons}
	\end{center}
\end{figure}

\begin{figure}[htbp]
\begin{center}
	\includegraphics[width=0.75\textwidth]{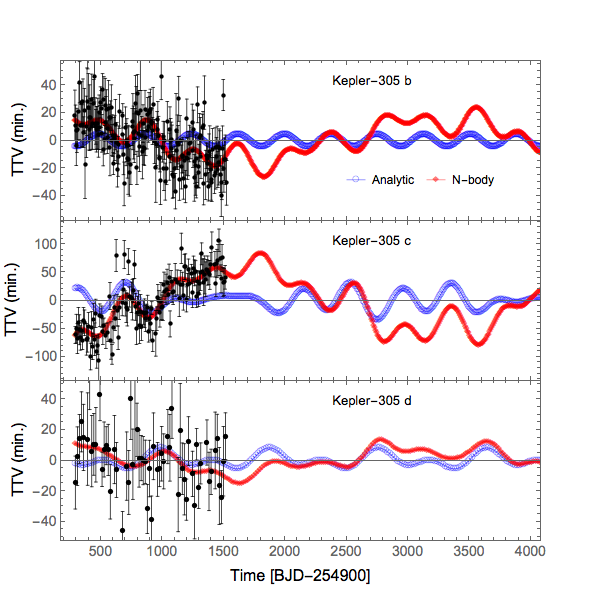}
\caption{
TTVs the outer three planets of the Kepler-305 system. The observed data are shown as black points.
The best fit $N$-body and analytic solutions are shown in red and blue, respectively, and are plotted beyond the
time baseline covered by the observed data.
The $N$-body solution shows a long-term variability not captured by the best-fit analytic solution.
 }
	\label{fig:kep305ttv}
	\end{center}
\end{figure}

{\it Kepler-305 03,b,c*,d*} (Fig. \ref{fig:kep305cons}):
	Kepler-305 b and c are near a 3:2 MMR and  planets c and d are near  2:1 MMR.
	The inferred masses and eccentricities agree well with the constraints 
	of the measured fundamental TTVs of planets b,c, and d.
	The mass of planets c and d are classified as robust based on the $N$-body MCMC results, though the
	analytic constraints shown in Figure \ref{fig:kep305cons} do not reveal what is responsible for breaking the 
	mass-eccentricity degeneracy.
	Figure \ref{fig:kep305ttv} shows that the $N$-body TTV solution exhibits a clear long-term variability not captured by the analytic model.
	We are unable to identify the origin of this variability after searching unsuccessfully for higher order MMRs and 3-body resonances among the
	planets. 
	Both the b/c and c/d pairs are fairly close to resonance ($\Delta=0.007$ and $\Delta=0.009$, respectively)
	and the variability could possibly be caused by resonance effects not captured in the analytic model.
	We speculate that the dynamical origin of the long-term variability breaks the mass-eccentricity degeneracy for this system.

\subsection{\ref{sec:Degenerate}: Degenerate Systems}
\label{sec:Degenerate}
\begin{figure}[htbp]
 	\begin{center}
 	\includegraphics[width=0.9\textwidth]{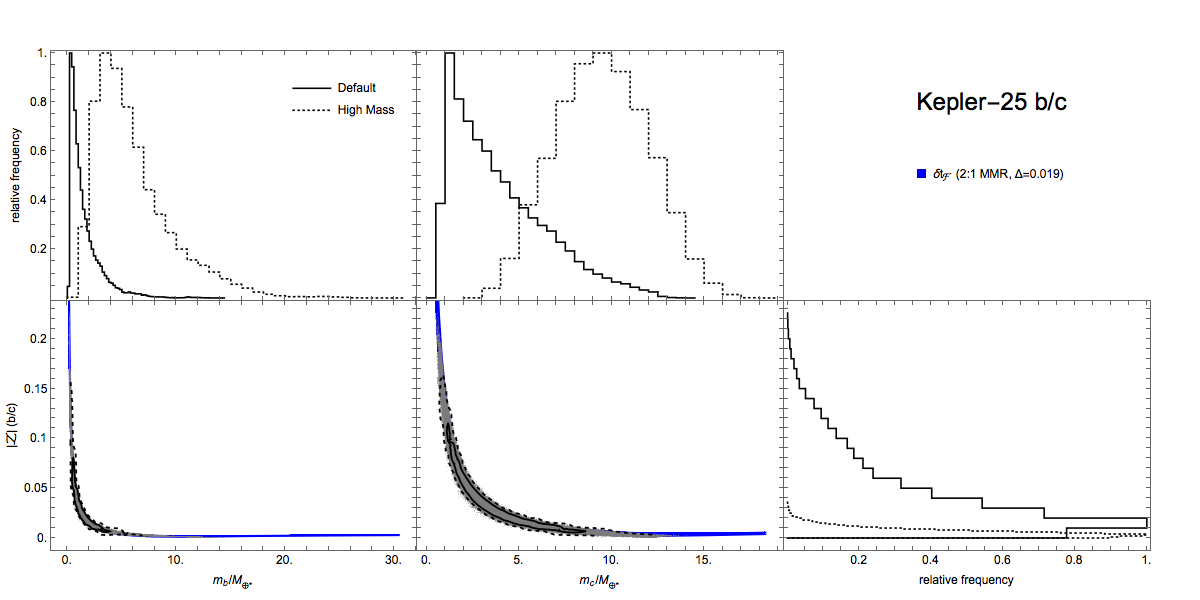}
 \caption{
 Constraint plots for Kepler-25 (see Figure \ref{fig:kep11cons} for description). 
 The purple region shows the 68\% confidence regions derived from an MCMC simulation using the analytic model 
 with priors based on RV mass measurements \citep{Marcy:2014hr}.
 }
 	\label{fig:kep25cons}
 	\end{center}
 \end{figure}

\begin{figure}[htbp]
 	\begin{center}
 	\includegraphics[width=0.9\textwidth]{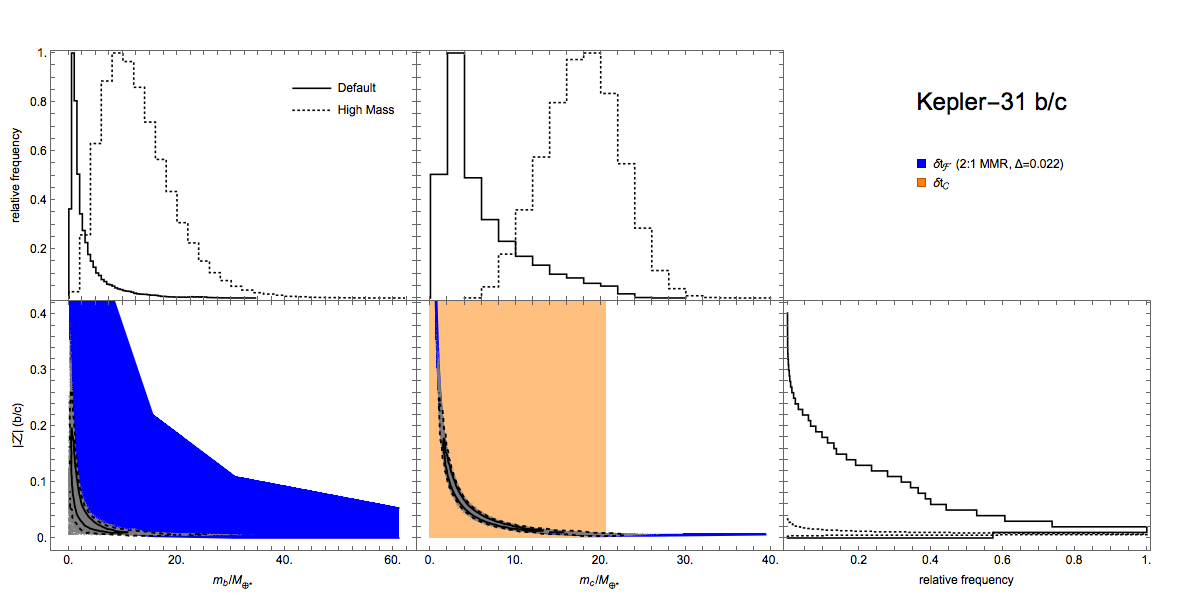}
 	\includegraphics[width=0.9\textwidth]{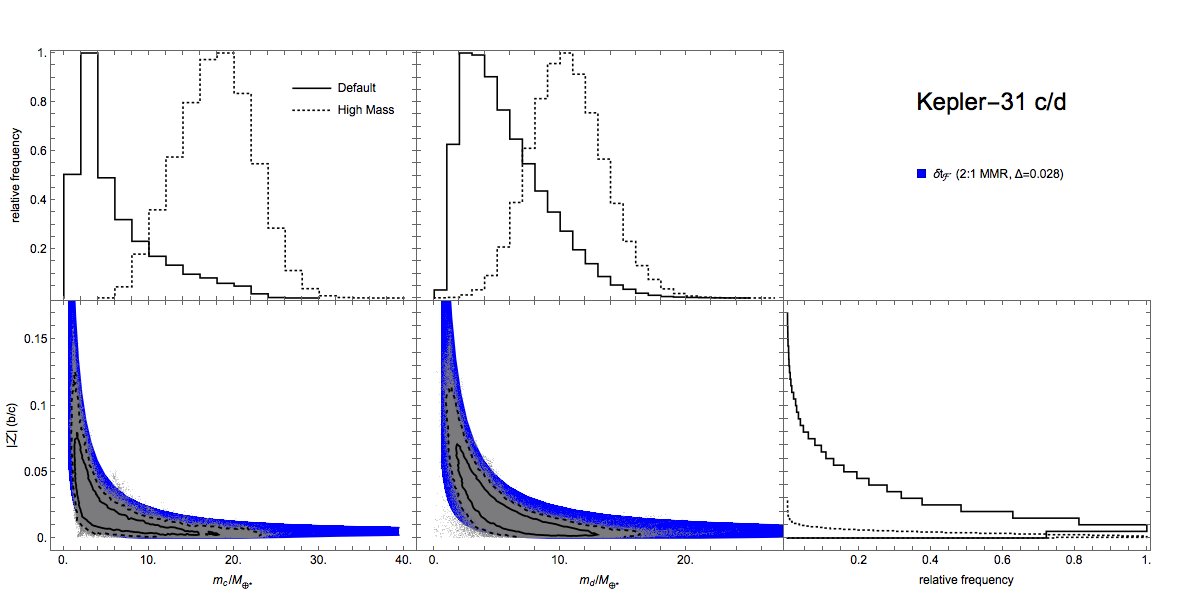}
 \caption{
 Constraint plots for Kepler-31 (see Figure \ref{fig:kep11cons} for description).
 }
 	\label{fig:kep31cons}
 	\end{center}
 \end{figure}

\begin{figure}[htbp]
 	\begin{center}
 	\includegraphics[width=0.9\textwidth]{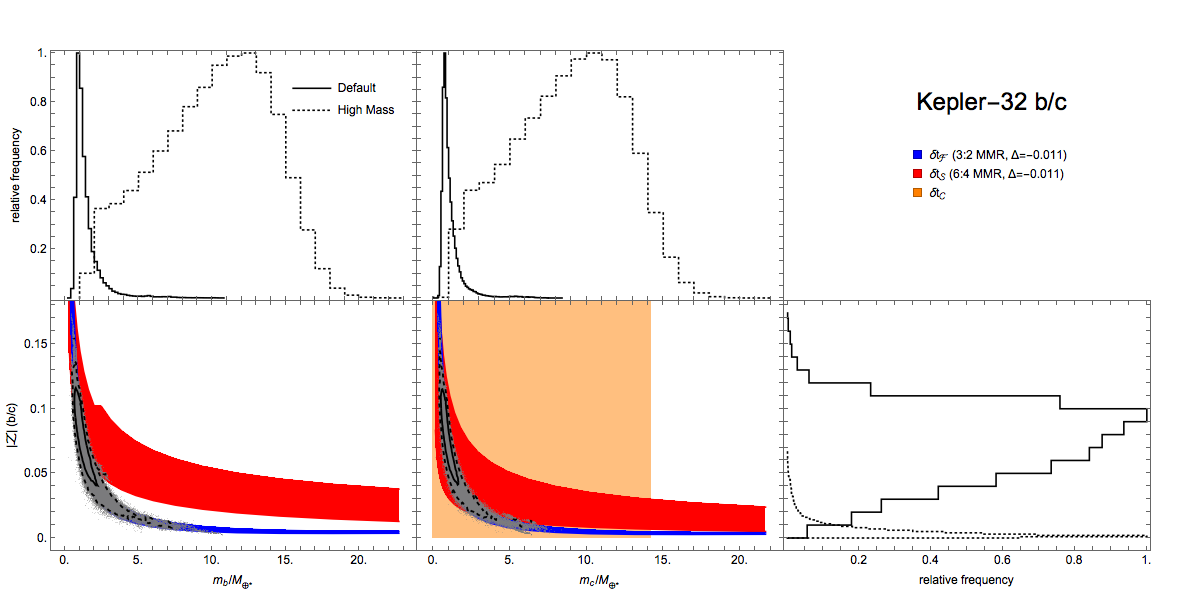}
 \caption{
 Constraint plots for Kepler-32 (see Figure \ref{fig:kep11cons} for description). 
 }
 	\label{fig:kep32cons}
 	\end{center}
 \end{figure}

\begin{figure}[htbp]
	\begin{center}
    \includegraphics[width=0.9\textwidth]{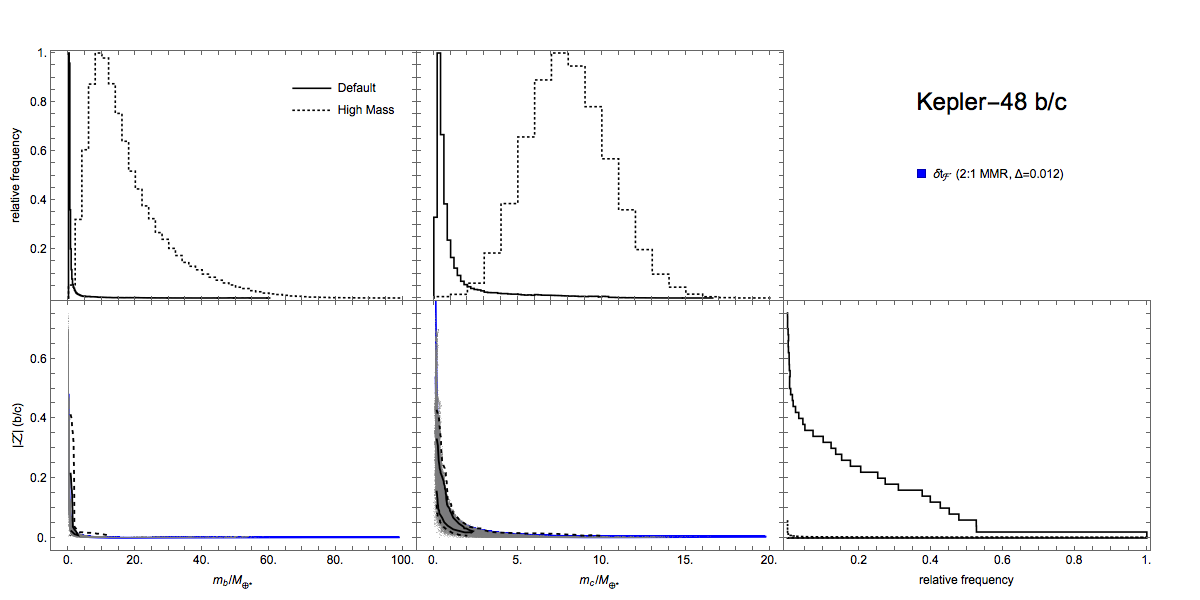}
\caption{
Constraint plots for Kepler-48 (see Figure \ref{fig:kep11cons} for description).
}
\label{fig:kep48cons}
\end{center}
\end{figure}

\begin{figure}[htbp]
	\begin{center}
    \includegraphics[width=0.9\textwidth]{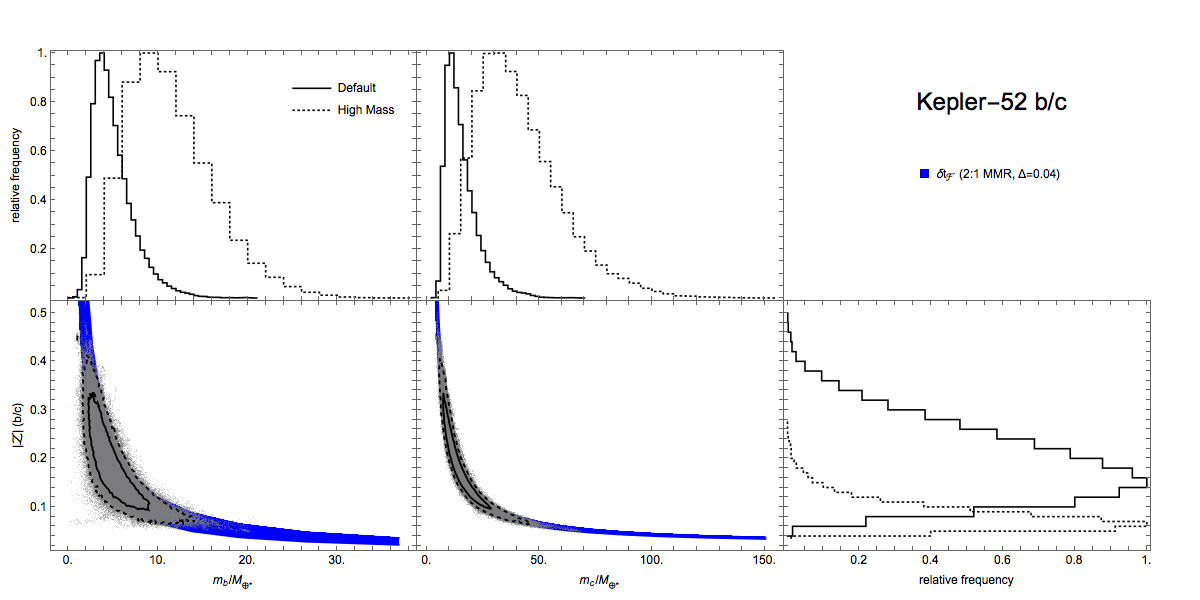}
	\includegraphics[width=0.9\textwidth]{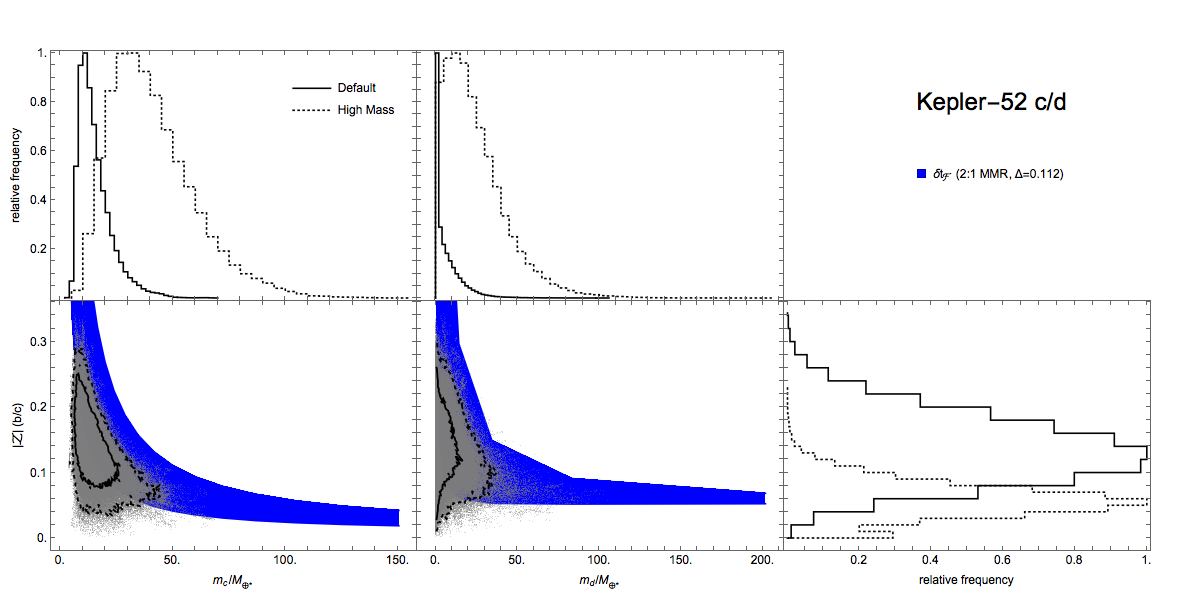}
\caption{
Constraint plots for Kepler-52 (see Figure \ref{fig:kep11cons} for description).
}
\label{fig:kep52cons}
\end{center}
\end{figure}

\begin{figure}[htbp]
 	\begin{center}
 	\includegraphics[width=0.9\textwidth]{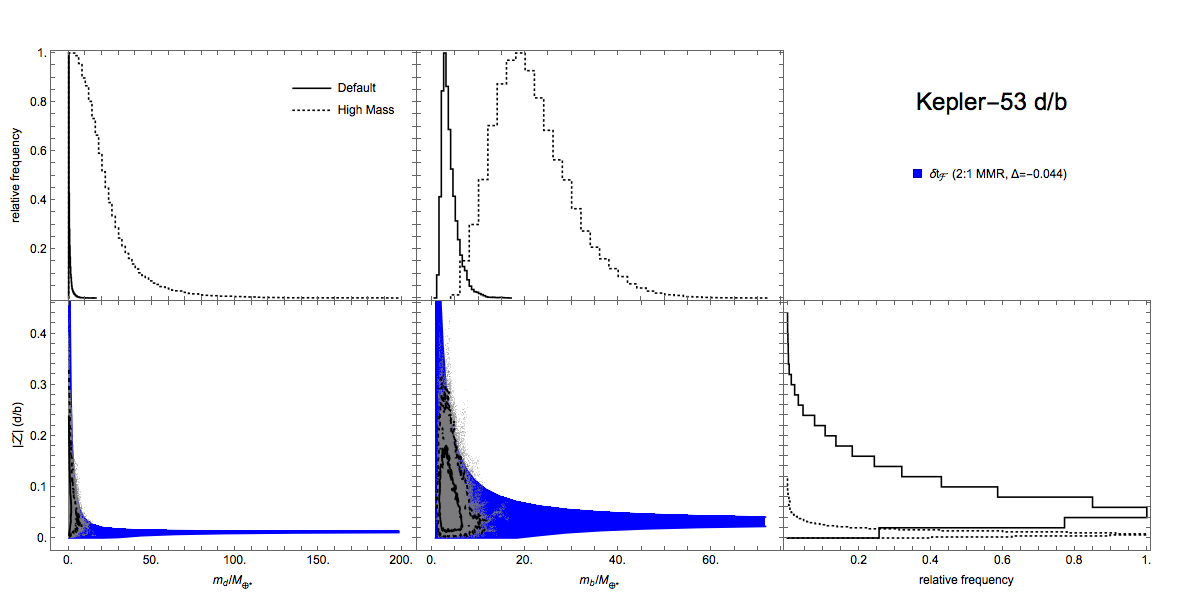}
	 \includegraphics[width=0.9\textwidth]{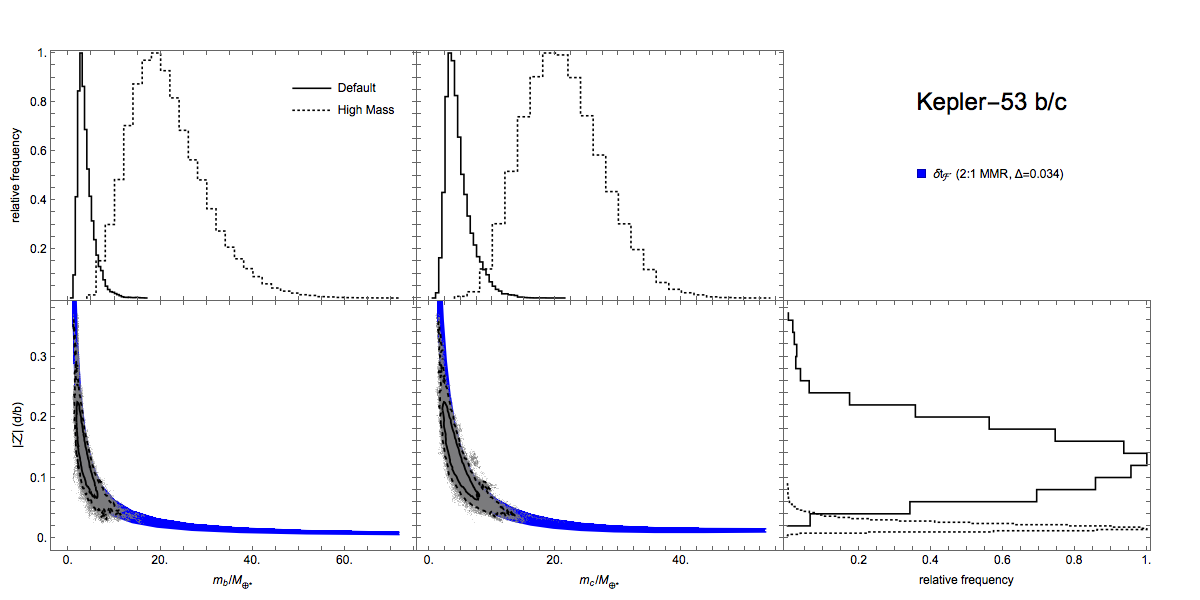}
 \caption{
 Constraint plots for Kepler-53 (see Figure \ref{fig:kep11cons} for description). 
 }
 	\label{fig:kep53cons}
 	\end{center}
 \end{figure}

\begin{figure}[htbp]
 	\begin{center}
 	\includegraphics[width=0.9\textwidth]{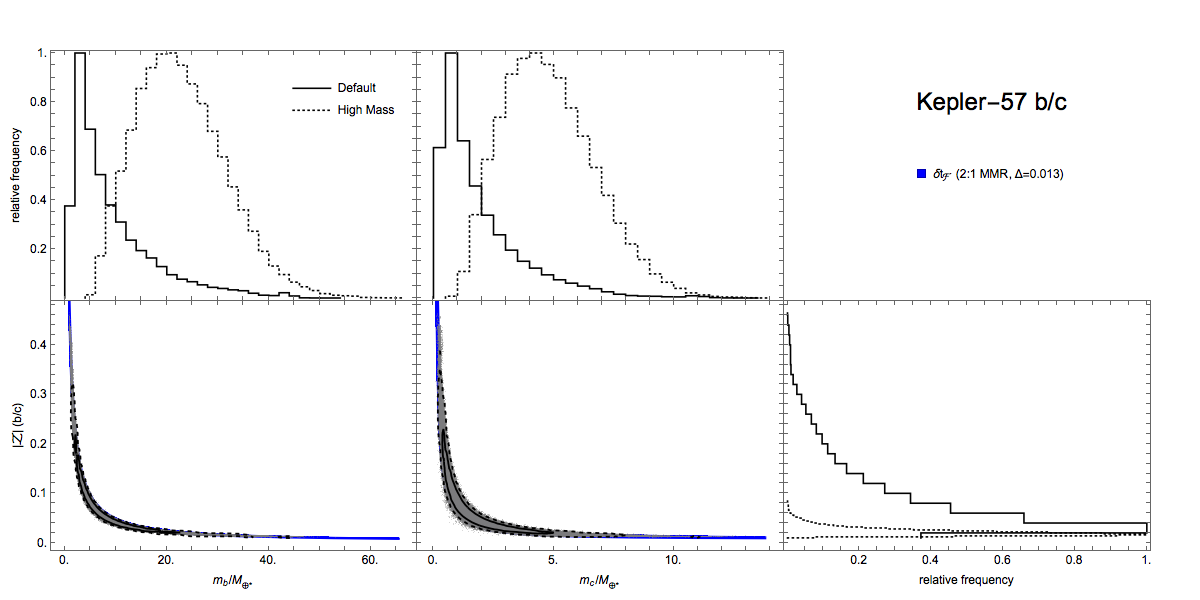}
 \caption{
 Constraint plots for Kepler-57 (see Figure \ref{fig:kep11cons} for description). 
 }
 	\label{fig:kep57cons}
 	\end{center}
 \end{figure}

\begin{figure}[htbp]
 	\begin{center}
 	\includegraphics[width=0.9\textwidth]{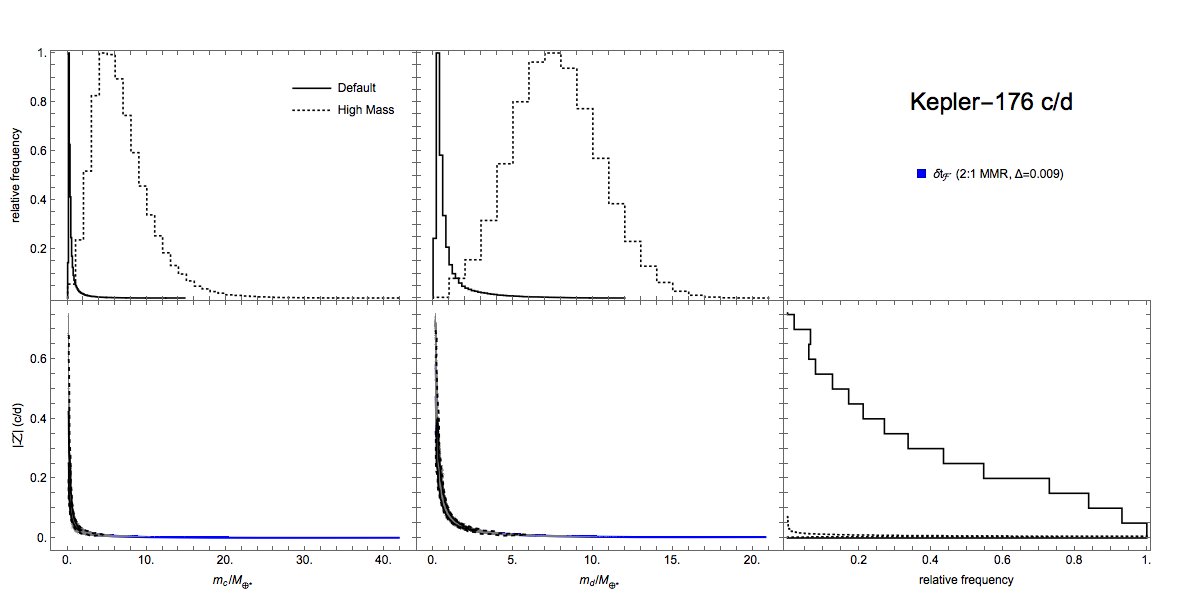}
 \caption{
 Constraint plots for Kepler-176 (see Figure \ref{fig:kep11cons} for description). 
 }
 	\label{fig:kep176cons}
 	\end{center}
 \end{figure}

\begin{figure}[htbp]
 	\begin{center}
	 \includegraphics[width=0.45\textwidth]{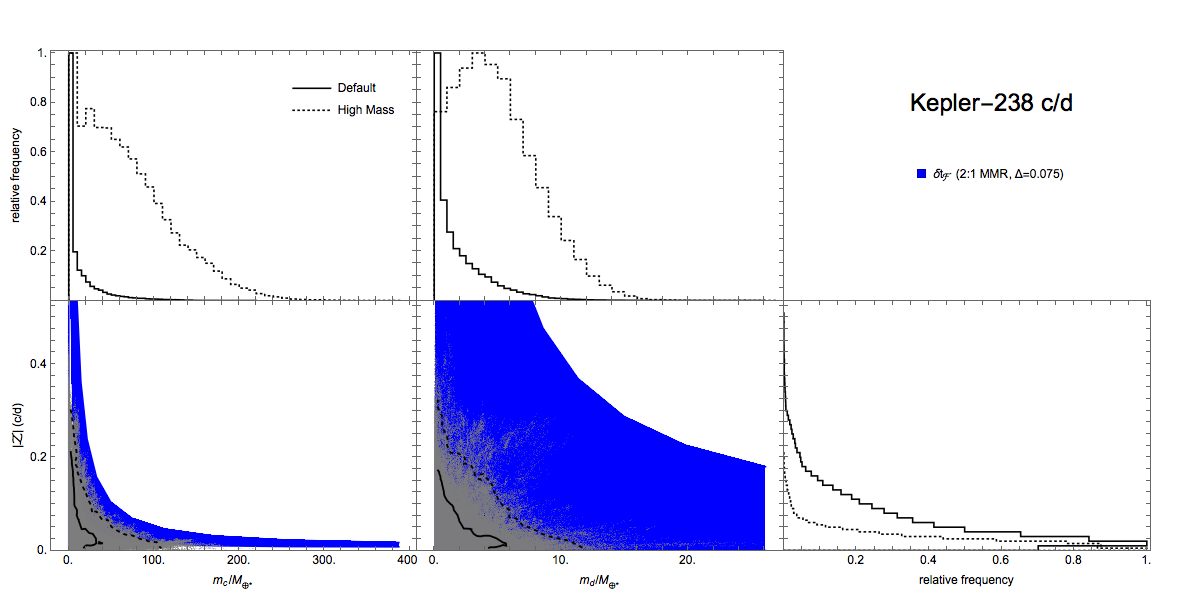}
	 \includegraphics[width=0.45\textwidth]{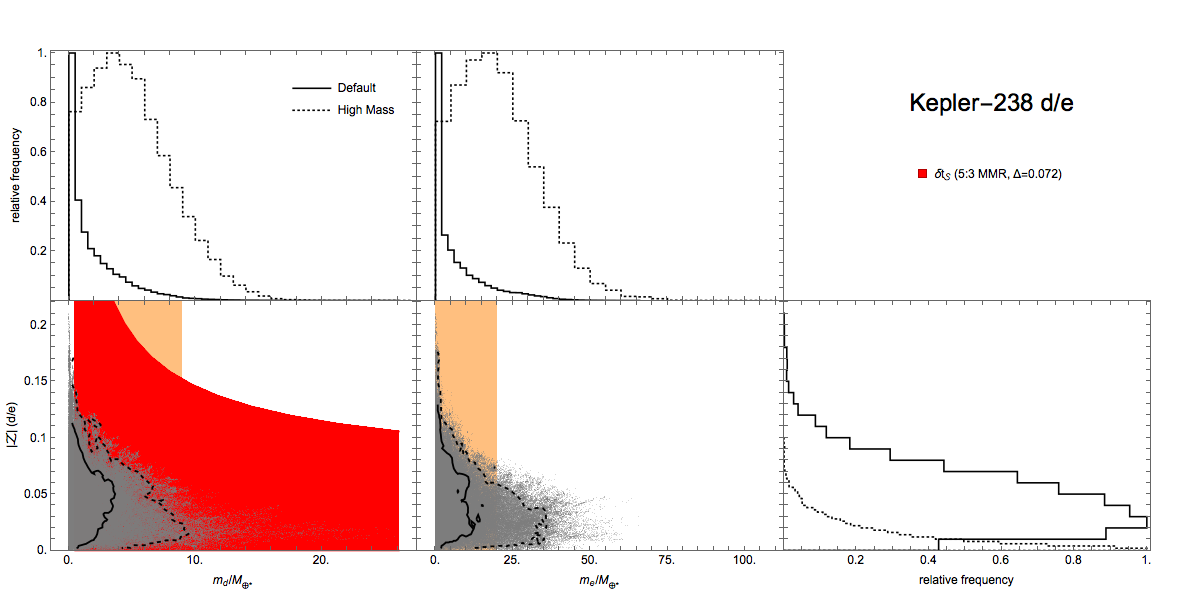}
 	\includegraphics[width=0.45\textwidth]{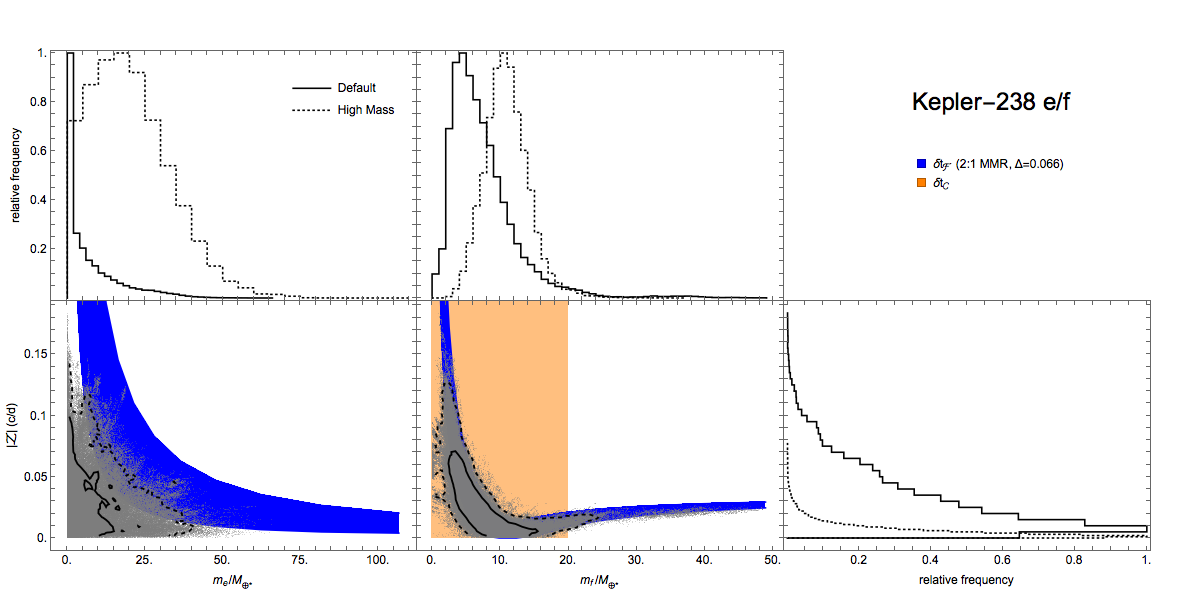}
 \caption{
 Constraint plots for Kepler-238 (see Figure \ref{fig:kep11cons} for description). 
 The combined eccentricity of the e/f pair has been corrected by subtracting the forced component.
 }
 	\label{fig:kep238cons}
 	\end{center}
 \end{figure}

\begin{figure}[htbp]
 	\begin{center}
 	\includegraphics[width=0.9\textwidth]{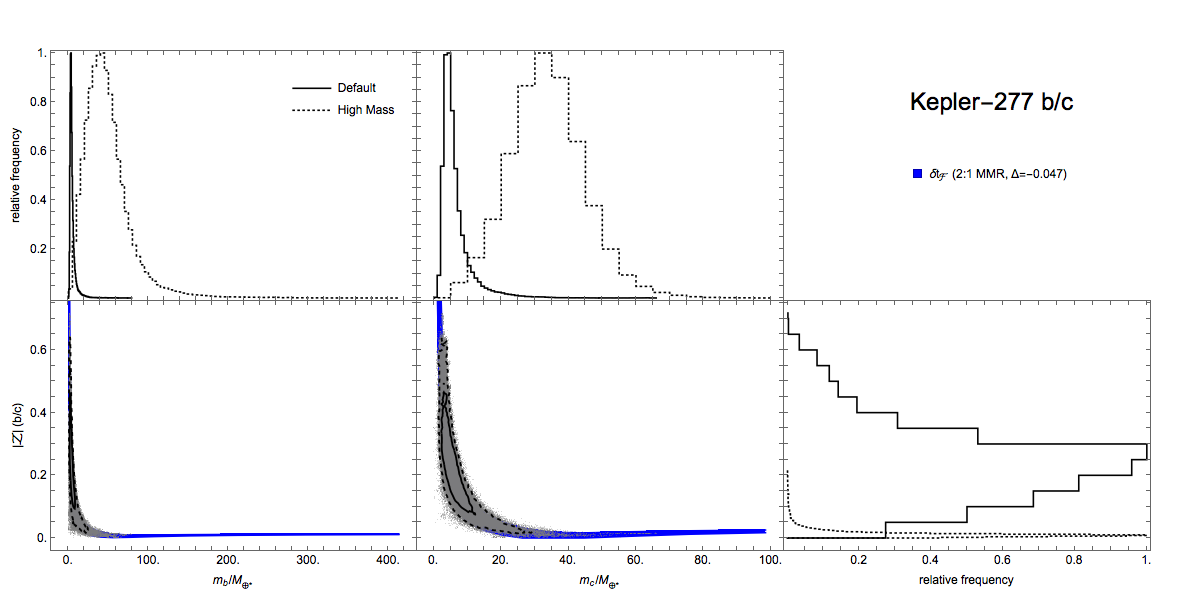}
 \caption{
 Constraint plots for Kepler-277 (see Figure \ref{fig:kep11cons} for description). 
 }
 	\label{fig:kep277cons}
 	\end{center}
 \end{figure}

\begin{figure}[htbp]
 	\begin{center}
 	\includegraphics[width=0.9\textwidth]{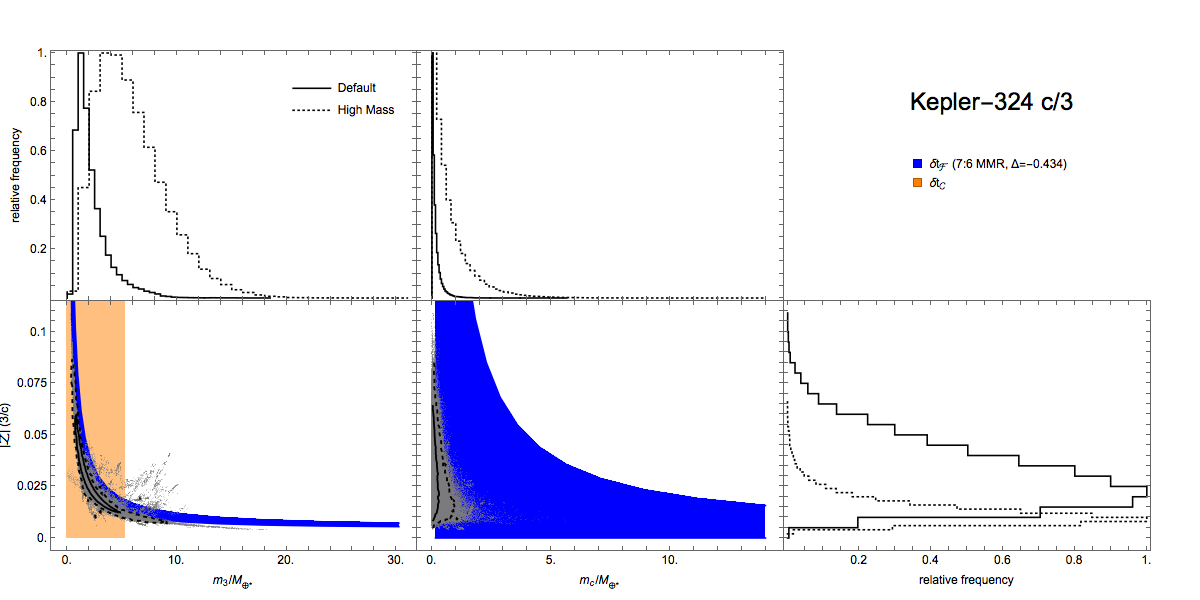}
 \caption{
 Constraint plot for Kepler-324 (see Figure \ref{fig:kep11cons} for description). 
 }
 	\label{fig:kep324cons}
 	\end{center}
 \end{figure}

\begin{figure}[htbp]
 	\begin{center}
 	\includegraphics[width=0.9\textwidth]{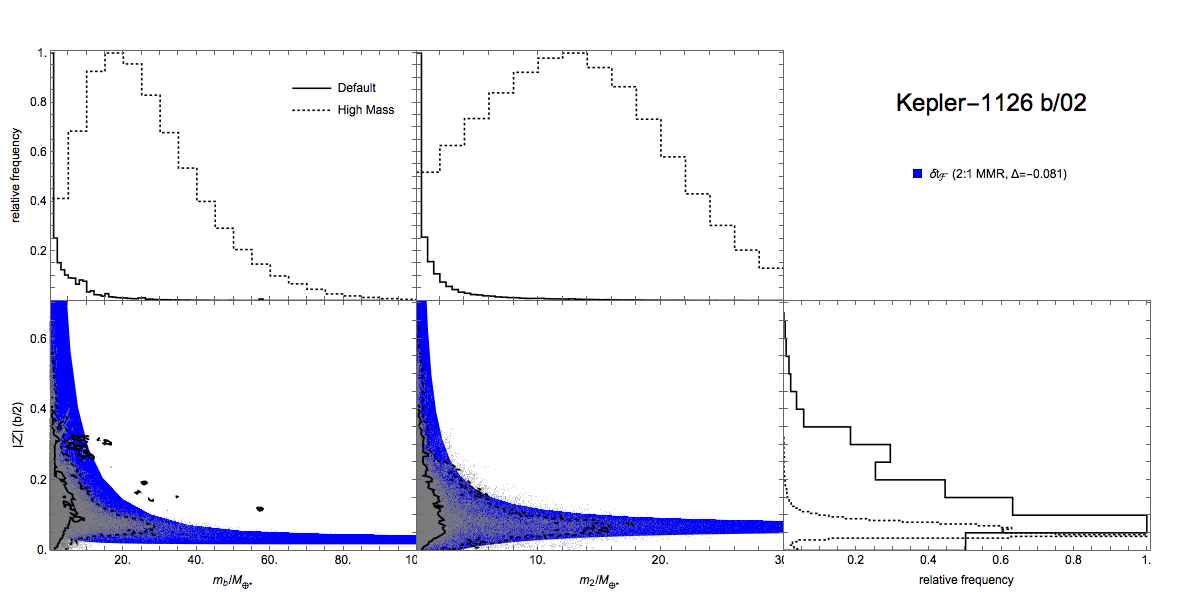}
 \caption{
 Constraint plot for Kepler-1126 (see Figure \ref{fig:kep11cons} for description). 
 }
 	\label{fig:kep1126cons}
 	\end{center}
 \end{figure}

A number of the TTV systems we analyze exhibit strong mass-eccentricity degeneracy.
The constraint plots are shown for each degenerate system:
Kepler-25, 
 Kepler-31, 
 Kepler-32, 
 Kepler-48, 
  Kepler-52, 
 Kepler-53, 
 Kepler-57, 
 Kepler-176, 
 Kepler-238, 
 Kepler-277, 
Kepler-324, 
and Kepler-1126 
in Figures \ref{fig:kep25cons}--\ref{fig:kep1126cons}.

\citet{JontofHutter:2016ch} also analyze the TTVs of Kepler-57 b and c and similarly find that the planets' masses and eccentricities
are poorly constrained by the TTVs.

\begin{figure}[htbp]
\begin{center}
 \includegraphics[width=0.45\textwidth]{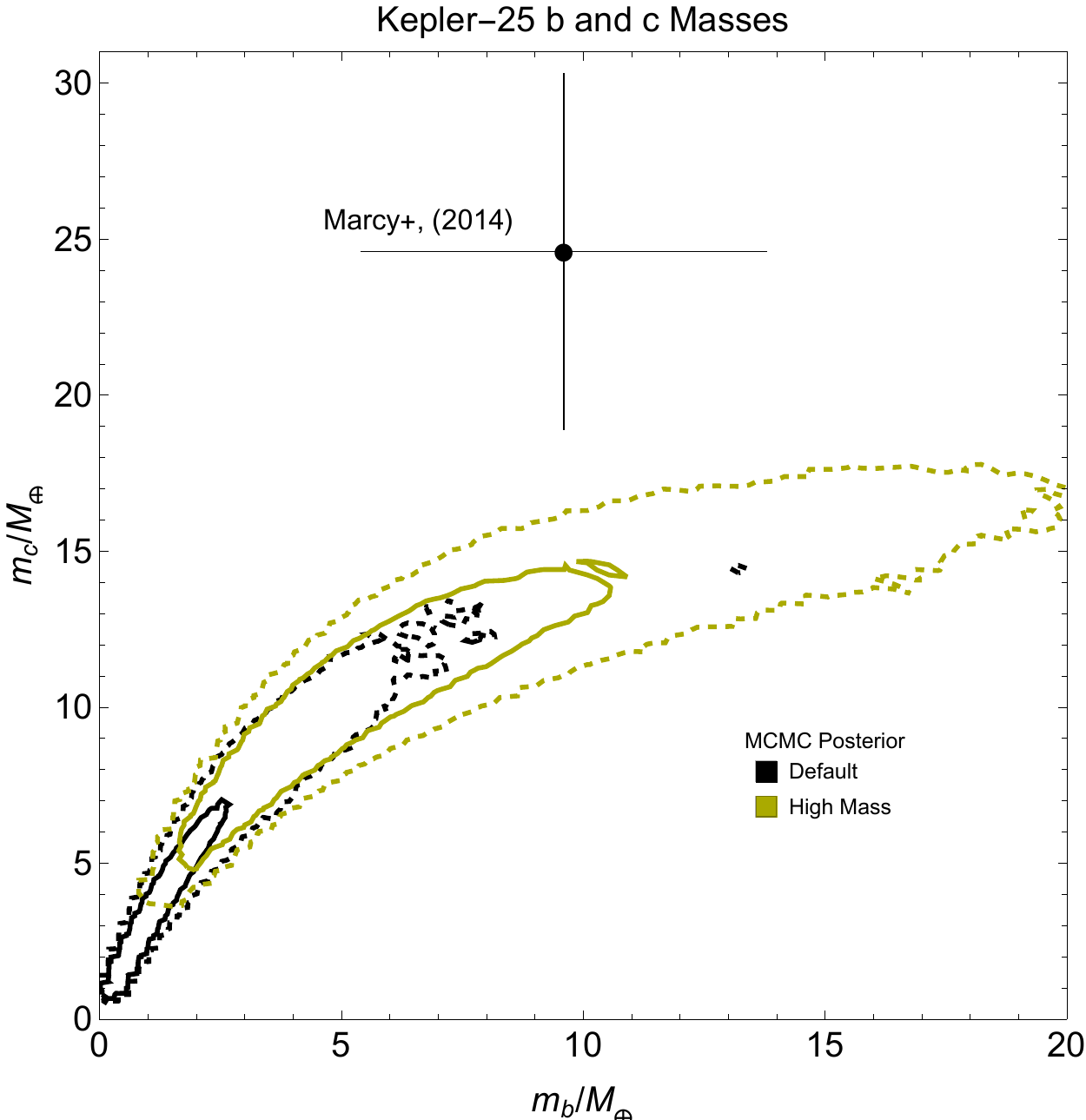}
 \includegraphics[width=0.45\textwidth]{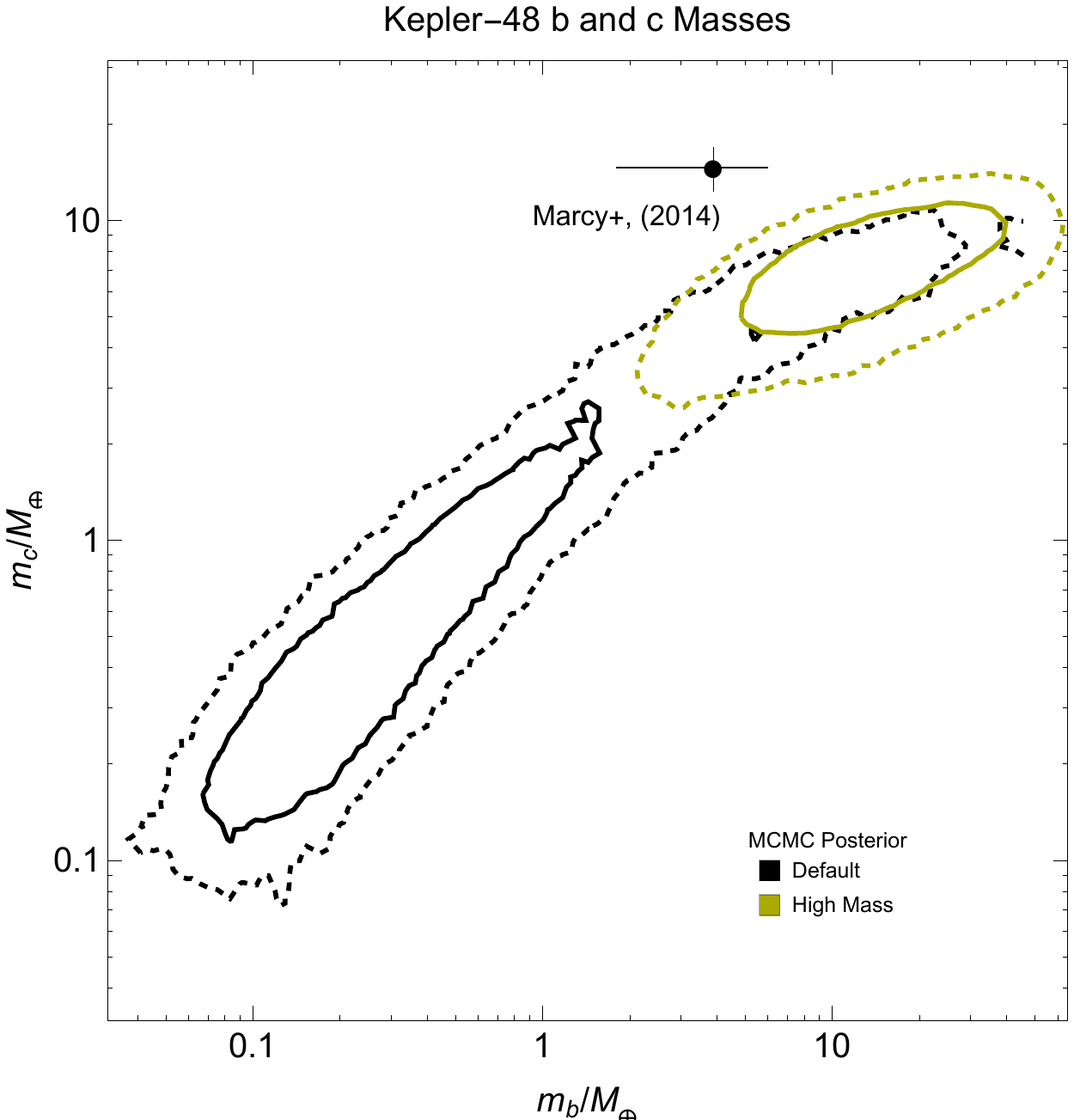}
\caption{
Contours show 68\%  (solid) and and 95\% (dashed) credible regions in planet mass from the default (black) and high mass (yellow)
$N$-body MCMC posteriors for Kepler-25 and Kepler-48.
The black points with error bars show the RV measurements of \citet{Marcy:2014hr}.
The planet-star mass ratios from the $N$-body MCMC posteriors have been multiplied by the best-fit stellar masses, $M_*=1.19  M_\sun$ (Kepler-25)
and $M_*=0.88 M_\sun $ (Kepler-48), found by \citet{Marcy:2014hr}.
}
\label{fig:rv_versus_ttv}
\end{center}
\end{figure}

\citet{Marcy:2014hr} measure RV planet masses for two of the degenerate systems, 
Kepler-25 b and c: $[m_b,m_c]$=$[9.6\pm4.2,24.6\pm5.7]M_\earth$
and Kepler-48 b and c: $[m_b,m_c]$=$[3.9\pm2.1,14.6\pm2.3]M_\earth$.
\citet{Marcy:2014hr}'s mass measurements are inconsistent with the TTVs of both the Kepler-25 and Kepler-48 systems.
Despite the fact that the TTVs suffer a strong mass-eccentricity degeneracies, 
the fundamental TTVs constrain the planets' mass ratios in both systems.
Figure \ref{fig:rv_versus_ttv} compares the RV mass measurements of \citet{Marcy:2014hr} with the $N$-body MCMC posteriors for Kepler-25 and Kepler-48 systems.
In both systems, \citet{Marcy:2014hr} find outer planet masses that are larger than the TTV observations support.  
\citet{Marcy:2014hr} report additional planets at distant orbital separations in both the Kepler-25 and Kepler-48 systems.
Kepler-25 hosts a non-transiting $\sim90~M_\earth$ planet with a period of $123$ days and
Kepler-48 hosts an additional transiting $\sim8~M_\earth$ planet on a $\sim$43 day orbit
as well as non-transiting $\sim 2~M_J$ planet  on a $\sim 1000$ day orbit.
We confirm via $N$-body calculations that these additional planets have negligible effects on the TTVs of Kepler-25 b/c and Kepler-48 b/c, as expected given their large orbital separations.

\clearpage


%

%


 \clearpage
 

%

%

\clearpage

\clearpage


\clearpage 

\clearpage

  \clearpage



%
%


\end{document}

%% file: masstable.tex
Kepler-9 b$^{*}$ & 19.243 & $8.2^{+1.0}_{-0.7}$ & $1.0^{+0.1}_{-0.1}$ & $43.5^{+2.7}_{-3.3}$ & $0.4^{+0.1}_{-0.1}$ & $43.4^{+2.7}_{-3.2}$ & $0.4^{+0.1}_{-0.1}$& \citetalias{Holman:2010db,2014arXiv1403.1372D,Borsato:2014it} \\ 
Kepler-9 c$^{*}$ & 38.969 & $8.3^{+0.8}_{-0.9}$ & ... & $29.9^{+1.8}_{-2.3}$ & $0.3^{+0.1}_{-0.1}$ & $29.9^{+1.9}_{-2.2}$ & $0.3^{+0.1}_{-0.1}$& \citetalias{Holman:2010db,2014arXiv1403.1372D,Borsato:2014it} \\ 
\hline 
Kepler-11 b & 10.304 & $1.9^{+0.1}_{-0.1}$ & $0.9^{+0.1}_{-0.1}$ & $0.7^{+0.3}_{-0.2}$ & $0.6^{+0.3}_{-0.2}$ & $1.2^{+0.6}_{-0.5}$ & $1.0^{+0.5}_{-0.5}$& \citetalias{Lissauer:2011el,2013ApJ...770..131L,Borsato:2014it} \\ 
Kepler-11 c & 13.025 & $3.0^{+0.2}_{-0.2}$ & ... & $1.8^{+0.9}_{-0.5}$ & $0.4^{+0.2}_{-0.1}$ & $3.4^{+1.4}_{-1.5}$ & $0.7^{+0.4}_{-0.3}$& \citetalias{Lissauer:2011el,2013ApJ...770..131L,Borsato:2014it} \\ 
Kepler-11 d$^{*}$ & 22.687 & $3.3^{+0.2}_{-0.2}$ & ... & $6.8^{+0.7}_{-0.8}$ & $1.0^{+0.2}_{-0.2}$ & $6.9^{+0.8}_{-0.8}$ & $1.0^{+0.3}_{-0.2}$& \citetalias{Lissauer:2011el,2013ApJ...770..131L,Borsato:2014it} \\ 
Kepler-11 e$^{*}$ & 31.995 & $4.0^{+0.2}_{-0.3}$ & ... & $6.7^{+1.2}_{-1.0}$ & $0.6^{+0.2}_{-0.1}$ & $7.2^{+1.1}_{-1.0}$ & $0.7^{+0.1}_{-0.1}$& \citetalias{Lissauer:2011el,2013ApJ...770..131L,Borsato:2014it} \\ 
Kepler-11 f$^{*}$ & 46.686 & $2.6^{+0.2}_{-0.2}$ & ... & $1.7^{+0.5}_{-0.4}$ & $0.5^{+0.2}_{-0.2}$ & $1.9^{+0.5}_{-0.4}$ & $0.6^{+0.2}_{-0.2}$& \citetalias{Lissauer:2011el,2013ApJ...770..131L,Borsato:2014it} \\ 
\hline 
Kepler-18 c & 7.642 & $5.0^{+0.3}_{-0.3}$ & $0.9^{+0.1}_{-0.02}$ & $12.9^{+5.6}_{-6.6}$ & $0.5^{+0.3}_{-0.3}$ & $21.6^{+3.2}_{-4.0}$ & $1.0^{+0.2}_{-0.3}$& \citetalias{Cochran:2011kg} \\ 
Kepler-18 d$^{*}$ & 14.859 & $6.0^{+0.4}_{-0.4}$ & ... & $14.9^{+1.8}_{-4.2}$ & $0.3^{+0.1}_{-0.1}$ & $16.2^{+1.3}_{-1.5}$ & $0.4^{+0.1}_{-0.1}$& \citetalias{Cochran:2011kg} \\ 
\hline 
Kepler-23 b & 7.107 & $1.8^{+0.1}_{-0.1}$ & $1.0^{+0.1}_{-0.1}$ & $1.3^{+1.3}_{-0.5}$ & $1.2^{+1.4}_{-0.5}$ & $4.7^{+1.9}_{-1.9}$ & $4.2^{+2.3}_{-1.6}$&-- \\ 
Kepler-23 c & 10.742 & $3.2^{+0.2}_{-0.2}$ & ... & $2.2^{+2.8}_{-0.9}$ & $0.3^{+0.5}_{-0.1}$ & $9.1^{+3.4}_{-3.9}$ & $1.3^{+0.8}_{-0.4}$&-- \\ 
Kepler-23 d & 15.274 & $2.3^{+0.1}_{-0.1}$ & ... & $<2.4$ & $<1.1$ & $4.9^{+3.6}_{-3.5}$ & $2.1^{+1.8}_{-1.5}$&-- \\ 
\hline 
Kepler-24 b & 8.145 & $2.7^{+0.6}_{-0.5}$ & $0.9^{+0.1}_{-0.1}$ & $2.0^{+1.6}_{-0.7}$ & $0.5^{+0.7}_{-0.3}$ & $7.5^{+2.3}_{-2.4}$ & $1.6^{+1.8}_{-1.0}$&-- \\ 
Kepler-24 c\tnote{a} & 12.333 & $3.2^{+1.7}_{-1.8}$ & ... & $1.7^{+1.2}_{-0.7}$ & $0.1^{+0.8}_{-0.1}$ & $5.2^{+2.0}_{-1.4}$ & $0.2^{+2.0}_{-0.2}$&-- \\ 
Kepler-24 e & 18.999 & $2.6^{+0.5}_{-0.5}$ & ... & $<0.7$ & $<0.2$ & $6.8^{+7.1}_{-6.4}$ & $0.3^{+4.6}_{-0.3}$&-- \\ 
\hline 
Kepler-25 b & 6.239 & $2.5^{+0.1}_{-0.1}$ & $1.1^{+0.1}_{-0.1}$ & $0.4^{+1.5}_{-0.2}$ & $0.1^{+0.5}_{-0.1}$ & $4.0^{+4.0}_{-2.0}$ & $1.3^{+1.4}_{-0.6}$& \citetalias{Marcy:2014hr} \\ 
Kepler-25 c & 12.720 & $4.4^{+0.2}_{-0.2}$ & ... & $1.4^{+4.0}_{-0.6}$ & $0.1^{+0.3}_{-0.05}$ & $10.0^{+3.5}_{-2.5}$ & $0.7^{+0.2}_{-0.2}$& \citetalias{Marcy:2014hr} \\ 
\hline 
Kepler-26 b & 12.283 & $2.9^{+0.2}_{-0.2}$ & $0.5^{+0.04}_{-0.03}$ & $5.1^{+0.6}_{-0.7}$ & $1.1^{+0.2}_{-0.2}$ & $4.2^{+0.6}_{-0.6}$ & $1.0^{+0.2}_{-0.2}$& \citetalias{Hadden:2016ki,JontofHutter:2016ch} \\ 
Kepler-26 c$^{*}$ & 17.251 & $2.5^{+0.2}_{-0.2}$ & ... & $6.1^{+0.7}_{-0.7}$ & $2.1^{+0.5}_{-0.4}$ & $5.6^{+0.8}_{-0.6}$ & $1.9^{+0.5}_{-0.4}$& \citetalias{Hadden:2016ki,JontofHutter:2016ch} \\ 
\hline 
K00841.03 & 6.546 & $2.4^{+0.2}_{-0.2}$ & $0.9^{+0.03}_{-0.1}$ & $<9.8$ & $<3.9$ & $239.1^{+101.4}_{-101.4}$ & $82.2^{+49.5}_{-39.1}$&-- \\ 
Kepler-27 b & 15.335 & $4.9^{+0.4}_{-0.3}$ & ... & $3.6^{+1.7}_{-1.6}$ & $0.1^{+0.1}_{-0.1}$ & $7.6^{+2.6}_{-2.1}$ & $0.4^{+0.1}_{-0.1}$&-- \\ 
Kepler-27 c & 31.330 & $7.0^{+0.5}_{-0.5}$ & ... & $4.3^{+2.0}_{-1.6}$ & $0.1^{+0.04}_{-0.02}$ & $8.5^{+2.0}_{-2.0}$ & $0.1^{+0.04}_{-0.04}$&-- \\ 
\hline 
Kepler-28 b & 5.912 & $1.8^{+0.1}_{-0.1}$ & $0.6^{+0.03}_{-0.04}$ & $1.2^{+0.8}_{-0.5}$ & $0.9^{+0.9}_{-0.3}$ & $3.1^{+2.7}_{-1.4}$ & $2.8^{+2.4}_{-1.4}$&-- \\ 
Kepler-28 c & 8.986 & $1.7^{+0.2}_{-0.1}$ & ... & $1.0^{+0.7}_{-0.3}$ & $1.0^{+0.8}_{-0.4}$ & $2.7^{+2.5}_{-1.3}$ & $2.7^{+2.9}_{-1.3}$&-- \\ 
\hline 
Kepler-29 b\tnote{b} & 10.339 & $3.3^{+0.2}_{-0.2}$ & $1.0^{+0.05}_{-0.1}$ & $0.2^{+1.9}_{-0.1}$ & $0.03^{+0.3}_{-0.02}$ & $3.6^{+1.3}_{-1.2}$ & $0.5^{+0.2}_{-0.2}$& \citetalias{JontofHutter:2016ch,2016arXiv160909135M} \\ 
Kepler-29 c\tnote{b} & 13.287 & $2.7^{+0.2}_{-0.2}$ & ... & $0.2^{+1.8}_{-0.1}$ & $0.1^{+0.5}_{-0.03}$ & $3.4^{+1.2}_{-1.2}$ & $0.9^{+0.4}_{-0.3}$& \citetalias{JontofHutter:2016ch,2016arXiv160909135M} \\ 
\hline 
Kepler-30 b$^{*}$ & 29.323 & $1.9^{+0.2}_{-0.2}$ & $1.0^{+0.04}_{-0.1}$ & $8.8^{+0.6}_{-0.5}$ & $6.7^{+1.7}_{-1.6}$ & $8.9^{+0.5}_{-0.6}$ & $6.6^{+1.8}_{-1.5}$& \citetalias{2012Natur.487..449S} \\ 
Kepler-30 c$^{*}$ & 60.332 & $13.1^{+0.9}_{-0.8}$ & ... & $527.7^{+35.2}_{-29.7}$ & $1.3^{+0.2}_{-0.2}$ & $529.9^{+32.7}_{-32.2}$ & $1.3^{+0.2}_{-0.2}$& \citetalias{2012Natur.487..449S} \\ 
\hline 
Kepler-31 b & 20.860 & $5.6^{+1.0}_{-1.0}$ & $1.0^{+0.1}_{-0.1}$ & $0.7^{+2.4}_{-0.6}$ & $0.02^{+0.1}_{-0.02}$ & $10.0^{+5.8}_{-5.5}$ & $0.3^{+0.3}_{-0.2}$&-- \\ 
Kepler-31 c & 42.634 & $5.3^{+0.9}_{-1.0}$ & ... & $2.2^{+4.6}_{-1.2}$ & $0.1^{+0.2}_{-0.1}$ & $17.6^{+3.9}_{-5.1}$ & $0.6^{+0.5}_{-0.3}$&-- \\ 
Kepler-31 d & 87.647 & $4.2^{+0.8}_{-0.8}$ & ... & $2.8^{+4.2}_{-1.4}$ & $0.2^{+0.4}_{-0.2}$ & $10.0^{+2.7}_{-3.0}$ & $0.7^{+0.4}_{-0.4}$&-- \\ 
\hline 
Kepler-32 b & 5.901 & $2.0^{+0.2}_{-0.1}$ & $0.5^{+0.04}_{-0.04}$ & $0.5^{+0.3}_{-0.1}$ & $0.3^{+0.2}_{-0.1}$ & $5.5^{+2.0}_{-2.2}$ & $3.2^{+1.4}_{-1.5}$&-- \\ 
Kepler-32 c & 8.752 & $1.9^{+0.2}_{-0.2}$ & ... & $0.4^{+0.2}_{-0.1}$ & $0.3^{+0.2}_{-0.1}$ & $5.1^{+1.5}_{-2.5}$ & $3.1^{+1.6}_{-1.6}$&-- \\ 
\hline 
Kepler-33 c & 13.176 & $2.7^{+0.5}_{-0.6}$ & $1.1^{+0.1}_{-0.1}$ & $<0.8$ & $<0.3$ & $<4.6$ & $<1.7$& \citetalias{Hadden:2016ki} \\ 
Kepler-33 d$^{*}$ & 21.776 & $4.6^{+1.0}_{-0.9}$ & ... & $4.1^{+1.7}_{-2.0}$ & $0.2^{+0.2}_{-0.2}$ & $5.0^{+1.8}_{-1.7}$ & $0.3^{+0.2}_{-0.2}$& \citetalias{Hadden:2016ki} \\ 
Kepler-33 e$^{*}$ & 31.784 & $3.5^{+0.8}_{-0.7}$ & ... & $5.5^{+1.2}_{-1.1}$ & $0.8^{+0.4}_{-0.4}$ & $5.5^{+1.1}_{-1.1}$ & $0.8^{+0.4}_{-0.4}$& \citetalias{Hadden:2016ki} \\ 
Kepler-33 f$^{*}$ & 41.028 & $4.0^{+0.8}_{-0.9}$ & ... & $9.6^{+1.7}_{-1.8}$ & $0.9^{+0.5}_{-0.5}$ & $9.8^{+1.7}_{-1.7}$ & $0.9^{+0.6}_{-0.4}$& \citetalias{Hadden:2016ki} \\ 
\hline 
Kepler-36 b$^{*}$ & 13.848 & $1.5^{+0.1}_{-0.1}$ & $1.0^{+0.04}_{-0.04}$ & $3.9^{+0.2}_{-0.2}$ & $7.0^{+1.0}_{-1.1}$ & $4.0^{+0.2}_{-0.2}$ & $7.0^{+1.0}_{-1.1}$& \citetalias{Carter:2012gq} \\ 
Kepler-36 c$^{*}$ & 16.233 & $3.8^{+0.2}_{-0.2}$ & ... & $7.5^{+0.3}_{-0.3}$ & $0.8^{+0.1}_{-0.1}$ & $7.4^{+0.3}_{-0.3}$ & $0.8^{+0.1}_{-0.1}$& \citetalias{Carter:2012gq} \\ 
\hline 
Kepler-48 b & 4.778 & $1.9^{+0.2}_{-0.1}$ & $0.9^{+0.05}_{-0.04}$ & $0.1^{+0.4}_{-0.1}$ & $0.1^{+0.3}_{-0.1}$ & $8.1^{+10.0}_{-5.0}$ & $5.9^{+8.6}_{-3.8}$& \citetalias{Marcy:2014hr} \\ 
Kepler-48 c & 9.674 & $2.7^{+0.2}_{-0.2}$ & ... & $0.2^{+0.6}_{-0.1}$ & $0.1^{+0.2}_{-0.04}$ & $6.6^{+2.3}_{-2.0}$ & $1.8^{+0.8}_{-0.6}$& \citetalias{Marcy:2014hr} \\ 
\hline 
Kepler-49 b$^{*}$ & 7.204 & $2.3^{+0.2}_{-0.2}$ & $0.5^{+0.04}_{-0.04}$ & $8.0^{+1.9}_{-1.6}$ & $3.4^{+0.9}_{-0.9}$ & $7.8^{+1.4}_{-1.0}$ & $3.5^{+0.6}_{-0.8}$& \citetalias{JontofHutter:2016ch} \\ 
Kepler-49 c$^{*}$ & 10.913 & $2.8^{+0.3}_{-0.3}$ & ... & $5.9^{+1.5}_{-1.5}$ & $1.3^{+0.6}_{-0.5}$ & $5.7^{+1.1}_{-0.9}$ & $1.4^{+0.4}_{-0.5}$& \citetalias{JontofHutter:2016ch} \\ 
\hline 
Kepler-51 b$^{*}$ & 45.155 & $8.6^{+1.7}_{-1.5}$ & $1.1^{+0.1}_{-0.1}$ & $2.3^{+1.7}_{-1.6}$ & $0.01^{+0.02}_{-0.01}$ & $3.4^{+2.0}_{-1.6}$ & $0.02^{+0.03}_{-0.02}$& \citetalias{Masuda:2014dm} \\ 
Kepler-51 c$^{*}$\tnote{c} & 85.316 & $6.4^{+1.7}_{-1.5}$ & ... & $3.9^{+0.8}_{-0.8}$ & $0.1^{+0.1}_{-0.04}$ & $4.3^{+0.8}_{-0.9}$ & $0.1^{+0.1}_{-0.05}$& \citetalias{Masuda:2014dm} \\ 
Kepler-51 d$^{*}$ & 130.180 & $11.8^{+2.3}_{-2.1}$ & ... & $6.2^{+1.6}_{-1.5}$ & $0.02^{+0.01}_{-0.01}$ & $7.2^{+1.6}_{-1.6}$ & $0.03^{+0.01}_{-0.02}$& \citetalias{Masuda:2014dm} \\ 
\hline 
Kepler-52 b & 7.877 & $2.3^{+0.2}_{-0.1}$ & $0.6^{+0.03}_{-0.04}$ & $2.3^{+1.5}_{-0.8}$ & $0.9^{+0.7}_{-0.4}$ & $5.9^{+3.2}_{-2.5}$ & $2.6^{+1.3}_{-1.2}$&-- \\ 
Kepler-52 c & 16.385 & $2.1^{+0.1}_{-0.1}$ & ... & $6.3^{+5.0}_{-2.0}$ & $3.9^{+3.0}_{-1.4}$ & $18.3^{+14.2}_{-7.2}$ & $11.2^{+8.3}_{-4.8}$&-- \\ 
Kepler-52 d & 36.445 & $2.1^{+0.1}_{-0.1}$ & ... & $<5.0$ & $<3.0$ & $<18.5$ & $<11.1$&-- \\ 
\hline 
Kepler-53 d & 9.752 & $2.3^{+0.3}_{-0.3}$ & $1.0^{+0.1}_{-0.1}$ & $<0.8$ & $<0.4$ & $<20.7$ & $<9.5$&-- \\ 
Kepler-53 b & 18.649 & $3.4^{+0.4}_{-0.4}$ & ... & $2.8^{+1.9}_{-0.9}$ & $0.4^{+0.3}_{-0.2}$ & $18.3^{+9.4}_{-6.6}$ & $2.7^{+1.5}_{-1.5}$&-- \\ 
Kepler-53 c & 38.558 & $3.6^{+0.5}_{-0.4}$ & ... & $3.5^{+2.3}_{-1.1}$ & $0.4^{+0.4}_{-0.2}$ & $20.5^{+6.7}_{-6.5}$ & $2.2^{+1.3}_{-1.0}$&-- \\ 
\hline 
Kepler-54 b & 8.011 & $2.3^{+0.4}_{-0.3}$ & $0.5^{+0.04}_{-0.1}$ & $0.8^{+1.0}_{-0.3}$ & $0.3^{+0.5}_{-0.2}$ & $2.6^{+0.8}_{-0.8}$ & $1.1^{+0.5}_{-0.5}$&-- \\ 
Kepler-54 c & 12.071 & $1.3^{+0.1}_{-0.1}$ & ... & $0.7^{+0.9}_{-0.3}$ & $2.1^{+2.5}_{-1.0}$ & $2.3^{+0.7}_{-0.7}$ & $5.6^{+2.9}_{-1.9}$&-- \\ 
\hline 
Kepler-55 b & 27.954 & $1.9^{+0.2}_{-0.1}$ & $0.7^{+0.03}_{-0.04}$ & $1.3^{+2.1}_{-0.6}$ & $1.0^{+1.3}_{-0.7}$ & $9.6^{+5.8}_{-4.9}$ & $6.2^{+4.0}_{-4.5}$&-- \\ 
Kepler-55 c\tnote{d} & 42.142 & $2.5^{+0.3}_{-0.3}$ & ... & $1.8^{+2.8}_{-0.8}$ & $0.6^{+1.1}_{-0.3}$ & $12.1^{+9.1}_{-5.9}$ & $3.7^{+3.9}_{-2.0}$&-- \\ 
\hline 
Kepler-56 b$^{*}$ & 10.501 & $5.1^{+0.5}_{-0.7}$ & $1.4^{+0.2}_{-0.3}$ & $32.0^{+11.2}_{-10.5}$ & $1.4^{+0.5}_{-0.5}$ & $38.5^{+9.8}_{-10.0}$ & $1.7^{+0.5}_{-0.5}$& \citetalias{Huber:2013ha} \\ 
Kepler-56 c$^{*}$ & 21.405 & $10.6^{+1.4}_{-1.3}$ & ... & $137.6^{+44.1}_{-50.3}$ & $0.6^{+0.2}_{-0.2}$ & $160.1^{+44.2}_{-40.3}$ & $0.7^{+0.2}_{-0.2}$& \citetalias{Huber:2013ha} \\ 
\hline 
Kepler-57 b & 5.729 & $2.6^{+0.3}_{-0.3}$ & $0.9^{+0.1}_{-0.1}$ & $1.8^{+7.5}_{-0.8}$ & $0.6^{+2.4}_{-0.4}$ & $16.5^{+9.3}_{-6.0}$ & $4.6^{+3.9}_{-2.1}$& \citetalias{JontofHutter:2016ch} \\ 
Kepler-57 c & 11.609 & $2.9^{+0.6}_{-0.5}$ & ... & $0.4^{+1.5}_{-0.2}$ & $0.0^{+0.0}_{0.0}$ & $3.4^{+2.0}_{-1.4}$ & $0.0^{+0.0}_{0.0}$& \citetalias{JontofHutter:2016ch} \\ 
\hline 
Kepler-58 b & 10.219 & $2.8^{+0.5}_{-0.6}$ & $1.0^{+0.2}_{-0.1}$ & $2.9^{+2.1}_{-1.1}$ & $0.8^{+1.1}_{-0.6}$ & $11.7^{+10.7}_{-7.2}$ & $2.3^{+5.4}_{-2.0}$&-- \\ 
Kepler-58 c & 15.573 & $2.7^{+0.6}_{-0.5}$ & ... & $2.3^{+1.8}_{-0.9}$ & $0.7^{+0.8}_{-0.6}$ & $10.3^{+8.1}_{-6.7}$ & $2.1^{+4.3}_{-1.8}$&-- \\ 
\hline 
Kepler-60 b$^{*}$ & 7.133 & $1.6^{+0.3}_{-0.3}$ & $1.0^{+0.1}_{-0.1}$ & $3.7^{+0.6}_{-0.6}$ & $6.0^{+2.7}_{-3.1}$ & $4.4^{+0.6}_{-0.5}$ & $7.2^{+2.9}_{-3.9}$& \citetalias{Gozdziewski:2015ws,JontofHutter:2016ch} \\ 
Kepler-60 c$^{*}$ & 8.919 & $1.8^{+0.3}_{-0.3}$ & ... & $2.0^{+0.3}_{-0.5}$ & $1.7^{+1.1}_{-0.9}$ & $2.7^{+0.7}_{-0.6}$ & $2.6^{+1.7}_{-1.4}$& \citetalias{Gozdziewski:2015ws,JontofHutter:2016ch} \\ 
Kepler-60 d$^{*}$ & 11.899 & $1.6^{+0.3}_{-0.3}$ & ... & $3.9^{+0.7}_{-0.6}$ & $6.0^{+3.0}_{-3.3}$ & $4.0^{+0.8}_{-0.8}$ & $5.7^{+3.5}_{-3.2}$& \citetalias{Gozdziewski:2015ws,JontofHutter:2016ch} \\ 
\hline 
Kepler-79 b & 13.485 & $3.5^{+0.7}_{-0.5}$ & $1.1^{+0.1}_{-0.1}$ & $7.3^{+6.7}_{-3.3}$ & $0.8^{+1.2}_{-0.5}$ & $22.3^{+12.2}_{-9.3}$ & $2.6^{+2.6}_{-1.6}$& \citetalias{2014ApJ...785...15J} \\ 
Kepler-79 c & 27.402 & $3.8^{+0.7}_{-0.6}$ & ... & $5.8^{+3.0}_{-1.8}$ & $0.6^{+0.5}_{-0.3}$ & $11.6^{+2.4}_{-2.3}$ & $1.2^{+0.6}_{-0.6}$& \citetalias{2014ApJ...785...15J} \\ 
Kepler-79 d & 52.091 & $7.3^{+1.4}_{-1.1}$ & ... & $5.8^{+1.7}_{-1.8}$ & $0.1^{+0.1}_{-0.04}$ & $7.9^{+2.0}_{-2.1}$ & $0.1^{+0.1}_{-0.1}$& \citetalias{2014ApJ...785...15J} \\ 
Kepler-79 e$^{*}$ & 81.064 & $2.9^{+0.5}_{-0.4}$ & ... & $3.4^{+1.0}_{-0.8}$ & $0.8^{+0.5}_{-0.4}$ & $4.2^{+0.9}_{-0.9}$ & $1.0^{+0.6}_{-0.5}$& \citetalias{2014ApJ...785...15J} \\ 
\hline 
Kepler-80 d$^{*}$ & 3.072 & $1.3^{+0.1}_{-0.1}$ & $0.6^{+0.03}_{-0.04}$ & $3.7^{+0.8}_{-0.6}$ & $10.1^{+2.4}_{-2.5}$ & $4.4^{+0.6}_{-0.7}$ & $11.5^{+2.5}_{-2.6}$& \citetalias{MacDonald:2016vo} \\ 
Kepler-80 e$^{*}$ & 4.645 & $1.3^{+0.1}_{-0.1}$ & ... & $2.1^{+0.7}_{-0.7}$ & $4.6^{+1.8}_{-1.7}$ & $2.6^{+0.8}_{-0.7}$ & $6.2^{+1.9}_{-2.0}$& \citetalias{MacDonald:2016vo} \\ 
Kepler-80 b & 7.054 & $2.2^{+0.1}_{-0.1}$ & ... & $1.4^{+1.2}_{-0.5}$ & $0.8^{+0.5}_{-0.3}$ & $3.2^{+0.8}_{-1.0}$ & $1.5^{+0.4}_{-0.5}$& \citetalias{MacDonald:2016vo} \\ 
Kepler-80 c & 9.522 & $2.3^{+0.1}_{-0.1}$ & ... & $1.7^{+0.8}_{-0.5}$ & $0.8^{+0.3}_{-0.3}$ & $2.7^{+0.8}_{-0.6}$ & $1.2^{+0.4}_{-0.3}$& \citetalias{MacDonald:2016vo} \\ 
\hline 
Kepler-81 b & 5.955 & $2.4^{+0.1}_{-0.1}$ & $0.6^{+0.04}_{-0.03}$ & $0.2^{+1.7}_{-0.1}$ & $0.1^{+0.6}_{-0.04}$ & $8.2^{+3.4}_{-3.8}$ & $2.9^{+1.6}_{-1.3}$&-- \\ 
Kepler-81 c & 12.040 & $2.3^{+0.1}_{-0.1}$ & ... & $0.1^{+1.1}_{-0.1}$ & $0.05^{+0.5}_{-0.03}$ & $3.9^{+1.0}_{-1.2}$ & $1.7^{+0.6}_{-0.5}$&-- \\ 
\hline 
Kepler-84 d & 4.225 & $1.5^{+0.2}_{-0.2}$ & $1.0^{+0.1}_{-0.1}$ & $<1.7$ & $<3.1$ & $<41.3$ & $<77.5$&-- \\ 
Kepler-84 b & 8.726 & $2.5^{+0.4}_{-0.4}$ & ... & $5.0^{+5.7}_{-2.9}$ & $1.6^{+2.5}_{-1.1}$ & $18.6^{+5.7}_{-6.8}$ & $5.7^{+3.8}_{-3.3}$&-- \\ 
Kepler-84 c & 12.883 & $2.8^{+0.7}_{-0.5}$ & ... & $<3.6$ & $<0.8$ & $6.4^{+5.3}_{-3.0}$ & $1.1^{+1.6}_{-0.8}$&-- \\ 
Kepler-84 e & 27.435 & $2.6^{+0.4}_{-0.4}$ & ... & $<35.2$ & $<11.8$ & $46.1^{+16.9}_{-15.7}$ & $13.1^{+9.7}_{-7.2}$&-- \\ 
Kepler-84 f & 44.551 & $2.3^{+0.4}_{-0.3}$ & ... & $<3.4$ & $<1.6$ & $<11.9$ & $<5.7$&-- \\ 
\hline 
Kepler-85 b & 8.305 & $1.8^{+0.2}_{-0.2}$ & $0.9^{+0.1}_{-0.05}$ & $0.7^{+1.9}_{-0.4}$ & $0.6^{+1.8}_{-0.4}$ & $5.7^{+2.1}_{-2.2}$ & $4.8^{+2.8}_{-2.2}$&-- \\ 
Kepler-85 c & 12.514 & $1.9^{+0.2}_{-0.2}$ & ... & $1.0^{+2.2}_{-0.6}$ & $0.8^{+2.0}_{-0.5}$ & $7.4^{+2.4}_{-2.8}$ & $6.0^{+2.5}_{-2.9}$&-- \\ 
Kepler-85 d & 17.913 & $1.3^{+0.1}_{-0.1}$ & ... & $<1.2$ & $<3.0$ & $<6.2$ & $<15.7$&-- \\ 
Kepler-85 e$^{*}$ & 25.215 & $1.2^{+0.1}_{-0.1}$ & ... & $0.6^{+0.5}_{-0.4}$ & $1.6^{+1.6}_{-1.2}$ & $0.8^{+0.4}_{-0.4}$ & $2.2^{+1.3}_{-1.3}$&-- \\ 
\hline 
Kepler-89 c$^{*}$ & 10.424 & $3.8^{+0.7}_{-0.6}$ & $1.1^{+0.1}_{-0.1}$ & $7.8^{+3.0}_{-2.4}$ & $0.8^{+0.6}_{-0.4}$ & $9.6^{+3.1}_{-2.7}$ & $1.0^{+0.6}_{-0.6}$& \citetalias{2013ApJ...778..185M,2012ApJ...759L..36H,2013ApJ...768...14W} \\ 
Kepler-89 d$^{*}$ & 22.343 & $10.1^{+1.7}_{-1.6}$ & ... & $58.4^{+11.9}_{-10.6}$ & $0.3^{+0.2}_{-0.1}$ & $62.0^{+11.6}_{-9.0}$ & $0.3^{+0.2}_{-0.1}$& \citetalias{2013ApJ...778..185M,2012ApJ...759L..36H,2013ApJ...768...14W} \\ 
\hline 
K00115.03 & 3.436 & $0.5^{+0.1}_{-0.1}$ & $0.9^{+0.1}_{-0.1}$ & $<1.9$ & $<69.5$ & $1.8^{+2.5}_{-1.4}$ & $35.6^{+110.5}_{-35.3}$& \citetalias{JontofHutter:2016ch} \\ 
Kepler-105 b & 5.412 & $2.7^{+0.4}_{-0.4}$ & ... & $<2.8$ & $<0.8$ & $1.6^{+2.3}_{-1.5}$ & $0.3^{+0.8}_{-0.3}$& \citetalias{JontofHutter:2016ch} \\ 
Kepler-105 c$^{*}$ & 7.126 & $1.6^{+0.2}_{-0.3}$ & ... & $3.6^{+1.3}_{-1.3}$ & $2.7^{+3.9}_{-1.8}$ & $4.2^{+1.2}_{-1.3}$ & $5.5^{+3.5}_{-3.0}$& \citetalias{JontofHutter:2016ch} \\ 
\hline 
Kepler-114 b & 5.189 & $1.0^{+0.1}_{-0.1}$ & $0.6^{+0.03}_{-0.04}$ & $<1.0$ & $<5.4$ & $2.9^{+2.8}_{-1.9}$ & $13.9^{+16.4}_{-9.4}$&-- \\ 
Kepler-114 c & 8.041 & $1.3^{+0.1}_{-0.1}$ & ... & $0.8^{+0.6}_{-0.3}$ & $2.0^{+1.8}_{-0.6}$ & $2.4^{+0.6}_{-0.7}$ & $5.5^{+2.0}_{-1.5}$&-- \\ 
Kepler-114 d & 11.776 & $2.0^{+0.1}_{-0.1}$ & ... & $<0.4$ & $<0.2$ & $0.8^{+0.8}_{-0.8}$ & $0.4^{+0.6}_{-0.4}$&-- \\ 
\hline 
Kepler-122 e & 37.997 & $2.0^{+0.3}_{-0.3}$ & $1.0^{+0.1}_{-0.1}$ & $<0.2$ & $<0.1$ & $<0.6$ & $<0.5$&-- \\ 
Kepler-122 f & 56.261 & $2.3^{+0.3}_{-0.4}$ & ... & $2.1^{+2.5}_{-1.0}$ & $1.0^{+1.6}_{-0.7}$ & $5.3^{+3.6}_{-2.4}$ & $2.5^{+2.6}_{-1.6}$&-- \\ 
\hline 
Kepler-127 b & 14.436 & $1.4^{+0.1}_{-0.04}$ & $1.2^{+0.1}_{-0.04}$ & $0.8^{+2.3}_{-0.4}$ & $1.4^{+4.0}_{-0.7}$ & $4.6^{+3.3}_{-1.7}$ & $8.1^{+5.9}_{-3.1}$&-- \\ 
Kepler-127 c & 29.394 & $2.4^{+0.1}_{-0.1}$ & ... & $1.0^{+3.5}_{-0.6}$ & $0.4^{+1.2}_{-0.4}$ & $7.2^{+2.6}_{-2.8}$ & $2.6^{+1.1}_{-1.0}$&-- \\ 
Kepler-127 d & 48.630 & $2.6^{+0.1}_{-0.1}$ & ... & $6.1^{+3.1}_{-3.5}$ & $2.0^{+1.0}_{-1.2}$ & $10.1^{+2.6}_{-2.9}$ & $3.2^{+1.0}_{-0.9}$&-- \\ 
\hline 
Kepler-128 b & 15.090 & $1.4^{+0.1}_{-0.1}$ & $1.1^{+0.1}_{-0.05}$ & $0.7^{+0.4}_{-0.3}$ & $1.2^{+0.7}_{-0.5}$ & $3.8^{+2.9}_{-2.9}$ & $3.4^{+9.2}_{-1.8}$& \citetalias{Hadden:2016ki} \\ 
Kepler-128 c & 22.803 & $1.3^{+0.1}_{-0.1}$ & ... & $0.7^{+0.5}_{-0.2}$ & $1.6^{+1.2}_{-0.6}$ & $1.9^{+6.3}_{-0.8}$ & $10.7^{+8.2}_{-8.2}$& \citetalias{Hadden:2016ki} \\ 
\hline 
Kepler-138 b & 10.313 & $0.6^{+0.1}_{-0.1}$ & $0.5^{+0.04}_{-0.04}$ & $0.01^{+0.01}_{-0.01}$ & $0.4^{+0.4}_{-0.2}$ & $0.1^{+0.05}_{-0.03}$ & $1.3^{+1.4}_{-0.8}$& \citetalias{2015Natur.522..321J} \\ 
Kepler-138 c & 13.781 & $1.3^{+0.1}_{-0.1}$ & ... & $0.4^{+0.4}_{-0.1}$ & $0.8^{+0.9}_{-0.4}$ & $1.6^{+1.6}_{-0.8}$ & $4.3^{+3.0}_{-2.7}$& \citetalias{2015Natur.522..321J} \\ 
Kepler-138 d & 23.089 & $1.3^{+0.1}_{-0.1}$ & ... & $0.1^{+0.1}_{-0.04}$ & $0.3^{+0.3}_{-0.1}$ & $0.5^{+0.5}_{-0.3}$ & $1.4^{+1.2}_{-0.9}$& \citetalias{2015Natur.522..321J} \\ 
\hline 
Kepler-176 c & 12.759 & $2.5^{+0.1}_{-0.2}$ & $0.8^{+0.1}_{-0.03}$ & $0.1^{+0.3}_{-0.04}$ & $0.04^{+0.1}_{-0.02}$ & $3.8^{+3.2}_{-1.8}$ & $1.4^{+1.2}_{-0.7}$&-- \\ 
Kepler-176 d & 25.753 & $2.3^{+0.1}_{-0.2}$ & ... & $0.2^{+0.5}_{-0.1}$ & $0.1^{+0.2}_{-0.04}$ & $5.9^{+2.2}_{-2.2}$ & $2.4^{+1.4}_{-0.9}$&-- \\ 
\hline 
Kepler-177 b$^{*}$ & 36.855 & $3.1^{+0.4}_{-0.4}$ & $1.0^{+0.1}_{-0.1}$ & $5.4^{+1.0}_{-0.9}$ & $1.0^{+0.4}_{-0.4}$ & $5.6^{+0.9}_{-0.9}$ & $1.0^{+0.4}_{-0.4}$& \citetalias{JontofHutter:2016ch} \\ 
Kepler-177 c$^{*}$ & 49.412 & $7.3^{+0.9}_{-0.7}$ & ... & $13.5^{+2.6}_{-2.8}$ & $0.2^{+0.1}_{-0.1}$ & $13.5^{+3.0}_{-2.5}$ & $0.2^{+0.1}_{-0.1}$& \citetalias{JontofHutter:2016ch} \\ 
\hline 
Kepler-223 b$^{*}$ & 7.385 & $2.2^{+0.4}_{-0.4}$ & $1.1^{+0.1}_{-0.1}$ & $4.0^{+1.7}_{-2.1}$ & $1.4^{+1.2}_{-1.4}$ & $4.2^{+1.9}_{-1.6}$ & $2.0^{+1.4}_{-1.3}$& \citetalias{Mills:2016gi} \\ 
Kepler-223 c$^{*}$ & 9.848 & $2.6^{+0.5}_{-0.5}$ & ... & $12.4^{+2.8}_{-2.9}$ & $4.1^{+1.9}_{-2.5}$ & $10.8^{+2.8}_{-2.5}$ & $3.3^{+2.0}_{-1.9}$& \citetalias{Mills:2016gi} \\ 
Kepler-223 d$^{*}$ & 14.787 & $3.8^{+0.7}_{-0.7}$ & ... & $5.9^{+1.9}_{-1.9}$ & $0.6^{+0.4}_{-0.4}$ & $6.6^{+1.9}_{-1.8}$ & $0.7^{+0.4}_{-0.4}$& \citetalias{Mills:2016gi} \\ 
Kepler-223 e\tnote{e} & 19.725 & $2.9^{+0.6}_{-0.5}$ & ... & $<1.0$ & $<0.2$ & $<2.5$ & $<0.6$& \citetalias{Mills:2016gi} \\ 
\hline 
Kepler-238 c & 6.156 & $2.4^{+0.4}_{-0.5}$ & $1.1^{+0.1}_{-0.1}$ & $<13.8$ & $<5.4$ & $<94.8$ & $<40.8$&-- \\ 
Kepler-238 d & 13.233 & $3.2^{+0.5}_{-0.6}$ & ... & $<2.5$ & $<0.5$ & $3.8^{+3.1}_{-3.6}$ & $0.3^{+1.2}_{-0.3}$&-- \\ 
Kepler-238 e & 23.654 & $8.2^{+1.6}_{-1.5}$ & ... & $<8.9$ & $<0.1$ & $16.6^{+13.7}_{-13.7}$ & $0.1^{+0.3}_{-0.1}$&-- \\ 
Kepler-238 f & 50.446 & $2.8^{+0.5}_{-0.5}$ & ... & $4.5^{+5.4}_{-2.2}$ & $1.2^{+1.8}_{-0.9}$ & $11.0^{+4.1}_{-2.8}$ & $2.9^{+1.8}_{-1.8}$&-- \\ 
\hline 
Kepler-277 b & 17.324 & $2.9^{+0.5}_{-0.7}$ & $1.2^{+0.1}_{-0.1}$ & $3.5^{+2.8}_{-1.6}$ & $0.9^{+1.2}_{-0.6}$ & $46.7^{+27.0}_{-27.0}$ & $8.8^{+13.6}_{-6.7}$&-- \\ 
Kepler-277 c & 33.007 & $3.0^{+0.5}_{-0.7}$ & ... & $4.5^{+3.7}_{-2.0}$ & $1.0^{+1.3}_{-0.7}$ & $37.6^{+14.1}_{-12.8}$ & $7.8^{+6.2}_{-5.2}$&-- \\ 
\hline 
Kepler-279 c$^{*}$ & 35.735 & $5.5^{+1.1}_{-1.1}$ & $1.2^{+0.1}_{-0.1}$ & $7.4^{+2.2}_{-1.5}$ & $0.3^{+0.2}_{-0.2}$ & $8.7^{+2.3}_{-2.0}$ & $0.3^{+0.2}_{-0.2}$&-- \\ 
Kepler-279 d$^{*}$ & 54.414 & $4.3^{+1.0}_{-0.9}$ & ... & $4.5^{+1.2}_{-0.9}$ & $0.3^{+0.2}_{-0.2}$ & $5.2^{+1.2}_{-1.3}$ & $0.4^{+0.2}_{-0.2}$&-- \\ 
\hline 
K01563.03 & 3.205 & $1.8^{+0.2}_{-0.1}$ & $0.8^{+0.04}_{-0.05}$ & $<8.7$ & $<7.5$ & $31.1^{+23.0}_{-24.7}$ & $22.2^{+24.1}_{-17.9}$&-- \\ 
Kepler-305 b & 5.487 & $3.1^{+0.2}_{-0.2}$ & ... & $2.7^{+2.9}_{-1.1}$ & $0.5^{+0.5}_{-0.3}$ & $7.5^{+2.4}_{-3.3}$ & $1.2^{+0.5}_{-0.6}$&-- \\ 
Kepler-305 c & 8.291 & $2.7^{+0.2}_{-0.2}$ & ... & $1.4^{+1.2}_{-0.7}$ & $0.4^{+0.3}_{-0.3}$ & $3.4^{+1.5}_{-1.5}$ & $0.9^{+0.5}_{-0.4}$&-- \\ 
Kepler-305 d$^{*}$ & 16.739 & $2.7^{+0.2}_{-0.2}$ & ... & $9.1^{+6.1}_{-3.8}$ & $2.5^{+1.8}_{-1.1}$ & $6.4^{+2.7}_{-2.2}$ & $1.6^{+1.0}_{-0.6}$&-- \\ 
\hline 
Kepler-307 b$^{*}$ & 10.416 & $3.0^{+0.3}_{-0.3}$ & $0.9^{+0.1}_{-0.1}$ & $8.8^{+0.9}_{-0.9}$ & $1.8^{+0.7}_{-0.6}$ & $8.8^{+0.8}_{-0.9}$ & $1.8^{+0.7}_{-0.6}$& \citetalias{Hadden:2016ki,JontofHutter:2016ch} \\ 
Kepler-307 c$^{*}$ & 13.084 & $2.7^{+0.3}_{-0.3}$ & ... & $3.9^{+0.7}_{-0.7}$ & $1.1^{+0.4}_{-0.5}$ & $4.0^{+0.7}_{-0.7}$ & $1.0^{+0.5}_{-0.3}$& \citetalias{Hadden:2016ki,JontofHutter:2016ch} \\ 
\hline 
Kepler-310 c & 56.476 & $2.9^{+0.3}_{-0.2}$ & $1.0^{+0.03}_{-0.1}$ & $<4.9$ & $<1.0$ & $5.9^{+4.1}_{-4.0}$ & $1.1^{+0.9}_{-0.8}$&-- \\ 
Kepler-310 d$^{*}$ & 92.874 & $2.1^{+0.2}_{-0.1}$ & ... & $7.0^{+3.4}_{-4.1}$ & $3.3^{+2.2}_{-2.0}$ & $8.4^{+2.9}_{-2.8}$ & $4.3^{+1.9}_{-1.6}$&-- \\ 
\hline 
K01831.03 & 34.207 & $1.2^{+0.1}_{-0.1}$ & $0.8^{+0.05}_{-0.04}$ & $1.0^{+1.2}_{-0.4}$ & $3.0^{+4.4}_{-1.5}$ & $3.5^{+2.8}_{-2.0}$ & $10.3^{+11.0}_{-6.0}$&-- \\ 
Kepler-324 c & 51.810 & $2.6^{+0.6}_{-0.2}$ & ... & $<0.1$ & $<0.03$ & $<0.7$ & $<0.2$&-- \\ 
\hline 
Kepler-345 b$^{*}$ & 7.416 & $0.8^{+0.1}_{-0.1}$ & $0.7^{+0.03}_{-0.04}$ & $0.5^{+0.3}_{-0.3}$ & $5.2^{+3.3}_{-3.4}$ & $0.9^{+0.8}_{-0.4}$ & $8.6^{+7.4}_{-4.7}$&-- \\ 
Kepler-345 c$^{*}$ & 9.387 & $1.3^{+0.1}_{-0.1}$ & ... & $2.2^{+0.9}_{-0.9}$ & $4.8^{+2.4}_{-2.2}$ & $2.8^{+2.7}_{-1.1}$ & $6.9^{+6.0}_{-3.5}$&-- \\ 
\hline 
Kepler-359 c$^{*}$ & 57.693 & $4.8^{+1.0}_{-0.9}$ & $1.1^{+0.2}_{-0.1}$ & $2.9^{+2.4}_{-1.9}$ & $0.1^{+0.2}_{-0.1}$ & $5.1^{+2.3}_{-2.1}$ & $0.2^{+0.2}_{-0.1}$&-- \\ 
Kepler-359 d$^{*}$\tnote{f} & 77.083 & $4.6^{+0.9}_{-0.9}$ & ... & $2.7^{+2.5}_{-1.5}$ & $0.1^{+0.2}_{-0.1}$ & $4.4^{+2.4}_{-1.3}$ & $0.3^{+0.2}_{-0.2}$&-- \\ 
\hline 
Kepler-396 b & 42.994 & $3.3^{+0.3}_{-0.4}$ & $0.9^{+0.1}_{-0.1}$ & $1.1^{+0.5}_{-0.2}$ & $0.2^{+0.1}_{-0.1}$ & $3.0^{+0.7}_{-1.3}$ & $0.4^{+0.3}_{-0.2}$&-- \\ 
Kepler-396 c & 88.508 & $5.6^{+0.5}_{-0.6}$ & ... & $1.1^{+0.6}_{-0.2}$ & $0.04^{+0.03}_{-0.02}$ & $3.1^{+0.9}_{-1.3}$ & $0.1^{+0.1}_{-0.03}$&-- \\ 
\hline 
Kepler-444 b & 3.600 & $0.4^{+0.03}_{-0.03}$ & $0.7^{+0.1}_{-0.03}$ & $<1.8$ & $<172.9$ & $<3.2$ & $<306.8$&-- \\ 
Kepler-444 c & 4.546 & $0.5^{+0.04}_{-0.03}$ & ... & $<0.6$ & $<30.1$ & $<1.2$ & $<58.8$&-- \\ 
Kepler-444 d$^{*}$ & 6.189 & $0.5^{+0.04}_{-0.04}$ & ... & $0.2^{+0.5}_{-0.1}$ & $7.9^{+20.6}_{-5.7}$ & $0.6^{+0.3}_{-0.3}$ & $24.6^{+14.7}_{-12.7}$&-- \\ 
Kepler-444 e$^{*}$ & 7.743 & $0.5^{+0.03}_{-0.04}$ & ... & $0.1^{+0.2}_{-0.1}$ & $4.8^{+6.5}_{-3.9}$ & $0.3^{+0.2}_{-0.2}$ & $9.8^{+7.2}_{-5.9}$&-- \\ 
Kepler-444 f & 9.741 & $0.6^{+0.04}_{-0.04}$ & ... & $<0.5$ & $<10.5$ & $0.5^{+0.6}_{-0.5}$ & $9.5^{+13.3}_{-9.3}$&-- \\ 
\hline 
Kepler-526 b & 5.459 & $1.9^{+0.2}_{-0.2}$ & $1.2^{+0.1}_{-0.1}$ & $<0.6$ & $<0.5$ & $<3.9$ & $<3.1$&-- \\ 
K00332.02 & 6.867 & $0.8^{+0.1}_{-0.1}$ & ... & $0.5^{+0.4}_{-0.3}$ & $5.5^{+6.0}_{-3.1}$ & $1.8^{+1.3}_{-1.2}$ & $16.4^{+22.8}_{-10.7}$&-- \\ 
\hline 
K00427.01 & 24.615 & $4.3^{+0.5}_{-0.5}$ & $0.9^{+0.1}_{-0.03}$ & $<8.0$ & $<0.6$ & $11.9^{+8.9}_{-8.0}$ & $0.7^{+0.7}_{-0.6}$&-- \\ 
Kepler-549 b$^{*}$ & 42.950 & $3.1^{+0.4}_{-0.4}$ & ... & $11.0^{+4.2}_{-3.2}$ & $1.9^{+1.4}_{-0.8}$ & $12.1^{+2.7}_{-3.0}$ & $2.2^{+1.0}_{-1.0}$&-- \\ 
\hline 
Kepler-1126 b\tnote{g} & 108.590 & $1.8^{+0.2}_{-0.2}$ & $0.8^{+0.1}_{-0.03}$ & $<3.9$ & $<4.1$ & $15.2^{+13.6}_{-10.5}$ & $15.2^{+15.5}_{-13.0}$&-- \\ 
K02162.02\tnote{g} & 199.670 & $1.7^{+0.3}_{-0.2}$ & ... & $<1.1$ & $<1.3$ & $10.2^{+5.9}_{-6.7}$ & $9.0^{+8.5}_{-8.5}$&-- \\ 

%% file: ecctable.tex
Kepler-9 b/c	 & 2:1	 & 0.0126	 & $0.083_{-0.001}^{+0.001}$	 & $0.083_{-0.001}^{+0.001}$	 & $0.083_{-0.001}^{+0.001}$	 & $0.083_{-0.001}^{+0.001}$ \\ 
Kepler-11 b/c	 & 5:4	 & 0.0113	 & $0.028_{-0.01}^{+0.006}$	 & $0.028_{-0.01}^{+0.006}$	 & $0.01_{-0.004}^{+0.008}$	 & $0.01_{-0.004}^{+0.008}$ \\ 
Kepler-11 c/d	 & 5:3	 & 0.0451	 & $0.013_{-0.009}^{+0.015}$	 & $0.001_{-0.008}^{+0.023}$	 & $0.001_{-0.001}^{+0.008}$	 & $0.00_{-0.002}^{+0.009}$ \\ 
Kepler-11 d/e	 & 7:5	 & 0.0074	 & $0.009_{-0.001}^{+0.001}$	 & $0.009_{-0.002}^{+0.001}$	 & $0.009_{-0.001}^{+0.001}$	 & $0.009_{-0.001}^{+0.002}$ \\ 
Kepler-11 e/f	 & 3:2	 & -0.0272	 & $0.018_{-0.004}^{+0.005}$	 & $0.018_{-0.004}^{+0.006}$	 & $0.016_{-0.004}^{+0.005}$	 & $0.016_{-0.004}^{+0.004}$ \\ 
Kepler-18 c/d	 & 2:1	 & -0.0278	 & $0.001_{-0.001}^{+0.005}$	 & $0.001_{-0.002}^{+0.006}$	 & $0.002_{-0.001}^{+0.0004}$	 & $0.002_{-0.001}^{+0.001}$ \\ 
Kepler-23 b/c	 & 3:2	 & 0.0077	 & $0.017_{-0.006}^{+0.026}$	 & $0.025_{-0.013}^{+0.018}$	 & $0.011_{-0.003}^{+0.005}$	 & $0.011_{-0.003}^{+0.005}$ \\ 
Kepler-23 c/d	 & 7:5	 & 0.0156	 & $0.021_{-0.014}^{+0.013}$	 & $0.002_{-0.018}^{+0.022}$	 & $0.009_{-0.009}^{+0.003}$	 & $0.009_{-0.011}^{+0.004}$ \\ 
Kepler-24 b/c	 & 3:2	 & 0.0095	 & $0.035_{-0.014}^{+0.025}$	 & $0.035_{-0.014}^{+0.025}$	 & $0.014_{-0.004}^{+0.006}$	 & $0.014_{-0.004}^{+0.006}$ \\ 
Kepler-24 c/e	 & 3:2	 & 0.0269	 & $0.038_{-0.018}^{+0.014}$	 & $0.037_{-0.024}^{+0.013}$	 & $0.006_{-0.006}^{+0.006}$	 & $0.001_{-0.001}^{+0.011}$ \\ 
Kepler-25 b/c	 & 2:1	 & 0.0195	 & $0.009_{-0.008}^{+0.043}$	 & $0.009_{-0.008}^{+0.044}$	 & $0.002_{-0.001}^{+0.005}$	 & $0.003_{-0.003}^{+0.004}$ \\ 
Kepler-26 b/c	 & 7:5	 & 0.0032	 & $0.013_{-0.001}^{+0.001}$	 & $0.013_{-0.001}^{+0.001}$	 & $0.01_{-0.001}^{+0.001}$	 & $0.01_{-0.001}^{+0.001}$ \\ 
Kepler-27 03/b	 & 2:1	 & 0.1713	 & $0.009_{-0.008}^{+0.082}$	 & $-0.006_{-0.033}^{+0.071}$	 & $0.006_{-0.002}^{+0.003}$	 & $0.005_{-0.003}^{+0.003}$ \\ 
Kepler-27 b/c	 & 2:1	 & 0.0215	 & $0.034_{-0.013}^{+0.022}$	 & $0.034_{-0.012}^{+0.022}$	 & $0.019_{-0.005}^{+0.007}$	 & $0.02_{-0.005}^{+0.007}$ \\ 
Kepler-28 b/c	 & 3:2	 & 0.0132	 & $0.038_{-0.019}^{+0.017}$	 & $0.038_{-0.018}^{+0.018}$	 & $0.005_{-0.001}^{+0.012}$	 & $0.008_{-0.005}^{+0.009}$ \\ 
Kepler-29 b/c	 & 9:7	 & -0.0005	 & $0.014_{-0.003}^{+0.014}$	 & $0.013_{-0.003}^{+0.014}$	 & $0.012_{-0.001}^{+0.002}$	 & $0.012_{-0.001}^{+0.002}$ \\ 
Kepler-30 b/c	 & 2:1	 & 0.0287	 & $0.039_{-0.0003}^{+0.0003}$	 & $0.039_{-0.0003}^{+0.0003}$	 & $0.039_{-0.0003}^{+0.0003}$	 & $0.039_{-0.0003}^{+0.0003}$ \\ 
Kepler-31 b/c	 & 2:1	 & 0.0219	 & $0.008_{-0.005}^{+0.082}$	 & $0.01_{-0.009}^{+0.08}$	 & $0.005_{-0.001}^{+0.004}$	 & $0.005_{-0.002}^{+0.003}$ \\ 
Kepler-31 c/d	 & 2:1	 & 0.0279	 & $0.007_{-0.005}^{+0.024}$	 & $0.004_{-0.005}^{+0.024}$	 & $0.001_{-0.001}^{+0.003}$	 & $0.001_{-0.001}^{+0.003}$ \\ 
Kepler-32 b/c	 & 3:2	 & -0.0113	 & $0.096_{-0.043}^{+0.011}$	 & $0.094_{-0.044}^{+0.01}$	 & $0.004_{-0.001}^{+0.004}$	 & $0.004_{-0.001}^{+0.004}$ \\ 
Kepler-33 c/d	 & 5:3	 & -0.0084	 & $0.029_{-0.015}^{+0.014}$	 & $0.024_{-0.018}^{+0.021}$	 & $0.016_{-0.008}^{+0.009}$	 & $0.016_{-0.01}^{+0.009}$ \\ 
Kepler-33 d/e	 & 3:2	 & -0.0269	 & $0.008_{-0.004}^{+0.004}$	 & $0.007_{-0.003}^{+0.005}$	 & $0.009_{-0.004}^{+0.003}$	 & $0.008_{-0.004}^{+0.004}$ \\ 
Kepler-33 e/f	 & 9:7	 & 0.004	 & $0.006_{-0.002}^{+0.002}$	 & $0.006_{-0.002}^{+0.002}$	 & $0.006_{-0.002}^{+0.002}$	 & $0.006_{-0.003}^{+0.002}$ \\ 
Kepler-36 b/c	 & 7:6	 & 0.0048	 & $0.02_{-0.0005}^{+0.0004}$	 & $0.02_{-0.0004}^{+0.0005}$	 & $0.02_{-0.0003}^{+0.0003}$	 & $0.02_{-0.0004}^{+0.0003}$ \\ 
Kepler-48 b/c	 & 2:1	 & 0.0123	 & $0.003_{-0.003}^{+0.16}$	 & $0.08_{-0.082}^{+0.074}$	 & $0.001_{-0.001}^{+0.001}$	 & $0.0004_{-0.0005}^{+0.0005}$ \\ 
Kepler-49 b/c	 & 3:2	 & 0.0099	 & $0.003_{-0.0004}^{+0.001}$	 & $0.003_{-0.001}^{+0.001}$	 & $0.004_{-0.0004}^{+0.0004}$	 & $0.004_{-0.001}^{+0.0004}$ \\ 
Kepler-51 b/c	 & 2:1	 & -0.0553	 & $0.041_{-0.011}^{+0.014}$	 & $0.041_{-0.011}^{+0.014}$	 & $0.033_{-0.009}^{+0.012}$	 & $0.033_{-0.009}^{+0.012}$ \\ 
Kepler-51 c/d	 & 3:2	 & 0.0172	 & $0.004_{-0.001}^{+0.002}$	 & $0.004_{-0.002}^{+0.002}$	 & $0.004_{-0.002}^{+0.001}$	 & $0.004_{-0.002}^{+0.001}$ \\ 
Kepler-52 b/c	 & 2:1	 & 0.04	 & $0.151_{-0.053}^{+0.091}$	 & $0.154_{-0.056}^{+0.084}$	 & $0.066_{-0.019}^{+0.029}$	 & $0.066_{-0.019}^{+0.029}$ \\ 
Kepler-52 c/d	 & 2:1	 & 0.1122	 & $0.129_{-0.052}^{+0.051}$	 & $0.115_{-0.043}^{+0.06}$	 & $0.056_{-0.026}^{+0.026}$	 & $0.052_{-0.025}^{+0.028}$ \\ 
Kepler-53 b/c	 & 2:1	 & 0.0338	 & $0.133_{-0.062}^{+0.042}$	 & $0.123_{-0.055}^{+0.049}$	 & $0.017_{-0.005}^{+0.009}$	 & $0.018_{-0.006}^{+0.007}$ \\ 
Kepler-53 d/b	 & 2:1	 & -0.0438	 & $0.055_{-0.039}^{+0.052}$	 & $0.043_{-0.073}^{+0.046}$	 & $0.007_{-0.005}^{+0.008}$	 & $0.005_{-0.005}^{+0.007}$ \\ 
Kepler-54 b/c	 & 3:2	 & 0.0046	 & $0.016_{-0.005}^{+0.019}$	 & $0.016_{-0.005}^{+0.019}$	 & $0.011_{-0.002}^{+0.005}$	 & $0.011_{-0.002}^{+0.005}$ \\ 
Kepler-55 b/c	 & 3:2	 & 0.005	 & $0.036_{-0.023}^{+0.02}$	 & $0.037_{-0.024}^{+0.019}$	 & $0.002_{-0.001}^{+0.007}$	 & $0.005_{-0.005}^{+0.006}$ \\ 
Kepler-56 b/c	 & 2:1	 & 0.0192	 & $0.03_{-0.004}^{+0.008}$	 & $0.03_{-0.005}^{+0.008}$	 & $0.028_{-0.003}^{+0.004}$	 & $0.028_{-0.003}^{+0.004}$ \\ 
Kepler-57 b/c	 & 2:1	 & 0.0131	 & $0.024_{-0.012}^{+0.076}$	 & $0.023_{-0.015}^{+0.075}$	 & $0.016_{-0.004}^{+0.009}$	 & $0.016_{-0.004}^{+0.008}$ \\ 
Kepler-58 b/c	 & 3:2	 & 0.016	 & $0.067_{-0.027}^{+0.023}$	 & $0.066_{-0.027}^{+0.023}$	 & $0.009_{-0.003}^{+0.011}$	 & $0.009_{-0.003}^{+0.011}$ \\ 
Kepler-60 b/c	 & 5:4	 & 0.0003	 & $0.03_{-0.003}^{+0.004}$	 & $0.03_{-0.004}^{+0.004}$	 & $0.026_{-0.002}^{+0.002}$	 & $0.026_{-0.002}^{+0.002}$ \\ 
Kepler-60 c/d	 & 4:3	 & 0.0006	 & $0.072_{-0.007}^{+0.008}$	 & $0.072_{-0.007}^{+0.008}$	 & $0.007_{-0.002}^{+0.002}$	 & $0.007_{-0.002}^{+0.002}$ \\ 
Kepler-79 b/c	 & 2:1	 & 0.0161	 & $0.007_{-0.003}^{+0.008}$	 & $0.008_{-0.005}^{+0.007}$	 & $0.002_{-0.0005}^{+0.002}$	 & $0.003_{-0.001}^{+0.002}$ \\ 
Kepler-79 c/d	 & 2:1	 & -0.0495	 & $0.02_{-0.012}^{+0.015}$	 & $0.018_{-0.011}^{+0.016}$	 & $0.005_{-0.003}^{+0.005}$	 & $0.005_{-0.004}^{+0.005}$ \\ 
Kepler-79 d/e	 & 3:2	 & 0.0375	 & $0.013_{-0.005}^{+0.007}$	 & $0.013_{-0.006}^{+0.007}$	 & $0.009_{-0.004}^{+0.005}$	 & $0.009_{-0.004}^{+0.005}$ \\ 
Kepler-80 b/c	 & 4:3	 & 0.0124	 & $0.009_{-0.005}^{+0.009}$	 & $0.008_{-0.005}^{+0.009}$	 & $0.004_{-0.002}^{+0.004}$	 & $0.004_{-0.002}^{+0.004}$ \\ 
Kepler-80 d/e	 & 3:2	 & 0.0081	 & $0.004_{-0.001}^{+0.002}$	 & $0.004_{-0.002}^{+0.002}$	 & $0.004_{-0.002}^{+0.001}$	 & $0.004_{-0.002}^{+0.001}$ \\ 
Kepler-80 e/b	 & 3:2	 & 0.0123	 & $0.002_{-0.001}^{+0.003}$	 & $0.001_{-0.002}^{+0.004}$	 & $0.002_{-0.001}^{+0.002}$	 & $0.002_{-0.001}^{+0.002}$ \\ 
Kepler-81 b/c	 & 2:1	 & 0.0109	 & $0.007_{-0.005}^{+0.135}$	 & $0.004_{-0.005}^{+0.124}$	 & $0.004_{-0.001}^{+0.002}$	 & $0.004_{-0.001}^{+0.002}$ \\ 
Kepler-84 b/c	 & 3:2	 & -0.0157	 & $0.008_{-0.006}^{+0.029}$	 & $0.008_{-0.007}^{+0.028}$	 & $0.003_{-0.002}^{+0.004}$	 & $0.003_{-0.003}^{+0.004}$ \\ 
Kepler-84 c/e	 & 2:1	 & 0.0648	 & $0.019_{-0.015}^{+0.036}$	 & $0.018_{-0.019}^{+0.034}$	 & $0.009_{-0.005}^{+0.007}$	 & $0.009_{-0.004}^{+0.007}$ \\ 
Kepler-84 d/b	 & 2:1	 & 0.0328	 & $0.005_{-0.005}^{+0.011}$	 & $-0.0002_{-0.005}^{+0.013}$	 & $0.001_{-0.001}^{+0.005}$	 & $-0.001_{-0.002}^{+0.006}$ \\ 
Kepler-84 e/f	 & 5:3	 & -0.0257	 & $0.01_{-0.008}^{+0.015}$	 & $0.005_{-0.017}^{+0.017}$	 & $0.002_{-0.001}^{+0.007}$	 & $0.001_{-0.003}^{+0.007}$ \\ 
Kepler-85 b/c	 & 3:2	 & 0.0045	 & $0.002_{-0.001}^{+0.014}$	 & $0.002_{-0.002}^{+0.014}$	 & $0.001_{-0.0004}^{+0.001}$	 & $0.001_{-0.001}^{+0.001}$ \\ 
Kepler-85 c/d	 & 7:5	 & 0.0225	 & $0.016_{-0.011}^{+0.013}$	 & $0.005_{-0.008}^{+0.02}$	 & $0.006_{-0.004}^{+0.004}$	 & $0.006_{-0.006}^{+0.003}$ \\ 
Kepler-85 d/e	 & 7:5	 & 0.0054	 & $0.016_{-0.012}^{+0.018}$	 & $0.001_{-0.01}^{+0.03}$	 & $0.001_{-0.001}^{+0.009}$	 & $0.0002_{-0.003}^{+0.007}$ \\ 
Kepler-89 c/d	 & 2:1	 & 0.0717	 & $0.015_{-0.003}^{+0.005}$	 & $0.015_{-0.003}^{+0.005}$	 & $0.014_{-0.002}^{+0.003}$	 & $0.014_{-0.002}^{+0.003}$ \\ 
Kepler-105 03/b	 & 3:2	 & 0.0501	 & $0.035_{-0.02}^{+0.018}$	 & $0.027_{-0.015}^{+0.016}$	 & $0.011_{-0.007}^{+0.011}$	 & $0.008_{-0.005}^{+0.011}$ \\ 
Kepler-105 b/c	 & 4:3	 & -0.0125	 & $0.01_{-0.002}^{+0.002}$	 & $0.01_{-0.002}^{+0.003}$	 & $0.01_{-0.001}^{+0.002}$	 & $0.01_{-0.001}^{+0.002}$ \\ 
Kepler-114 b/c	 & 3:2	 & 0.0332	 & $0.011_{-0.009}^{+0.03}$	 & $0.007_{-0.014}^{+0.03}$	 & $0.001_{-0.001}^{+0.005}$	 & $0.001_{-0.003}^{+0.004}$ \\ 
Kepler-114 c/d	 & 3:2	 & -0.0237	 & $0.015_{-0.009}^{+0.02}$	 & $0.015_{-0.01}^{+0.019}$	 & $0.002_{-0.002}^{+0.004}$	 & $0.0001_{-0.001}^{+0.005}$ \\ 
Kepler-122 e/f	 & 3:2	 & -0.0129	 & $0.033_{-0.018}^{+0.015}$	 & $0.028_{-0.012}^{+0.023}$	 & $0.017_{-0.005}^{+0.009}$	 & $0.017_{-0.005}^{+0.009}$ \\ 
Kepler-127 b/c	 & 2:1	 & 0.0181	 & $0.027_{-0.012}^{+0.038}$	 & $0.027_{-0.011}^{+0.037}$	 & $0.017_{-0.006}^{+0.007}$	 & $0.017_{-0.006}^{+0.007}$ \\ 
Kepler-127 c/d	 & 5:3	 & -0.0073	 & $0.023_{-0.005}^{+0.008}$	 & $0.023_{-0.004}^{+0.008}$	 & $0.02_{-0.002}^{+0.003}$	 & $0.02_{-0.003}^{+0.003}$ \\ 
Kepler-128 b/c	 & 3:2	 & 0.0075	 & $0.084_{-0.033}^{+0.031}$	 & $0.088_{-0.037}^{+0.025}$	 & $0.008_{-0.003}^{+0.014}$	 & $0.008_{-0.003}^{+0.014}$ \\ 
Kepler-138 b/c	 & 4:3	 & 0.0022	 & $0.006_{-0.004}^{+0.004}$	 & $0.006_{-0.004}^{+0.004}$	 & $0.001_{-0.0003}^{+0.001}$	 & $0.001_{-0.001}^{+0.001}$ \\ 
Kepler-138 c/d	 & 5:3	 & 0.0052	 & $0.08_{-0.021}^{+0.022}$	 & $0.084_{-0.023}^{+0.02}$	 & $0.035_{-0.008}^{+0.011}$	 & $0.035_{-0.009}^{+0.011}$ \\ 
Kepler-176 c/d	 & 2:1	 & 0.0092	 & $0.13_{-0.125}^{+0.113}$	 & $0.018_{-0.014}^{+0.215}$	 & $0.004_{-0.002}^{+0.004}$	 & $0.004_{-0.002}^{+0.004}$ \\ 
Kepler-177 b/c	 & 4:3	 & 0.0056	 & $0.002_{-0.0003}^{+0.0002}$	 & $0.002_{-0.0003}^{+0.0002}$	 & $0.002_{-0.0002}^{+0.0002}$	 & $0.002_{-0.0002}^{+0.0002}$ \\ 
Kepler-223 b/c	 & 4:3	 & 0.0002	 & $0.073_{-0.014}^{+0.014}$	 & $0.072_{-0.013}^{+0.015}$	 & $0.071_{-0.013}^{+0.016}$	 & $0.071_{-0.013}^{+0.016}$ \\ 
Kepler-223 c/d	 & 3:2	 & 0.001	 & $0.003_{-0.002}^{+0.004}$	 & $0.002_{-0.003}^{+0.003}$	 & $0.005_{-0.003}^{+0.006}$	 & $0.005_{-0.005}^{+0.006}$ \\ 
Kepler-223 d/e	 & 4:3	 & 0.0005	 & $0.021_{-0.008}^{+0.01}$	 & $0.021_{-0.01}^{+0.01}$	 & $0.023_{-0.005}^{+0.007}$	 & $0.022_{-0.005}^{+0.008}$ \\ 
Kepler-238 c/d	 & 2:1	 & 0.0749	 & $0.012_{-0.012}^{+0.061}$	 & $0.003_{-0.028}^{+0.068}$	 & $0.012_{-0.012}^{+0.014}$	 & $0.01_{-0.012}^{+0.016}$ \\ 
Kepler-238 d/e	 & 5:3	 & 0.0724	 & $0.037_{-0.027}^{+0.025}$	 & $0.013_{-0.024}^{+0.041}$	 & $0.001_{-0.001}^{+0.015}$	 & $0.0001_{-0.007}^{+0.011}$ \\ 
Kepler-238 e/f	 & 2:1	 & 0.0663	 & $0.007_{-0.007}^{+0.032}$	 & $0.013_{-0.022}^{+0.029}$	 & $0.001_{-0.001}^{+0.006}$	 & $0.0003_{-0.004}^{+0.006}$ \\ 
Kepler-277 b/c	 & 2:1	 & -0.0474	 & $0.259_{-0.173}^{+0.06}$	 & $0.118_{-0.053}^{+0.179}$	 & $0.006_{-0.005}^{+0.006}$	 & $0.002_{-0.005}^{+0.009}$ \\ 
Kepler-279 c/d	 & 3:2	 & 0.0151	 & $0.056_{-0.009}^{+0.009}$	 & $0.056_{-0.009}^{+0.009}$	 & $0.051_{-0.01}^{+0.008}$	 & $0.051_{-0.009}^{+0.008}$ \\ 
Kepler-305 03/b	 & 5:3	 & 0.0271	 & $0.015_{-0.012}^{+0.027}$	 & $0.003_{-0.019}^{+0.025}$	 & $0.002_{-0.002}^{+0.007}$	 & $0.002_{-0.007}^{+0.004}$ \\ 
Kepler-305 b/c	 & 3:2	 & 0.0073	 & $0.006_{-0.003}^{+0.006}$	 & $0.006_{-0.003}^{+0.006}$	 & $0.003_{-0.001}^{+0.002}$	 & $0.003_{-0.001}^{+0.002}$ \\ 
Kepler-305 c/d	 & 2:1	 & 0.0095	 & $0.006_{-0.002}^{+0.003}$	 & $0.005_{-0.002}^{+0.004}$	 & $0.007_{-0.002}^{+0.002}$	 & $0.008_{-0.003}^{+0.002}$ \\ 
Kepler-307 b/c	 & 5:4	 & 0.005	 & $0.003_{-0.0002}^{+0.0002}$	 & $0.003_{-0.0002}^{+0.0002}$	 & $0.003_{-0.0001}^{+0.0001}$	 & $0.003_{-0.0001}^{+0.0001}$ \\ 
Kepler-310 c/d	 & 5:3	 & -0.0133	 & $0.026_{-0.014}^{+0.008}$	 & $0.026_{-0.035}^{+0.006}$	 & $0.017_{-0.008}^{+0.012}$	 & $0.008_{-0.005}^{+0.005}$ \\ 
Kepler-324 03/c	 & 3:2	 & 0.0097	 & $0.019_{-0.008}^{+0.019}$	 & $0.019_{-0.009}^{+0.02}$	 & $0.009_{-0.002}^{+0.006}$	 & $0.009_{-0.002}^{+0.006}$ \\ 
Kepler-345 b/c	 & 5:4	 & 0.0127	 & $0.024_{-0.005}^{+0.005}$	 & $0.025_{-0.005}^{+0.004}$	 & $0.007_{-0.002}^{+0.016}$	 & $0.007_{-0.002}^{+0.016}$ \\ 
Kepler-359 c/d	 & 4:3	 & 0.0021	 & $0.007_{-0.004}^{+0.011}$	 & $0.005_{-0.005}^{+0.007}$	 & $0.006_{-0.003}^{+0.005}$	 & $0.006_{-0.006}^{+0.004}$ \\ 
Kepler-396 b/c	 & 2:1	 & 0.0293	 & $0.229_{-0.072}^{+0.113}$	 & $0.244_{-0.087}^{+0.097}$	 & $0.088_{-0.029}^{+0.045}$	 & $0.089_{-0.025}^{+0.049}$ \\ 
Kepler-444 b/c	 & 5:4	 & 0.0102	 & $0.003_{-0.002}^{+0.008}$	 & $0.00_{-0.003}^{+0.009}$	 & $0.001_{-0.001}^{+0.003}$	 & $0.00_{-0.002}^{+0.003}$ \\ 
Kepler-444 c/d	 & 4:3	 & 0.0212	 & $0.003_{-0.002}^{+0.011}$	 & $0.0004_{-0.007}^{+0.009}$	 & $0.002_{-0.002}^{+0.002}$	 & $0.001_{-0.003}^{+0.002}$ \\ 
Kepler-444 d/e	 & 5:4	 & 0.0009	 & $0.001_{-0.0002}^{+0.0004}$	 & $0.001_{-0.0002}^{+0.0005}$	 & $0.001_{-0.0002}^{+0.0002}$	 & $0.001_{-0.0002}^{+0.0002}$ \\ 
Kepler-444 e/f	 & 5:4	 & 0.0063	 & $0.003_{-0.002}^{+0.003}$	 & $0.002_{-0.004}^{+0.003}$	 & $0.003_{-0.002}^{+0.001}$	 & $0.002_{-0.002}^{+0.002}$ \\ 
Kepler-526 b/02	 & 5:4	 & 0.0064	 & $0.036_{-0.013}^{+0.007}$	 & $0.036_{-0.014}^{+0.006}$	 & $0.008_{-0.003}^{+0.006}$	 & $0.008_{-0.003}^{+0.006}$ \\ 
Kepler-549 01/b	 & 5:3	 & 0.0469	 & $0.015_{-0.007}^{+0.006}$	 & $0.013_{-0.009}^{+0.006}$	 & $0.001_{-0.001}^{+0.012}$	 & $-0.0002_{-0.001}^{+0.012}$ \\ 
Kepler-1126 b/02	 & 2:1	 & -0.0807	 & $0.071_{-0.055}^{+0.11}$	 & $0.07_{-0.11}^{+0.124}$	 & $0.062_{-0.025}^{+0.011}$	 & $0.057_{-0.02}^{+0.016}$

%% file: rvtable.tex
Kepler-4 b	&	3.21	&	$24.5^{+3.8}_{-3.8}$	&	$4.0^{+0.2}_{-0.2}$	&	\citet{2010Sci...327..977B}\\ 
CoRoT-7 b	&	0.85	&	$5.7^{+0.9}_{-0.9}$	&	$1.6^{+0.1}_{-0.1}$	&	\citet{2014AandA...569A..74B}\\ 
CoRoT-8 b	&	6.21	&	$69.9^{+9.5}_{-9.5}$	&	$6.4^{+0.2}_{-0.2}$	&	\citet{2010AandA...520A..66B}\\ 
Kepler-10 b	&	0.84	&	$4.6^{+1.3}_{-1.5}$	&	$1.5^{+0.05}_{-0.03}$	&	\citet{2015ApJ...804..150E}\\ 
Kepler-10 c	&	45.29	&	$17.2^{+1.9}_{-1.9}$	&	$2.4^{+0.1}_{-0.04}$	&	\citet{2014ApJ...789..154D}\\ 
HAT-P-11 b	&	4.89	&	$25.7^{+2.9}_{-2.9}$	&	$4.7^{+0.2}_{-0.2}$	&	\citet{2010ApJ...710.1724B}\\ 
Kepler-20 b	&	3.70	&	$8.7^{+2.1}_{-2.2}$	&	$1.9^{+0.1}_{-0.2}$	&	\citet{2012ApJ...749...15G}\\ 
Kepler-20 c	&	10.85	&	$16.1^{+3.3}_{-3.7}$	&	$3.1^{+0.2}_{-0.3}$	&	\citet{2012ApJ...749...15G}\\ 
CoRoT-22 b	&	9.76	&	$12.2^{+14.}_{-8.8}$	&	$4.9^{+0.2}_{-0.4}$	&	\citet{2014MNRAS.444.2783M}\\ 
CoRoT-24 c	&	11.76	&	$28.0^{+11.}_{-11.}$	&	$5.0^{+0.5}_{-0.5}$	&	\citet{2014AandA...567A.112A}\\ 
HAT-P-26 b	&	4.23	&	$18.8^{+2.2}_{-2.2}$	&	$6.3^{+0.8}_{-0.4}$	&	\citet{Hartman:2011ei}\\ 
Kepler-48 d	&	42.90	&	$7.9^{+4.6}_{-4.6}$	&	$2.0^{+0.1}_{-0.1}$	&	\citet{Marcy:2014hr}\\ 
Kepler-68 b	&	5.40	&	$8.3^{+2.2}_{-2.4}$	&	$2.3^{+0.1}_{-0.1}$	&	\citet{Gilliland:2013ep}\\ 
Kepler-68 c	&	9.61	&	$4.8^{+2.5}_{-3.6}$	&	$1.0^{+0.04}_{-0.04}$	&	\citet{Gilliland:2013ep}\\ 
Kepler-78 b	&	0.36	&	$1.9^{+0.4}_{-0.2}$	&	$1.2^{+0.2}_{-0.1}$	&	\citet{Pepe:2013hs}\\ 
Kepler-93 b	&	4.73	&	$4.0^{+0.7}_{-0.7}$	&	$1.5^{+0.02}_{-0.02}$	&	\citet{Dressing:2015je}\\ 
Kepler-94 b	&	2.51	&	$10.8^{+1.4}_{-1.4}$	&	$3.5^{+0.1}_{-0.1}$	&	\citet{Marcy:2014hr}\\ 
Kepler-95 b	&	11.52	&	$13.0^{+2.9}_{-2.9}$	&	$3.4^{+0.1}_{-0.1}$	&	\citet{Marcy:2014hr}\\ 
Kepler-96 b	&	16.24	&	$8.5^{+3.4}_{-3.4}$	&	$2.7^{+0.2}_{-0.2}$	&	\citet{Marcy:2014hr}\\ 
Kepler-97 b	&	2.59	&	$3.5^{+1.9}_{-1.9}$	&	$1.5^{+0.1}_{-0.1}$	&	\citet{Marcy:2014hr}\\ 
Kepler-98 b	&	1.54	&	$3.5^{+1.6}_{-1.6}$	&	$2.0^{+0.2}_{-0.2}$	&	\citet{Marcy:2014hr}\\ 
Kepler-99 b	&	4.60	&	$6.2^{+1.3}_{-1.3}$	&	$1.5^{+0.1}_{-0.1}$	&	\citet{Marcy:2014hr}\\ 
Kepler-100 b	&	6.89	&	$7.3^{+3.2}_{-3.2}$	&	$1.3^{+0.04}_{-0.04}$	&	\citet{Marcy:2014hr}\\ 
Kepler-102 d	&	10.31	&	$3.8^{+1.8}_{-1.8}$	&	$1.2^{+0.04}_{-0.04}$	&	\citet{Marcy:2014hr}\\ 
Kepler-102 e	&	16.15	&	$8.9^{+2.}_{-2.}$	&	$2.2^{+0.1}_{-0.1}$	&	\citet{Marcy:2014hr}\\ 
Kepler-106 c	&	13.57	&	$10.4^{+3.2}_{-3.2}$	&	$2.5^{+0.3}_{-0.3}$	&	\citet{Marcy:2014hr}\\ 
Kepler-106 e	&	43.84	&	$11.2^{+5.8}_{-5.8}$	&	$2.6^{+0.3}_{-0.3}$	&	\citet{Marcy:2014hr}\\ 
Kepler-113 b	&	4.75	&	$11.7^{+4.2}_{-4.2}$	&	$1.8^{+0.05}_{-0.05}$	&	\citet{Marcy:2014hr}\\ 
Kepler-131 b	&	16.09	&	$16.1^{+3.5}_{-3.5}$	&	$2.4^{+0.2}_{-0.2}$	&	\citet{Marcy:2014hr}\\ 
Kepler-131 c	&	25.52	&	$8.2^{+5.9}_{-5.9}$	&	$0.8^{+0.1}_{-0.1}$	&	\citet{Marcy:2014hr}\\ 
KOI-142 b	&	10.95	&	$8.7^{+2.5}_{-2.5}$	&	$3.8^{+0.4}_{-0.4}$	&	\citet{Nesvorny:2013cb}\\ 
Kepler-406 b	&	2.43	&	$6.3^{+1.4}_{-1.4}$	&	$1.4^{+0.03}_{-0.03}$	&	\citet{Marcy:2014hr}\\ 
Kepler-406 c	&	4.62	&	$2.7^{+1.8}_{-1.8}$	&	$0.8^{+0.03}_{-0.03}$	&	\citet{Marcy:2014hr}\\ 
Kepler-413 b	&	66.26	&	$67.0^{+22.}_{-21.}$	&	$4.3^{+0.1}_{-0.1}$	&	\citet{2014ApJ...784...14K}\\ 
Kepler-454 b	&	10.57	&	$6.8^{+1.4}_{-1.4}$	&	$2.4^{+0.1}_{-0.1}$	&	\citet{2016ApJ...816...95G}\\ 
GJ 1132 b	&	1.63	&	$1.6^{+0.6}_{-0.6}$	&	$1.2^{+0.1}_{-0.1}$	&	\citet{BertaThompson:2015el}\\ 
GJ 1214 b	&	1.58	&	$6.3^{+0.9}_{-0.9}$	&	$2.8^{+0.2}_{-0.2}$	&	\citet{Harpsoe:2013ki}\\ 
HIP 116454 b	&	9.12	&	$11.8^{+1.3}_{-1.3}$	&	$2.5^{+0.2}_{-0.2}$	&	\citet{Vanderburg:2015fx}\\ 
K2-38 b	&	4.02	&	$12.0^{+2.9}_{-2.9}$	&	$1.6^{+0.2}_{-0.2}$	&	\citet{Sinukoff:2016cf}\\ 
K2-38 c	&	10.56	&	$9.9^{+4.6}_{-4.6}$	&	$2.4^{+0.3}_{-0.3}$	&	\citet{Sinukoff:2016cf}\\ 
BD+20 594 b	&	41.69	&	$16.3^{+6.}_{-6.1}$	&	$2.2^{+0.1}_{-0.1}$	&	\citet{Espinoza:2016wq}